\documentclass[apjl,iop]{emulateapj}
\pdfoutput=1

\usepackage{times}

\usepackage[breaklinks=true,
  colorlinks=true,
  urlcolor=blue,
  citecolor=blue,
  linkcolor=blue]{hyperref}
\newcommand{\forloop}[5][1]%
{%
\setcounter{#2}{#3}%
\ifthenelse{#4}%
	{%
	#5%
	\addtocounter{#2}{#1}%
	\forloop[#1]{#2}{\value{#2}}{#4}{#5}%
	}%
	{%
	}%
}%

\newcommand{\ctbd}[1]{}

\newcommand{\Lc}{Light curve}

\newcommand{\masy}{\ensuremath{\rm mas\,yr^{-1}}}
\newcommand{\kms}{\ensuremath{\rm km\,s^{-1}}}
\newcommand{\ms}{\ensuremath{\rm m\,s^{-1}}}

\newcommand{\gcmc}{\ensuremath{\rm g\,cm^{-3}}}

\newcommand{\vsini}{\ensuremath{v \sin{i}}}
\newcommand{\feh}{\ensuremath{\rm [Fe/H]}}

\newcommand{\vmac}{\ensuremath{v_{\rm mac}}}
\newcommand{\vmic}{\ensuremath{v_{\rm mic}}}

\newcommand{\rsun}{\ensuremath{R_\sun}}
\newcommand{\msun}{\ensuremath{M_\sun}}
\newcommand{\lsun}{\ensuremath{L_\sun}}

\newcommand{\rstar}{\ensuremath{R_\star}}
\newcommand{\mstar}{\ensuremath{M_\star}}
\newcommand{\lstar}{\ensuremath{L_\star}}

\newcommand{\teffstar}{\ensuremath{T_{\rm eff\star}}}
\newcommand{\rhostar}{\ensuremath{\rho_\star}}
\newcommand{\loggstar}{\ensuremath{\log{g_{\star}}}}

\newcommand{\rpl}{\ensuremath{R_{p}}}
\newcommand{\mpl}{\ensuremath{M_{p}}}

\newcommand{\rhopl}{\ensuremath{\rho_{p}}}

\newcommand{\arstar}{\ensuremath{a/\rstar}}
\newcommand{\zrstar}{\ensuremath{\zeta/\rstar}}

\newcommand{\rjup}{\ensuremath{R_{\rm J}}}
\newcommand{\mjup}{\ensuremath{M_{\rm J}}}

\newcommand{\hatcurhtrxxxxxA}{HATS606-019}                      
\newcommand{\hatcurfieldxxxxxA}{\ensuremath{string}}            
\newcommand{\hatcurCCraxxxxxA}{\ensuremath{09^{\mathrm h}49^{\mathrm m}37.63{\mathrm s}}}                     
\newcommand{\hatcurCCdecxxxxxA}{\ensuremath{-33{\arcdeg}13{\arcmin}06.6{\arcsec}}}                    
\newcommand{\hatcurCCmagxxxxxA}{13.030}                         
\newcommand{\hatcurCCtwomassxxxxxA}{2MASS~J09493761-3313065}     
\newcommand{\hatcurCCgscxxxxxA}{GSC~7172-01459}                 
\newcommand{\hatcurCCtassmvxxxxxA}{\ensuremath{13.030\pm0.060}} 
\newcommand{\hatcurCCtassmvshortxxxxxA}{\ensuremath{13.0}}      
\newcommand{\hatcurCCtassmBxxxxxA}{\ensuremath{13.691\pm0.050}} 
\newcommand{\hatcurCCtassmBshortxxxxxA}{\ensuremath{13.7}}      
\newcommand{\hatcurCCtassmIxxxxxA}{\ensuremath{100\pm100}}      
\newcommand{\hatcurCCtassmIshortxxxxxA}{\ensuremath{100.0}}     
\newcommand{\hatcurCCtassmgxxxxxA}{\ensuremath{13.36\pm0.12}}   
\newcommand{\hatcurCCtassmgshortxxxxxA}{\ensuremath{13.4}}      
\newcommand{\hatcurCCtassmrxxxxxA}{\ensuremath{12.832\pm0.020}} 
\newcommand{\hatcurCCtassmrshortxxxxxA}{\ensuremath{12.8}}      
\newcommand{\hatcurCCtassmixxxxxA}{\ensuremath{12.804\pm0.070}} 
\newcommand{\hatcurCCtassmishortxxxxxA}{\ensuremath{12.8}}      
\newcommand{\hatcurCCtwomassJmagxxxxxA}{\ensuremath{11.811\pm0.023}} 
\newcommand{\hatcurCCtwomassHmagxxxxxA}{\ensuremath{11.553\pm0.022}} 
\newcommand{\hatcurCCtwomassKmagxxxxxA}{\ensuremath{11.510\pm0.019}} 
\newcommand{\hatcurCCcitJmagxxxxxA}{\ensuremath{11.832\pm0.023}} 
\newcommand{\hatcurCCcitHmagxxxxxA}{\ensuremath{11.548\pm0.022}} 
\newcommand{\hatcurCCcitKmagxxxxxA}{\ensuremath{11.534\pm0.019}} 
\newcommand{\hatcurCCbbJmagxxxxxA}{\ensuremath{11.875\pm0.025}} 
\newcommand{\hatcurCCbbHmagxxxxxA}{\ensuremath{11.569\pm0.023}} 
\newcommand{\hatcurCCbbKmagxxxxxA}{\ensuremath{11.554\pm0.019}} 
\newcommand{\hatcurCCesoJmagxxxxxA}{\ensuremath{11.876\pm0.026}} 
\newcommand{\hatcurCCesoHmagxxxxxA}{\ensuremath{11.563\pm0.025}} 
\newcommand{\hatcurCCesoKmagxxxxxA}{\ensuremath{11.553\pm0.020}} 
\newcommand{\hatcurCCesoJHmagxxxxxA}{\ensuremath{0.313\pm0.034}} 
\newcommand{\hatcurCCesoJKmagxxxxxA}{\ensuremath{0.324\pm0.032}} 
\newcommand{\hatcurCCesoHKmagxxxxxA}{\ensuremath{0.010\pm0.032}} 
\newcommand{\hatcurLCdipxxxxxA}{\ensuremath{0.0}}               
\newcommand{\hatcurLCrprstarxxxxxA}{\ensuremath{0.0973\pm0.0021}} 
\newcommand{\hatcurLCbsqxxxxxA}{\ensuremath{0.100_{-0.068}^{+0.084}}} 
\newcommand{\hatcurLCimpxxxxxA}{\ensuremath{0.32_{-0.14}^{+0.11}}} 
\newcommand{\hatcurLCzetaxxxxxA}{\ensuremath{13.271\pm0.093}}   
\newcommand{\hatcurLCdurxxxxxA}{\ensuremath{0.1669\pm0.0019}}   
\newcommand{\hatcurLCdurshortxxxxxA}{\ensuremath{0.1669}}       
\newcommand{\hatcurLCdurhrxxxxxA}{\ensuremath{4.005\pm0.046}}   
\newcommand{\hatcurLCdurhrshortxxxxxA}{\ensuremath{4.005}}      
\newcommand{\hatcurLCqxxxxxA}{\ensuremath{0.03650\pm0.00042}}   
\newcommand{\hatcurLCqshortxxxxxA}{\ensuremath{0.036}}          
\newcommand{\hatcurLCingdurxxxxxA}{\ensuremath{0.0161\pm0.0017}} 
\newcommand{\hatcurLCPxxxxxA}{\ensuremath{4.569671\pm0.000011}} 
\newcommand{\hatcurLCPprecxxxxxA}{\ensuremath{4.5696711}}       
\newcommand{\hatcurLCPshortxxxxxA}{\ensuremath{4.5697}}         
\newcommand{\hatcurLCTxxxxxA}{\ensuremath{2456692.02112\pm0.00050}} 
\newcommand{\hatcurLCTAxxxxxA}{\ensuremath{2455572.4516\pm0.0027}} 
\newcommand{\hatcurLCTBxxxxxA}{\ensuremath{2456737.71786\pm0.00055}} 
\newcommand{\hatcurLChatnetmAxxxxxA}{\ensuremath{12.769560\pm0.000098}} 
\newcommand{\hatcurLCiblendAxxxxxA}{\ensuremath{0.868\pm0.052}} 
\newcommand{\hatcurLChatnetmBxxxxxA}{\ensuremath{12.769610\pm0.000090}} 
\newcommand{\hatcurLCiblendBxxxxxA}{\ensuremath{0.857\pm0.047}} 
\newcommand{\hatcurLCrhoxxxxxA}{\ensuremath{0.697\pm0.081}}     
\newcommand{\hatcurSMEiteffxxxxxA}{\ensuremath{5812\pm74}}      
\newcommand{\hatcurSMEizfehxxxxxA}{\ensuremath{0.210\pm0.040}}  
\newcommand{\hatcurSMEizfehshortxxxxxA}{\ensuremath{0.21}}      
\newcommand{\hatcurSMEiloggxxxxxA}{\ensuremath{4.15\pm0.11}}    
\newcommand{\hatcurSMEivsinxxxxxA}{\ensuremath{4.85\pm0.50}}    
\newcommand{\hatcurSMEivmacxxxxxA}{\ensuremath{0.0}}            
\newcommand{\hatcurSMEivmicxxxxxA}{\ensuremath{0.0}}            
\newcommand{\hatcurSMEiiteffxxxxxA}{\ensuremath{5896\pm77}}     
\newcommand{\hatcurSMEiizfehxxxxxA}{\ensuremath{0.240\pm0.050}} 
\newcommand{\hatcurSMEiizfehshortxxxxxA}{\ensuremath{0.24}}     
\newcommand{\hatcurSMEiiloggxxxxxA}{\ensuremath{4\pm0}}         
\newcommand{\hatcurSMEiivsinxxxxxA}{\ensuremath{4.79\pm0.50}}   
\newcommand{\hatcurLBizxxxxxA}{\ensuremath{0.1948}}             
\newcommand{\hatcurLBiizxxxxxA}{\ensuremath{0.3415}}            
\newcommand{\hatcurLBiixxxxxA}{\ensuremath{0.2566}}             
\newcommand{\hatcurLBiiixxxxxA}{\ensuremath{0.3437}}            
\newcommand{\hatcurLBiIxxxxxA}{\ensuremath{0.2353}}             
\newcommand{\hatcurLBiiIxxxxxA}{\ensuremath{0.3442}}            
\newcommand{\hatcurLBigxxxxxA}{\ensuremath{0.5443}}             
\newcommand{\hatcurLBiigxxxxxA}{\ensuremath{0.2452}}            
\newcommand{\hatcurLBirxxxxxA}{\ensuremath{0.3468}}             
\newcommand{\hatcurLBiirxxxxxA}{\ensuremath{0.3369}}            
\newcommand{\hatcurLBiRxxxxxA}{\ensuremath{0.3216}}             
\newcommand{\hatcurLBiiRxxxxxA}{\ensuremath{0.3400}}            
\newcommand{\hatcurLBikepxxxxxA}{\ensuremath{0.1000}}           
\newcommand{\hatcurLBiikepxxxxxA}{\ensuremath{0.1000}}          
\newcommand{\hatcurISOmxxxxxA}{\ensuremath{1.188\pm0.035}}      
\newcommand{\hatcurISOmshortxxxxxA}{\ensuremath{1.19}}          
\newcommand{\hatcurISOmlongxxxxxA}{\ensuremath{1.188\pm0.035}}  
\newcommand{\hatcurISOrxxxxxA}{\ensuremath{1.336_{-0.051}^{+0.081}}} 
\newcommand{\hatcurISOrshortxxxxxA}{\ensuremath{1.34}}          
\newcommand{\hatcurISOrlongxxxxxA}{\ensuremath{1.336_{-0.051}^{+0.081}}} 
\newcommand{\hatcurISOrhoxxxxxA}{\ensuremath{0.697\pm0.081}}    
\newcommand{\hatcurISOrholongxxxxxA}{\ensuremath{0.697\pm0.081}} 
\newcommand{\hatcurISOloggxxxxxA}{\ensuremath{4.260\pm0.034}}   
\newcommand{\hatcurISOlumxxxxxA}{\ensuremath{1.94_{-0.19}^{+0.26}}} 
\newcommand{\hatcurISOlumshortxxxxxA}{\ensuremath{1.94}}        
\newcommand{\hatcurISOmvxxxxxA}{\ensuremath{4.08\pm0.13}}       
\newcommand{\hatcurISOvixxxxxA}{\ensuremath{0.664\pm0.023}}     
\newcommand{\hatcurISOagexxxxxA}{\ensuremath{4.05\pm0.67}}      
\newcommand{\hatcurISOsigmaxxxxxA}{\ensuremath{0.000200\pm0.000060}} 
\newcommand{\hatcurISOMJxxxxxA}{\ensuremath{2.99\pm0.11}}       
\newcommand{\hatcurISOMHxxxxxA}{\ensuremath{2.68\pm0.11}}       
\newcommand{\hatcurISOMKxxxxxA}{\ensuremath{2.63\pm0.11}}       
\newcommand{\hatcurISOJKxxxxxA}{\ensuremath{0.370\pm0.010}}     
\newcommand{\hatcurISOspecxxxxxA}{G}                            
\newcommand{\hatcurRVKxxxxxA}{\ensuremath{34.0\pm5.4}}          
\newcommand{\hatcurRVrkxxxxxA}{\ensuremath{0\pm0}}              
\newcommand{\hatcurRVrhxxxxxA}{\ensuremath{0\pm0}}              
\newcommand{\hatcurRVkxxxxxA}{\ensuremath{0\pm0}}               
\newcommand{\hatcurRVhxxxxxA}{\ensuremath{0\pm0}}               
\newcommand{\hatcurRVtronexxxxxA}{\ensuremath{0\pm0}}           
\newcommand{\hatcurRVtrtwoxxxxxA}{\ensuremath{0\pm0}}           
\newcommand{\hatcurRVgammaAxxxxxA}{\ensuremath{27543.1\pm7.1}}  
\newcommand{\hatcurRVjitterAxxxxxA}{\ensuremath{19.3\pm6.1}}    
\newcommand{\hatcurRVjittertwosiglimAxxxxxA}{\ensuremath{<31.3}} 
\newcommand{\hatcurRVfitrmsAxxxxxA}{\ensuremath{0.0}}           
\newcommand{\hatcurRVgammaBxxxxxA}{\ensuremath{27443\pm11}}     
\newcommand{\hatcurRVjitterBxxxxxA}{\ensuremath{0.0\pm4.1}}     
\newcommand{\hatcurRVjittertwosiglimBxxxxxA}{\ensuremath{<9.6}} 
\newcommand{\hatcurRVfitrmsBxxxxxA}{\ensuremath{0.0}}           
\newcommand{\hatcurRVgammaCxxxxxA}{\ensuremath{-4.5\pm5.2}}     
\newcommand{\hatcurRVjitterCxxxxxA}{\ensuremath{15.5\pm5.3}}    
\newcommand{\hatcurRVjittertwosiglimCxxxxxA}{\ensuremath{<25.8}} 
\newcommand{\hatcurRVfitrmsCxxxxxA}{\ensuremath{0.0}}           
\newcommand{\hatcurRVeccenxxxxxA}{\ensuremath{0\pm0}}           
\newcommand{\hatcurRVeccentwosiglimxxxxxA}{\ensuremath{<0.000}} 
\newcommand{\hatcurRVomegaxxxxxA}{\ensuremath{0\pm0}}           
\newcommand{\hatcurPPixxxxxA}{\ensuremath{88.02\pm0.82}}        
\newcommand{\hatcurPPgxxxxxA}{\ensuremath{4.69\pm0.95}}         
\newcommand{\hatcurPPloggxxxxxA}{\ensuremath{2.671\pm0.087}}    
\newcommand{\hatcurPParxxxxxA}{\ensuremath{9.16_{-0.45}^{+0.32}}} 
\newcommand{\hatcurPParelxxxxxA}{\ensuremath{0.05707\pm0.00055}} 
\newcommand{\hatcurPPrhoxxxxxA}{\ensuremath{0.184_{-0.039}^{+0.052}}} 
\newcommand{\hatcurPPmxxxxxA}{\ensuremath{0.312\pm0.050}}       
\newcommand{\hatcurPPmshortxxxxxA}{\ensuremath{0.31}}           
\newcommand{\hatcurPPmlongxxxxxA}{\ensuremath{0.312\pm0.050}}   
\newcommand{\hatcurPPmexxxxxA}{\ensuremath{99\pm16}}            
\newcommand{\hatcurPPmeshortxxxxxA}{\ensuremath{99.2}}          
\newcommand{\hatcurPPmelongxxxxxA}{\ensuremath{99\pm16}}        
\newcommand{\hatcurPPrxxxxxA}{\ensuremath{1.258_{-0.058}^{+0.107}}} 
\newcommand{\hatcurPPrshortxxxxxA}{\ensuremath{1.26}}           
\newcommand{\hatcurPPrlongxxxxxA}{\ensuremath{1.258_{-0.058}^{+0.107}}} 
\newcommand{\hatcurPPrexxxxxA}{\ensuremath{14.10_{-0.66}^{+1.20}}} 
\newcommand{\hatcurPPreshortxxxxxA}{\ensuremath{14.1}}          
\newcommand{\hatcurPPrelongxxxxxA}{\ensuremath{14.10_{-0.66}^{+1.20}}} 
\newcommand{\hatcurPPmrcorrxxxxxA}{\ensuremath{0.06}}           
\newcommand{\hatcurPPteffxxxxxA}{\ensuremath{1378\pm34}}        
\newcommand{\hatcurPPthetaxxxxxA}{\ensuremath{0.0234\pm0.0041}} 
\newcommand{\hatcurPPfluxperixxxxxA}{\ensuremath{8.14_{-0.68}^{+0.94}}} 
\newcommand{\hatcurPPfluxperidimxxxxxA}{\ensuremath{8}}         
\newcommand{\hatcurPPfluxapxxxxxA}{\ensuremath{8.14_{-0.68}^{+0.94}}} 
\newcommand{\hatcurPPfluxapdimxxxxxA}{\ensuremath{8}}           
\newcommand{\hatcurPPfluxavgxxxxxA}{\ensuremath{8.14_{-0.68}^{+0.94}}} 
\newcommand{\hatcurPPfluxavgdimxxxxxA}{\ensuremath{8}}          
\newcommand{\hatcurPPfluxavglogxxxxxA}{\ensuremath{8.911\pm0.042}} 
\newcommand{\hatcurXsecphasexxxxxA}{\ensuremath{0\pm0}}         
\newcommand{\hatcurXsecondaryxxxxxA}{\ensuremath{2456694.30595\pm0.00050}} 
\newcommand{\hatcurXsecdurxxxxxA}{\ensuremath{0.1669\pm0.0019}} 
\newcommand{\hatcurXsecingdurxxxxxA}{\ensuremath{0.0161\pm0.0017}} 
\newcommand{\hatcurPPphiconjxxxxxA}{\ensuremath{0\pm0}}         
\newcommand{\hatcurPPperixxxxxA}{\ensuremath{2456690.87870\pm0.00050}} 
\newcommand{\hatcurPPaequivxxxxxA}{\ensuremath{0.0409\pm0.0020}} 
\newcommand{\hatcurPPtcircxxxxxA}{\ensuremath{230\pm75}}        
\newcommand{\hatcurPPtinfallxxxxxA}{\ensuremath{10500_{-2400}^{+3200}}} 
\newcommand{\hatcurXdistxxxxxA}{\ensuremath{609_{-26}^{+38}}}   
\newcommand{\hatcurXAvxxxxxA}{\ensuremath{0.011_{-0.011}^{+0.088}}} 
\newcommand{\hatcurXdistredxxxxxA}{\ensuremath{601_{-25}^{+37}}} 
\newcommand{\hatcurXEBVxxxxxA}{\ensuremath{0.0030_{-0.0030}^{+0.0290}}} 
\newcommand{\hatcurXmvisoredxxxxxA}{\ensuremath{13.018\pm0.043}} 
\newcommand{\hatcurXmiisoredxxxxxA}{\ensuremath{12.334\pm0.020}} 
\newcommand{\hatcurXmjisoredxxxxxA}{\ensuremath{11.901\pm0.014}} 
\newcommand{\hatcurXmhisoredxxxxxA}{\ensuremath{11.582\pm0.014}} 
\newcommand{\hatcurXmkisoredxxxxxA}{\ensuremath{11.527\pm0.014}} 
\newcommand{\hatcurXviisoredxxxxxA}{\ensuremath{0.684\pm0.026}} 
\newcommand{\hatcurXvkisoredxxxxxA}{\ensuremath{1.490\pm0.046}} 
\newcommand{\hatcurXjhisoredxxxxxA}{\ensuremath{0.3180\pm0.0098}} 
\newcommand{\hatcurXjkisoredxxxxxA}{\ensuremath{0.373\pm0.011}} 
\newcommand{\hatcurCCpmraxxxxxA}{\ensuremath{-4.6\pm1.3}}       
\newcommand{\hatcurCCpmdecxxxxxA}{\ensuremath{2.0\pm1.3}}       
\newcommand{\hatcurCCpmxxxxxA}{\ensuremath{5.0\pm1.8}}          

\newcommand{\hatcurhtrxxxxxB}{HATS700-026}                      
\newcommand{\hatcurfieldxxxxxB}{\ensuremath{string}}            
\newcommand{\hatcurCCraxxxxxB}{\ensuremath{13^{\mathrm h}12^{\mathrm m}32.04{\mathrm s}}}                     
\newcommand{\hatcurCCdecxxxxxB}{\ensuremath{-45{\arcdeg}35{\arcmin}26.0{\arcsec}}}                    
\newcommand{\hatcurCCmagxxxxxB}{13.765}                         
\newcommand{\hatcurCCtwomassxxxxxB}{2MASS~J13123190-4535259}     
\newcommand{\hatcurCCgscxxxxxB}{GSC~8247-02184}                 
\newcommand{\hatcurCCtassmvxxxxxB}{\ensuremath{13.765\pm0.050}} 
\newcommand{\hatcurCCtassmvshortxxxxxB}{\ensuremath{13.8}}      
\newcommand{\hatcurCCtassmBxxxxxB}{\ensuremath{14.581\pm0.060}} 
\newcommand{\hatcurCCtassmBshortxxxxxB}{\ensuremath{14.6}}      
\newcommand{\hatcurCCtassmIxxxxxB}{\ensuremath{nff\pmnff}}      
\newcommand{\hatcurCCtassmIshortxxxxxB}{\ensuremath{0.0}}       
\newcommand{\hatcurCCtassmgxxxxxB}{\ensuremath{14.126\pm0.010}} 
\newcommand{\hatcurCCtassmgshortxxxxxB}{\ensuremath{14.1}}      
\newcommand{\hatcurCCtassmrxxxxxB}{\ensuremath{13.478\pm0.010}} 
\newcommand{\hatcurCCtassmrshortxxxxxB}{\ensuremath{13.5}}      
\newcommand{\hatcurCCtassmixxxxxB}{\ensuremath{13.317\pm0.020}} 
\newcommand{\hatcurCCtassmishortxxxxxB}{\ensuremath{13.3}}      
\newcommand{\hatcurCCtwomassJmagxxxxxB}{\ensuremath{12.351\pm0.024}} 
\newcommand{\hatcurCCtwomassHmagxxxxxB}{\ensuremath{11.951\pm0.022}} 
\newcommand{\hatcurCCtwomassKmagxxxxxB}{\ensuremath{11.864\pm0.023}} 
\newcommand{\hatcurCCcitJmagxxxxxB}{\ensuremath{12.362\pm0.024}} 
\newcommand{\hatcurCCcitHmagxxxxxB}{\ensuremath{11.945\pm0.023}} 
\newcommand{\hatcurCCcitKmagxxxxxB}{\ensuremath{11.888\pm0.023}} 
\newcommand{\hatcurCCbbJmagxxxxxB}{\ensuremath{12.421\pm0.026}} 
\newcommand{\hatcurCCbbHmagxxxxxB}{\ensuremath{11.967\pm0.023}} 
\newcommand{\hatcurCCbbKmagxxxxxB}{\ensuremath{11.908\pm0.023}} 
\newcommand{\hatcurCCesoJmagxxxxxB}{\ensuremath{12.424\pm0.028}} 
\newcommand{\hatcurCCesoHmagxxxxxB}{\ensuremath{11.963\pm0.027}} 
\newcommand{\hatcurCCesoKmagxxxxxB}{\ensuremath{11.906\pm0.024}} 
\newcommand{\hatcurCCesoJHmagxxxxxB}{\ensuremath{0.461\pm0.036}} 
\newcommand{\hatcurCCesoJKmagxxxxxB}{\ensuremath{0.518\pm0.036}} 
\newcommand{\hatcurCCesoHKmagxxxxxB}{\ensuremath{0.057\pm0.036}} 
\newcommand{\hatcurLCdipxxxxxB}{\ensuremath{8.3}}               
\newcommand{\hatcurLCrprstarxxxxxB}{\ensuremath{0.0898\pm0.0022}} 
\newcommand{\hatcurLCbsqxxxxxB}{\ensuremath{0.303_{-0.077}^{+0.085}}} 
\newcommand{\hatcurLCimpxxxxxB}{\ensuremath{0.551_{-0.075}^{+0.072}}} 
\newcommand{\hatcurLCzetaxxxxxB}{\ensuremath{22.01\pm0.24}}     
\newcommand{\hatcurLCdurxxxxxB}{\ensuremath{0.1022\pm0.0016}}   
\newcommand{\hatcurLCdurshortxxxxxB}{\ensuremath{0.1022}}       
\newcommand{\hatcurLCdurhrxxxxxB}{\ensuremath{2.453\pm0.040}}   
\newcommand{\hatcurLCdurhrshortxxxxxB}{\ensuremath{2.453}}      
\newcommand{\hatcurLCqxxxxxB}{\ensuremath{0.02690\pm0.00044}}   
\newcommand{\hatcurLCqshortxxxxxB}{\ensuremath{0.027}}          
\newcommand{\hatcurLCingdurxxxxxB}{\ensuremath{0.0117\pm0.0016}} 
\newcommand{\hatcurLCPxxxxxB}{\ensuremath{3.7992969\pm0.0000079}} 
\newcommand{\hatcurLCPprecxxxxxB}{\ensuremath{3.7992969}}       
\newcommand{\hatcurLCPshortxxxxxB}{\ensuremath{3.7993}}         
\newcommand{\hatcurLCTxxxxxB}{\ensuremath{2457121.16516\pm0.00055}} 
\newcommand{\hatcurLCTAxxxxxB}{\ensuremath{2455681.2316\pm0.0031}} 
\newcommand{\hatcurLCTBxxxxxB}{\ensuremath{2457170.55603\pm0.00055}} 
\newcommand{\hatcurLChatnetmAxxxxxB}{\ensuremath{13.63335\pm0.00022}} 
\newcommand{\hatcurLCiblendAxxxxxB}{\ensuremath{0.62\pm0.17}}   
\newcommand{\hatcurLChatnetmBxxxxxB}{\ensuremath{13.62982\pm0.00012}} 
\newcommand{\hatcurLCiblendBxxxxxB}{\ensuremath{0.876\pm0.074}} 
\newcommand{\hatcurLCrhoxxxxxB}{\ensuremath{1.98\pm0.48}}       
\newcommand{\hatcurSMEiteffxxxxxB}{\ensuremath{5485\pm96}}      
\newcommand{\hatcurSMEizfehxxxxxB}{\ensuremath{0.120\pm0.070}}  
\newcommand{\hatcurSMEizfehshortxxxxxB}{\ensuremath{0.12}}      
\newcommand{\hatcurSMEiloggxxxxxB}{\ensuremath{4.56\pm0.17}}    
\newcommand{\hatcurSMEivsinxxxxxB}{\ensuremath{1.4\pm1.2}}      
\newcommand{\hatcurSMEivmacxxxxxB}{\ensuremath{0.0}}            
\newcommand{\hatcurSMEivmicxxxxxB}{\ensuremath{0.0}}            
\newcommand{\hatcurSMEiiteffxxxxxB}{\ensuremath{5406\pm49}}     
\newcommand{\hatcurSMEiizfehxxxxxB}{\ensuremath{0.030\pm0.050}} 
\newcommand{\hatcurSMEiizfehshortxxxxxB}{\ensuremath{0.030}}    
\newcommand{\hatcurSMEiiloggxxxxxB}{\ensuremath{4.493\pm0.046}} 
\newcommand{\hatcurSMEiivsinxxxxxB}{\ensuremath{1.51\pm0.89}}   
\newcommand{\hatcurLBizxxxxxB}{\ensuremath{0.2609}}             
\newcommand{\hatcurLBiizxxxxxB}{\ensuremath{0.3012}}            
\newcommand{\hatcurLBiixxxxxB}{\ensuremath{0.3320}}             
\newcommand{\hatcurLBiiixxxxxB}{\ensuremath{0.2933}}            
\newcommand{\hatcurLBiIxxxxxB}{\ensuremath{0.3085}}             
\newcommand{\hatcurLBiiIxxxxxB}{\ensuremath{0.2962}}            
\newcommand{\hatcurLBigxxxxxB}{\ensuremath{0.6542}}             
\newcommand{\hatcurLBiigxxxxxB}{\ensuremath{0.1580}}            
\newcommand{\hatcurLBirxxxxxB}{\ensuremath{0.4366}}             
\newcommand{\hatcurLBiirxxxxxB}{\ensuremath{0.2753}}            
\newcommand{\hatcurLBiRxxxxxB}{\ensuremath{0.4077}}             
\newcommand{\hatcurLBiiRxxxxxB}{\ensuremath{0.2811}}            
\newcommand{\hatcurLBikepxxxxxB}{\ensuremath{0.1000}}           
\newcommand{\hatcurLBiikepxxxxxB}{\ensuremath{0.1000}}          
\newcommand{\hatcurISOmxxxxxB}{\ensuremath{0.910\pm0.026}}      
\newcommand{\hatcurISOmshortxxxxxB}{\ensuremath{0.91}}          
\newcommand{\hatcurISOmlongxxxxxB}{\ensuremath{0.910\pm0.026}}  
\newcommand{\hatcurISOrxxxxxB}{\ensuremath{0.892_{-0.040}^{+0.057}}} 
\newcommand{\hatcurISOrshortxxxxxB}{\ensuremath{0.89}}          
\newcommand{\hatcurISOrlongxxxxxB}{\ensuremath{0.892_{-0.040}^{+0.057}}} 
\newcommand{\hatcurISOrhoxxxxxB}{\ensuremath{1.81\pm0.30}}      
\newcommand{\hatcurISOrholongxxxxxB}{\ensuremath{1.81\pm0.30}}  
\newcommand{\hatcurISOloggxxxxxB}{\ensuremath{4.497\pm0.052}}   
\newcommand{\hatcurISOlumxxxxxB}{\ensuremath{0.612_{-0.066}^{+0.087}}} 
\newcommand{\hatcurISOlumshortxxxxxB}{\ensuremath{0.61}}        
\newcommand{\hatcurISOmvxxxxxB}{\ensuremath{5.44\pm0.14}}       
\newcommand{\hatcurISOvixxxxxB}{\ensuremath{0.807\pm0.013}}     
\newcommand{\hatcurISOagexxxxxB}{\ensuremath{6.4\pm3.4}}        
\newcommand{\hatcurISOsigmaxxxxxB}{\ensuremath{0.000500\pm0.000099}} 
\newcommand{\hatcurISOMJxxxxxB}{\ensuremath{4.10\pm0.13}}       
\newcommand{\hatcurISOMHxxxxxB}{\ensuremath{3.68\pm0.12}}       
\newcommand{\hatcurISOMKxxxxxB}{\ensuremath{3.62\pm0.12}}       
\newcommand{\hatcurISOJKxxxxxB}{\ensuremath{0.490\pm0.010}}     
\newcommand{\hatcurISOspecxxxxxB}{G}                            
\newcommand{\hatcurRVKxxxxxB}{\ensuremath{37.9\pm4.8}}          
\newcommand{\hatcurRVrkxxxxxB}{\ensuremath{0\pm0}}              
\newcommand{\hatcurRVrhxxxxxB}{\ensuremath{0\pm0}}              
\newcommand{\hatcurRVkxxxxxB}{\ensuremath{0\pm0}}               
\newcommand{\hatcurRVhxxxxxB}{\ensuremath{0\pm0}}               
\newcommand{\hatcurRVtronexxxxxB}{\ensuremath{0\pm0}}           
\newcommand{\hatcurRVtrtwoxxxxxB}{\ensuremath{0\pm0}}           
\newcommand{\hatcurRVgammaxxxxxB}{\ensuremath{22116.9\pm4.0}}   
\newcommand{\hatcurRVjitterxxxxxB}{\ensuremath{0.0\pm3.5}}      
\newcommand{\hatcurRVjittertwosiglimxxxxxB}{\ensuremath{<8.6}}  
\newcommand{\hatcurRVfitrmsxxxxxB}{\ensuremath{.1fym}}          %
\newcommand{\hatcurRVeccenxxxxxB}{\ensuremath{0\pm0}}           
\newcommand{\hatcurRVeccentwosiglimxxxxxB}{\ensuremath{<0.000}} 
\newcommand{\hatcurRVomegaxxxxxB}{\ensuremath{0\pm0}}           
\newcommand{\hatcurPPixxxxxB}{\ensuremath{87.16\pm0.54}}        
\newcommand{\hatcurPPgxxxxxB}{\ensuremath{11.2\pm2.1}}          
\newcommand{\hatcurPPloggxxxxxB}{\ensuremath{3.050\pm0.084}}    
\newcommand{\hatcurPParxxxxxB}{\ensuremath{11.14\pm0.63}}       
\newcommand{\hatcurPParelxxxxxB}{\ensuremath{0.04619\pm0.00044}} 
\newcommand{\hatcurPPrhoxxxxxB}{\ensuremath{0.73\pm0.18}}       
\newcommand{\hatcurPPmxxxxxB}{\ensuremath{0.273\pm0.035}}       
\newcommand{\hatcurPPmshortxxxxxB}{\ensuremath{0.27}}           
\newcommand{\hatcurPPmlongxxxxxB}{\ensuremath{0.273\pm0.035}}   
\newcommand{\hatcurPPmexxxxxB}{\ensuremath{87\pm11}}            
\newcommand{\hatcurPPmeshortxxxxxB}{\ensuremath{86.9}}          
\newcommand{\hatcurPPmelongxxxxxB}{\ensuremath{87\pm11}}        
\newcommand{\hatcurPPrxxxxxB}{\ensuremath{0.776\pm0.055}}       
\newcommand{\hatcurPPrshortxxxxxB}{\ensuremath{0.78}}           
\newcommand{\hatcurPPrlongxxxxxB}{\ensuremath{0.776\pm0.055}}   
\newcommand{\hatcurPPrexxxxxB}{\ensuremath{8.70\pm0.62}}        
\newcommand{\hatcurPPreshortxxxxxB}{\ensuremath{8.7}}           
\newcommand{\hatcurPPrelongxxxxxB}{\ensuremath{8.70\pm0.62}}    
\newcommand{\hatcurPPmrcorrxxxxxB}{\ensuremath{-0.03}}          
\newcommand{\hatcurPPteffxxxxxB}{\ensuremath{1147\pm36}}        
\newcommand{\hatcurPPthetaxxxxxB}{\ensuremath{0.0355\pm0.0050}} 
\newcommand{\hatcurPPfluxperixxxxxB}{\ensuremath{3.91_{-0.43}^{+0.59}}} 
\newcommand{\hatcurPPfluxperidimxxxxxB}{\ensuremath{8}}         
\newcommand{\hatcurPPfluxapxxxxxB}{\ensuremath{3.91_{-0.43}^{+0.59}}} 
\newcommand{\hatcurPPfluxapdimxxxxxB}{\ensuremath{8}}           
\newcommand{\hatcurPPfluxavgxxxxxB}{\ensuremath{3.91_{-0.43}^{+0.59}}} 
\newcommand{\hatcurPPfluxavgdimxxxxxB}{\ensuremath{8}}          
\newcommand{\hatcurPPfluxavglogxxxxxB}{\ensuremath{8.592\pm0.054}} 
\newcommand{\hatcurXsecphasexxxxxB}{\ensuremath{0\pm0}}         
\newcommand{\hatcurXsecondaryxxxxxB}{\ensuremath{2457123.06481\pm0.00055}} 
\newcommand{\hatcurXsecdurxxxxxB}{\ensuremath{0.1022\pm0.0016}} 
\newcommand{\hatcurXsecingdurxxxxxB}{\ensuremath{0.0117\pm0.0016}} 
\newcommand{\hatcurPPphiconjxxxxxB}{\ensuremath{0\pm0}}         
\newcommand{\hatcurPPperixxxxxB}{\ensuremath{2457120.21533\pm0.00055}} 
\newcommand{\hatcurPPaequivxxxxxB}{\ensuremath{0.0591\pm0.0036}} 
\newcommand{\hatcurPPtcircxxxxxB}{\ensuremath{870\pm340}}       
\newcommand{\hatcurPPtinfallxxxxxB}{\ensuremath{21000\pm6300}}  
\newcommand{\hatcurXdistxxxxxB}{\ensuremath{455_{-23}^{+30}}}   
\newcommand{\hatcurXAvxxxxxB}{\ensuremath{0.044_{-0.044}^{+0.078}}} 
\newcommand{\hatcurXdistredxxxxxB}{\ensuremath{454_{-22}^{+30}}} 
\newcommand{\hatcurXEBVxxxxxB}{\ensuremath{0.014_{-0.014}^{+0.025}}} 
\newcommand{\hatcurXmvisoredxxxxxB}{\ensuremath{13.781\pm0.040}} 
\newcommand{\hatcurXmiisoredxxxxxB}{\ensuremath{12.947\pm0.023}} 
\newcommand{\hatcurXmjisoredxxxxxB}{\ensuremath{12.403\pm0.015}} 
\newcommand{\hatcurXmhisoredxxxxxB}{\ensuremath{11.978\pm0.015}} 
\newcommand{\hatcurXmkisoredxxxxxB}{\ensuremath{11.905\pm0.016}} 
\newcommand{\hatcurXviisoredxxxxxB}{\ensuremath{0.831_{-0.019}^{+0.033}}} 
\newcommand{\hatcurXvkisoredxxxxxB}{\ensuremath{1.875\pm0.045}} 
\newcommand{\hatcurXjhisoredxxxxxB}{\ensuremath{0.4250\pm0.0091}} 
\newcommand{\hatcurXjkisoredxxxxxB}{\ensuremath{0.498\pm0.011}} 
\newcommand{\hatcurCCpmraxxxxxB}{\ensuremath{-15.2\pm1.8}}      
\newcommand{\hatcurCCpmdecxxxxxB}{\ensuremath{-3.8\pm1.8}}      
\newcommand{\hatcurCCpmxxxxxB}{\ensuremath{15.7\pm2.5}}         

\newcommand{\hatcurhtrxxxxxC}{HATS777-021}                      
\newcommand{\hatcurfieldxxxxxC}{\ensuremath{string}}            
\newcommand{\hatcurCCraxxxxxC}{\ensuremath{18^{\mathrm h}40^{\mathrm m}44.40{\mathrm s}}}                     
\newcommand{\hatcurCCdecxxxxxC}{\ensuremath{-58{\arcdeg}27{\arcmin}33.3{\arcsec}}}                    
\newcommand{\hatcurCCmagxxxxxC}{12.191}                         
\newcommand{\hatcurCCtwomassxxxxxC}{2MASS~J18404426-5827332}     
\newcommand{\hatcurCCgscxxxxxC}{GSC~8770-00400}                 
\newcommand{\hatcurCCtassmvxxxxxC}{\ensuremath{12.191\pm0.020}} 
\newcommand{\hatcurCCtassmvshortxxxxxC}{\ensuremath{12.2}}      
\newcommand{\hatcurCCtassmBxxxxxC}{\ensuremath{13.018\pm0.010}} 
\newcommand{\hatcurCCtassmBshortxxxxxC}{\ensuremath{13.0}}      
\newcommand{\hatcurCCtassmIxxxxxC}{\ensuremath{nff\pmnff}}      
\newcommand{\hatcurCCtassmIshortxxxxxC}{\ensuremath{0.0}}       
\newcommand{\hatcurCCtassmgxxxxxC}{\ensuremath{12.595\pm0.010}} 
\newcommand{\hatcurCCtassmgshortxxxxxC}{\ensuremath{12.6}}      
\newcommand{\hatcurCCtassmrxxxxxC}{\ensuremath{11.954\pm0.010}} 
\newcommand{\hatcurCCtassmrshortxxxxxC}{\ensuremath{12.0}}      
\newcommand{\hatcurCCtassmixxxxxC}{\ensuremath{11.833\pm0.040}} 
\newcommand{\hatcurCCtassmishortxxxxxC}{\ensuremath{11.8}}      
\newcommand{\hatcurCCtwomassJmagxxxxxC}{\ensuremath{10.912\pm0.020}} 
\newcommand{\hatcurCCtwomassHmagxxxxxC}{\ensuremath{10.562\pm0.022}} 
\newcommand{\hatcurCCtwomassKmagxxxxxC}{\ensuremath{10.526\pm0.023}} 
\newcommand{\hatcurCCcitJmagxxxxxC}{\ensuremath{10.928\pm0.020}} 
\newcommand{\hatcurCCcitHmagxxxxxC}{\ensuremath{10.558\pm0.023}} 
\newcommand{\hatcurCCcitKmagxxxxxC}{\ensuremath{10.550\pm0.023}} 
\newcommand{\hatcurCCbbJmagxxxxxC}{\ensuremath{10.979\pm0.022}} 
\newcommand{\hatcurCCbbHmagxxxxxC}{\ensuremath{10.578\pm0.023}} 
\newcommand{\hatcurCCbbKmagxxxxxC}{\ensuremath{10.570\pm0.023}} 
\newcommand{\hatcurCCesoJmagxxxxxC}{\ensuremath{10.981\pm0.024}} 
\newcommand{\hatcurCCesoHmagxxxxxC}{\ensuremath{10.571\pm0.026}} 
\newcommand{\hatcurCCesoKmagxxxxxC}{\ensuremath{10.569\pm0.024}} 
\newcommand{\hatcurCCesoJHmagxxxxxC}{\ensuremath{0.409\pm0.033}} 
\newcommand{\hatcurCCesoJKmagxxxxxC}{\ensuremath{0.413\pm0.033}} 
\newcommand{\hatcurCCesoHKmagxxxxxC}{\ensuremath{0.003\pm0.035}} 
\newcommand{\hatcurLCdipxxxxxC}{\ensuremath{11.1}}              
\newcommand{\hatcurLCrprstarxxxxxC}{\ensuremath{0.113\pm0.011}} 
\newcommand{\hatcurLCbsqxxxxxC}{\ensuremath{0.728_{-0.024}^{+0.059}}} 
\newcommand{\hatcurLCimpxxxxxC}{\ensuremath{0.853_{-0.014}^{+0.034}}} 
\newcommand{\hatcurLCzetaxxxxxC}{\ensuremath{33.41_{-0.84}^{+1.52}}} 
\newcommand{\hatcurLCdurxxxxxC}{\ensuremath{0.0824\pm0.0026}}   
\newcommand{\hatcurLCdurshortxxxxxC}{\ensuremath{0.0824}}       
\newcommand{\hatcurLCdurhrxxxxxC}{\ensuremath{1.978\pm0.062}}   
\newcommand{\hatcurLCdurhrshortxxxxxC}{\ensuremath{1.978}}      
\newcommand{\hatcurLCqxxxxxC}{\ensuremath{0.02320\pm0.00073}}   
\newcommand{\hatcurLCqshortxxxxxC}{\ensuremath{0.023}}          
\newcommand{\hatcurLCingdurxxxxxC}{\ensuremath{0.027\pm0.034}}  
\newcommand{\hatcurLCPxxxxxC}{\ensuremath{3.5543973\pm0.0000058}} 
\newcommand{\hatcurLCPprecxxxxxC}{\ensuremath{3.5543973}}       
\newcommand{\hatcurLCPshortxxxxxC}{\ensuremath{3.5544}}         
\newcommand{\hatcurLCTxxxxxC}{\ensuremath{2457109.22538\pm0.00057}} 
\newcommand{\hatcurLCTAxxxxxC}{\ensuremath{2455680.3576\pm0.0022}} 
\newcommand{\hatcurLCTBxxxxxC}{\ensuremath{2457219.41170\pm0.00064}} 
\newcommand{\hatcurLChatnetmAxxxxxC}{\ensuremath{12.07052\pm0.00010}} 
\newcommand{\hatcurLCiblendAxxxxxC}{\ensuremath{0.757\pm0.070}} 
\newcommand{\hatcurLChatnetmBxxxxxC}{\ensuremath{12.070690\pm0.000091}} 
\newcommand{\hatcurLCiblendBxxxxxC}{\ensuremath{0.741\pm0.059}} 
\newcommand{\hatcurLCrhoxxxxxC}{\ensuremath{1.55\pm0.38}}       
\newcommand{\hatcurSMEiteffxxxxxC}{\ensuremath{5678\pm97}}      
\newcommand{\hatcurSMEizfehxxxxxC}{\ensuremath{0.290\pm0.060}}  
\newcommand{\hatcurSMEizfehshortxxxxxC}{\ensuremath{0.29}}      
\newcommand{\hatcurSMEiloggxxxxxC}{\ensuremath{4.55\pm0.14}}    
\newcommand{\hatcurSMEivsinxxxxxC}{\ensuremath{2.43\pm0.60}}    
\newcommand{\hatcurSMEivmacxxxxxC}{\ensuremath{0.0}}            
\newcommand{\hatcurSMEivmicxxxxxC}{\ensuremath{0.0}}            
\newcommand{\hatcurSMEiiteffxxxxxC}{\ensuremath{5695\pm67}}     
\newcommand{\hatcurSMEiizfehxxxxxC}{\ensuremath{0.300\pm0.040}} 
\newcommand{\hatcurSMEiizfehshortxxxxxC}{\ensuremath{0.30}}     
\newcommand{\hatcurSMEiiloggxxxxxC}{\ensuremath{4.460\pm0.050}} 
\newcommand{\hatcurSMEiivsinxxxxxC}{\ensuremath{2.63\pm0.55}}   
\newcommand{\hatcurLBizxxxxxC}{\ensuremath{0.2266}}             
\newcommand{\hatcurLBiizxxxxxC}{\ensuremath{0.3265}}            
\newcommand{\hatcurLBiixxxxxC}{\ensuremath{0.2962}}             
\newcommand{\hatcurLBiiixxxxxC}{\ensuremath{0.3225}}            
\newcommand{\hatcurLBiIxxxxxC}{\ensuremath{0.2730}}             
\newcommand{\hatcurLBiiIxxxxxC}{\ensuremath{0.3245}}            
\newcommand{\hatcurLBigxxxxxC}{\ensuremath{0.6056}}             
\newcommand{\hatcurLBiigxxxxxC}{\ensuremath{0.1986}}            
\newcommand{\hatcurLBirxxxxxC}{\ensuremath{0.3952}}             
\newcommand{\hatcurLBiirxxxxxC}{\ensuremath{0.3072}}            
\newcommand{\hatcurLBiRxxxxxC}{\ensuremath{0.3677}}             
\newcommand{\hatcurLBiiRxxxxxC}{\ensuremath{0.3125}}            
\newcommand{\hatcurLBikepxxxxxC}{\ensuremath{0.1000}}           
\newcommand{\hatcurLBiikepxxxxxC}{\ensuremath{0.1000}}          
\newcommand{\hatcurISOmxxxxxC}{\ensuremath{1.080\pm0.026}}      
\newcommand{\hatcurISOmshortxxxxxC}{\ensuremath{1.08}}          
\newcommand{\hatcurISOmlongxxxxxC}{\ensuremath{1.080\pm0.026}}  
\newcommand{\hatcurISOrxxxxxC}{\ensuremath{1.021_{-0.036}^{+0.089}}} 
\newcommand{\hatcurISOrshortxxxxxC}{\ensuremath{1.02}}          
\newcommand{\hatcurISOrlongxxxxxC}{\ensuremath{1.021_{-0.036}^{+0.089}}} 
\newcommand{\hatcurISOrhoxxxxxC}{\ensuremath{1.42_{-0.33}^{+0.16}}} 
\newcommand{\hatcurISOrholongxxxxxC}{\ensuremath{1.42_{-0.33}^{+0.16}}} 
\newcommand{\hatcurISOloggxxxxxC}{\ensuremath{4.453\pm0.055}}   
\newcommand{\hatcurISOlumxxxxxC}{\ensuremath{0.98_{-0.10}^{+0.18}}} 
\newcommand{\hatcurISOlumshortxxxxxC}{\ensuremath{0.98}}        
\newcommand{\hatcurISOmvxxxxxC}{\ensuremath{4.85\pm0.17}}       
\newcommand{\hatcurISOvixxxxxC}{\ensuremath{0.734\pm0.021}}     
\newcommand{\hatcurISOagexxxxxC}{\ensuremath{2.3\pm1.7}}        
\newcommand{\hatcurISOsigmaxxxxxC}{\ensuremath{0.00050\pm0.00011}} 
\newcommand{\hatcurISOMJxxxxxC}{\ensuremath{3.67\pm0.15}}       
\newcommand{\hatcurISOMHxxxxxC}{\ensuremath{3.32\pm0.15}}       
\newcommand{\hatcurISOMKxxxxxC}{\ensuremath{3.26\pm0.15}}       
\newcommand{\hatcurISOJKxxxxxC}{\ensuremath{0.34\pm0.15}}       
\newcommand{\hatcurISOspecxxxxxC}{G}                            
\newcommand{\hatcurRVKxxxxxC}{\ensuremath{41.6\pm4.3}}          
\newcommand{\hatcurRVrkxxxxxC}{\ensuremath{0\pm0}}              
\newcommand{\hatcurRVrhxxxxxC}{\ensuremath{0\pm0}}              
\newcommand{\hatcurRVkxxxxxC}{\ensuremath{0\pm0}}               
\newcommand{\hatcurRVhxxxxxC}{\ensuremath{0\pm0}}               
\newcommand{\hatcurRVtronexxxxxC}{\ensuremath{0\pm0}}           
\newcommand{\hatcurRVtrtwoxxxxxC}{\ensuremath{0\pm0}}           
\newcommand{\hatcurRVgammaAxxxxxC}{\ensuremath{32041.2\pm8.7}}  
\newcommand{\hatcurRVjitterAxxxxxC}{\ensuremath{20.0\pm10.0}}   
\newcommand{\hatcurRVjittertwosiglimAxxxxxC}{\ensuremath{<37.4}} 
\newcommand{\hatcurRVfitrmsAxxxxxC}{\ensuremath{0.0}}           
\newcommand{\hatcurRVgammaBxxxxxC}{\ensuremath{32057.5\pm6.8}}  
\newcommand{\hatcurRVjitterBxxxxxC}{\ensuremath{0.00\pm0.27}}   
\newcommand{\hatcurRVjittertwosiglimBxxxxxC}{\ensuremath{<0.4}} 
\newcommand{\hatcurRVfitrmsBxxxxxC}{\ensuremath{0.0}}           
\newcommand{\hatcurRVgammaCxxxxxC}{\ensuremath{17.3\pm4.0}}     
\newcommand{\hatcurRVjitterCxxxxxC}{\ensuremath{7.3\pm2.9}}     
\newcommand{\hatcurRVjittertwosiglimCxxxxxC}{\ensuremath{<13.6}} 
\newcommand{\hatcurRVfitrmsCxxxxxC}{\ensuremath{0.0}}           
\newcommand{\hatcurRVeccenxxxxxC}{\ensuremath{0\pm0}}           
\newcommand{\hatcurRVeccentwosiglimxxxxxC}{\ensuremath{<0.000}} 
\newcommand{\hatcurRVomegaxxxxxC}{\ensuremath{0\pm0}}           
\newcommand{\hatcurPPixxxxxC}{\ensuremath{85.04_{-0.65}^{+0.23}}} 
\newcommand{\hatcurPPgxxxxxC}{\ensuremath{6.5\pm1.5}}           
\newcommand{\hatcurPPloggxxxxxC}{\ensuremath{2.811_{-0.108}^{+0.068}}} 
\newcommand{\hatcurPParxxxxxC}{\ensuremath{9.84_{-0.83}^{+0.35}}} 
\newcommand{\hatcurPParelxxxxxC}{\ensuremath{0.04676\pm0.00038}} 
\newcommand{\hatcurPPrhoxxxxxC}{\ensuremath{0.290_{-0.086}^{+0.063}}} 
\newcommand{\hatcurPPmxxxxxC}{\ensuremath{0.332_{-0.030}^{+0.040}}} 
\newcommand{\hatcurPPmshortxxxxxC}{\ensuremath{0.33}}           
\newcommand{\hatcurPPmlongxxxxxC}{\ensuremath{0.332_{-0.030}^{+0.040}}} 
\newcommand{\hatcurPPmexxxxxC}{\ensuremath{105.4_{-9.4}^{+12.7}}} 
\newcommand{\hatcurPPmeshortxxxxxC}{\ensuremath{105.4}}         
\newcommand{\hatcurPPmelongxxxxxC}{\ensuremath{105.4_{-9.4}^{+12.7}}} 
\newcommand{\hatcurPPrxxxxxC}{\ensuremath{1.123_{-0.054}^{+0.147}}} 
\newcommand{\hatcurPPrshortxxxxxC}{\ensuremath{1.12}}           
\newcommand{\hatcurPPrlongxxxxxC}{\ensuremath{1.123_{-0.054}^{+0.147}}} 
\newcommand{\hatcurPPrexxxxxC}{\ensuremath{12.59_{-0.61}^{+1.65}}} 
\newcommand{\hatcurPPreshortxxxxxC}{\ensuremath{12.6}}          
\newcommand{\hatcurPPrelongxxxxxC}{\ensuremath{12.59_{-0.61}^{+1.65}}} 
\newcommand{\hatcurPPmrcorrxxxxxC}{\ensuremath{-0.03}}          
\newcommand{\hatcurPPteffxxxxxC}{\ensuremath{1284_{-31}^{+55}}} 
\newcommand{\hatcurPPthetaxxxxxC}{\ensuremath{0.0253\pm0.0041}} 
\newcommand{\hatcurPPfluxperixxxxxC}{\ensuremath{6.13_{-0.58}^{+1.11}}} 
\newcommand{\hatcurPPfluxperidimxxxxxC}{\ensuremath{8}}         
\newcommand{\hatcurPPfluxapxxxxxC}{\ensuremath{6.13_{-0.58}^{+1.11}}} 
\newcommand{\hatcurPPfluxapdimxxxxxC}{\ensuremath{8}}           
\newcommand{\hatcurPPfluxavgxxxxxC}{\ensuremath{6.13_{-0.58}^{+1.11}}} 
\newcommand{\hatcurPPfluxavgdimxxxxxC}{\ensuremath{8}}          
\newcommand{\hatcurPPfluxavglogxxxxxC}{\ensuremath{8.787_{-0.043}^{+0.072}}} 
\newcommand{\hatcurXsecphasexxxxxC}{\ensuremath{0\pm0}}         
\newcommand{\hatcurXsecondaryxxxxxC}{\ensuremath{2457111.00258\pm0.00057}} 
\newcommand{\hatcurXsecdurxxxxxC}{\ensuremath{0.0824\pm0.0069}} 
\newcommand{\hatcurXsecingdurxxxxxC}{\ensuremath{0.0269\pm0.0094}} 
\newcommand{\hatcurPPphiconjxxxxxC}{\ensuremath{0\pm0}}         
\newcommand{\hatcurPPperixxxxxC}{\ensuremath{2457108.33678\pm0.00057}} 
\newcommand{\hatcurPPaequivxxxxxC}{\ensuremath{0.0472_{-0.0038}^{+0.0024}}} 
\newcommand{\hatcurPPtcircxxxxxC}{\ensuremath{144_{-65}^{+49}}} 
\newcommand{\hatcurPPtinfallxxxxxC}{\ensuremath{9800_{-3200}^{+2300}}} 
\newcommand{\hatcurXdistxxxxxC}{\ensuremath{289_{-12}^{+24}}}   
\newcommand{\hatcurXAvxxxxxC}{\ensuremath{0.047\pm0.049}}       
\newcommand{\hatcurXdistredxxxxxC}{\ensuremath{286_{-12}^{+24}}} 
\newcommand{\hatcurXEBVxxxxxC}{\ensuremath{0.015_{-0.015}^{+0.020}}} 
\newcommand{\hatcurXmvisoredxxxxxC}{\ensuremath{12.195\pm0.020}} 
\newcommand{\hatcurXmiisoredxxxxxC}{\ensuremath{11.433\pm0.014}} 
\newcommand{\hatcurXmjisoredxxxxxC}{\ensuremath{10.964\pm0.013}} 
\newcommand{\hatcurXmhisoredxxxxxC}{\ensuremath{10.609\pm0.016}} 
\newcommand{\hatcurXmkisoredxxxxxC}{\ensuremath{10.548\pm0.017}} 
\newcommand{\hatcurXviisoredxxxxxC}{\ensuremath{0.762\pm0.015}} 
\newcommand{\hatcurXvkisoredxxxxxC}{\ensuremath{1.647\pm0.028}} 
\newcommand{\hatcurXjhisoredxxxxxC}{\ensuremath{0.3540\pm0.0093}} 
\newcommand{\hatcurXjkisoredxxxxxC}{\ensuremath{0.4150\pm0.0097}} 
\newcommand{\hatcurCCpmraxxxxxC}{\ensuremath{-26.1\pm1.7}}      
\newcommand{\hatcurCCpmdecxxxxxC}{\ensuremath{-29.4\pm1.7}}     
\newcommand{\hatcurCCpmxxxxxC}{\ensuremath{39.3\pm2.4}}         

\newcommand{\hatcurCCbbHmag}[1]{\ifnum#1=19 %
\hatcurCCbbHmagxxxxxA
\else
\ifnum#1=20 %
\hatcurCCbbHmagxxxxxB
\else
\ifnum#1=21 %
\hatcurCCbbHmagxxxxxC
\else
??????\fi
\fi
\fi
}
\newcommand{\hatcurCCbbJmag}[1]{\ifnum#1=19 %
\hatcurCCbbJmagxxxxxA
\else
\ifnum#1=20 %
\hatcurCCbbJmagxxxxxB
\else
\ifnum#1=21 %
\hatcurCCbbJmagxxxxxC
\else
??????\fi
\fi
\fi
}
\newcommand{\hatcurCCbbKmag}[1]{\ifnum#1=19 %
\hatcurCCbbKmagxxxxxA
\else
\ifnum#1=20 %
\hatcurCCbbKmagxxxxxB
\else
\ifnum#1=21 %
\hatcurCCbbKmagxxxxxC
\else
??????\fi
\fi
\fi
}
\newcommand{\hatcurCCcitHmag}[1]{\ifnum#1=19 %
\hatcurCCcitHmagxxxxxA
\else
\ifnum#1=20 %
\hatcurCCcitHmagxxxxxB
\else
\ifnum#1=21 %
\hatcurCCcitHmagxxxxxC
\else
??????\fi
\fi
\fi
}
\newcommand{\hatcurCCcitJmag}[1]{\ifnum#1=19 %
\hatcurCCcitJmagxxxxxA
\else
\ifnum#1=20 %
\hatcurCCcitJmagxxxxxB
\else
\ifnum#1=21 %
\hatcurCCcitJmagxxxxxC
\else
??????\fi
\fi
\fi
}
\newcommand{\hatcurCCcitKmag}[1]{\ifnum#1=19 %
\hatcurCCcitKmagxxxxxA
\else
\ifnum#1=20 %
\hatcurCCcitKmagxxxxxB
\else
\ifnum#1=21 %
\hatcurCCcitKmagxxxxxC
\else
??????\fi
\fi
\fi
}
\newcommand{\hatcurCCdec}[1]{\ifnum#1=19 %
\hatcurCCdecxxxxxA
\else
\ifnum#1=20 %
\hatcurCCdecxxxxxB
\else
\ifnum#1=21 %
\hatcurCCdecxxxxxC
\else
??????\fi
\fi
\fi
}
\newcommand{\hatcurCCesoHKmag}[1]{\ifnum#1=19 %
\hatcurCCesoHKmagxxxxxA
\else
\ifnum#1=20 %
\hatcurCCesoHKmagxxxxxB
\else
\ifnum#1=21 %
\hatcurCCesoHKmagxxxxxC
\else
??????\fi
\fi
\fi
}
\newcommand{\hatcurCCesoHmag}[1]{\ifnum#1=19 %
\hatcurCCesoHmagxxxxxA
\else
\ifnum#1=20 %
\hatcurCCesoHmagxxxxxB
\else
\ifnum#1=21 %
\hatcurCCesoHmagxxxxxC
\else
??????\fi
\fi
\fi
}
\newcommand{\hatcurCCesoJHmag}[1]{\ifnum#1=19 %
\hatcurCCesoJHmagxxxxxA
\else
\ifnum#1=20 %
\hatcurCCesoJHmagxxxxxB
\else
\ifnum#1=21 %
\hatcurCCesoJHmagxxxxxC
\else
??????\fi
\fi
\fi
}
\newcommand{\hatcurCCesoJKmag}[1]{\ifnum#1=19 %
\hatcurCCesoJKmagxxxxxA
\else
\ifnum#1=20 %
\hatcurCCesoJKmagxxxxxB
\else
\ifnum#1=21 %
\hatcurCCesoJKmagxxxxxC
\else
??????\fi
\fi
\fi
}
\newcommand{\hatcurCCesoJmag}[1]{\ifnum#1=19 %
\hatcurCCesoJmagxxxxxA
\else
\ifnum#1=20 %
\hatcurCCesoJmagxxxxxB
\else
\ifnum#1=21 %
\hatcurCCesoJmagxxxxxC
\else
??????\fi
\fi
\fi
}
\newcommand{\hatcurCCesoKmag}[1]{\ifnum#1=19 %
\hatcurCCesoKmagxxxxxA
\else
\ifnum#1=20 %
\hatcurCCesoKmagxxxxxB
\else
\ifnum#1=21 %
\hatcurCCesoKmagxxxxxC
\else
??????\fi
\fi
\fi
}
\newcommand{\hatcurCCgsc}[1]{\ifnum#1=19 %
\hatcurCCgscxxxxxA
\else
\ifnum#1=20 %
\hatcurCCgscxxxxxB
\else
\ifnum#1=21 %
\hatcurCCgscxxxxxC
\else
??????\fi
\fi
\fi
}
\newcommand{\hatcurCCmag}[1]{\ifnum#1=19 %
\hatcurCCmagxxxxxA
\else
\ifnum#1=20 %
\hatcurCCmagxxxxxB
\else
\ifnum#1=21 %
\hatcurCCmagxxxxxC
\else
??????\fi
\fi
\fi
}
\newcommand{\hatcurCCpm}[1]{\ifnum#1=19 %
\hatcurCCpmxxxxxA
\else
\ifnum#1=20 %
\hatcurCCpmxxxxxB
\else
\ifnum#1=21 %
\hatcurCCpmxxxxxC
\else
??????\fi
\fi
\fi
}
\newcommand{\hatcurCCpmdec}[1]{\ifnum#1=19 %
\hatcurCCpmdecxxxxxA
\else
\ifnum#1=20 %
\hatcurCCpmdecxxxxxB
\else
\ifnum#1=21 %
\hatcurCCpmdecxxxxxC
\else
??????\fi
\fi
\fi
}
\newcommand{\hatcurCCpmra}[1]{\ifnum#1=19 %
\hatcurCCpmraxxxxxA
\else
\ifnum#1=20 %
\hatcurCCpmraxxxxxB
\else
\ifnum#1=21 %
\hatcurCCpmraxxxxxC
\else
??????\fi
\fi
\fi
}
\newcommand{\hatcurCCra}[1]{\ifnum#1=19 %
\hatcurCCraxxxxxA
\else
\ifnum#1=20 %
\hatcurCCraxxxxxB
\else
\ifnum#1=21 %
\hatcurCCraxxxxxC
\else
??????\fi
\fi
\fi
}
\newcommand{\hatcurCCtassmB}[1]{\ifnum#1=19 %
\hatcurCCtassmBxxxxxA
\else
\ifnum#1=20 %
\hatcurCCtassmBxxxxxB
\else
\ifnum#1=21 %
\hatcurCCtassmBxxxxxC
\else
??????\fi
\fi
\fi
}
\newcommand{\hatcurCCtassmBshort}[1]{\ifnum#1=19 %
\hatcurCCtassmBshortxxxxxA
\else
\ifnum#1=20 %
\hatcurCCtassmBshortxxxxxB
\else
\ifnum#1=21 %
\hatcurCCtassmBshortxxxxxC
\else
??????\fi
\fi
\fi
}
\newcommand{\hatcurCCtassmg}[1]{\ifnum#1=19 %
\hatcurCCtassmgxxxxxA
\else
\ifnum#1=20 %
\hatcurCCtassmgxxxxxB
\else
\ifnum#1=21 %
\hatcurCCtassmgxxxxxC
\else
??????\fi
\fi
\fi
}
\newcommand{\hatcurCCtassmgshort}[1]{\ifnum#1=19 %
\hatcurCCtassmgshortxxxxxA
\else
\ifnum#1=20 %
\hatcurCCtassmgshortxxxxxB
\else
\ifnum#1=21 %
\hatcurCCtassmgshortxxxxxC
\else
??????\fi
\fi
\fi
}
\newcommand{\hatcurCCtassmi}[1]{\ifnum#1=19 %
\hatcurCCtassmixxxxxA
\else
\ifnum#1=20 %
\hatcurCCtassmixxxxxB
\else
\ifnum#1=21 %
\hatcurCCtassmixxxxxC
\else
??????\fi
\fi
\fi
}
\newcommand{\hatcurCCtassmI}[1]{\ifnum#1=19 %
\hatcurCCtassmIxxxxxA
\else
\ifnum#1=20 %
\hatcurCCtassmIxxxxxB
\else
\ifnum#1=21 %
\hatcurCCtassmIxxxxxC
\else
??????\fi
\fi
\fi
}
\newcommand{\hatcurCCtassmishort}[1]{\ifnum#1=19 %
\hatcurCCtassmishortxxxxxA
\else
\ifnum#1=20 %
\hatcurCCtassmishortxxxxxB
\else
\ifnum#1=21 %
\hatcurCCtassmishortxxxxxC
\else
??????\fi
\fi
\fi
}
\newcommand{\hatcurCCtassmIshort}[1]{\ifnum#1=19 %
\hatcurCCtassmIshortxxxxxA
\else
\ifnum#1=20 %
\hatcurCCtassmIshortxxxxxB
\else
\ifnum#1=21 %
\hatcurCCtassmIshortxxxxxC
\else
??????\fi
\fi
\fi
}
\newcommand{\hatcurCCtassmr}[1]{\ifnum#1=19 %
\hatcurCCtassmrxxxxxA
\else
\ifnum#1=20 %
\hatcurCCtassmrxxxxxB
\else
\ifnum#1=21 %
\hatcurCCtassmrxxxxxC
\else
??????\fi
\fi
\fi
}
\newcommand{\hatcurCCtassmrshort}[1]{\ifnum#1=19 %
\hatcurCCtassmrshortxxxxxA
\else
\ifnum#1=20 %
\hatcurCCtassmrshortxxxxxB
\else
\ifnum#1=21 %
\hatcurCCtassmrshortxxxxxC
\else
??????\fi
\fi
\fi
}
\newcommand{\hatcurCCtassmv}[1]{\ifnum#1=19 %
\hatcurCCtassmvxxxxxA
\else
\ifnum#1=20 %
\hatcurCCtassmvxxxxxB
\else
\ifnum#1=21 %
\hatcurCCtassmvxxxxxC
\else
??????\fi
\fi
\fi
}
\newcommand{\hatcurCCtassmvshort}[1]{\ifnum#1=19 %
\hatcurCCtassmvshortxxxxxA
\else
\ifnum#1=20 %
\hatcurCCtassmvshortxxxxxB
\else
\ifnum#1=21 %
\hatcurCCtassmvshortxxxxxC
\else
??????\fi
\fi
\fi
}
\newcommand{\hatcurCCtwomass}[1]{\ifnum#1=19 %
\hatcurCCtwomassxxxxxA
\else
\ifnum#1=20 %
\hatcurCCtwomassxxxxxB
\else
\ifnum#1=21 %
\hatcurCCtwomassxxxxxC
\else
??????\fi
\fi
\fi
}
\newcommand{\hatcurCCtwomassHmag}[1]{\ifnum#1=19 %
\hatcurCCtwomassHmagxxxxxA
\else
\ifnum#1=20 %
\hatcurCCtwomassHmagxxxxxB
\else
\ifnum#1=21 %
\hatcurCCtwomassHmagxxxxxC
\else
??????\fi
\fi
\fi
}
\newcommand{\hatcurCCtwomassJmag}[1]{\ifnum#1=19 %
\hatcurCCtwomassJmagxxxxxA
\else
\ifnum#1=20 %
\hatcurCCtwomassJmagxxxxxB
\else
\ifnum#1=21 %
\hatcurCCtwomassJmagxxxxxC
\else
??????\fi
\fi
\fi
}
\newcommand{\hatcurCCtwomassKmag}[1]{\ifnum#1=19 %
\hatcurCCtwomassKmagxxxxxA
\else
\ifnum#1=20 %
\hatcurCCtwomassKmagxxxxxB
\else
\ifnum#1=21 %
\hatcurCCtwomassKmagxxxxxC
\else
??????\fi
\fi
\fi
}
\newcommand{\hatcurfield}[1]{\ifnum#1=19 %
\hatcurfieldxxxxxA
\else
\ifnum#1=20 %
\hatcurfieldxxxxxB
\else
\ifnum#1=21 %
\hatcurfieldxxxxxC
\else
??????\fi
\fi
\fi
}
\newcommand{\hatcurhtr}[1]{\ifnum#1=19 %
\hatcurhtrxxxxxA
\else
\ifnum#1=20 %
\hatcurhtrxxxxxB
\else
\ifnum#1=21 %
\hatcurhtrxxxxxC
\else
??????\fi
\fi
\fi
}
\newcommand{\hatcurISOage}[1]{\ifnum#1=19 %
\hatcurISOagexxxxxA
\else
\ifnum#1=20 %
\hatcurISOagexxxxxB
\else
\ifnum#1=21 %
\hatcurISOagexxxxxC
\else
??????\fi
\fi
\fi
}
\newcommand{\hatcurISOJK}[1]{\ifnum#1=19 %
\hatcurISOJKxxxxxA
\else
\ifnum#1=20 %
\hatcurISOJKxxxxxB
\else
\ifnum#1=21 %
\hatcurISOJKxxxxxC
\else
??????\fi
\fi
\fi
}
\newcommand{\hatcurISOlogg}[1]{\ifnum#1=19 %
\hatcurISOloggxxxxxA
\else
\ifnum#1=20 %
\hatcurISOloggxxxxxB
\else
\ifnum#1=21 %
\hatcurISOloggxxxxxC
\else
??????\fi
\fi
\fi
}
\newcommand{\hatcurISOlum}[1]{\ifnum#1=19 %
\hatcurISOlumxxxxxA
\else
\ifnum#1=20 %
\hatcurISOlumxxxxxB
\else
\ifnum#1=21 %
\hatcurISOlumxxxxxC
\else
??????\fi
\fi
\fi
}
\newcommand{\hatcurISOlumshort}[1]{\ifnum#1=19 %
\hatcurISOlumshortxxxxxA
\else
\ifnum#1=20 %
\hatcurISOlumshortxxxxxB
\else
\ifnum#1=21 %
\hatcurISOlumshortxxxxxC
\else
??????\fi
\fi
\fi
}
\newcommand{\hatcurISOm}[1]{\ifnum#1=19 %
\hatcurISOmxxxxxA
\else
\ifnum#1=20 %
\hatcurISOmxxxxxB
\else
\ifnum#1=21 %
\hatcurISOmxxxxxC
\else
??????\fi
\fi
\fi
}
\newcommand{\hatcurISOMH}[1]{\ifnum#1=19 %
\hatcurISOMHxxxxxA
\else
\ifnum#1=20 %
\hatcurISOMHxxxxxB
\else
\ifnum#1=21 %
\hatcurISOMHxxxxxC
\else
??????\fi
\fi
\fi
}
\newcommand{\hatcurISOMJ}[1]{\ifnum#1=19 %
\hatcurISOMJxxxxxA
\else
\ifnum#1=20 %
\hatcurISOMJxxxxxB
\else
\ifnum#1=21 %
\hatcurISOMJxxxxxC
\else
??????\fi
\fi
\fi
}
\newcommand{\hatcurISOMK}[1]{\ifnum#1=19 %
\hatcurISOMKxxxxxA
\else
\ifnum#1=20 %
\hatcurISOMKxxxxxB
\else
\ifnum#1=21 %
\hatcurISOMKxxxxxC
\else
??????\fi
\fi
\fi
}
\newcommand{\hatcurISOmlong}[1]{\ifnum#1=19 %
\hatcurISOmlongxxxxxA
\else
\ifnum#1=20 %
\hatcurISOmlongxxxxxB
\else
\ifnum#1=21 %
\hatcurISOmlongxxxxxC
\else
??????\fi
\fi
\fi
}
\newcommand{\hatcurISOmshort}[1]{\ifnum#1=19 %
\hatcurISOmshortxxxxxA
\else
\ifnum#1=20 %
\hatcurISOmshortxxxxxB
\else
\ifnum#1=21 %
\hatcurISOmshortxxxxxC
\else
??????\fi
\fi
\fi
}
\newcommand{\hatcurISOmv}[1]{\ifnum#1=19 %
\hatcurISOmvxxxxxA
\else
\ifnum#1=20 %
\hatcurISOmvxxxxxB
\else
\ifnum#1=21 %
\hatcurISOmvxxxxxC
\else
??????\fi
\fi
\fi
}
\newcommand{\hatcurISOr}[1]{\ifnum#1=19 %
\hatcurISOrxxxxxA
\else
\ifnum#1=20 %
\hatcurISOrxxxxxB
\else
\ifnum#1=21 %
\hatcurISOrxxxxxC
\else
??????\fi
\fi
\fi
}
\newcommand{\hatcurISOrho}[1]{\ifnum#1=19 %
\hatcurISOrhoxxxxxA
\else
\ifnum#1=20 %
\hatcurISOrhoxxxxxB
\else
\ifnum#1=21 %
\hatcurISOrhoxxxxxC
\else
??????\fi
\fi
\fi
}
\newcommand{\hatcurISOrholong}[1]{\ifnum#1=19 %
\hatcurISOrholongxxxxxA
\else
\ifnum#1=20 %
\hatcurISOrholongxxxxxB
\else
\ifnum#1=21 %
\hatcurISOrholongxxxxxC
\else
??????\fi
\fi
\fi
}
\newcommand{\hatcurISOrlong}[1]{\ifnum#1=19 %
\hatcurISOrlongxxxxxA
\else
\ifnum#1=20 %
\hatcurISOrlongxxxxxB
\else
\ifnum#1=21 %
\hatcurISOrlongxxxxxC
\else
??????\fi
\fi
\fi
}
\newcommand{\hatcurISOrshort}[1]{\ifnum#1=19 %
\hatcurISOrshortxxxxxA
\else
\ifnum#1=20 %
\hatcurISOrshortxxxxxB
\else
\ifnum#1=21 %
\hatcurISOrshortxxxxxC
\else
??????\fi
\fi
\fi
}
\newcommand{\hatcurISOsigma}[1]{\ifnum#1=19 %
\hatcurISOsigmaxxxxxA
\else
\ifnum#1=20 %
\hatcurISOsigmaxxxxxB
\else
\ifnum#1=21 %
\hatcurISOsigmaxxxxxC
\else
??????\fi
\fi
\fi
}
\newcommand{\hatcurISOspec}[1]{\ifnum#1=19 %
\hatcurISOspecxxxxxA
\else
\ifnum#1=20 %
\hatcurISOspecxxxxxB
\else
\ifnum#1=21 %
\hatcurISOspecxxxxxC
\else
??????\fi
\fi
\fi
}
\newcommand{\hatcurISOvi}[1]{\ifnum#1=19 %
\hatcurISOvixxxxxA
\else
\ifnum#1=20 %
\hatcurISOvixxxxxB
\else
\ifnum#1=21 %
\hatcurISOvixxxxxC
\else
??????\fi
\fi
\fi
}
\newcommand{\hatcurLBig}[1]{\ifnum#1=19 %
\hatcurLBigxxxxxA
\else
\ifnum#1=20 %
\hatcurLBigxxxxxB
\else
\ifnum#1=21 %
\hatcurLBigxxxxxC
\else
??????\fi
\fi
\fi
}
\newcommand{\hatcurLBii}[1]{\ifnum#1=19 %
\hatcurLBiixxxxxA
\else
\ifnum#1=20 %
\hatcurLBiixxxxxB
\else
\ifnum#1=21 %
\hatcurLBiixxxxxC
\else
??????\fi
\fi
\fi
}
\newcommand{\hatcurLBiI}[1]{\ifnum#1=19 %
\hatcurLBiIxxxxxA
\else
\ifnum#1=20 %
\hatcurLBiIxxxxxB
\else
\ifnum#1=21 %
\hatcurLBiIxxxxxC
\else
??????\fi
\fi
\fi
}
\newcommand{\hatcurLBiig}[1]{\ifnum#1=19 %
\hatcurLBiigxxxxxA
\else
\ifnum#1=20 %
\hatcurLBiigxxxxxB
\else
\ifnum#1=21 %
\hatcurLBiigxxxxxC
\else
??????\fi
\fi
\fi
}
\newcommand{\hatcurLBiii}[1]{\ifnum#1=19 %
\hatcurLBiiixxxxxA
\else
\ifnum#1=20 %
\hatcurLBiiixxxxxB
\else
\ifnum#1=21 %
\hatcurLBiiixxxxxC
\else
??????\fi
\fi
\fi
}
\newcommand{\hatcurLBiiI}[1]{\ifnum#1=19 %
\hatcurLBiiIxxxxxA
\else
\ifnum#1=20 %
\hatcurLBiiIxxxxxB
\else
\ifnum#1=21 %
\hatcurLBiiIxxxxxC
\else
??????\fi
\fi
\fi
}
\newcommand{\hatcurLBiikep}[1]{\ifnum#1=19 %
\hatcurLBiikepxxxxxA
\else
\ifnum#1=20 %
\hatcurLBiikepxxxxxB
\else
\ifnum#1=21 %
\hatcurLBiikepxxxxxC
\else
??????\fi
\fi
\fi
}
\newcommand{\hatcurLBiir}[1]{\ifnum#1=19 %
\hatcurLBiirxxxxxA
\else
\ifnum#1=20 %
\hatcurLBiirxxxxxB
\else
\ifnum#1=21 %
\hatcurLBiirxxxxxC
\else
??????\fi
\fi
\fi
}
\newcommand{\hatcurLBiiR}[1]{\ifnum#1=19 %
\hatcurLBiiRxxxxxA
\else
\ifnum#1=20 %
\hatcurLBiiRxxxxxB
\else
\ifnum#1=21 %
\hatcurLBiiRxxxxxC
\else
??????\fi
\fi
\fi
}
\newcommand{\hatcurLBiiz}[1]{\ifnum#1=19 %
\hatcurLBiizxxxxxA
\else
\ifnum#1=20 %
\hatcurLBiizxxxxxB
\else
\ifnum#1=21 %
\hatcurLBiizxxxxxC
\else
??????\fi
\fi
\fi
}
\newcommand{\hatcurLBikep}[1]{\ifnum#1=19 %
\hatcurLBikepxxxxxA
\else
\ifnum#1=20 %
\hatcurLBikepxxxxxB
\else
\ifnum#1=21 %
\hatcurLBikepxxxxxC
\else
??????\fi
\fi
\fi
}
\newcommand{\hatcurLBir}[1]{\ifnum#1=19 %
\hatcurLBirxxxxxA
\else
\ifnum#1=20 %
\hatcurLBirxxxxxB
\else
\ifnum#1=21 %
\hatcurLBirxxxxxC
\else
??????\fi
\fi
\fi
}
\newcommand{\hatcurLBiR}[1]{\ifnum#1=19 %
\hatcurLBiRxxxxxA
\else
\ifnum#1=20 %
\hatcurLBiRxxxxxB
\else
\ifnum#1=21 %
\hatcurLBiRxxxxxC
\else
??????\fi
\fi
\fi
}
\newcommand{\hatcurLBiz}[1]{\ifnum#1=19 %
\hatcurLBizxxxxxA
\else
\ifnum#1=20 %
\hatcurLBizxxxxxB
\else
\ifnum#1=21 %
\hatcurLBizxxxxxC
\else
??????\fi
\fi
\fi
}
\newcommand{\hatcurLCbsq}[1]{\ifnum#1=19 %
\hatcurLCbsqxxxxxA
\else
\ifnum#1=20 %
\hatcurLCbsqxxxxxB
\else
\ifnum#1=21 %
\hatcurLCbsqxxxxxC
\else
??????\fi
\fi
\fi
}
\newcommand{\hatcurLCdip}[1]{\ifnum#1=19 %
\hatcurLCdipxxxxxA
\else
\ifnum#1=20 %
\hatcurLCdipxxxxxB
\else
\ifnum#1=21 %
\hatcurLCdipxxxxxC
\else
??????\fi
\fi
\fi
}
\newcommand{\hatcurLCdur}[1]{\ifnum#1=19 %
\hatcurLCdurxxxxxA
\else
\ifnum#1=20 %
\hatcurLCdurxxxxxB
\else
\ifnum#1=21 %
\hatcurLCdurxxxxxC
\else
??????\fi
\fi
\fi
}
\newcommand{\hatcurLCdurhr}[1]{\ifnum#1=19 %
\hatcurLCdurhrxxxxxA
\else
\ifnum#1=20 %
\hatcurLCdurhrxxxxxB
\else
\ifnum#1=21 %
\hatcurLCdurhrxxxxxC
\else
??????\fi
\fi
\fi
}
\newcommand{\hatcurLCdurhrshort}[1]{\ifnum#1=19 %
\hatcurLCdurhrshortxxxxxA
\else
\ifnum#1=20 %
\hatcurLCdurhrshortxxxxxB
\else
\ifnum#1=21 %
\hatcurLCdurhrshortxxxxxC
\else
??????\fi
\fi
\fi
}
\newcommand{\hatcurLCdurshort}[1]{\ifnum#1=19 %
\hatcurLCdurshortxxxxxA
\else
\ifnum#1=20 %
\hatcurLCdurshortxxxxxB
\else
\ifnum#1=21 %
\hatcurLCdurshortxxxxxC
\else
??????\fi
\fi
\fi
}
\newcommand{\hatcurLChatnetmA}[1]{\ifnum#1=19 %
\hatcurLChatnetmAxxxxxA
\else
\ifnum#1=20 %
\hatcurLChatnetmAxxxxxB
\else
\ifnum#1=21 %
\hatcurLChatnetmAxxxxxC
\else
??????\fi
\fi
\fi
}
\newcommand{\hatcurLChatnetmB}[1]{\ifnum#1=19 %
\hatcurLChatnetmBxxxxxA
\else
\ifnum#1=20 %
\hatcurLChatnetmBxxxxxB
\else
\ifnum#1=21 %
\hatcurLChatnetmBxxxxxC
\else
??????\fi
\fi
\fi
}
\newcommand{\hatcurLCiblendA}[1]{\ifnum#1=19 %
\hatcurLCiblendAxxxxxA
\else
\ifnum#1=20 %
\hatcurLCiblendAxxxxxB
\else
\ifnum#1=21 %
\hatcurLCiblendAxxxxxC
\else
??????\fi
\fi
\fi
}
\newcommand{\hatcurLCiblendB}[1]{\ifnum#1=19 %
\hatcurLCiblendBxxxxxA
\else
\ifnum#1=20 %
\hatcurLCiblendBxxxxxB
\else
\ifnum#1=21 %
\hatcurLCiblendBxxxxxC
\else
??????\fi
\fi
\fi
}
\newcommand{\hatcurLCimp}[1]{\ifnum#1=19 %
\hatcurLCimpxxxxxA
\else
\ifnum#1=20 %
\hatcurLCimpxxxxxB
\else
\ifnum#1=21 %
\hatcurLCimpxxxxxC
\else
??????\fi
\fi
\fi
}
\newcommand{\hatcurLCingdur}[1]{\ifnum#1=19 %
\hatcurLCingdurxxxxxA
\else
\ifnum#1=20 %
\hatcurLCingdurxxxxxB
\else
\ifnum#1=21 %
\hatcurLCingdurxxxxxC
\else
??????\fi
\fi
\fi
}
\newcommand{\hatcurLCP}[1]{\ifnum#1=19 %
\hatcurLCPxxxxxA
\else
\ifnum#1=20 %
\hatcurLCPxxxxxB
\else
\ifnum#1=21 %
\hatcurLCPxxxxxC
\else
??????\fi
\fi
\fi
}
\newcommand{\hatcurLCPprec}[1]{\ifnum#1=19 %
\hatcurLCPprecxxxxxA
\else
\ifnum#1=20 %
\hatcurLCPprecxxxxxB
\else
\ifnum#1=21 %
\hatcurLCPprecxxxxxC
\else
??????\fi
\fi
\fi
}
\newcommand{\hatcurLCPshort}[1]{\ifnum#1=19 %
\hatcurLCPshortxxxxxA
\else
\ifnum#1=20 %
\hatcurLCPshortxxxxxB
\else
\ifnum#1=21 %
\hatcurLCPshortxxxxxC
\else
??????\fi
\fi
\fi
}
\newcommand{\hatcurLCq}[1]{\ifnum#1=19 %
\hatcurLCqxxxxxA
\else
\ifnum#1=20 %
\hatcurLCqxxxxxB
\else
\ifnum#1=21 %
\hatcurLCqxxxxxC
\else
??????\fi
\fi
\fi
}
\newcommand{\hatcurLCqshort}[1]{\ifnum#1=19 %
\hatcurLCqshortxxxxxA
\else
\ifnum#1=20 %
\hatcurLCqshortxxxxxB
\else
\ifnum#1=21 %
\hatcurLCqshortxxxxxC
\else
??????\fi
\fi
\fi
}
\newcommand{\hatcurLCrho}[1]{\ifnum#1=19 %
\hatcurLCrhoxxxxxA
\else
\ifnum#1=20 %
\hatcurLCrhoxxxxxB
\else
\ifnum#1=21 %
\hatcurLCrhoxxxxxC
\else
??????\fi
\fi
\fi
}
\newcommand{\hatcurLCrprstar}[1]{\ifnum#1=19 %
\hatcurLCrprstarxxxxxA
\else
\ifnum#1=20 %
\hatcurLCrprstarxxxxxB
\else
\ifnum#1=21 %
\hatcurLCrprstarxxxxxC
\else
??????\fi
\fi
\fi
}
\newcommand{\hatcurLCT}[1]{\ifnum#1=19 %
\hatcurLCTxxxxxA
\else
\ifnum#1=20 %
\hatcurLCTxxxxxB
\else
\ifnum#1=21 %
\hatcurLCTxxxxxC
\else
??????\fi
\fi
\fi
}
\newcommand{\hatcurLCTA}[1]{\ifnum#1=19 %
\hatcurLCTAxxxxxA
\else
\ifnum#1=20 %
\hatcurLCTAxxxxxB
\else
\ifnum#1=21 %
\hatcurLCTAxxxxxC
\else
??????\fi
\fi
\fi
}
\newcommand{\hatcurLCTB}[1]{\ifnum#1=19 %
\hatcurLCTBxxxxxA
\else
\ifnum#1=20 %
\hatcurLCTBxxxxxB
\else
\ifnum#1=21 %
\hatcurLCTBxxxxxC
\else
??????\fi
\fi
\fi
}
\newcommand{\hatcurLCzeta}[1]{\ifnum#1=19 %
\hatcurLCzetaxxxxxA
\else
\ifnum#1=20 %
\hatcurLCzetaxxxxxB
\else
\ifnum#1=21 %
\hatcurLCzetaxxxxxC
\else
??????\fi
\fi
\fi
}
\newcommand{\hatcurPPaequiv}[1]{\ifnum#1=19 %
\hatcurPPaequivxxxxxA
\else
\ifnum#1=20 %
\hatcurPPaequivxxxxxB
\else
\ifnum#1=21 %
\hatcurPPaequivxxxxxC
\else
??????\fi
\fi
\fi
}
\newcommand{\hatcurPPar}[1]{\ifnum#1=19 %
\hatcurPParxxxxxA
\else
\ifnum#1=20 %
\hatcurPParxxxxxB
\else
\ifnum#1=21 %
\hatcurPParxxxxxC
\else
??????\fi
\fi
\fi
}
\newcommand{\hatcurPParel}[1]{\ifnum#1=19 %
\hatcurPParelxxxxxA
\else
\ifnum#1=20 %
\hatcurPParelxxxxxB
\else
\ifnum#1=21 %
\hatcurPParelxxxxxC
\else
??????\fi
\fi
\fi
}
\newcommand{\hatcurPPfluxap}[1]{\ifnum#1=19 %
\hatcurPPfluxapxxxxxA
\else
\ifnum#1=20 %
\hatcurPPfluxapxxxxxB
\else
\ifnum#1=21 %
\hatcurPPfluxapxxxxxC
\else
??????\fi
\fi
\fi
}
\newcommand{\hatcurPPfluxapdim}[1]{\ifnum#1=19 %
\hatcurPPfluxapdimxxxxxA
\else
\ifnum#1=20 %
\hatcurPPfluxapdimxxxxxB
\else
\ifnum#1=21 %
\hatcurPPfluxapdimxxxxxC
\else
??????\fi
\fi
\fi
}
\newcommand{\hatcurPPfluxavg}[1]{\ifnum#1=19 %
\hatcurPPfluxavgxxxxxA
\else
\ifnum#1=20 %
\hatcurPPfluxavgxxxxxB
\else
\ifnum#1=21 %
\hatcurPPfluxavgxxxxxC
\else
??????\fi
\fi
\fi
}
\newcommand{\hatcurPPfluxavgdim}[1]{\ifnum#1=19 %
\hatcurPPfluxavgdimxxxxxA
\else
\ifnum#1=20 %
\hatcurPPfluxavgdimxxxxxB
\else
\ifnum#1=21 %
\hatcurPPfluxavgdimxxxxxC
\else
??????\fi
\fi
\fi
}
\newcommand{\hatcurPPfluxavglog}[1]{\ifnum#1=19 %
\hatcurPPfluxavglogxxxxxA
\else
\ifnum#1=20 %
\hatcurPPfluxavglogxxxxxB
\else
\ifnum#1=21 %
\hatcurPPfluxavglogxxxxxC
\else
??????\fi
\fi
\fi
}
\newcommand{\hatcurPPfluxperi}[1]{\ifnum#1=19 %
\hatcurPPfluxperixxxxxA
\else
\ifnum#1=20 %
\hatcurPPfluxperixxxxxB
\else
\ifnum#1=21 %
\hatcurPPfluxperixxxxxC
\else
??????\fi
\fi
\fi
}
\newcommand{\hatcurPPfluxperidim}[1]{\ifnum#1=19 %
\hatcurPPfluxperidimxxxxxA
\else
\ifnum#1=20 %
\hatcurPPfluxperidimxxxxxB
\else
\ifnum#1=21 %
\hatcurPPfluxperidimxxxxxC
\else
??????\fi
\fi
\fi
}
\newcommand{\hatcurPPg}[1]{\ifnum#1=19 %
\hatcurPPgxxxxxA
\else
\ifnum#1=20 %
\hatcurPPgxxxxxB
\else
\ifnum#1=21 %
\hatcurPPgxxxxxC
\else
??????\fi
\fi
\fi
}
\newcommand{\hatcurPPi}[1]{\ifnum#1=19 %
\hatcurPPixxxxxA
\else
\ifnum#1=20 %
\hatcurPPixxxxxB
\else
\ifnum#1=21 %
\hatcurPPixxxxxC
\else
??????\fi
\fi
\fi
}
\newcommand{\hatcurPPlogg}[1]{\ifnum#1=19 %
\hatcurPPloggxxxxxA
\else
\ifnum#1=20 %
\hatcurPPloggxxxxxB
\else
\ifnum#1=21 %
\hatcurPPloggxxxxxC
\else
??????\fi
\fi
\fi
}
\newcommand{\hatcurPPm}[1]{\ifnum#1=19 %
\hatcurPPmxxxxxA
\else
\ifnum#1=20 %
\hatcurPPmxxxxxB
\else
\ifnum#1=21 %
\hatcurPPmxxxxxC
\else
??????\fi
\fi
\fi
}
\newcommand{\hatcurPPme}[1]{\ifnum#1=19 %
\hatcurPPmexxxxxA
\else
\ifnum#1=20 %
\hatcurPPmexxxxxB
\else
\ifnum#1=21 %
\hatcurPPmexxxxxC
\else
??????\fi
\fi
\fi
}
\newcommand{\hatcurPPmelong}[1]{\ifnum#1=19 %
\hatcurPPmelongxxxxxA
\else
\ifnum#1=20 %
\hatcurPPmelongxxxxxB
\else
\ifnum#1=21 %
\hatcurPPmelongxxxxxC
\else
??????\fi
\fi
\fi
}
\newcommand{\hatcurPPmeshort}[1]{\ifnum#1=19 %
\hatcurPPmeshortxxxxxA
\else
\ifnum#1=20 %
\hatcurPPmeshortxxxxxB
\else
\ifnum#1=21 %
\hatcurPPmeshortxxxxxC
\else
??????\fi
\fi
\fi
}
\newcommand{\hatcurPPmlong}[1]{\ifnum#1=19 %
\hatcurPPmlongxxxxxA
\else
\ifnum#1=20 %
\hatcurPPmlongxxxxxB
\else
\ifnum#1=21 %
\hatcurPPmlongxxxxxC
\else
??????\fi
\fi
\fi
}
\newcommand{\hatcurPPmrcorr}[1]{\ifnum#1=19 %
\hatcurPPmrcorrxxxxxA
\else
\ifnum#1=20 %
\hatcurPPmrcorrxxxxxB
\else
\ifnum#1=21 %
\hatcurPPmrcorrxxxxxC
\else
??????\fi
\fi
\fi
}
\newcommand{\hatcurPPmshort}[1]{\ifnum#1=19 %
\hatcurPPmshortxxxxxA
\else
\ifnum#1=20 %
\hatcurPPmshortxxxxxB
\else
\ifnum#1=21 %
\hatcurPPmshortxxxxxC
\else
??????\fi
\fi
\fi
}
\newcommand{\hatcurPPperi}[1]{\ifnum#1=19 %
\hatcurPPperixxxxxA
\else
\ifnum#1=20 %
\hatcurPPperixxxxxB
\else
\ifnum#1=21 %
\hatcurPPperixxxxxC
\else
??????\fi
\fi
\fi
}
\newcommand{\hatcurPPphiconj}[1]{\ifnum#1=19 %
\hatcurPPphiconjxxxxxA
\else
\ifnum#1=20 %
\hatcurPPphiconjxxxxxB
\else
\ifnum#1=21 %
\hatcurPPphiconjxxxxxC
\else
??????\fi
\fi
\fi
}
\newcommand{\hatcurPPr}[1]{\ifnum#1=19 %
\hatcurPPrxxxxxA
\else
\ifnum#1=20 %
\hatcurPPrxxxxxB
\else
\ifnum#1=21 %
\hatcurPPrxxxxxC
\else
??????\fi
\fi
\fi
}
\newcommand{\hatcurPPre}[1]{\ifnum#1=19 %
\hatcurPPrexxxxxA
\else
\ifnum#1=20 %
\hatcurPPrexxxxxB
\else
\ifnum#1=21 %
\hatcurPPrexxxxxC
\else
??????\fi
\fi
\fi
}
\newcommand{\hatcurPPrelong}[1]{\ifnum#1=19 %
\hatcurPPrelongxxxxxA
\else
\ifnum#1=20 %
\hatcurPPrelongxxxxxB
\else
\ifnum#1=21 %
\hatcurPPrelongxxxxxC
\else
??????\fi
\fi
\fi
}
\newcommand{\hatcurPPreshort}[1]{\ifnum#1=19 %
\hatcurPPreshortxxxxxA
\else
\ifnum#1=20 %
\hatcurPPreshortxxxxxB
\else
\ifnum#1=21 %
\hatcurPPreshortxxxxxC
\else
??????\fi
\fi
\fi
}
\newcommand{\hatcurPPrho}[1]{\ifnum#1=19 %
\hatcurPPrhoxxxxxA
\else
\ifnum#1=20 %
\hatcurPPrhoxxxxxB
\else
\ifnum#1=21 %
\hatcurPPrhoxxxxxC
\else
??????\fi
\fi
\fi
}
\newcommand{\hatcurPPrlong}[1]{\ifnum#1=19 %
\hatcurPPrlongxxxxxA
\else
\ifnum#1=20 %
\hatcurPPrlongxxxxxB
\else
\ifnum#1=21 %
\hatcurPPrlongxxxxxC
\else
??????\fi
\fi
\fi
}
\newcommand{\hatcurPPrshort}[1]{\ifnum#1=19 %
\hatcurPPrshortxxxxxA
\else
\ifnum#1=20 %
\hatcurPPrshortxxxxxB
\else
\ifnum#1=21 %
\hatcurPPrshortxxxxxC
\else
??????\fi
\fi
\fi
}
\newcommand{\hatcurPPtcirc}[1]{\ifnum#1=19 %
\hatcurPPtcircxxxxxA
\else
\ifnum#1=20 %
\hatcurPPtcircxxxxxB
\else
\ifnum#1=21 %
\hatcurPPtcircxxxxxC
\else
??????\fi
\fi
\fi
}
\newcommand{\hatcurPPteff}[1]{\ifnum#1=19 %
\hatcurPPteffxxxxxA
\else
\ifnum#1=20 %
\hatcurPPteffxxxxxB
\else
\ifnum#1=21 %
\hatcurPPteffxxxxxC
\else
??????\fi
\fi
\fi
}
\newcommand{\hatcurPPtheta}[1]{\ifnum#1=19 %
\hatcurPPthetaxxxxxA
\else
\ifnum#1=20 %
\hatcurPPthetaxxxxxB
\else
\ifnum#1=21 %
\hatcurPPthetaxxxxxC
\else
??????\fi
\fi
\fi
}
\newcommand{\hatcurPPtinfall}[1]{\ifnum#1=19 %
\hatcurPPtinfallxxxxxA
\else
\ifnum#1=20 %
\hatcurPPtinfallxxxxxB
\else
\ifnum#1=21 %
\hatcurPPtinfallxxxxxC
\else
??????\fi
\fi
\fi
}
\newcommand{\hatcurRVeccen}[1]{\ifnum#1=19 %
\hatcurRVeccenxxxxxA
\else
\ifnum#1=20 %
\hatcurRVeccenxxxxxB
\else
\ifnum#1=21 %
\hatcurRVeccenxxxxxC
\else
??????\fi
\fi
\fi
}
\newcommand{\hatcurRVeccentwosiglim}[1]{\ifnum#1=19 %
\hatcurRVeccentwosiglimxxxxxA
\else
\ifnum#1=20 %
\hatcurRVeccentwosiglimxxxxxB
\else
\ifnum#1=21 %
\hatcurRVeccentwosiglimxxxxxC
\else
??????\fi
\fi
\fi
}
\newcommand{\hatcurRVfitrms}[1]{\ifnum#1=20 %
\hatcurRVfitrmsxxxxxB
\else
??????\fi
}
\newcommand{\hatcurRVfitrmsA}[1]{\ifnum#1=19 %
\hatcurRVfitrmsAxxxxxA
\else
\ifnum#1=21 %
\hatcurRVfitrmsAxxxxxC
\else
??????\fi
\fi
}
\newcommand{\hatcurRVfitrmsB}[1]{\ifnum#1=19 %
\hatcurRVfitrmsBxxxxxA
\else
\ifnum#1=21 %
\hatcurRVfitrmsBxxxxxC
\else
??????\fi
\fi
}
\newcommand{\hatcurRVfitrmsC}[1]{\ifnum#1=19 %
\hatcurRVfitrmsCxxxxxA
\else
\ifnum#1=21 %
\hatcurRVfitrmsCxxxxxC
\else
??????\fi
\fi
}
\newcommand{\hatcurRVgamma}[1]{\ifnum#1=20 %
\hatcurRVgammaxxxxxB
\else
??????\fi
}
\newcommand{\hatcurRVgammaA}[1]{\ifnum#1=19 %
\hatcurRVgammaAxxxxxA
\else
\ifnum#1=21 %
\hatcurRVgammaAxxxxxC
\else
??????\fi
\fi
}
\newcommand{\hatcurRVgammaB}[1]{\ifnum#1=19 %
\hatcurRVgammaBxxxxxA
\else
\ifnum#1=21 %
\hatcurRVgammaBxxxxxC
\else
??????\fi
\fi
}
\newcommand{\hatcurRVgammaC}[1]{\ifnum#1=19 %
\hatcurRVgammaCxxxxxA
\else
\ifnum#1=21 %
\hatcurRVgammaCxxxxxC
\else
??????\fi
\fi
}
\newcommand{\hatcurRVh}[1]{\ifnum#1=19 %
\hatcurRVhxxxxxA
\else
\ifnum#1=20 %
\hatcurRVhxxxxxB
\else
\ifnum#1=21 %
\hatcurRVhxxxxxC
\else
??????\fi
\fi
\fi
}
\newcommand{\hatcurRVjitter}[1]{\ifnum#1=20 %
\hatcurRVjitterxxxxxB
\else
??????\fi
}
\newcommand{\hatcurRVjitterA}[1]{\ifnum#1=19 %
\hatcurRVjitterAxxxxxA
\else
\ifnum#1=21 %
\hatcurRVjitterAxxxxxC
\else
??????\fi
\fi
}
\newcommand{\hatcurRVjitterB}[1]{\ifnum#1=19 %
\hatcurRVjitterBxxxxxA
\else
\ifnum#1=21 %
\hatcurRVjitterBxxxxxC
\else
??????\fi
\fi
}
\newcommand{\hatcurRVjitterC}[1]{\ifnum#1=19 %
\hatcurRVjitterCxxxxxA
\else
\ifnum#1=21 %
\hatcurRVjitterCxxxxxC
\else
??????\fi
\fi
}
\newcommand{\hatcurRVjittertwosiglim}[1]{\ifnum#1=20 %
\hatcurRVjittertwosiglimxxxxxB
\else
??????\fi
}
\newcommand{\hatcurRVjittertwosiglimA}[1]{\ifnum#1=19 %
\hatcurRVjittertwosiglimAxxxxxA
\else
\ifnum#1=21 %
\hatcurRVjittertwosiglimAxxxxxC
\else
??????\fi
\fi
}
\newcommand{\hatcurRVjittertwosiglimB}[1]{\ifnum#1=19 %
\hatcurRVjittertwosiglimBxxxxxA
\else
\ifnum#1=21 %
\hatcurRVjittertwosiglimBxxxxxC
\else
??????\fi
\fi
}
\newcommand{\hatcurRVjittertwosiglimC}[1]{\ifnum#1=19 %
\hatcurRVjittertwosiglimCxxxxxA
\else
\ifnum#1=21 %
\hatcurRVjittertwosiglimCxxxxxC
\else
??????\fi
\fi
}
\newcommand{\hatcurRVk}[1]{\ifnum#1=19 %
\hatcurRVkxxxxxA
\else
\ifnum#1=20 %
\hatcurRVkxxxxxB
\else
\ifnum#1=21 %
\hatcurRVkxxxxxC
\else
??????\fi
\fi
\fi
}
\newcommand{\hatcurRVK}[1]{\ifnum#1=19 %
\hatcurRVKxxxxxA
\else
\ifnum#1=20 %
\hatcurRVKxxxxxB
\else
\ifnum#1=21 %
\hatcurRVKxxxxxC
\else
??????\fi
\fi
\fi
}
\newcommand{\hatcurRVomega}[1]{\ifnum#1=19 %
\hatcurRVomegaxxxxxA
\else
\ifnum#1=20 %
\hatcurRVomegaxxxxxB
\else
\ifnum#1=21 %
\hatcurRVomegaxxxxxC
\else
??????\fi
\fi
\fi
}
\newcommand{\hatcurRVrh}[1]{\ifnum#1=19 %
\hatcurRVrhxxxxxA
\else
\ifnum#1=20 %
\hatcurRVrhxxxxxB
\else
\ifnum#1=21 %
\hatcurRVrhxxxxxC
\else
??????\fi
\fi
\fi
}
\newcommand{\hatcurRVrk}[1]{\ifnum#1=19 %
\hatcurRVrkxxxxxA
\else
\ifnum#1=20 %
\hatcurRVrkxxxxxB
\else
\ifnum#1=21 %
\hatcurRVrkxxxxxC
\else
??????\fi
\fi
\fi
}
\newcommand{\hatcurRVtrone}[1]{\ifnum#1=19 %
\hatcurRVtronexxxxxA
\else
\ifnum#1=20 %
\hatcurRVtronexxxxxB
\else
\ifnum#1=21 %
\hatcurRVtronexxxxxC
\else
??????\fi
\fi
\fi
}
\newcommand{\hatcurRVtrtwo}[1]{\ifnum#1=19 %
\hatcurRVtrtwoxxxxxA
\else
\ifnum#1=20 %
\hatcurRVtrtwoxxxxxB
\else
\ifnum#1=21 %
\hatcurRVtrtwoxxxxxC
\else
??????\fi
\fi
\fi
}
\newcommand{\hatcurSMEiilogg}[1]{\ifnum#1=19 %
\hatcurSMEiiloggxxxxxA
\else
\ifnum#1=20 %
\hatcurSMEiiloggxxxxxB
\else
\ifnum#1=21 %
\hatcurSMEiiloggxxxxxC
\else
??????\fi
\fi
\fi
}
\newcommand{\hatcurSMEiiteff}[1]{\ifnum#1=19 %
\hatcurSMEiiteffxxxxxA
\else
\ifnum#1=20 %
\hatcurSMEiiteffxxxxxB
\else
\ifnum#1=21 %
\hatcurSMEiiteffxxxxxC
\else
??????\fi
\fi
\fi
}
\newcommand{\hatcurSMEiivsin}[1]{\ifnum#1=19 %
\hatcurSMEiivsinxxxxxA
\else
\ifnum#1=20 %
\hatcurSMEiivsinxxxxxB
\else
\ifnum#1=21 %
\hatcurSMEiivsinxxxxxC
\else
??????\fi
\fi
\fi
}
\newcommand{\hatcurSMEiizfeh}[1]{\ifnum#1=19 %
\hatcurSMEiizfehxxxxxA
\else
\ifnum#1=20 %
\hatcurSMEiizfehxxxxxB
\else
\ifnum#1=21 %
\hatcurSMEiizfehxxxxxC
\else
??????\fi
\fi
\fi
}
\newcommand{\hatcurSMEiizfehshort}[1]{\ifnum#1=19 %
\hatcurSMEiizfehshortxxxxxA
\else
\ifnum#1=20 %
\hatcurSMEiizfehshortxxxxxB
\else
\ifnum#1=21 %
\hatcurSMEiizfehshortxxxxxC
\else
??????\fi
\fi
\fi
}
\newcommand{\hatcurSMEilogg}[1]{\ifnum#1=19 %
\hatcurSMEiloggxxxxxA
\else
\ifnum#1=20 %
\hatcurSMEiloggxxxxxB
\else
\ifnum#1=21 %
\hatcurSMEiloggxxxxxC
\else
??????\fi
\fi
\fi
}
\newcommand{\hatcurSMEiteff}[1]{\ifnum#1=19 %
\hatcurSMEiteffxxxxxA
\else
\ifnum#1=20 %
\hatcurSMEiteffxxxxxB
\else
\ifnum#1=21 %
\hatcurSMEiteffxxxxxC
\else
??????\fi
\fi
\fi
}
\newcommand{\hatcurSMEivmac}[1]{\ifnum#1=19 %
\hatcurSMEivmacxxxxxA
\else
\ifnum#1=20 %
\hatcurSMEivmacxxxxxB
\else
\ifnum#1=21 %
\hatcurSMEivmacxxxxxC
\else
??????\fi
\fi
\fi
}
\newcommand{\hatcurSMEivmic}[1]{\ifnum#1=19 %
\hatcurSMEivmicxxxxxA
\else
\ifnum#1=20 %
\hatcurSMEivmicxxxxxB
\else
\ifnum#1=21 %
\hatcurSMEivmicxxxxxC
\else
??????\fi
\fi
\fi
}
\newcommand{\hatcurSMEivsin}[1]{\ifnum#1=19 %
\hatcurSMEivsinxxxxxA
\else
\ifnum#1=20 %
\hatcurSMEivsinxxxxxB
\else
\ifnum#1=21 %
\hatcurSMEivsinxxxxxC
\else
??????\fi
\fi
\fi
}
\newcommand{\hatcurSMEizfeh}[1]{\ifnum#1=19 %
\hatcurSMEizfehxxxxxA
\else
\ifnum#1=20 %
\hatcurSMEizfehxxxxxB
\else
\ifnum#1=21 %
\hatcurSMEizfehxxxxxC
\else
??????\fi
\fi
\fi
}
\newcommand{\hatcurSMEizfehshort}[1]{\ifnum#1=19 %
\hatcurSMEizfehshortxxxxxA
\else
\ifnum#1=20 %
\hatcurSMEizfehshortxxxxxB
\else
\ifnum#1=21 %
\hatcurSMEizfehshortxxxxxC
\else
??????\fi
\fi
\fi
}
\newcommand{\hatcurXAv}[1]{\ifnum#1=19 %
\hatcurXAvxxxxxA
\else
\ifnum#1=20 %
\hatcurXAvxxxxxB
\else
\ifnum#1=21 %
\hatcurXAvxxxxxC
\else
??????\fi
\fi
\fi
}
\newcommand{\hatcurXdist}[1]{\ifnum#1=19 %
\hatcurXdistxxxxxA
\else
\ifnum#1=20 %
\hatcurXdistxxxxxB
\else
\ifnum#1=21 %
\hatcurXdistxxxxxC
\else
??????\fi
\fi
\fi
}
\newcommand{\hatcurXdistred}[1]{\ifnum#1=19 %
\hatcurXdistredxxxxxA
\else
\ifnum#1=20 %
\hatcurXdistredxxxxxB
\else
\ifnum#1=21 %
\hatcurXdistredxxxxxC
\else
??????\fi
\fi
\fi
}
\newcommand{\hatcurXEBV}[1]{\ifnum#1=19 %
\hatcurXEBVxxxxxA
\else
\ifnum#1=20 %
\hatcurXEBVxxxxxB
\else
\ifnum#1=21 %
\hatcurXEBVxxxxxC
\else
??????\fi
\fi
\fi
}
\newcommand{\hatcurXjhisored}[1]{\ifnum#1=19 %
\hatcurXjhisoredxxxxxA
\else
\ifnum#1=20 %
\hatcurXjhisoredxxxxxB
\else
\ifnum#1=21 %
\hatcurXjhisoredxxxxxC
\else
??????\fi
\fi
\fi
}
\newcommand{\hatcurXjkisored}[1]{\ifnum#1=19 %
\hatcurXjkisoredxxxxxA
\else
\ifnum#1=20 %
\hatcurXjkisoredxxxxxB
\else
\ifnum#1=21 %
\hatcurXjkisoredxxxxxC
\else
??????\fi
\fi
\fi
}
\newcommand{\hatcurXmhisored}[1]{\ifnum#1=19 %
\hatcurXmhisoredxxxxxA
\else
\ifnum#1=20 %
\hatcurXmhisoredxxxxxB
\else
\ifnum#1=21 %
\hatcurXmhisoredxxxxxC
\else
??????\fi
\fi
\fi
}
\newcommand{\hatcurXmiisored}[1]{\ifnum#1=19 %
\hatcurXmiisoredxxxxxA
\else
\ifnum#1=20 %
\hatcurXmiisoredxxxxxB
\else
\ifnum#1=21 %
\hatcurXmiisoredxxxxxC
\else
??????\fi
\fi
\fi
}
\newcommand{\hatcurXmjisored}[1]{\ifnum#1=19 %
\hatcurXmjisoredxxxxxA
\else
\ifnum#1=20 %
\hatcurXmjisoredxxxxxB
\else
\ifnum#1=21 %
\hatcurXmjisoredxxxxxC
\else
??????\fi
\fi
\fi
}
\newcommand{\hatcurXmkisored}[1]{\ifnum#1=19 %
\hatcurXmkisoredxxxxxA
\else
\ifnum#1=20 %
\hatcurXmkisoredxxxxxB
\else
\ifnum#1=21 %
\hatcurXmkisoredxxxxxC
\else
??????\fi
\fi
\fi
}
\newcommand{\hatcurXmvisored}[1]{\ifnum#1=19 %
\hatcurXmvisoredxxxxxA
\else
\ifnum#1=20 %
\hatcurXmvisoredxxxxxB
\else
\ifnum#1=21 %
\hatcurXmvisoredxxxxxC
\else
??????\fi
\fi
\fi
}
\newcommand{\hatcurXsecdur}[1]{\ifnum#1=19 %
\hatcurXsecdurxxxxxA
\else
\ifnum#1=20 %
\hatcurXsecdurxxxxxB
\else
\ifnum#1=21 %
\hatcurXsecdurxxxxxC
\else
??????\fi
\fi
\fi
}
\newcommand{\hatcurXsecingdur}[1]{\ifnum#1=19 %
\hatcurXsecingdurxxxxxA
\else
\ifnum#1=20 %
\hatcurXsecingdurxxxxxB
\else
\ifnum#1=21 %
\hatcurXsecingdurxxxxxC
\else
??????\fi
\fi
\fi
}
\newcommand{\hatcurXsecondary}[1]{\ifnum#1=19 %
\hatcurXsecondaryxxxxxA
\else
\ifnum#1=20 %
\hatcurXsecondaryxxxxxB
\else
\ifnum#1=21 %
\hatcurXsecondaryxxxxxC
\else
??????\fi
\fi
\fi
}
\newcommand{\hatcurXsecphase}[1]{\ifnum#1=19 %
\hatcurXsecphasexxxxxA
\else
\ifnum#1=20 %
\hatcurXsecphasexxxxxB
\else
\ifnum#1=21 %
\hatcurXsecphasexxxxxC
\else
??????\fi
\fi
\fi
}
\newcommand{\hatcurXviisored}[1]{\ifnum#1=19 %
\hatcurXviisoredxxxxxA
\else
\ifnum#1=20 %
\hatcurXviisoredxxxxxB
\else
\ifnum#1=21 %
\hatcurXviisoredxxxxxC
\else
??????\fi
\fi
\fi
}
\newcommand{\hatcurXvkisored}[1]{\ifnum#1=19 %
\hatcurXvkisoredxxxxxA
\else
\ifnum#1=20 %
\hatcurXvkisoredxxxxxB
\else
\ifnum#1=21 %
\hatcurXvkisoredxxxxxC
\else
??????\fi
\fi
\fi
}

\newcommand{\hatcurhtreccenxxxxxA}{HATS606-019}                      
\newcommand{\hatcurfieldeccenxxxxxA}{\ensuremath{string}}            
\newcommand{\hatcurCCraeccenxxxxxA}{\ensuremath{09^{\mathrm h}49^{\mathrm m}37.63{\mathrm s}}}                     
\newcommand{\hatcurCCdececcenxxxxxA}{\ensuremath{-33{\arcdeg}13{\arcmin}06.6{\arcsec}}}                    
\newcommand{\hatcurCCmageccenxxxxxA}{13.030}                         
\newcommand{\hatcurCCtwomasseccenxxxxxA}{2MASS~J09493761-3313065}     
\newcommand{\hatcurCCgsceccenxxxxxA}{GSC~7172-01459}                 
\newcommand{\hatcurCCtassmveccenxxxxxA}{\ensuremath{13.030\pm0.060}} 
\newcommand{\hatcurCCtassmvshorteccenxxxxxA}{\ensuremath{13.0}}      
\newcommand{\hatcurCCtassmBeccenxxxxxA}{\ensuremath{13.691\pm0.050}} 
\newcommand{\hatcurCCtassmBshorteccenxxxxxA}{\ensuremath{13.7}}      
\newcommand{\hatcurCCtassmIeccenxxxxxA}{\ensuremath{100\pm100}}      
\newcommand{\hatcurCCtassmIshorteccenxxxxxA}{\ensuremath{100.0}}     
\newcommand{\hatcurCCtassmgeccenxxxxxA}{\ensuremath{13.36\pm0.12}}   
\newcommand{\hatcurCCtassmgshorteccenxxxxxA}{\ensuremath{13.4}}      
\newcommand{\hatcurCCtassmreccenxxxxxA}{\ensuremath{12.832\pm0.020}} 
\newcommand{\hatcurCCtassmrshorteccenxxxxxA}{\ensuremath{12.8}}      
\newcommand{\hatcurCCtassmieccenxxxxxA}{\ensuremath{12.804\pm0.070}} 
\newcommand{\hatcurCCtassmishorteccenxxxxxA}{\ensuremath{12.8}}      
\newcommand{\hatcurCCtwomassJmageccenxxxxxA}{\ensuremath{11.811\pm0.023}} 
\newcommand{\hatcurCCtwomassHmageccenxxxxxA}{\ensuremath{11.553\pm0.022}} 
\newcommand{\hatcurCCtwomassKmageccenxxxxxA}{\ensuremath{11.510\pm0.019}} 
\newcommand{\hatcurCCcitJmageccenxxxxxA}{\ensuremath{11.832\pm0.023}} 
\newcommand{\hatcurCCcitHmageccenxxxxxA}{\ensuremath{11.548\pm0.022}} 
\newcommand{\hatcurCCcitKmageccenxxxxxA}{\ensuremath{11.534\pm0.019}} 
\newcommand{\hatcurCCbbJmageccenxxxxxA}{\ensuremath{11.875\pm0.025}} 
\newcommand{\hatcurCCbbHmageccenxxxxxA}{\ensuremath{11.569\pm0.023}} 
\newcommand{\hatcurCCbbKmageccenxxxxxA}{\ensuremath{11.554\pm0.019}} 
\newcommand{\hatcurCCesoJmageccenxxxxxA}{\ensuremath{11.876\pm0.026}} 
\newcommand{\hatcurCCesoHmageccenxxxxxA}{\ensuremath{11.563\pm0.025}} 
\newcommand{\hatcurCCesoKmageccenxxxxxA}{\ensuremath{11.553\pm0.020}} 
\newcommand{\hatcurCCesoJHmageccenxxxxxA}{\ensuremath{0.313\pm0.034}} 
\newcommand{\hatcurCCesoJKmageccenxxxxxA}{\ensuremath{0.324\pm0.032}} 
\newcommand{\hatcurCCesoHKmageccenxxxxxA}{\ensuremath{0.010\pm0.032}} 
\newcommand{\hatcurLCdipeccenxxxxxA}{\ensuremath{0.0}}               
\newcommand{\hatcurLCrprstareccenxxxxxA}{\ensuremath{0.0976\pm0.0021}} 
\newcommand{\hatcurLCbsqeccenxxxxxA}{\ensuremath{0.105_{-0.068}^{+0.080}}} 
\newcommand{\hatcurLCimpeccenxxxxxA}{\ensuremath{0.32_{-0.13}^{+0.11}}} 
\newcommand{\hatcurLCzetaeccenxxxxxA}{\ensuremath{13.32\pm0.11}}     
\newcommand{\hatcurLCdureccenxxxxxA}{\ensuremath{0.1666\pm0.0020}}   
\newcommand{\hatcurLCdurshorteccenxxxxxA}{\ensuremath{0.1666}}       
\newcommand{\hatcurLCdurhreccenxxxxxA}{\ensuremath{3.998\pm0.047}}   
\newcommand{\hatcurLCdurhrshorteccenxxxxxA}{\ensuremath{3.998}}      
\newcommand{\hatcurLCqeccenxxxxxA}{\ensuremath{0.03650\pm0.00043}}   
\newcommand{\hatcurLCqshorteccenxxxxxA}{\ensuremath{0.036}}          
\newcommand{\hatcurLCingdureccenxxxxxA}{\ensuremath{0.0164\pm0.0016}} 
\newcommand{\hatcurLCPeccenxxxxxA}{\ensuremath{4.569673\pm0.000010}} 
\newcommand{\hatcurLCPprececcenxxxxxA}{\ensuremath{4.5696734}}       
\newcommand{\hatcurLCPshorteccenxxxxxA}{\ensuremath{4.5697}}         
\newcommand{\hatcurLCTeccenxxxxxA}{\ensuremath{2456660.03375\pm0.00059}} 
\newcommand{\hatcurLCTAeccenxxxxxA}{\ensuremath{2455572.4516\pm0.0024}} 
\newcommand{\hatcurLCTBeccenxxxxxA}{\ensuremath{2456737.71821\pm0.00063}} 
\newcommand{\hatcurLChatnetmAeccenxxxxxA}{\ensuremath{12.769560\pm0.000095}} 
\newcommand{\hatcurLCiblendAeccenxxxxxA}{\ensuremath{0.858\pm0.047}} 
\newcommand{\hatcurLChatnetmBeccenxxxxxA}{\ensuremath{12.769610\pm0.000077}} 
\newcommand{\hatcurLCiblendBeccenxxxxxA}{\ensuremath{0.854\pm0.048}} 
\newcommand{\hatcurLCrhoeccenxxxxxA}{\ensuremath{0.34_{-0.11}^{+0.15}}} 
\newcommand{\hatcurSMEiteffeccenxxxxxA}{\ensuremath{5812\pm74}}      
\newcommand{\hatcurSMEizfeheccenxxxxxA}{\ensuremath{0.210\pm0.040}}  
\newcommand{\hatcurSMEizfehshorteccenxxxxxA}{\ensuremath{0.21}}      
\newcommand{\hatcurSMEiloggeccenxxxxxA}{\ensuremath{4.15\pm0.11}}    
\newcommand{\hatcurSMEivsineccenxxxxxA}{\ensuremath{4.85\pm0.50}}    
\newcommand{\hatcurSMEivmaceccenxxxxxA}{\ensuremath{0.0}}            
\newcommand{\hatcurSMEivmiceccenxxxxxA}{\ensuremath{0.0}}            
\newcommand{\hatcurSMEiiteffeccenxxxxxA}{\ensuremath{5896\pm77}}     
\newcommand{\hatcurSMEiizfeheccenxxxxxA}{\ensuremath{0.240\pm0.050}} 
\newcommand{\hatcurSMEiizfehshorteccenxxxxxA}{\ensuremath{0.24}}     
\newcommand{\hatcurSMEiiloggeccenxxxxxA}{\ensuremath{4\pm0}}         
\newcommand{\hatcurSMEiivsineccenxxxxxA}{\ensuremath{4.79\pm0.50}}   
\newcommand{\hatcurLBizeccenxxxxxA}{\ensuremath{0.1948}}             
\newcommand{\hatcurLBiizeccenxxxxxA}{\ensuremath{0.3415}}            
\newcommand{\hatcurLBiieccenxxxxxA}{\ensuremath{0.2566}}             
\newcommand{\hatcurLBiiieccenxxxxxA}{\ensuremath{0.3437}}            
\newcommand{\hatcurLBiIeccenxxxxxA}{\ensuremath{0.2353}}             
\newcommand{\hatcurLBiiIeccenxxxxxA}{\ensuremath{0.3442}}            
\newcommand{\hatcurLBigeccenxxxxxA}{\ensuremath{0.5443}}             
\newcommand{\hatcurLBiigeccenxxxxxA}{\ensuremath{0.2452}}            
\newcommand{\hatcurLBireccenxxxxxA}{\ensuremath{0.3468}}             
\newcommand{\hatcurLBiireccenxxxxxA}{\ensuremath{0.3369}}            
\newcommand{\hatcurLBiReccenxxxxxA}{\ensuremath{0.3216}}             
\newcommand{\hatcurLBiiReccenxxxxxA}{\ensuremath{0.3400}}            
\newcommand{\hatcurLBikepeccenxxxxxA}{\ensuremath{0.1000}}           
\newcommand{\hatcurLBiikepeccenxxxxxA}{\ensuremath{0.1000}}          
\newcommand{\hatcurISOmeccenxxxxxA}{\ensuremath{1.303\pm0.083}}      
\newcommand{\hatcurISOmshorteccenxxxxxA}{\ensuremath{1.30}}          
\newcommand{\hatcurISOmlongeccenxxxxxA}{\ensuremath{1.303\pm0.083}}  
\newcommand{\hatcurISOreccenxxxxxA}{\ensuremath{1.75\pm0.25}}        
\newcommand{\hatcurISOrshorteccenxxxxxA}{\ensuremath{1.75}}          
\newcommand{\hatcurISOrlongeccenxxxxxA}{\ensuremath{1.75\pm0.25}}    
\newcommand{\hatcurISOrhoeccenxxxxxA}{\ensuremath{0.34\pm0.14}}      
\newcommand{\hatcurISOrholongeccenxxxxxA}{\ensuremath{0.34\pm0.14}}  
\newcommand{\hatcurISOloggeccenxxxxxA}{\ensuremath{4.07\pm0.10}}     
\newcommand{\hatcurISOlumeccenxxxxxA}{\ensuremath{3.31_{-0.81}^{+1.15}}} 
\newcommand{\hatcurISOlumshorteccenxxxxxA}{\ensuremath{3.31}}        
\newcommand{\hatcurISOmveccenxxxxxA}{\ensuremath{3.50\pm0.32}}       
\newcommand{\hatcurISOvieccenxxxxxA}{\ensuremath{0.664\pm0.023}}     
\newcommand{\hatcurISOageeccenxxxxxA}{\ensuremath{3.94_{-0.50}^{+0.96}}} 
\newcommand{\hatcurISOsigmaeccenxxxxxA}{\ensuremath{0.000200\pm0.000051}} 
\newcommand{\hatcurISOMJeccenxxxxxA}{\ensuremath{2.42\pm0.31}}       
\newcommand{\hatcurISOMHeccenxxxxxA}{\ensuremath{2.10\pm0.31}}       
\newcommand{\hatcurISOMKeccenxxxxxA}{\ensuremath{2.05\pm0.31}}       
\newcommand{\hatcurISOJKeccenxxxxxA}{\ensuremath{0.370\pm0.020}}     
\newcommand{\hatcurISOspececcenxxxxxA}{G}                            
\newcommand{\hatcurRVKeccenxxxxxA}{\ensuremath{46.1\pm8.0}}          
\newcommand{\hatcurRVrkeccenxxxxxA}{\ensuremath{0.39\pm0.13}}        
\newcommand{\hatcurRVrheccenxxxxxA}{\ensuremath{0.37\pm0.16}}        
\newcommand{\hatcurRVkeccenxxxxxA}{\ensuremath{0.209\pm0.084}}       
\newcommand{\hatcurRVheccenxxxxxA}{\ensuremath{0.20\pm0.11}}         
\newcommand{\hatcurRVtroneeccenxxxxxA}{\ensuremath{0\pm0}}           
\newcommand{\hatcurRVtrtwoeccenxxxxxA}{\ensuremath{0\pm0}}           
\newcommand{\hatcurRVgammaAeccenxxxxxA}{\ensuremath{27551.6\pm7.2}}  
\newcommand{\hatcurRVjitterAeccenxxxxxA}{\ensuremath{15.6\pm6.2}}    
\newcommand{\hatcurRVjittertwosiglimAeccenxxxxxA}{\ensuremath{<25.9}} 
\newcommand{\hatcurRVfitrmsAeccenxxxxxA}{\ensuremath{0.0}}           
\newcommand{\hatcurRVgammaBeccenxxxxxA}{\ensuremath{27443\pm10}}     
\newcommand{\hatcurRVjitterBeccenxxxxxA}{\ensuremath{0.0\pm5.5}}     
\newcommand{\hatcurRVjittertwosiglimBeccenxxxxxA}{\ensuremath{<13.2}} 
\newcommand{\hatcurRVfitrmsBeccenxxxxxA}{\ensuremath{0.0}}           
\newcommand{\hatcurRVgammaCeccenxxxxxA}{\ensuremath{1.8\pm4.9}}      
\newcommand{\hatcurRVjitterCeccenxxxxxA}{\ensuremath{11.4\pm3.6}}    
\newcommand{\hatcurRVjittertwosiglimCeccenxxxxxA}{\ensuremath{<18.3}} 
\newcommand{\hatcurRVfitrmsCeccenxxxxxA}{\ensuremath{0.0}}           
\newcommand{\hatcurRVecceneccenxxxxxA}{\ensuremath{0.30\pm0.10}}     
\newcommand{\hatcurRVeccentwosiglimeccenxxxxxA}{\ensuremath{<0.492}} 
\newcommand{\hatcurRVomegaeccenxxxxxA}{\ensuremath{44\pm50}}         
\newcommand{\hatcurPPieccenxxxxxA}{\ensuremath{86.6\pm1.7}}          
\newcommand{\hatcurPPgeccenxxxxxA}{\ensuremath{3.82\pm0.89}}         
\newcommand{\hatcurPPloggeccenxxxxxA}{\ensuremath{2.58\pm0.10}}      
\newcommand{\hatcurPPareccenxxxxxA}{\ensuremath{7.24\pm0.90}}        
\newcommand{\hatcurPPareleccenxxxxxA}{\ensuremath{0.0589\pm0.0013}}  
\newcommand{\hatcurPPrhoeccenxxxxxA}{\ensuremath{0.116\pm0.042}}     
\newcommand{\hatcurPPmeccenxxxxxA}{\ensuremath{0.427\pm0.071}}       
\newcommand{\hatcurPPmshorteccenxxxxxA}{\ensuremath{0.43}}           
\newcommand{\hatcurPPmlongeccenxxxxxA}{\ensuremath{0.427\pm0.071}}   
\newcommand{\hatcurPPmeeccenxxxxxA}{\ensuremath{136\pm23}}           
\newcommand{\hatcurPPmeshorteccenxxxxxA}{\ensuremath{135.7}}         
\newcommand{\hatcurPPmelongeccenxxxxxA}{\ensuremath{136\pm23}}       
\newcommand{\hatcurPPreccenxxxxxA}{\ensuremath{1.66_{-0.21}^{+0.27}}} 
\newcommand{\hatcurPPrshorteccenxxxxxA}{\ensuremath{1.66}}           
\newcommand{\hatcurPPrlongeccenxxxxxA}{\ensuremath{1.66_{-0.21}^{+0.27}}} 
\newcommand{\hatcurPPreeccenxxxxxA}{\ensuremath{18.6_{-2.3}^{+3.1}}} 
\newcommand{\hatcurPPreshorteccenxxxxxA}{\ensuremath{18.6}}          
\newcommand{\hatcurPPrelongeccenxxxxxA}{\ensuremath{18.6_{-2.3}^{+3.1}}} 
\newcommand{\hatcurPPmrcorreccenxxxxxA}{\ensuremath{0.55}}           
\newcommand{\hatcurPPteffeccenxxxxxA}{\ensuremath{1570\pm110}}       
\newcommand{\hatcurPPthetaeccenxxxxxA}{\ensuremath{0.0231\pm0.0036}} 
\newcommand{\hatcurPPfluxperieccenxxxxxA}{\ensuremath{2.65_{-0.98}^{+2.07}}} 
\newcommand{\hatcurPPfluxperidimeccenxxxxxA}{\ensuremath{9}}         
\newcommand{\hatcurPPfluxapeccenxxxxxA}{\ensuremath{7.8\pm1.2}}      
\newcommand{\hatcurPPfluxapdimeccenxxxxxA}{\ensuremath{8}}           
\newcommand{\hatcurPPfluxavgeccenxxxxxA}{\ensuremath{1.37_{-0.32}^{+0.49}}} 
\newcommand{\hatcurPPfluxavgdimeccenxxxxxA}{\ensuremath{9}}          
\newcommand{\hatcurPPfluxavglogeccenxxxxxA}{\ensuremath{9.14\pm0.12}} 
\newcommand{\hatcurXsecphaseeccenxxxxxA}{\ensuremath{0.635\pm0.055}} 
\newcommand{\hatcurXsecondaryeccenxxxxxA}{\ensuremath{2456662.93\pm0.25}} 
\newcommand{\hatcurXsecdureccenxxxxxA}{\ensuremath{0.228\pm0.042}}   
\newcommand{\hatcurXsecingdureccenxxxxxA}{\ensuremath{0.027\pm0.022}} 
\newcommand{\hatcurPPphiconjeccenxxxxxA}{\ensuremath{0.065_{-0.030}^{+0.043}}} 
\newcommand{\hatcurPPperieccenxxxxxA}{\ensuremath{2456659.73\pm0.31}} 
\newcommand{\hatcurPPaequiveccenxxxxxA}{\ensuremath{0.0324\pm0.0041}} 
\newcommand{\hatcurPPtcirceccenxxxxxA}{\ensuremath{48_{-33}^{+64}}}  
\newcommand{\hatcurPPtinfalleccenxxxxxA}{\ensuremath{2600_{-1300}^{+2300}}} 
\newcommand{\hatcurXdisteccenxxxxxA}{\ensuremath{800\pm120}}         
\newcommand{\hatcurXAveccenxxxxxA}{\ensuremath{0.012_{-0.012}^{+0.091}}} 
\newcommand{\hatcurXdistredeccenxxxxxA}{\ensuremath{780\pm110}}      
\newcommand{\hatcurXEBVeccenxxxxxA}{\ensuremath{0.0040_{-0.0040}^{+0.0290}}} 
\newcommand{\hatcurXmvisoredeccenxxxxxA}{\ensuremath{13.019\pm0.044}} 
\newcommand{\hatcurXmiisoredeccenxxxxxA}{\ensuremath{12.334\pm0.021}} 
\newcommand{\hatcurXmjisoredeccenxxxxxA}{\ensuremath{11.900\pm0.015}} 
\newcommand{\hatcurXmhisoredeccenxxxxxA}{\ensuremath{11.584\pm0.014}} 
\newcommand{\hatcurXmkisoredeccenxxxxxA}{\ensuremath{11.527\pm0.014}} 
\newcommand{\hatcurXviisoredeccenxxxxxA}{\ensuremath{0.684\pm0.026}} 
\newcommand{\hatcurXvkisoredeccenxxxxxA}{\ensuremath{1.491\pm0.047}} 
\newcommand{\hatcurXjhisoredeccenxxxxxA}{\ensuremath{0.316\pm0.010}} 
\newcommand{\hatcurXjkisoredeccenxxxxxA}{\ensuremath{0.373\pm0.012}} 
\newcommand{\hatcurCCpmraeccenxxxxxA}{\ensuremath{-4.6\pm1.3}}       
\newcommand{\hatcurCCpmdececcenxxxxxA}{\ensuremath{2.0\pm1.3}}       
\newcommand{\hatcurCCpmeccenxxxxxA}{\ensuremath{5.0\pm1.8}}          

\newcommand{\hatcurhtreccenxxxxxB}{HATS700-026}                      
\newcommand{\hatcurfieldeccenxxxxxB}{\ensuremath{string}}            
\newcommand{\hatcurCCraeccenxxxxxB}{\ensuremath{13^{\mathrm h}12^{\mathrm m}32.04{\mathrm s}}}                     
\newcommand{\hatcurCCdececcenxxxxxB}{\ensuremath{-45{\arcdeg}35{\arcmin}26.0{\arcsec}}}                    
\newcommand{\hatcurCCmageccenxxxxxB}{13.765}                         
\newcommand{\hatcurCCtwomasseccenxxxxxB}{2MASS~13123190-4535259}     
\newcommand{\hatcurCCgsceccenxxxxxB}{GSC~8247-02184}                 
\newcommand{\hatcurCCtassmveccenxxxxxB}{\ensuremath{13.765\pm0.050}} 
\newcommand{\hatcurCCtassmvshorteccenxxxxxB}{\ensuremath{13.8}}      
\newcommand{\hatcurCCtassmBeccenxxxxxB}{\ensuremath{14.581\pm0.060}} 
\newcommand{\hatcurCCtassmBshorteccenxxxxxB}{\ensuremath{14.6}}      
\newcommand{\hatcurCCtassmIeccenxxxxxB}{\ensuremath{nff\pmnff}}      
\newcommand{\hatcurCCtassmIshorteccenxxxxxB}{\ensuremath{0.0}}       
\newcommand{\hatcurCCtassmgeccenxxxxxB}{\ensuremath{14.126\pm0.010}} 
\newcommand{\hatcurCCtassmgshorteccenxxxxxB}{\ensuremath{14.1}}      
\newcommand{\hatcurCCtassmreccenxxxxxB}{\ensuremath{13.478\pm0.010}} 
\newcommand{\hatcurCCtassmrshorteccenxxxxxB}{\ensuremath{13.5}}      
\newcommand{\hatcurCCtassmieccenxxxxxB}{\ensuremath{13.317\pm0.020}} 
\newcommand{\hatcurCCtassmishorteccenxxxxxB}{\ensuremath{13.3}}      
\newcommand{\hatcurCCtwomassJmageccenxxxxxB}{\ensuremath{12.351\pm0.024}} 
\newcommand{\hatcurCCtwomassHmageccenxxxxxB}{\ensuremath{11.951\pm0.022}} 
\newcommand{\hatcurCCtwomassKmageccenxxxxxB}{\ensuremath{11.864\pm0.023}} 
\newcommand{\hatcurCCcitJmageccenxxxxxB}{\ensuremath{12.362\pm0.024}} 
\newcommand{\hatcurCCcitHmageccenxxxxxB}{\ensuremath{11.945\pm0.023}} 
\newcommand{\hatcurCCcitKmageccenxxxxxB}{\ensuremath{11.888\pm0.023}} 
\newcommand{\hatcurCCbbJmageccenxxxxxB}{\ensuremath{12.421\pm0.026}} 
\newcommand{\hatcurCCbbHmageccenxxxxxB}{\ensuremath{11.967\pm0.023}} 
\newcommand{\hatcurCCbbKmageccenxxxxxB}{\ensuremath{11.908\pm0.023}} 
\newcommand{\hatcurCCesoJmageccenxxxxxB}{\ensuremath{12.424\pm0.028}} 
\newcommand{\hatcurCCesoHmageccenxxxxxB}{\ensuremath{11.963\pm0.027}} 
\newcommand{\hatcurCCesoKmageccenxxxxxB}{\ensuremath{11.906\pm0.024}} 
\newcommand{\hatcurCCesoJHmageccenxxxxxB}{\ensuremath{0.461\pm0.036}} 
\newcommand{\hatcurCCesoJKmageccenxxxxxB}{\ensuremath{0.518\pm0.036}} 
\newcommand{\hatcurCCesoHKmageccenxxxxxB}{\ensuremath{0.057\pm0.036}} 
\newcommand{\hatcurLCdipeccenxxxxxB}{\ensuremath{8.3}}               
\newcommand{\hatcurLCrprstareccenxxxxxB}{\ensuremath{0.0889\pm0.0029}} 
\newcommand{\hatcurLCbsqeccenxxxxxB}{\ensuremath{0.28_{-0.13}^{+0.11}}} 
\newcommand{\hatcurLCimpeccenxxxxxB}{\ensuremath{0.525_{-0.147}^{+0.098}}} 
\newcommand{\hatcurLCzetaeccenxxxxxB}{\ensuremath{22.10\pm0.26}}     
\newcommand{\hatcurLCdureccenxxxxxB}{\ensuremath{0.1015\pm0.0023}}   
\newcommand{\hatcurLCdurshorteccenxxxxxB}{\ensuremath{0.1015}}       
\newcommand{\hatcurLCdurhreccenxxxxxB}{\ensuremath{2.437\pm0.055}}   
\newcommand{\hatcurLCdurhrshorteccenxxxxxB}{\ensuremath{2.437}}      
\newcommand{\hatcurLCqeccenxxxxxB}{\ensuremath{0.02670\pm0.00061}}   
\newcommand{\hatcurLCqshorteccenxxxxxB}{\ensuremath{0.027}}          
\newcommand{\hatcurLCingdureccenxxxxxB}{\ensuremath{0.0111\pm0.0023}} 
\newcommand{\hatcurLCPeccenxxxxxB}{\ensuremath{3.7992961\pm0.0000078}} 
\newcommand{\hatcurLCPprececcenxxxxxB}{\ensuremath{3.7992961}}       
\newcommand{\hatcurLCPshorteccenxxxxxB}{\ensuremath{3.7993}}         
\newcommand{\hatcurLCTeccenxxxxxB}{\ensuremath{2457098.36911\pm0.00059}} 
\newcommand{\hatcurLCTAeccenxxxxxB}{\ensuremath{2455681.2317\pm0.0029}} 
\newcommand{\hatcurLCTBeccenxxxxxB}{\ensuremath{2457170.55572\pm0.00061}} 
\newcommand{\hatcurLChatnetmAeccenxxxxxB}{\ensuremath{13.63339\pm0.00021}} 
\newcommand{\hatcurLCiblendAeccenxxxxxB}{\ensuremath{0.65\pm0.20}}   
\newcommand{\hatcurLChatnetmBeccenxxxxxB}{\ensuremath{13.62984\pm0.00010}} 
\newcommand{\hatcurLCiblendBeccenxxxxxB}{\ensuremath{0.894\pm0.084}} 
\newcommand{\hatcurLCrhoeccenxxxxxB}{\ensuremath{2.2_{-1.3}^{+4.7}}} 
\newcommand{\hatcurSMEiteffeccenxxxxxB}{\ensuremath{5485\pm96}}      
\newcommand{\hatcurSMEizfeheccenxxxxxB}{\ensuremath{0.120\pm0.070}}  
\newcommand{\hatcurSMEizfehshorteccenxxxxxB}{\ensuremath{0.12}}      
\newcommand{\hatcurSMEiloggeccenxxxxxB}{\ensuremath{4.56\pm0.17}}    
\newcommand{\hatcurSMEivsineccenxxxxxB}{\ensuremath{1.4\pm1.2}}      
\newcommand{\hatcurSMEivmaceccenxxxxxB}{\ensuremath{0.0}}            
\newcommand{\hatcurSMEivmiceccenxxxxxB}{\ensuremath{0.0}}            
\newcommand{\hatcurSMEiiteffeccenxxxxxB}{\ensuremath{5406\pm49}}     
\newcommand{\hatcurSMEiizfeheccenxxxxxB}{\ensuremath{0.030\pm0.050}} 
\newcommand{\hatcurSMEiizfehshorteccenxxxxxB}{\ensuremath{0.030}}    
\newcommand{\hatcurSMEiiloggeccenxxxxxB}{\ensuremath{4.493\pm0.046}} 
\newcommand{\hatcurSMEiivsineccenxxxxxB}{\ensuremath{1.51\pm0.89}}   
\newcommand{\hatcurLBizeccenxxxxxB}{\ensuremath{0.2609}}             
\newcommand{\hatcurLBiizeccenxxxxxB}{\ensuremath{0.3012}}            
\newcommand{\hatcurLBiieccenxxxxxB}{\ensuremath{0.3320}}             
\newcommand{\hatcurLBiiieccenxxxxxB}{\ensuremath{0.2933}}            
\newcommand{\hatcurLBiIeccenxxxxxB}{\ensuremath{0.3085}}             
\newcommand{\hatcurLBiiIeccenxxxxxB}{\ensuremath{0.2962}}            
\newcommand{\hatcurLBigeccenxxxxxB}{\ensuremath{0.6542}}             
\newcommand{\hatcurLBiigeccenxxxxxB}{\ensuremath{0.1580}}            
\newcommand{\hatcurLBireccenxxxxxB}{\ensuremath{0.4366}}             
\newcommand{\hatcurLBiireccenxxxxxB}{\ensuremath{0.2753}}            
\newcommand{\hatcurLBiReccenxxxxxB}{\ensuremath{0.4077}}             
\newcommand{\hatcurLBiiReccenxxxxxB}{\ensuremath{0.2811}}            
\newcommand{\hatcurLBikepeccenxxxxxB}{\ensuremath{0.1000}}           
\newcommand{\hatcurLBiikepeccenxxxxxB}{\ensuremath{0.1000}}          
\newcommand{\hatcurISOmeccenxxxxxB}{\ensuremath{0.912\pm0.031}}      
\newcommand{\hatcurISOmshorteccenxxxxxB}{\ensuremath{0.91}}          
\newcommand{\hatcurISOmlongeccenxxxxxB}{\ensuremath{0.912\pm0.031}}  
\newcommand{\hatcurISOreccenxxxxxB}{\ensuremath{0.903_{-0.049}^{+0.096}}} 
\newcommand{\hatcurISOrshorteccenxxxxxB}{\ensuremath{0.90}}          
\newcommand{\hatcurISOrlongeccenxxxxxB}{\ensuremath{0.903_{-0.049}^{+0.096}}} 
\newcommand{\hatcurISOrhoeccenxxxxxB}{\ensuremath{1.73\pm0.46}}      
\newcommand{\hatcurISOrholongeccenxxxxxB}{\ensuremath{1.73\pm0.46}}  
\newcommand{\hatcurISOloggeccenxxxxxB}{\ensuremath{4.48\pm0.11}}     
\newcommand{\hatcurISOlumeccenxxxxxB}{\ensuremath{0.627_{-0.077}^{+0.154}}} 
\newcommand{\hatcurISOlumshorteccenxxxxxB}{\ensuremath{0.63}}        
\newcommand{\hatcurISOmveccenxxxxxB}{\ensuremath{5.41\pm0.32}}       
\newcommand{\hatcurISOvieccenxxxxxB}{\ensuremath{0.805\pm0.013}}     
\newcommand{\hatcurISOageeccenxxxxxB}{\ensuremath{7.2\pm3.7}}        
\newcommand{\hatcurISOsigmaeccenxxxxxB}{\ensuremath{0.0006\pm0.0017}} 
\newcommand{\hatcurISOMJeccenxxxxxB}{\ensuremath{4.08\pm0.31}}       
\newcommand{\hatcurISOMHeccenxxxxxB}{\ensuremath{3.66\pm0.30}}       
\newcommand{\hatcurISOMKeccenxxxxxB}{\ensuremath{3.59\pm0.30}}       
\newcommand{\hatcurISOJKeccenxxxxxB}{\ensuremath{0.44\pm0.14}}       
\newcommand{\hatcurISOspececcenxxxxxB}{G}                            
\newcommand{\hatcurRVKeccenxxxxxB}{\ensuremath{40\pm20}}             
\newcommand{\hatcurRVrkeccenxxxxxB}{\ensuremath{-0.22\pm0.29}}       
\newcommand{\hatcurRVrheccenxxxxxB}{\ensuremath{0.06\pm0.24}}        
\newcommand{\hatcurRVkeccenxxxxxB}{\ensuremath{-0.063_{-0.256}^{+0.095}}} 
\newcommand{\hatcurRVheccenxxxxxB}{\ensuremath{0.009_{-0.068}^{+0.125}}} 
\newcommand{\hatcurRVtroneeccenxxxxxB}{\ensuremath{0\pm0}}           
\newcommand{\hatcurRVtrtwoeccenxxxxxB}{\ensuremath{0\pm0}}           
\newcommand{\hatcurRVgammaeccenxxxxxB}{\ensuremath{22116.8\pm6.6}}   
\newcommand{\hatcurRVjittereccenxxxxxB}{\ensuremath{0.0\pm3.4}}      
\newcommand{\hatcurRVjittertwosiglimeccenxxxxxB}{\ensuremath{<8.8}}  
\newcommand{\hatcurRVfitrmseccenxxxxxB}{\ensuremath{.1fym}}          %
\newcommand{\hatcurRVecceneccenxxxxxB}{\ensuremath{0.15\pm0.16}}     
\newcommand{\hatcurRVeccentwosiglimeccenxxxxxB}{\ensuremath{<0.501}} 
\newcommand{\hatcurRVomegaeccenxxxxxB}{\ensuremath{167\pm75}}        
\newcommand{\hatcurPPieccenxxxxxB}{\ensuremath{87.2\pm1.6}}          
\newcommand{\hatcurPPgeccenxxxxxB}{\ensuremath{11.8\pm3.7}}          
\newcommand{\hatcurPPloggeccenxxxxxB}{\ensuremath{3.07\pm0.12}}      
\newcommand{\hatcurPPareccenxxxxxB}{\ensuremath{10.96_{-1.08}^{+0.72}}} 
\newcommand{\hatcurPPareleccenxxxxxB}{\ensuremath{0.04621\pm0.00052}} 
\newcommand{\hatcurPPrhoeccenxxxxxB}{\ensuremath{0.77\pm0.26}}       
\newcommand{\hatcurPPmeccenxxxxxB}{\ensuremath{0.289_{-0.041}^{+0.128}}} 
\newcommand{\hatcurPPmshorteccenxxxxxB}{\ensuremath{0.29}}           
\newcommand{\hatcurPPmlongeccenxxxxxB}{\ensuremath{0.289_{-0.041}^{+0.128}}} 
\newcommand{\hatcurPPmeeccenxxxxxB}{\ensuremath{92_{-13}^{+41}}}     
\newcommand{\hatcurPPmeshorteccenxxxxxB}{\ensuremath{91.7}}          
\newcommand{\hatcurPPmelongeccenxxxxxB}{\ensuremath{92_{-13}^{+41}}} 
\newcommand{\hatcurPPreccenxxxxxB}{\ensuremath{0.784_{-0.055}^{+0.088}}} 
\newcommand{\hatcurPPrshorteccenxxxxxB}{\ensuremath{0.78}}           
\newcommand{\hatcurPPrlongeccenxxxxxB}{\ensuremath{0.784_{-0.055}^{+0.088}}} 
\newcommand{\hatcurPPreeccenxxxxxB}{\ensuremath{8.79_{-0.62}^{+0.99}}} 
\newcommand{\hatcurPPreshorteccenxxxxxB}{\ensuremath{8.8}}           
\newcommand{\hatcurPPrelongeccenxxxxxB}{\ensuremath{8.79_{-0.62}^{+0.99}}} 
\newcommand{\hatcurPPmrcorreccenxxxxxB}{\ensuremath{0.56}}           
\newcommand{\hatcurPPteffeccenxxxxxB}{\ensuremath{1160_{-40}^{+76}}} 
\newcommand{\hatcurPPthetaeccenxxxxxB}{\ensuremath{0.0372_{-0.0056}^{+0.0091}}} 
\newcommand{\hatcurPPfluxperieccenxxxxxB}{\ensuremath{5.6_{-1.5}^{+5.8}}} 
\newcommand{\hatcurPPfluxperidimeccenxxxxxB}{\ensuremath{8}}         
\newcommand{\hatcurPPfluxapeccenxxxxxB}{\ensuremath{3.08\pm0.72}}    
\newcommand{\hatcurPPfluxapdimeccenxxxxxB}{\ensuremath{8}}           
\newcommand{\hatcurPPfluxavgeccenxxxxxB}{\ensuremath{4.08_{-0.53}^{+1.19}}} 
\newcommand{\hatcurPPfluxavgdimeccenxxxxxB}{\ensuremath{8}}          
\newcommand{\hatcurPPfluxavglogeccenxxxxxB}{\ensuremath{8.611_{-0.060}^{+0.111}}} 
\newcommand{\hatcurXsecphaseeccenxxxxxB}{\ensuremath{0.46\pm0.11}}   
\newcommand{\hatcurXsecondaryeccenxxxxxB}{\ensuremath{2457100.11\pm0.42}} 
\newcommand{\hatcurXsecdureccenxxxxxB}{\ensuremath{0.103\pm0.026}}   
\newcommand{\hatcurXsecingdureccenxxxxxB}{\ensuremath{0.011\pm0.014}} 
\newcommand{\hatcurPPphiconjeccenxxxxxB}{\ensuremath{-0.10\pm0.19}}  
\newcommand{\hatcurPPperieccenxxxxxB}{\ensuremath{2457098.76\pm0.73}} 
\newcommand{\hatcurPPaequiveccenxxxxxB}{\ensuremath{0.0584_{-0.0062}^{+0.0040}}} 
\newcommand{\hatcurPPtcirceccenxxxxxB}{\ensuremath{720\pm390}}       
\newcommand{\hatcurPPtinfalleccenxxxxxB}{\ensuremath{18000\pm8700}}  
\newcommand{\hatcurXdisteccenxxxxxB}{\ensuremath{460_{-26}^{+50}}}   
\newcommand{\hatcurXAveccenxxxxxB}{\ensuremath{0.049_{-0.049}^{+0.080}}} 
\newcommand{\hatcurXdistredeccenxxxxxB}{\ensuremath{459_{-26}^{+50}}} 
\newcommand{\hatcurXEBVeccenxxxxxB}{\ensuremath{0.016_{-0.016}^{+0.026}}} 
\newcommand{\hatcurXmvisoredeccenxxxxxB}{\ensuremath{13.781\pm0.041}} 
\newcommand{\hatcurXmiisoredeccenxxxxxB}{\ensuremath{12.945\pm0.023}} 
\newcommand{\hatcurXmjisoredeccenxxxxxB}{\ensuremath{12.402\pm0.015}} 
\newcommand{\hatcurXmhisoredeccenxxxxxB}{\ensuremath{11.979\pm0.015}} 
\newcommand{\hatcurXmkisoredeccenxxxxxB}{\ensuremath{11.906\pm0.016}} 
\newcommand{\hatcurXviisoredeccenxxxxxB}{\ensuremath{0.832_{-0.020}^{+0.034}}} 
\newcommand{\hatcurXvkisoredeccenxxxxxB}{\ensuremath{1.875\pm0.046}} 
\newcommand{\hatcurXjhisoredeccenxxxxxB}{\ensuremath{0.4240\pm0.0097}} 
\newcommand{\hatcurXjkisoredeccenxxxxxB}{\ensuremath{0.497\pm0.011}} 
\newcommand{\hatcurCCpmraeccenxxxxxB}{\ensuremath{-15.2\pm1.8}}      
\newcommand{\hatcurCCpmdececcenxxxxxB}{\ensuremath{-3.8\pm1.8}}      
\newcommand{\hatcurCCpmeccenxxxxxB}{\ensuremath{15.7\pm2.5}}         

\newcommand{\hatcurhtreccenxxxxxC}{HATS777-021}                      
\newcommand{\hatcurfieldeccenxxxxxC}{\ensuremath{string}}            
\newcommand{\hatcurCCraeccenxxxxxC}{\ensuremath{18^{\mathrm h}40^{\mathrm m}44.40{\mathrm s}}}                     
\newcommand{\hatcurCCdececcenxxxxxC}{\ensuremath{-58{\arcdeg}27{\arcmin}33.3{\arcsec}}}                    
\newcommand{\hatcurCCmageccenxxxxxC}{12.191}                         
\newcommand{\hatcurCCtwomasseccenxxxxxC}{2MASS~18404426-5827332}     
\newcommand{\hatcurCCgsceccenxxxxxC}{GSC~8770-00400}                 
\newcommand{\hatcurCCtassmveccenxxxxxC}{\ensuremath{12.191\pm0.020}} 
\newcommand{\hatcurCCtassmvshorteccenxxxxxC}{\ensuremath{12.2}}      
\newcommand{\hatcurCCtassmBeccenxxxxxC}{\ensuremath{13.018\pm0.010}} 
\newcommand{\hatcurCCtassmBshorteccenxxxxxC}{\ensuremath{13.0}}      
\newcommand{\hatcurCCtassmIeccenxxxxxC}{\ensuremath{nff\pmnff}}      
\newcommand{\hatcurCCtassmIshorteccenxxxxxC}{\ensuremath{0.0}}       
\newcommand{\hatcurCCtassmgeccenxxxxxC}{\ensuremath{12.595\pm0.010}} 
\newcommand{\hatcurCCtassmgshorteccenxxxxxC}{\ensuremath{12.6}}      
\newcommand{\hatcurCCtassmreccenxxxxxC}{\ensuremath{11.954\pm0.010}} 
\newcommand{\hatcurCCtassmrshorteccenxxxxxC}{\ensuremath{12.0}}      
\newcommand{\hatcurCCtassmieccenxxxxxC}{\ensuremath{11.833\pm0.040}} 
\newcommand{\hatcurCCtassmishorteccenxxxxxC}{\ensuremath{11.8}}      
\newcommand{\hatcurCCtwomassJmageccenxxxxxC}{\ensuremath{10.912\pm0.020}} 
\newcommand{\hatcurCCtwomassHmageccenxxxxxC}{\ensuremath{10.562\pm0.022}} 
\newcommand{\hatcurCCtwomassKmageccenxxxxxC}{\ensuremath{10.526\pm0.023}} 
\newcommand{\hatcurCCcitJmageccenxxxxxC}{\ensuremath{10.928\pm0.020}} 
\newcommand{\hatcurCCcitHmageccenxxxxxC}{\ensuremath{10.558\pm0.023}} 
\newcommand{\hatcurCCcitKmageccenxxxxxC}{\ensuremath{10.550\pm0.023}} 
\newcommand{\hatcurCCbbJmageccenxxxxxC}{\ensuremath{10.979\pm0.022}} 
\newcommand{\hatcurCCbbHmageccenxxxxxC}{\ensuremath{10.578\pm0.023}} 
\newcommand{\hatcurCCbbKmageccenxxxxxC}{\ensuremath{10.570\pm0.023}} 
\newcommand{\hatcurCCesoJmageccenxxxxxC}{\ensuremath{10.981\pm0.024}} 
\newcommand{\hatcurCCesoHmageccenxxxxxC}{\ensuremath{10.571\pm0.026}} 
\newcommand{\hatcurCCesoKmageccenxxxxxC}{\ensuremath{10.569\pm0.024}} 
\newcommand{\hatcurCCesoJHmageccenxxxxxC}{\ensuremath{0.409\pm0.033}} 
\newcommand{\hatcurCCesoJKmageccenxxxxxC}{\ensuremath{0.413\pm0.033}} 
\newcommand{\hatcurCCesoHKmageccenxxxxxC}{\ensuremath{0.003\pm0.035}} 
\newcommand{\hatcurLCdipeccenxxxxxC}{\ensuremath{11.1}}              
\newcommand{\hatcurLCrprstareccenxxxxxC}{\ensuremath{0.1133\pm0.0063}} 
\newcommand{\hatcurLCbsqeccenxxxxxC}{\ensuremath{0.731_{-0.030}^{+0.071}}} 
\newcommand{\hatcurLCimpeccenxxxxxC}{\ensuremath{0.855_{-0.018}^{+0.040}}} 
\newcommand{\hatcurLCzetaeccenxxxxxC}{\ensuremath{33.69_{-0.97}^{+1.78}}} 
\newcommand{\hatcurLCdureccenxxxxxC}{\ensuremath{0.0822\pm0.0028}}   
\newcommand{\hatcurLCdurshorteccenxxxxxC}{\ensuremath{0.0822}}       
\newcommand{\hatcurLCdurhreccenxxxxxC}{\ensuremath{1.973\pm0.067}}   
\newcommand{\hatcurLCdurhrshorteccenxxxxxC}{\ensuremath{1.973}}      
\newcommand{\hatcurLCqeccenxxxxxC}{\ensuremath{0.02310\pm0.00078}}   
\newcommand{\hatcurLCqshorteccenxxxxxC}{\ensuremath{0.023}}          
\newcommand{\hatcurLCingdureccenxxxxxC}{\ensuremath{0.028\pm0.039}}  
\newcommand{\hatcurLCPeccenxxxxxC}{\ensuremath{3.5543983\pm0.0000045}} 
\newcommand{\hatcurLCPprececcenxxxxxC}{\ensuremath{3.5543983}}       
\newcommand{\hatcurLCPshorteccenxxxxxC}{\ensuremath{3.5544}}         
\newcommand{\hatcurLCTeccenxxxxxC}{\ensuremath{2457034.58277\pm0.00063}} 
\newcommand{\hatcurLCTAeccenxxxxxC}{\ensuremath{2455680.3574\pm0.0019}} 
\newcommand{\hatcurLCTBeccenxxxxxC}{\ensuremath{2457219.41153\pm0.00065}} 
\newcommand{\hatcurLChatnetmAeccenxxxxxC}{\ensuremath{12.07053\pm0.00012}} 
\newcommand{\hatcurLCiblendAeccenxxxxxC}{\ensuremath{0.780\pm0.096}} 
\newcommand{\hatcurLChatnetmBeccenxxxxxC}{\ensuremath{12.070680\pm0.000058}} 
\newcommand{\hatcurLCiblendBeccenxxxxxC}{\ensuremath{0.767\pm0.070}} 
\newcommand{\hatcurLCrhoeccenxxxxxC}{\ensuremath{1.80_{-0.50}^{+0.73}}} 
\newcommand{\hatcurSMEiteffeccenxxxxxC}{\ensuremath{5678\pm97}}      
\newcommand{\hatcurSMEizfeheccenxxxxxC}{\ensuremath{0.290\pm0.060}}  
\newcommand{\hatcurSMEizfehshorteccenxxxxxC}{\ensuremath{0.29}}      
\newcommand{\hatcurSMEiloggeccenxxxxxC}{\ensuremath{4.55\pm0.14}}    
\newcommand{\hatcurSMEivsineccenxxxxxC}{\ensuremath{2.43\pm0.60}}    
\newcommand{\hatcurSMEivmaceccenxxxxxC}{\ensuremath{0.0}}            
\newcommand{\hatcurSMEivmiceccenxxxxxC}{\ensuremath{0.0}}            
\newcommand{\hatcurSMEiiteffeccenxxxxxC}{\ensuremath{5695\pm67}}     
\newcommand{\hatcurSMEiizfeheccenxxxxxC}{\ensuremath{0.300\pm0.040}} 
\newcommand{\hatcurSMEiizfehshorteccenxxxxxC}{\ensuremath{0.30}}     
\newcommand{\hatcurSMEiiloggeccenxxxxxC}{\ensuremath{4.460\pm0.050}} 
\newcommand{\hatcurSMEiivsineccenxxxxxC}{\ensuremath{2.63\pm0.55}}   
\newcommand{\hatcurLBizeccenxxxxxC}{\ensuremath{0.2266}}             
\newcommand{\hatcurLBiizeccenxxxxxC}{\ensuremath{0.3265}}            
\newcommand{\hatcurLBiieccenxxxxxC}{\ensuremath{0.2962}}             
\newcommand{\hatcurLBiiieccenxxxxxC}{\ensuremath{0.3225}}            
\newcommand{\hatcurLBiIeccenxxxxxC}{\ensuremath{0.2730}}             
\newcommand{\hatcurLBiiIeccenxxxxxC}{\ensuremath{0.3245}}            
\newcommand{\hatcurLBigeccenxxxxxC}{\ensuremath{0.6056}}             
\newcommand{\hatcurLBiigeccenxxxxxC}{\ensuremath{0.1986}}            
\newcommand{\hatcurLBireccenxxxxxC}{\ensuremath{0.3952}}             
\newcommand{\hatcurLBiireccenxxxxxC}{\ensuremath{0.3072}}            
\newcommand{\hatcurLBiReccenxxxxxC}{\ensuremath{0.3677}}             
\newcommand{\hatcurLBiiReccenxxxxxC}{\ensuremath{0.3125}}            
\newcommand{\hatcurLBikepeccenxxxxxC}{\ensuremath{0.1000}}           
\newcommand{\hatcurLBiikepeccenxxxxxC}{\ensuremath{0.1000}}          
\newcommand{\hatcurISOmeccenxxxxxC}{\ensuremath{1.078\pm0.027}}      
\newcommand{\hatcurISOmshorteccenxxxxxC}{\ensuremath{1.08}}          
\newcommand{\hatcurISOmlongeccenxxxxxC}{\ensuremath{1.078\pm0.027}}  
\newcommand{\hatcurISOreccenxxxxxC}{\ensuremath{1.032_{-0.042}^{+0.060}}} 
\newcommand{\hatcurISOrshorteccenxxxxxC}{\ensuremath{1.03}}          
\newcommand{\hatcurISOrlongeccenxxxxxC}{\ensuremath{1.032_{-0.042}^{+0.060}}} 
\newcommand{\hatcurISOrhoeccenxxxxxC}{\ensuremath{1.38\pm0.21}}      
\newcommand{\hatcurISOrholongeccenxxxxxC}{\ensuremath{1.38\pm0.21}}  
\newcommand{\hatcurISOloggeccenxxxxxC}{\ensuremath{4.443\pm0.049}}   
\newcommand{\hatcurISOlumeccenxxxxxC}{\ensuremath{1.01\pm0.16}}      
\newcommand{\hatcurISOlumshorteccenxxxxxC}{\ensuremath{1.01}}        
\newcommand{\hatcurISOmveccenxxxxxC}{\ensuremath{4.83\pm0.16}}       
\newcommand{\hatcurISOvieccenxxxxxC}{\ensuremath{0.733\pm0.021}}     
\newcommand{\hatcurISOageeccenxxxxxC}{\ensuremath{2.7\pm1.6}}        
\newcommand{\hatcurISOsigmaeccenxxxxxC}{\ensuremath{0.00050\pm0.00012}} 
\newcommand{\hatcurISOMJeccenxxxxxC}{\ensuremath{3.64\pm0.14}}       
\newcommand{\hatcurISOMHeccenxxxxxC}{\ensuremath{3.29\pm0.13}}       
\newcommand{\hatcurISOMKeccenxxxxxC}{\ensuremath{3.24\pm0.13}}       
\newcommand{\hatcurISOJKeccenxxxxxC}{\ensuremath{0.30\pm0.18}}       
\newcommand{\hatcurISOspececcenxxxxxC}{G}                            
\newcommand{\hatcurRVKeccenxxxxxC}{\ensuremath{41.9\pm1.4}}          
\newcommand{\hatcurRVrkeccenxxxxxC}{\ensuremath{0.03\pm0.11}}        
\newcommand{\hatcurRVrheccenxxxxxC}{\ensuremath{-0.05_{-0.16}^{+0.28}}} 
\newcommand{\hatcurRVkeccenxxxxxC}{\ensuremath{0.006\pm0.026}}       
\newcommand{\hatcurRVheccenxxxxxC}{\ensuremath{-0.006_{-0.042}^{+0.062}}} 
\newcommand{\hatcurRVtroneeccenxxxxxC}{\ensuremath{0\pm0}}           
\newcommand{\hatcurRVtrtwoeccenxxxxxC}{\ensuremath{0\pm0}}           
\newcommand{\hatcurRVgammaAeccenxxxxxC}{\ensuremath{32040.7\pm8.5}}  
\newcommand{\hatcurRVjitterAeccenxxxxxC}{\ensuremath{20.2\pm8.6}}    
\newcommand{\hatcurRVjittertwosiglimAeccenxxxxxC}{\ensuremath{<33.8}} 
\newcommand{\hatcurRVfitrmsAeccenxxxxxC}{\ensuremath{0.0}}           
\newcommand{\hatcurRVgammaBeccenxxxxxC}{\ensuremath{32055.0\pm6.3}}  
\newcommand{\hatcurRVjitterBeccenxxxxxC}{\ensuremath{0.00\pm0.74}}   
\newcommand{\hatcurRVjittertwosiglimBeccenxxxxxC}{\ensuremath{<1.4}} 
\newcommand{\hatcurRVfitrmsBeccenxxxxxC}{\ensuremath{0.0}}           
\newcommand{\hatcurRVgammaCeccenxxxxxC}{\ensuremath{16.5\pm3.0}}     
\newcommand{\hatcurRVjitterCeccenxxxxxC}{\ensuremath{7.2\pm3.4}}     
\newcommand{\hatcurRVjittertwosiglimCeccenxxxxxC}{\ensuremath{<13.0}} 
\newcommand{\hatcurRVfitrmsCeccenxxxxxC}{\ensuremath{0.0}}           
\newcommand{\hatcurRVecceneccenxxxxxC}{\ensuremath{0.046\pm0.048}}   
\newcommand{\hatcurRVeccentwosiglimeccenxxxxxC}{\ensuremath{<0.149}} 
\newcommand{\hatcurRVomegaeccenxxxxxC}{\ensuremath{230\pm110}}       
\newcommand{\hatcurPPieccenxxxxxC}{\ensuremath{84.92\pm0.67}}        
\newcommand{\hatcurPPgeccenxxxxxC}{\ensuremath{6.3\pm1.0}}           
\newcommand{\hatcurPPloggeccenxxxxxC}{\ensuremath{2.800_{-0.080}^{+0.058}}} 
\newcommand{\hatcurPPareccenxxxxxC}{\ensuremath{9.73\pm0.53}}        
\newcommand{\hatcurPPareleccenxxxxxC}{\ensuremath{0.04674\pm0.00039}} 
\newcommand{\hatcurPPrhoeccenxxxxxC}{\ensuremath{0.278\pm0.067}}     
\newcommand{\hatcurPPmeccenxxxxxC}{\ensuremath{0.330\pm0.012}}       
\newcommand{\hatcurPPmshorteccenxxxxxC}{\ensuremath{0.33}}           
\newcommand{\hatcurPPmlongeccenxxxxxC}{\ensuremath{0.330\pm0.012}}   
\newcommand{\hatcurPPmeeccenxxxxxC}{\ensuremath{105.0\pm3.8}}        
\newcommand{\hatcurPPmeshorteccenxxxxxC}{\ensuremath{105.0}}         
\newcommand{\hatcurPPmelongeccenxxxxxC}{\ensuremath{105.0\pm3.8}}    
\newcommand{\hatcurPPreccenxxxxxC}{\ensuremath{1.136_{-0.074}^{+0.117}}} 
\newcommand{\hatcurPPrshorteccenxxxxxC}{\ensuremath{1.14}}           
\newcommand{\hatcurPPrlongeccenxxxxxC}{\ensuremath{1.136_{-0.074}^{+0.117}}} 
\newcommand{\hatcurPPreeccenxxxxxC}{\ensuremath{12.73_{-0.83}^{+1.31}}} 
\newcommand{\hatcurPPreshorteccenxxxxxC}{\ensuremath{12.7}}          
\newcommand{\hatcurPPrelongeccenxxxxxC}{\ensuremath{12.73_{-0.83}^{+1.31}}} 
\newcommand{\hatcurPPmrcorreccenxxxxxC}{\ensuremath{0.43}}           
\newcommand{\hatcurPPteffeccenxxxxxC}{\ensuremath{1292\pm44}}        
\newcommand{\hatcurPPthetaeccenxxxxxC}{\ensuremath{0.0249\pm0.0022}} 
\newcommand{\hatcurPPfluxperieccenxxxxxC}{\ensuremath{7.0\pm2.1}}    
\newcommand{\hatcurPPfluxperidimeccenxxxxxC}{\ensuremath{8}}         
\newcommand{\hatcurPPfluxapeccenxxxxxC}{\ensuremath{5.73\pm0.75}}    
\newcommand{\hatcurPPfluxapdimeccenxxxxxC}{\ensuremath{8}}           
\newcommand{\hatcurPPfluxavgeccenxxxxxC}{\ensuremath{6.30\pm0.95}}   
\newcommand{\hatcurPPfluxavgdimeccenxxxxxC}{\ensuremath{8}}          
\newcommand{\hatcurPPfluxavglogeccenxxxxxC}{\ensuremath{8.799\pm0.058}} 
\newcommand{\hatcurXsecphaseeccenxxxxxC}{\ensuremath{0.504\pm0.017}} 
\newcommand{\hatcurXsecondaryeccenxxxxxC}{\ensuremath{2457036.374\pm0.059}} 
\newcommand{\hatcurXsecdureccenxxxxxC}{\ensuremath{0.0846\pm0.0082}} 
\newcommand{\hatcurXsecingdureccenxxxxxC}{\ensuremath{0.026\pm0.011}} 
\newcommand{\hatcurPPphiconjeccenxxxxxC}{\ensuremath{0.04\pm0.32}}   
\newcommand{\hatcurPPperieccenxxxxxC}{\ensuremath{2457034.4\pm1.1}}  
\newcommand{\hatcurPPaequiveccenxxxxxC}{\ensuremath{0.0466\pm0.0030}} 
\newcommand{\hatcurPPtcirceccenxxxxxC}{\ensuremath{132\pm53}}        
\newcommand{\hatcurPPtinfalleccenxxxxxC}{\ensuremath{9700\pm2400}}   
\newcommand{\hatcurXdisteccenxxxxxC}{\ensuremath{293_{-14}^{+18}}}   
\newcommand{\hatcurXAveccenxxxxxC}{\ensuremath{0.049\pm0.051}}       
\newcommand{\hatcurXdistredeccenxxxxxC}{\ensuremath{289\pm19}}       
\newcommand{\hatcurXEBVeccenxxxxxC}{\ensuremath{0.016\pm0.016}}      
\newcommand{\hatcurXmvisoredeccenxxxxxC}{\ensuremath{12.195\pm0.020}} 
\newcommand{\hatcurXmiisoredeccenxxxxxC}{\ensuremath{11.433\pm0.014}} 
\newcommand{\hatcurXmjisoredeccenxxxxxC}{\ensuremath{10.963\pm0.013}} 
\newcommand{\hatcurXmhisoredeccenxxxxxC}{\ensuremath{10.609\pm0.016}} 
\newcommand{\hatcurXmkisoredeccenxxxxxC}{\ensuremath{10.548\pm0.017}} 
\newcommand{\hatcurXviisoredeccenxxxxxC}{\ensuremath{0.762\pm0.015}} 
\newcommand{\hatcurXvkisoredeccenxxxxxC}{\ensuremath{1.646\pm0.029}} 
\newcommand{\hatcurXjhisoredeccenxxxxxC}{\ensuremath{0.3540\pm0.0095}} 
\newcommand{\hatcurXjkisoredeccenxxxxxC}{\ensuremath{0.414\pm0.010}} 
\newcommand{\hatcurCCpmraeccenxxxxxC}{\ensuremath{-26.1\pm1.7}}      
\newcommand{\hatcurCCpmdececcenxxxxxC}{\ensuremath{-29.4\pm1.7}}     
\newcommand{\hatcurCCpmeccenxxxxxC}{\ensuremath{39.3\pm2.4}}         

\newcommand{\hatcurCCbbHmageccen}[1]{\ifnum#1=19 %
\hatcurCCbbHmageccenxxxxxA
\else
\ifnum#1=20 %
\hatcurCCbbHmageccenxxxxxB
\else
\ifnum#1=21 %
\hatcurCCbbHmageccenxxxxxC
\else
??????\fi
\fi
\fi
}
\newcommand{\hatcurCCbbJmageccen}[1]{\ifnum#1=19 %
\hatcurCCbbJmageccenxxxxxA
\else
\ifnum#1=20 %
\hatcurCCbbJmageccenxxxxxB
\else
\ifnum#1=21 %
\hatcurCCbbJmageccenxxxxxC
\else
??????\fi
\fi
\fi
}
\newcommand{\hatcurCCbbKmageccen}[1]{\ifnum#1=19 %
\hatcurCCbbKmageccenxxxxxA
\else
\ifnum#1=20 %
\hatcurCCbbKmageccenxxxxxB
\else
\ifnum#1=21 %
\hatcurCCbbKmageccenxxxxxC
\else
??????\fi
\fi
\fi
}
\newcommand{\hatcurCCcitHmageccen}[1]{\ifnum#1=19 %
\hatcurCCcitHmageccenxxxxxA
\else
\ifnum#1=20 %
\hatcurCCcitHmageccenxxxxxB
\else
\ifnum#1=21 %
\hatcurCCcitHmageccenxxxxxC
\else
??????\fi
\fi
\fi
}
\newcommand{\hatcurCCcitJmageccen}[1]{\ifnum#1=19 %
\hatcurCCcitJmageccenxxxxxA
\else
\ifnum#1=20 %
\hatcurCCcitJmageccenxxxxxB
\else
\ifnum#1=21 %
\hatcurCCcitJmageccenxxxxxC
\else
??????\fi
\fi
\fi
}
\newcommand{\hatcurCCcitKmageccen}[1]{\ifnum#1=19 %
\hatcurCCcitKmageccenxxxxxA
\else
\ifnum#1=20 %
\hatcurCCcitKmageccenxxxxxB
\else
\ifnum#1=21 %
\hatcurCCcitKmageccenxxxxxC
\else
??????\fi
\fi
\fi
}
\newcommand{\hatcurCCdececcen}[1]{\ifnum#1=19 %
\hatcurCCdececcenxxxxxA
\else
\ifnum#1=20 %
\hatcurCCdececcenxxxxxB
\else
\ifnum#1=21 %
\hatcurCCdececcenxxxxxC
\else
??????\fi
\fi
\fi
}
\newcommand{\hatcurCCesoHKmageccen}[1]{\ifnum#1=19 %
\hatcurCCesoHKmageccenxxxxxA
\else
\ifnum#1=20 %
\hatcurCCesoHKmageccenxxxxxB
\else
\ifnum#1=21 %
\hatcurCCesoHKmageccenxxxxxC
\else
??????\fi
\fi
\fi
}
\newcommand{\hatcurCCesoHmageccen}[1]{\ifnum#1=19 %
\hatcurCCesoHmageccenxxxxxA
\else
\ifnum#1=20 %
\hatcurCCesoHmageccenxxxxxB
\else
\ifnum#1=21 %
\hatcurCCesoHmageccenxxxxxC
\else
??????\fi
\fi
\fi
}
\newcommand{\hatcurCCesoJHmageccen}[1]{\ifnum#1=19 %
\hatcurCCesoJHmageccenxxxxxA
\else
\ifnum#1=20 %
\hatcurCCesoJHmageccenxxxxxB
\else
\ifnum#1=21 %
\hatcurCCesoJHmageccenxxxxxC
\else
??????\fi
\fi
\fi
}
\newcommand{\hatcurCCesoJKmageccen}[1]{\ifnum#1=19 %
\hatcurCCesoJKmageccenxxxxxA
\else
\ifnum#1=20 %
\hatcurCCesoJKmageccenxxxxxB
\else
\ifnum#1=21 %
\hatcurCCesoJKmageccenxxxxxC
\else
??????\fi
\fi
\fi
}
\newcommand{\hatcurCCesoJmageccen}[1]{\ifnum#1=19 %
\hatcurCCesoJmageccenxxxxxA
\else
\ifnum#1=20 %
\hatcurCCesoJmageccenxxxxxB
\else
\ifnum#1=21 %
\hatcurCCesoJmageccenxxxxxC
\else
??????\fi
\fi
\fi
}
\newcommand{\hatcurCCesoKmageccen}[1]{\ifnum#1=19 %
\hatcurCCesoKmageccenxxxxxA
\else
\ifnum#1=20 %
\hatcurCCesoKmageccenxxxxxB
\else
\ifnum#1=21 %
\hatcurCCesoKmageccenxxxxxC
\else
??????\fi
\fi
\fi
}
\newcommand{\hatcurCCgsceccen}[1]{\ifnum#1=19 %
\hatcurCCgsceccenxxxxxA
\else
\ifnum#1=20 %
\hatcurCCgsceccenxxxxxB
\else
\ifnum#1=21 %
\hatcurCCgsceccenxxxxxC
\else
??????\fi
\fi
\fi
}
\newcommand{\hatcurCCmageccen}[1]{\ifnum#1=19 %
\hatcurCCmageccenxxxxxA
\else
\ifnum#1=20 %
\hatcurCCmageccenxxxxxB
\else
\ifnum#1=21 %
\hatcurCCmageccenxxxxxC
\else
??????\fi
\fi
\fi
}
\newcommand{\hatcurCCpmdececcen}[1]{\ifnum#1=19 %
\hatcurCCpmdececcenxxxxxA
\else
\ifnum#1=20 %
\hatcurCCpmdececcenxxxxxB
\else
\ifnum#1=21 %
\hatcurCCpmdececcenxxxxxC
\else
??????\fi
\fi
\fi
}
\newcommand{\hatcurCCpmeccen}[1]{\ifnum#1=19 %
\hatcurCCpmeccenxxxxxA
\else
\ifnum#1=20 %
\hatcurCCpmeccenxxxxxB
\else
\ifnum#1=21 %
\hatcurCCpmeccenxxxxxC
\else
??????\fi
\fi
\fi
}
\newcommand{\hatcurCCpmraeccen}[1]{\ifnum#1=19 %
\hatcurCCpmraeccenxxxxxA
\else
\ifnum#1=20 %
\hatcurCCpmraeccenxxxxxB
\else
\ifnum#1=21 %
\hatcurCCpmraeccenxxxxxC
\else
??????\fi
\fi
\fi
}
\newcommand{\hatcurCCraeccen}[1]{\ifnum#1=19 %
\hatcurCCraeccenxxxxxA
\else
\ifnum#1=20 %
\hatcurCCraeccenxxxxxB
\else
\ifnum#1=21 %
\hatcurCCraeccenxxxxxC
\else
??????\fi
\fi
\fi
}
\newcommand{\hatcurCCtassmBeccen}[1]{\ifnum#1=19 %
\hatcurCCtassmBeccenxxxxxA
\else
\ifnum#1=20 %
\hatcurCCtassmBeccenxxxxxB
\else
\ifnum#1=21 %
\hatcurCCtassmBeccenxxxxxC
\else
??????\fi
\fi
\fi
}
\newcommand{\hatcurCCtassmBshorteccen}[1]{\ifnum#1=19 %
\hatcurCCtassmBshorteccenxxxxxA
\else
\ifnum#1=20 %
\hatcurCCtassmBshorteccenxxxxxB
\else
\ifnum#1=21 %
\hatcurCCtassmBshorteccenxxxxxC
\else
??????\fi
\fi
\fi
}
\newcommand{\hatcurCCtassmgeccen}[1]{\ifnum#1=19 %
\hatcurCCtassmgeccenxxxxxA
\else
\ifnum#1=20 %
\hatcurCCtassmgeccenxxxxxB
\else
\ifnum#1=21 %
\hatcurCCtassmgeccenxxxxxC
\else
??????\fi
\fi
\fi
}
\newcommand{\hatcurCCtassmgshorteccen}[1]{\ifnum#1=19 %
\hatcurCCtassmgshorteccenxxxxxA
\else
\ifnum#1=20 %
\hatcurCCtassmgshorteccenxxxxxB
\else
\ifnum#1=21 %
\hatcurCCtassmgshorteccenxxxxxC
\else
??????\fi
\fi
\fi
}
\newcommand{\hatcurCCtassmieccen}[1]{\ifnum#1=19 %
\hatcurCCtassmieccenxxxxxA
\else
\ifnum#1=20 %
\hatcurCCtassmieccenxxxxxB
\else
\ifnum#1=21 %
\hatcurCCtassmieccenxxxxxC
\else
??????\fi
\fi
\fi
}
\newcommand{\hatcurCCtassmIeccen}[1]{\ifnum#1=19 %
\hatcurCCtassmIeccenxxxxxA
\else
\ifnum#1=20 %
\hatcurCCtassmIeccenxxxxxB
\else
\ifnum#1=21 %
\hatcurCCtassmIeccenxxxxxC
\else
??????\fi
\fi
\fi
}
\newcommand{\hatcurCCtassmishorteccen}[1]{\ifnum#1=19 %
\hatcurCCtassmishorteccenxxxxxA
\else
\ifnum#1=20 %
\hatcurCCtassmishorteccenxxxxxB
\else
\ifnum#1=21 %
\hatcurCCtassmishorteccenxxxxxC
\else
??????\fi
\fi
\fi
}
\newcommand{\hatcurCCtassmIshorteccen}[1]{\ifnum#1=19 %
\hatcurCCtassmIshorteccenxxxxxA
\else
\ifnum#1=20 %
\hatcurCCtassmIshorteccenxxxxxB
\else
\ifnum#1=21 %
\hatcurCCtassmIshorteccenxxxxxC
\else
??????\fi
\fi
\fi
}
\newcommand{\hatcurCCtassmreccen}[1]{\ifnum#1=19 %
\hatcurCCtassmreccenxxxxxA
\else
\ifnum#1=20 %
\hatcurCCtassmreccenxxxxxB
\else
\ifnum#1=21 %
\hatcurCCtassmreccenxxxxxC
\else
??????\fi
\fi
\fi
}
\newcommand{\hatcurCCtassmrshorteccen}[1]{\ifnum#1=19 %
\hatcurCCtassmrshorteccenxxxxxA
\else
\ifnum#1=20 %
\hatcurCCtassmrshorteccenxxxxxB
\else
\ifnum#1=21 %
\hatcurCCtassmrshorteccenxxxxxC
\else
??????\fi
\fi
\fi
}
\newcommand{\hatcurCCtassmveccen}[1]{\ifnum#1=19 %
\hatcurCCtassmveccenxxxxxA
\else
\ifnum#1=20 %
\hatcurCCtassmveccenxxxxxB
\else
\ifnum#1=21 %
\hatcurCCtassmveccenxxxxxC
\else
??????\fi
\fi
\fi
}
\newcommand{\hatcurCCtassmvshorteccen}[1]{\ifnum#1=19 %
\hatcurCCtassmvshorteccenxxxxxA
\else
\ifnum#1=20 %
\hatcurCCtassmvshorteccenxxxxxB
\else
\ifnum#1=21 %
\hatcurCCtassmvshorteccenxxxxxC
\else
??????\fi
\fi
\fi
}
\newcommand{\hatcurCCtwomasseccen}[1]{\ifnum#1=19 %
\hatcurCCtwomasseccenxxxxxA
\else
\ifnum#1=20 %
\hatcurCCtwomasseccenxxxxxB
\else
\ifnum#1=21 %
\hatcurCCtwomasseccenxxxxxC
\else
??????\fi
\fi
\fi
}
\newcommand{\hatcurCCtwomassHmageccen}[1]{\ifnum#1=19 %
\hatcurCCtwomassHmageccenxxxxxA
\else
\ifnum#1=20 %
\hatcurCCtwomassHmageccenxxxxxB
\else
\ifnum#1=21 %
\hatcurCCtwomassHmageccenxxxxxC
\else
??????\fi
\fi
\fi
}
\newcommand{\hatcurCCtwomassJmageccen}[1]{\ifnum#1=19 %
\hatcurCCtwomassJmageccenxxxxxA
\else
\ifnum#1=20 %
\hatcurCCtwomassJmageccenxxxxxB
\else
\ifnum#1=21 %
\hatcurCCtwomassJmageccenxxxxxC
\else
??????\fi
\fi
\fi
}
\newcommand{\hatcurCCtwomassKmageccen}[1]{\ifnum#1=19 %
\hatcurCCtwomassKmageccenxxxxxA
\else
\ifnum#1=20 %
\hatcurCCtwomassKmageccenxxxxxB
\else
\ifnum#1=21 %
\hatcurCCtwomassKmageccenxxxxxC
\else
??????\fi
\fi
\fi
}
\newcommand{\hatcurfieldeccen}[1]{\ifnum#1=19 %
\hatcurfieldeccenxxxxxA
\else
\ifnum#1=20 %
\hatcurfieldeccenxxxxxB
\else
\ifnum#1=21 %
\hatcurfieldeccenxxxxxC
\else
??????\fi
\fi
\fi
}
\newcommand{\hatcurhtreccen}[1]{\ifnum#1=19 %
\hatcurhtreccenxxxxxA
\else
\ifnum#1=20 %
\hatcurhtreccenxxxxxB
\else
\ifnum#1=21 %
\hatcurhtreccenxxxxxC
\else
??????\fi
\fi
\fi
}
\newcommand{\hatcurISOageeccen}[1]{\ifnum#1=19 %
\hatcurISOageeccenxxxxxA
\else
\ifnum#1=20 %
\hatcurISOageeccenxxxxxB
\else
\ifnum#1=21 %
\hatcurISOageeccenxxxxxC
\else
??????\fi
\fi
\fi
}
\newcommand{\hatcurISOJKeccen}[1]{\ifnum#1=19 %
\hatcurISOJKeccenxxxxxA
\else
\ifnum#1=20 %
\hatcurISOJKeccenxxxxxB
\else
\ifnum#1=21 %
\hatcurISOJKeccenxxxxxC
\else
??????\fi
\fi
\fi
}
\newcommand{\hatcurISOloggeccen}[1]{\ifnum#1=19 %
\hatcurISOloggeccenxxxxxA
\else
\ifnum#1=20 %
\hatcurISOloggeccenxxxxxB
\else
\ifnum#1=21 %
\hatcurISOloggeccenxxxxxC
\else
??????\fi
\fi
\fi
}
\newcommand{\hatcurISOlumeccen}[1]{\ifnum#1=19 %
\hatcurISOlumeccenxxxxxA
\else
\ifnum#1=20 %
\hatcurISOlumeccenxxxxxB
\else
\ifnum#1=21 %
\hatcurISOlumeccenxxxxxC
\else
??????\fi
\fi
\fi
}
\newcommand{\hatcurISOlumshorteccen}[1]{\ifnum#1=19 %
\hatcurISOlumshorteccenxxxxxA
\else
\ifnum#1=20 %
\hatcurISOlumshorteccenxxxxxB
\else
\ifnum#1=21 %
\hatcurISOlumshorteccenxxxxxC
\else
??????\fi
\fi
\fi
}
\newcommand{\hatcurISOmeccen}[1]{\ifnum#1=19 %
\hatcurISOmeccenxxxxxA
\else
\ifnum#1=20 %
\hatcurISOmeccenxxxxxB
\else
\ifnum#1=21 %
\hatcurISOmeccenxxxxxC
\else
??????\fi
\fi
\fi
}
\newcommand{\hatcurISOMHeccen}[1]{\ifnum#1=19 %
\hatcurISOMHeccenxxxxxA
\else
\ifnum#1=20 %
\hatcurISOMHeccenxxxxxB
\else
\ifnum#1=21 %
\hatcurISOMHeccenxxxxxC
\else
??????\fi
\fi
\fi
}
\newcommand{\hatcurISOMJeccen}[1]{\ifnum#1=19 %
\hatcurISOMJeccenxxxxxA
\else
\ifnum#1=20 %
\hatcurISOMJeccenxxxxxB
\else
\ifnum#1=21 %
\hatcurISOMJeccenxxxxxC
\else
??????\fi
\fi
\fi
}
\newcommand{\hatcurISOMKeccen}[1]{\ifnum#1=19 %
\hatcurISOMKeccenxxxxxA
\else
\ifnum#1=20 %
\hatcurISOMKeccenxxxxxB
\else
\ifnum#1=21 %
\hatcurISOMKeccenxxxxxC
\else
??????\fi
\fi
\fi
}
\newcommand{\hatcurISOmlongeccen}[1]{\ifnum#1=19 %
\hatcurISOmlongeccenxxxxxA
\else
\ifnum#1=20 %
\hatcurISOmlongeccenxxxxxB
\else
\ifnum#1=21 %
\hatcurISOmlongeccenxxxxxC
\else
??????\fi
\fi
\fi
}
\newcommand{\hatcurISOmshorteccen}[1]{\ifnum#1=19 %
\hatcurISOmshorteccenxxxxxA
\else
\ifnum#1=20 %
\hatcurISOmshorteccenxxxxxB
\else
\ifnum#1=21 %
\hatcurISOmshorteccenxxxxxC
\else
??????\fi
\fi
\fi
}
\newcommand{\hatcurISOmveccen}[1]{\ifnum#1=19 %
\hatcurISOmveccenxxxxxA
\else
\ifnum#1=20 %
\hatcurISOmveccenxxxxxB
\else
\ifnum#1=21 %
\hatcurISOmveccenxxxxxC
\else
??????\fi
\fi
\fi
}
\newcommand{\hatcurISOreccen}[1]{\ifnum#1=19 %
\hatcurISOreccenxxxxxA
\else
\ifnum#1=20 %
\hatcurISOreccenxxxxxB
\else
\ifnum#1=21 %
\hatcurISOreccenxxxxxC
\else
??????\fi
\fi
\fi
}
\newcommand{\hatcurISOrhoeccen}[1]{\ifnum#1=19 %
\hatcurISOrhoeccenxxxxxA
\else
\ifnum#1=20 %
\hatcurISOrhoeccenxxxxxB
\else
\ifnum#1=21 %
\hatcurISOrhoeccenxxxxxC
\else
??????\fi
\fi
\fi
}
\newcommand{\hatcurISOrholongeccen}[1]{\ifnum#1=19 %
\hatcurISOrholongeccenxxxxxA
\else
\ifnum#1=20 %
\hatcurISOrholongeccenxxxxxB
\else
\ifnum#1=21 %
\hatcurISOrholongeccenxxxxxC
\else
??????\fi
\fi
\fi
}
\newcommand{\hatcurISOrlongeccen}[1]{\ifnum#1=19 %
\hatcurISOrlongeccenxxxxxA
\else
\ifnum#1=20 %
\hatcurISOrlongeccenxxxxxB
\else
\ifnum#1=21 %
\hatcurISOrlongeccenxxxxxC
\else
??????\fi
\fi
\fi
}
\newcommand{\hatcurISOrshorteccen}[1]{\ifnum#1=19 %
\hatcurISOrshorteccenxxxxxA
\else
\ifnum#1=20 %
\hatcurISOrshorteccenxxxxxB
\else
\ifnum#1=21 %
\hatcurISOrshorteccenxxxxxC
\else
??????\fi
\fi
\fi
}
\newcommand{\hatcurISOsigmaeccen}[1]{\ifnum#1=19 %
\hatcurISOsigmaeccenxxxxxA
\else
\ifnum#1=20 %
\hatcurISOsigmaeccenxxxxxB
\else
\ifnum#1=21 %
\hatcurISOsigmaeccenxxxxxC
\else
??????\fi
\fi
\fi
}
\newcommand{\hatcurISOspececcen}[1]{\ifnum#1=19 %
\hatcurISOspececcenxxxxxA
\else
\ifnum#1=20 %
\hatcurISOspececcenxxxxxB
\else
\ifnum#1=21 %
\hatcurISOspececcenxxxxxC
\else
??????\fi
\fi
\fi
}
\newcommand{\hatcurISOvieccen}[1]{\ifnum#1=19 %
\hatcurISOvieccenxxxxxA
\else
\ifnum#1=20 %
\hatcurISOvieccenxxxxxB
\else
\ifnum#1=21 %
\hatcurISOvieccenxxxxxC
\else
??????\fi
\fi
\fi
}
\newcommand{\hatcurLBigeccen}[1]{\ifnum#1=19 %
\hatcurLBigeccenxxxxxA
\else
\ifnum#1=20 %
\hatcurLBigeccenxxxxxB
\else
\ifnum#1=21 %
\hatcurLBigeccenxxxxxC
\else
??????\fi
\fi
\fi
}
\newcommand{\hatcurLBiieccen}[1]{\ifnum#1=19 %
\hatcurLBiieccenxxxxxA
\else
\ifnum#1=20 %
\hatcurLBiieccenxxxxxB
\else
\ifnum#1=21 %
\hatcurLBiieccenxxxxxC
\else
??????\fi
\fi
\fi
}
\newcommand{\hatcurLBiIeccen}[1]{\ifnum#1=19 %
\hatcurLBiIeccenxxxxxA
\else
\ifnum#1=20 %
\hatcurLBiIeccenxxxxxB
\else
\ifnum#1=21 %
\hatcurLBiIeccenxxxxxC
\else
??????\fi
\fi
\fi
}
\newcommand{\hatcurLBiigeccen}[1]{\ifnum#1=19 %
\hatcurLBiigeccenxxxxxA
\else
\ifnum#1=20 %
\hatcurLBiigeccenxxxxxB
\else
\ifnum#1=21 %
\hatcurLBiigeccenxxxxxC
\else
??????\fi
\fi
\fi
}
\newcommand{\hatcurLBiiieccen}[1]{\ifnum#1=19 %
\hatcurLBiiieccenxxxxxA
\else
\ifnum#1=20 %
\hatcurLBiiieccenxxxxxB
\else
\ifnum#1=21 %
\hatcurLBiiieccenxxxxxC
\else
??????\fi
\fi
\fi
}
\newcommand{\hatcurLBiiIeccen}[1]{\ifnum#1=19 %
\hatcurLBiiIeccenxxxxxA
\else
\ifnum#1=20 %
\hatcurLBiiIeccenxxxxxB
\else
\ifnum#1=21 %
\hatcurLBiiIeccenxxxxxC
\else
??????\fi
\fi
\fi
}
\newcommand{\hatcurLBiikepeccen}[1]{\ifnum#1=19 %
\hatcurLBiikepeccenxxxxxA
\else
\ifnum#1=20 %
\hatcurLBiikepeccenxxxxxB
\else
\ifnum#1=21 %
\hatcurLBiikepeccenxxxxxC
\else
??????\fi
\fi
\fi
}
\newcommand{\hatcurLBiireccen}[1]{\ifnum#1=19 %
\hatcurLBiireccenxxxxxA
\else
\ifnum#1=20 %
\hatcurLBiireccenxxxxxB
\else
\ifnum#1=21 %
\hatcurLBiireccenxxxxxC
\else
??????\fi
\fi
\fi
}
\newcommand{\hatcurLBiiReccen}[1]{\ifnum#1=19 %
\hatcurLBiiReccenxxxxxA
\else
\ifnum#1=20 %
\hatcurLBiiReccenxxxxxB
\else
\ifnum#1=21 %
\hatcurLBiiReccenxxxxxC
\else
??????\fi
\fi
\fi
}
\newcommand{\hatcurLBiizeccen}[1]{\ifnum#1=19 %
\hatcurLBiizeccenxxxxxA
\else
\ifnum#1=20 %
\hatcurLBiizeccenxxxxxB
\else
\ifnum#1=21 %
\hatcurLBiizeccenxxxxxC
\else
??????\fi
\fi
\fi
}
\newcommand{\hatcurLBikepeccen}[1]{\ifnum#1=19 %
\hatcurLBikepeccenxxxxxA
\else
\ifnum#1=20 %
\hatcurLBikepeccenxxxxxB
\else
\ifnum#1=21 %
\hatcurLBikepeccenxxxxxC
\else
??????\fi
\fi
\fi
}
\newcommand{\hatcurLBireccen}[1]{\ifnum#1=19 %
\hatcurLBireccenxxxxxA
\else
\ifnum#1=20 %
\hatcurLBireccenxxxxxB
\else
\ifnum#1=21 %
\hatcurLBireccenxxxxxC
\else
??????\fi
\fi
\fi
}
\newcommand{\hatcurLBiReccen}[1]{\ifnum#1=19 %
\hatcurLBiReccenxxxxxA
\else
\ifnum#1=20 %
\hatcurLBiReccenxxxxxB
\else
\ifnum#1=21 %
\hatcurLBiReccenxxxxxC
\else
??????\fi
\fi
\fi
}
\newcommand{\hatcurLBizeccen}[1]{\ifnum#1=19 %
\hatcurLBizeccenxxxxxA
\else
\ifnum#1=20 %
\hatcurLBizeccenxxxxxB
\else
\ifnum#1=21 %
\hatcurLBizeccenxxxxxC
\else
??????\fi
\fi
\fi
}
\newcommand{\hatcurLCbsqeccen}[1]{\ifnum#1=19 %
\hatcurLCbsqeccenxxxxxA
\else
\ifnum#1=20 %
\hatcurLCbsqeccenxxxxxB
\else
\ifnum#1=21 %
\hatcurLCbsqeccenxxxxxC
\else
??????\fi
\fi
\fi
}
\newcommand{\hatcurLCdipeccen}[1]{\ifnum#1=19 %
\hatcurLCdipeccenxxxxxA
\else
\ifnum#1=20 %
\hatcurLCdipeccenxxxxxB
\else
\ifnum#1=21 %
\hatcurLCdipeccenxxxxxC
\else
??????\fi
\fi
\fi
}
\newcommand{\hatcurLCdureccen}[1]{\ifnum#1=19 %
\hatcurLCdureccenxxxxxA
\else
\ifnum#1=20 %
\hatcurLCdureccenxxxxxB
\else
\ifnum#1=21 %
\hatcurLCdureccenxxxxxC
\else
??????\fi
\fi
\fi
}
\newcommand{\hatcurLCdurhreccen}[1]{\ifnum#1=19 %
\hatcurLCdurhreccenxxxxxA
\else
\ifnum#1=20 %
\hatcurLCdurhreccenxxxxxB
\else
\ifnum#1=21 %
\hatcurLCdurhreccenxxxxxC
\else
??????\fi
\fi
\fi
}
\newcommand{\hatcurLCdurhrshorteccen}[1]{\ifnum#1=19 %
\hatcurLCdurhrshorteccenxxxxxA
\else
\ifnum#1=20 %
\hatcurLCdurhrshorteccenxxxxxB
\else
\ifnum#1=21 %
\hatcurLCdurhrshorteccenxxxxxC
\else
??????\fi
\fi
\fi
}
\newcommand{\hatcurLCdurshorteccen}[1]{\ifnum#1=19 %
\hatcurLCdurshorteccenxxxxxA
\else
\ifnum#1=20 %
\hatcurLCdurshorteccenxxxxxB
\else
\ifnum#1=21 %
\hatcurLCdurshorteccenxxxxxC
\else
??????\fi
\fi
\fi
}
\newcommand{\hatcurLChatnetmAeccen}[1]{\ifnum#1=19 %
\hatcurLChatnetmAeccenxxxxxA
\else
\ifnum#1=20 %
\hatcurLChatnetmAeccenxxxxxB
\else
\ifnum#1=21 %
\hatcurLChatnetmAeccenxxxxxC
\else
??????\fi
\fi
\fi
}
\newcommand{\hatcurLChatnetmBeccen}[1]{\ifnum#1=19 %
\hatcurLChatnetmBeccenxxxxxA
\else
\ifnum#1=20 %
\hatcurLChatnetmBeccenxxxxxB
\else
\ifnum#1=21 %
\hatcurLChatnetmBeccenxxxxxC
\else
??????\fi
\fi
\fi
}
\newcommand{\hatcurLCiblendAeccen}[1]{\ifnum#1=19 %
\hatcurLCiblendAeccenxxxxxA
\else
\ifnum#1=20 %
\hatcurLCiblendAeccenxxxxxB
\else
\ifnum#1=21 %
\hatcurLCiblendAeccenxxxxxC
\else
??????\fi
\fi
\fi
}
\newcommand{\hatcurLCiblendBeccen}[1]{\ifnum#1=19 %
\hatcurLCiblendBeccenxxxxxA
\else
\ifnum#1=20 %
\hatcurLCiblendBeccenxxxxxB
\else
\ifnum#1=21 %
\hatcurLCiblendBeccenxxxxxC
\else
??????\fi
\fi
\fi
}
\newcommand{\hatcurLCimpeccen}[1]{\ifnum#1=19 %
\hatcurLCimpeccenxxxxxA
\else
\ifnum#1=20 %
\hatcurLCimpeccenxxxxxB
\else
\ifnum#1=21 %
\hatcurLCimpeccenxxxxxC
\else
??????\fi
\fi
\fi
}
\newcommand{\hatcurLCingdureccen}[1]{\ifnum#1=19 %
\hatcurLCingdureccenxxxxxA
\else
\ifnum#1=20 %
\hatcurLCingdureccenxxxxxB
\else
\ifnum#1=21 %
\hatcurLCingdureccenxxxxxC
\else
??????\fi
\fi
\fi
}
\newcommand{\hatcurLCPeccen}[1]{\ifnum#1=19 %
\hatcurLCPeccenxxxxxA
\else
\ifnum#1=20 %
\hatcurLCPeccenxxxxxB
\else
\ifnum#1=21 %
\hatcurLCPeccenxxxxxC
\else
??????\fi
\fi
\fi
}
\newcommand{\hatcurLCPprececcen}[1]{\ifnum#1=19 %
\hatcurLCPprececcenxxxxxA
\else
\ifnum#1=20 %
\hatcurLCPprececcenxxxxxB
\else
\ifnum#1=21 %
\hatcurLCPprececcenxxxxxC
\else
??????\fi
\fi
\fi
}
\newcommand{\hatcurLCPshorteccen}[1]{\ifnum#1=19 %
\hatcurLCPshorteccenxxxxxA
\else
\ifnum#1=20 %
\hatcurLCPshorteccenxxxxxB
\else
\ifnum#1=21 %
\hatcurLCPshorteccenxxxxxC
\else
??????\fi
\fi
\fi
}
\newcommand{\hatcurLCqeccen}[1]{\ifnum#1=19 %
\hatcurLCqeccenxxxxxA
\else
\ifnum#1=20 %
\hatcurLCqeccenxxxxxB
\else
\ifnum#1=21 %
\hatcurLCqeccenxxxxxC
\else
??????\fi
\fi
\fi
}
\newcommand{\hatcurLCqshorteccen}[1]{\ifnum#1=19 %
\hatcurLCqshorteccenxxxxxA
\else
\ifnum#1=20 %
\hatcurLCqshorteccenxxxxxB
\else
\ifnum#1=21 %
\hatcurLCqshorteccenxxxxxC
\else
??????\fi
\fi
\fi
}
\newcommand{\hatcurLCrhoeccen}[1]{\ifnum#1=19 %
\hatcurLCrhoeccenxxxxxA
\else
\ifnum#1=20 %
\hatcurLCrhoeccenxxxxxB
\else
\ifnum#1=21 %
\hatcurLCrhoeccenxxxxxC
\else
??????\fi
\fi
\fi
}
\newcommand{\hatcurLCrprstareccen}[1]{\ifnum#1=19 %
\hatcurLCrprstareccenxxxxxA
\else
\ifnum#1=20 %
\hatcurLCrprstareccenxxxxxB
\else
\ifnum#1=21 %
\hatcurLCrprstareccenxxxxxC
\else
??????\fi
\fi
\fi
}
\newcommand{\hatcurLCTAeccen}[1]{\ifnum#1=19 %
\hatcurLCTAeccenxxxxxA
\else
\ifnum#1=20 %
\hatcurLCTAeccenxxxxxB
\else
\ifnum#1=21 %
\hatcurLCTAeccenxxxxxC
\else
??????\fi
\fi
\fi
}
\newcommand{\hatcurLCTBeccen}[1]{\ifnum#1=19 %
\hatcurLCTBeccenxxxxxA
\else
\ifnum#1=20 %
\hatcurLCTBeccenxxxxxB
\else
\ifnum#1=21 %
\hatcurLCTBeccenxxxxxC
\else
??????\fi
\fi
\fi
}
\newcommand{\hatcurLCTeccen}[1]{\ifnum#1=19 %
\hatcurLCTeccenxxxxxA
\else
\ifnum#1=20 %
\hatcurLCTeccenxxxxxB
\else
\ifnum#1=21 %
\hatcurLCTeccenxxxxxC
\else
??????\fi
\fi
\fi
}
\newcommand{\hatcurLCzetaeccen}[1]{\ifnum#1=19 %
\hatcurLCzetaeccenxxxxxA
\else
\ifnum#1=20 %
\hatcurLCzetaeccenxxxxxB
\else
\ifnum#1=21 %
\hatcurLCzetaeccenxxxxxC
\else
??????\fi
\fi
\fi
}
\newcommand{\hatcurPPaequiveccen}[1]{\ifnum#1=19 %
\hatcurPPaequiveccenxxxxxA
\else
\ifnum#1=20 %
\hatcurPPaequiveccenxxxxxB
\else
\ifnum#1=21 %
\hatcurPPaequiveccenxxxxxC
\else
??????\fi
\fi
\fi
}
\newcommand{\hatcurPPareccen}[1]{\ifnum#1=19 %
\hatcurPPareccenxxxxxA
\else
\ifnum#1=20 %
\hatcurPPareccenxxxxxB
\else
\ifnum#1=21 %
\hatcurPPareccenxxxxxC
\else
??????\fi
\fi
\fi
}
\newcommand{\hatcurPPareleccen}[1]{\ifnum#1=19 %
\hatcurPPareleccenxxxxxA
\else
\ifnum#1=20 %
\hatcurPPareleccenxxxxxB
\else
\ifnum#1=21 %
\hatcurPPareleccenxxxxxC
\else
??????\fi
\fi
\fi
}
\newcommand{\hatcurPPfluxapdimeccen}[1]{\ifnum#1=19 %
\hatcurPPfluxapdimeccenxxxxxA
\else
\ifnum#1=20 %
\hatcurPPfluxapdimeccenxxxxxB
\else
\ifnum#1=21 %
\hatcurPPfluxapdimeccenxxxxxC
\else
??????\fi
\fi
\fi
}
\newcommand{\hatcurPPfluxapeccen}[1]{\ifnum#1=19 %
\hatcurPPfluxapeccenxxxxxA
\else
\ifnum#1=20 %
\hatcurPPfluxapeccenxxxxxB
\else
\ifnum#1=21 %
\hatcurPPfluxapeccenxxxxxC
\else
??????\fi
\fi
\fi
}
\newcommand{\hatcurPPfluxavgdimeccen}[1]{\ifnum#1=19 %
\hatcurPPfluxavgdimeccenxxxxxA
\else
\ifnum#1=20 %
\hatcurPPfluxavgdimeccenxxxxxB
\else
\ifnum#1=21 %
\hatcurPPfluxavgdimeccenxxxxxC
\else
??????\fi
\fi
\fi
}
\newcommand{\hatcurPPfluxavgeccen}[1]{\ifnum#1=19 %
\hatcurPPfluxavgeccenxxxxxA
\else
\ifnum#1=20 %
\hatcurPPfluxavgeccenxxxxxB
\else
\ifnum#1=21 %
\hatcurPPfluxavgeccenxxxxxC
\else
??????\fi
\fi
\fi
}
\newcommand{\hatcurPPfluxavglogeccen}[1]{\ifnum#1=19 %
\hatcurPPfluxavglogeccenxxxxxA
\else
\ifnum#1=20 %
\hatcurPPfluxavglogeccenxxxxxB
\else
\ifnum#1=21 %
\hatcurPPfluxavglogeccenxxxxxC
\else
??????\fi
\fi
\fi
}
\newcommand{\hatcurPPfluxperidimeccen}[1]{\ifnum#1=19 %
\hatcurPPfluxperidimeccenxxxxxA
\else
\ifnum#1=20 %
\hatcurPPfluxperidimeccenxxxxxB
\else
\ifnum#1=21 %
\hatcurPPfluxperidimeccenxxxxxC
\else
??????\fi
\fi
\fi
}
\newcommand{\hatcurPPfluxperieccen}[1]{\ifnum#1=19 %
\hatcurPPfluxperieccenxxxxxA
\else
\ifnum#1=20 %
\hatcurPPfluxperieccenxxxxxB
\else
\ifnum#1=21 %
\hatcurPPfluxperieccenxxxxxC
\else
??????\fi
\fi
\fi
}
\newcommand{\hatcurPPgeccen}[1]{\ifnum#1=19 %
\hatcurPPgeccenxxxxxA
\else
\ifnum#1=20 %
\hatcurPPgeccenxxxxxB
\else
\ifnum#1=21 %
\hatcurPPgeccenxxxxxC
\else
??????\fi
\fi
\fi
}
\newcommand{\hatcurPPieccen}[1]{\ifnum#1=19 %
\hatcurPPieccenxxxxxA
\else
\ifnum#1=20 %
\hatcurPPieccenxxxxxB
\else
\ifnum#1=21 %
\hatcurPPieccenxxxxxC
\else
??????\fi
\fi
\fi
}
\newcommand{\hatcurPPloggeccen}[1]{\ifnum#1=19 %
\hatcurPPloggeccenxxxxxA
\else
\ifnum#1=20 %
\hatcurPPloggeccenxxxxxB
\else
\ifnum#1=21 %
\hatcurPPloggeccenxxxxxC
\else
??????\fi
\fi
\fi
}
\newcommand{\hatcurPPmeccen}[1]{\ifnum#1=19 %
\hatcurPPmeccenxxxxxA
\else
\ifnum#1=20 %
\hatcurPPmeccenxxxxxB
\else
\ifnum#1=21 %
\hatcurPPmeccenxxxxxC
\else
??????\fi
\fi
\fi
}
\newcommand{\hatcurPPmeeccen}[1]{\ifnum#1=19 %
\hatcurPPmeeccenxxxxxA
\else
\ifnum#1=20 %
\hatcurPPmeeccenxxxxxB
\else
\ifnum#1=21 %
\hatcurPPmeeccenxxxxxC
\else
??????\fi
\fi
\fi
}
\newcommand{\hatcurPPmelongeccen}[1]{\ifnum#1=19 %
\hatcurPPmelongeccenxxxxxA
\else
\ifnum#1=20 %
\hatcurPPmelongeccenxxxxxB
\else
\ifnum#1=21 %
\hatcurPPmelongeccenxxxxxC
\else
??????\fi
\fi
\fi
}
\newcommand{\hatcurPPmeshorteccen}[1]{\ifnum#1=19 %
\hatcurPPmeshorteccenxxxxxA
\else
\ifnum#1=20 %
\hatcurPPmeshorteccenxxxxxB
\else
\ifnum#1=21 %
\hatcurPPmeshorteccenxxxxxC
\else
??????\fi
\fi
\fi
}
\newcommand{\hatcurPPmlongeccen}[1]{\ifnum#1=19 %
\hatcurPPmlongeccenxxxxxA
\else
\ifnum#1=20 %
\hatcurPPmlongeccenxxxxxB
\else
\ifnum#1=21 %
\hatcurPPmlongeccenxxxxxC
\else
??????\fi
\fi
\fi
}
\newcommand{\hatcurPPmrcorreccen}[1]{\ifnum#1=19 %
\hatcurPPmrcorreccenxxxxxA
\else
\ifnum#1=20 %
\hatcurPPmrcorreccenxxxxxB
\else
\ifnum#1=21 %
\hatcurPPmrcorreccenxxxxxC
\else
??????\fi
\fi
\fi
}
\newcommand{\hatcurPPmshorteccen}[1]{\ifnum#1=19 %
\hatcurPPmshorteccenxxxxxA
\else
\ifnum#1=20 %
\hatcurPPmshorteccenxxxxxB
\else
\ifnum#1=21 %
\hatcurPPmshorteccenxxxxxC
\else
??????\fi
\fi
\fi
}
\newcommand{\hatcurPPperieccen}[1]{\ifnum#1=19 %
\hatcurPPperieccenxxxxxA
\else
\ifnum#1=20 %
\hatcurPPperieccenxxxxxB
\else
\ifnum#1=21 %
\hatcurPPperieccenxxxxxC
\else
??????\fi
\fi
\fi
}
\newcommand{\hatcurPPphiconjeccen}[1]{\ifnum#1=19 %
\hatcurPPphiconjeccenxxxxxA
\else
\ifnum#1=20 %
\hatcurPPphiconjeccenxxxxxB
\else
\ifnum#1=21 %
\hatcurPPphiconjeccenxxxxxC
\else
??????\fi
\fi
\fi
}
\newcommand{\hatcurPPreccen}[1]{\ifnum#1=19 %
\hatcurPPreccenxxxxxA
\else
\ifnum#1=20 %
\hatcurPPreccenxxxxxB
\else
\ifnum#1=21 %
\hatcurPPreccenxxxxxC
\else
??????\fi
\fi
\fi
}
\newcommand{\hatcurPPreeccen}[1]{\ifnum#1=19 %
\hatcurPPreeccenxxxxxA
\else
\ifnum#1=20 %
\hatcurPPreeccenxxxxxB
\else
\ifnum#1=21 %
\hatcurPPreeccenxxxxxC
\else
??????\fi
\fi
\fi
}
\newcommand{\hatcurPPrelongeccen}[1]{\ifnum#1=19 %
\hatcurPPrelongeccenxxxxxA
\else
\ifnum#1=20 %
\hatcurPPrelongeccenxxxxxB
\else
\ifnum#1=21 %
\hatcurPPrelongeccenxxxxxC
\else
??????\fi
\fi
\fi
}
\newcommand{\hatcurPPreshorteccen}[1]{\ifnum#1=19 %
\hatcurPPreshorteccenxxxxxA
\else
\ifnum#1=20 %
\hatcurPPreshorteccenxxxxxB
\else
\ifnum#1=21 %
\hatcurPPreshorteccenxxxxxC
\else
??????\fi
\fi
\fi
}
\newcommand{\hatcurPPrhoeccen}[1]{\ifnum#1=19 %
\hatcurPPrhoeccenxxxxxA
\else
\ifnum#1=20 %
\hatcurPPrhoeccenxxxxxB
\else
\ifnum#1=21 %
\hatcurPPrhoeccenxxxxxC
\else
??????\fi
\fi
\fi
}
\newcommand{\hatcurPPrlongeccen}[1]{\ifnum#1=19 %
\hatcurPPrlongeccenxxxxxA
\else
\ifnum#1=20 %
\hatcurPPrlongeccenxxxxxB
\else
\ifnum#1=21 %
\hatcurPPrlongeccenxxxxxC
\else
??????\fi
\fi
\fi
}
\newcommand{\hatcurPPrshorteccen}[1]{\ifnum#1=19 %
\hatcurPPrshorteccenxxxxxA
\else
\ifnum#1=20 %
\hatcurPPrshorteccenxxxxxB
\else
\ifnum#1=21 %
\hatcurPPrshorteccenxxxxxC
\else
??????\fi
\fi
\fi
}
\newcommand{\hatcurPPtcirceccen}[1]{\ifnum#1=19 %
\hatcurPPtcirceccenxxxxxA
\else
\ifnum#1=20 %
\hatcurPPtcirceccenxxxxxB
\else
\ifnum#1=21 %
\hatcurPPtcirceccenxxxxxC
\else
??????\fi
\fi
\fi
}
\newcommand{\hatcurPPteffeccen}[1]{\ifnum#1=19 %
\hatcurPPteffeccenxxxxxA
\else
\ifnum#1=20 %
\hatcurPPteffeccenxxxxxB
\else
\ifnum#1=21 %
\hatcurPPteffeccenxxxxxC
\else
??????\fi
\fi
\fi
}
\newcommand{\hatcurPPthetaeccen}[1]{\ifnum#1=19 %
\hatcurPPthetaeccenxxxxxA
\else
\ifnum#1=20 %
\hatcurPPthetaeccenxxxxxB
\else
\ifnum#1=21 %
\hatcurPPthetaeccenxxxxxC
\else
??????\fi
\fi
\fi
}
\newcommand{\hatcurPPtinfalleccen}[1]{\ifnum#1=19 %
\hatcurPPtinfalleccenxxxxxA
\else
\ifnum#1=20 %
\hatcurPPtinfalleccenxxxxxB
\else
\ifnum#1=21 %
\hatcurPPtinfalleccenxxxxxC
\else
??????\fi
\fi
\fi
}
\newcommand{\hatcurRVecceneccen}[1]{\ifnum#1=19 %
\hatcurRVecceneccenxxxxxA
\else
\ifnum#1=20 %
\hatcurRVecceneccenxxxxxB
\else
\ifnum#1=21 %
\hatcurRVecceneccenxxxxxC
\else
??????\fi
\fi
\fi
}
\newcommand{\hatcurRVeccentwosiglimeccen}[1]{\ifnum#1=19 %
\hatcurRVeccentwosiglimeccenxxxxxA
\else
\ifnum#1=20 %
\hatcurRVeccentwosiglimeccenxxxxxB
\else
\ifnum#1=21 %
\hatcurRVeccentwosiglimeccenxxxxxC
\else
??????\fi
\fi
\fi
}
\newcommand{\hatcurRVfitrmsAeccen}[1]{\ifnum#1=19 %
\hatcurRVfitrmsAeccenxxxxxA
\else
\ifnum#1=21 %
\hatcurRVfitrmsAeccenxxxxxC
\else
??????\fi
\fi
}
\newcommand{\hatcurRVfitrmsBeccen}[1]{\ifnum#1=19 %
\hatcurRVfitrmsBeccenxxxxxA
\else
\ifnum#1=21 %
\hatcurRVfitrmsBeccenxxxxxC
\else
??????\fi
\fi
}
\newcommand{\hatcurRVfitrmsCeccen}[1]{\ifnum#1=19 %
\hatcurRVfitrmsCeccenxxxxxA
\else
\ifnum#1=21 %
\hatcurRVfitrmsCeccenxxxxxC
\else
??????\fi
\fi
}
\newcommand{\hatcurRVfitrmseccen}[1]{\ifnum#1=20 %
\hatcurRVfitrmseccenxxxxxB
\else
??????\fi
}
\newcommand{\hatcurRVgammaAeccen}[1]{\ifnum#1=19 %
\hatcurRVgammaAeccenxxxxxA
\else
\ifnum#1=21 %
\hatcurRVgammaAeccenxxxxxC
\else
??????\fi
\fi
}
\newcommand{\hatcurRVgammaBeccen}[1]{\ifnum#1=19 %
\hatcurRVgammaBeccenxxxxxA
\else
\ifnum#1=21 %
\hatcurRVgammaBeccenxxxxxC
\else
??????\fi
\fi
}
\newcommand{\hatcurRVgammaCeccen}[1]{\ifnum#1=19 %
\hatcurRVgammaCeccenxxxxxA
\else
\ifnum#1=21 %
\hatcurRVgammaCeccenxxxxxC
\else
??????\fi
\fi
}
\newcommand{\hatcurRVgammaeccen}[1]{\ifnum#1=20 %
\hatcurRVgammaeccenxxxxxB
\else
??????\fi
}
\newcommand{\hatcurRVheccen}[1]{\ifnum#1=19 %
\hatcurRVheccenxxxxxA
\else
\ifnum#1=20 %
\hatcurRVheccenxxxxxB
\else
\ifnum#1=21 %
\hatcurRVheccenxxxxxC
\else
??????\fi
\fi
\fi
}
\newcommand{\hatcurRVjitterAeccen}[1]{\ifnum#1=19 %
\hatcurRVjitterAeccenxxxxxA
\else
\ifnum#1=21 %
\hatcurRVjitterAeccenxxxxxC
\else
??????\fi
\fi
}
\newcommand{\hatcurRVjitterBeccen}[1]{\ifnum#1=19 %
\hatcurRVjitterBeccenxxxxxA
\else
\ifnum#1=21 %
\hatcurRVjitterBeccenxxxxxC
\else
??????\fi
\fi
}
\newcommand{\hatcurRVjitterCeccen}[1]{\ifnum#1=19 %
\hatcurRVjitterCeccenxxxxxA
\else
\ifnum#1=21 %
\hatcurRVjitterCeccenxxxxxC
\else
??????\fi
\fi
}
\newcommand{\hatcurRVjittereccen}[1]{\ifnum#1=20 %
\hatcurRVjittereccenxxxxxB
\else
??????\fi
}
\newcommand{\hatcurRVjittertwosiglimAeccen}[1]{\ifnum#1=19 %
\hatcurRVjittertwosiglimAeccenxxxxxA
\else
\ifnum#1=21 %
\hatcurRVjittertwosiglimAeccenxxxxxC
\else
??????\fi
\fi
}
\newcommand{\hatcurRVjittertwosiglimBeccen}[1]{\ifnum#1=19 %
\hatcurRVjittertwosiglimBeccenxxxxxA
\else
\ifnum#1=21 %
\hatcurRVjittertwosiglimBeccenxxxxxC
\else
??????\fi
\fi
}
\newcommand{\hatcurRVjittertwosiglimCeccen}[1]{\ifnum#1=19 %
\hatcurRVjittertwosiglimCeccenxxxxxA
\else
\ifnum#1=21 %
\hatcurRVjittertwosiglimCeccenxxxxxC
\else
??????\fi
\fi
}
\newcommand{\hatcurRVjittertwosiglimeccen}[1]{\ifnum#1=20 %
\hatcurRVjittertwosiglimeccenxxxxxB
\else
??????\fi
}
\newcommand{\hatcurRVkeccen}[1]{\ifnum#1=19 %
\hatcurRVkeccenxxxxxA
\else
\ifnum#1=20 %
\hatcurRVkeccenxxxxxB
\else
\ifnum#1=21 %
\hatcurRVkeccenxxxxxC
\else
??????\fi
\fi
\fi
}
\newcommand{\hatcurRVKeccen}[1]{\ifnum#1=19 %
\hatcurRVKeccenxxxxxA
\else
\ifnum#1=20 %
\hatcurRVKeccenxxxxxB
\else
\ifnum#1=21 %
\hatcurRVKeccenxxxxxC
\else
??????\fi
\fi
\fi
}
\newcommand{\hatcurRVomegaeccen}[1]{\ifnum#1=19 %
\hatcurRVomegaeccenxxxxxA
\else
\ifnum#1=20 %
\hatcurRVomegaeccenxxxxxB
\else
\ifnum#1=21 %
\hatcurRVomegaeccenxxxxxC
\else
??????\fi
\fi
\fi
}
\newcommand{\hatcurRVrheccen}[1]{\ifnum#1=19 %
\hatcurRVrheccenxxxxxA
\else
\ifnum#1=20 %
\hatcurRVrheccenxxxxxB
\else
\ifnum#1=21 %
\hatcurRVrheccenxxxxxC
\else
??????\fi
\fi
\fi
}
\newcommand{\hatcurRVrkeccen}[1]{\ifnum#1=19 %
\hatcurRVrkeccenxxxxxA
\else
\ifnum#1=20 %
\hatcurRVrkeccenxxxxxB
\else
\ifnum#1=21 %
\hatcurRVrkeccenxxxxxC
\else
??????\fi
\fi
\fi
}
\newcommand{\hatcurRVtroneeccen}[1]{\ifnum#1=19 %
\hatcurRVtroneeccenxxxxxA
\else
\ifnum#1=20 %
\hatcurRVtroneeccenxxxxxB
\else
\ifnum#1=21 %
\hatcurRVtroneeccenxxxxxC
\else
??????\fi
\fi
\fi
}
\newcommand{\hatcurRVtrtwoeccen}[1]{\ifnum#1=19 %
\hatcurRVtrtwoeccenxxxxxA
\else
\ifnum#1=20 %
\hatcurRVtrtwoeccenxxxxxB
\else
\ifnum#1=21 %
\hatcurRVtrtwoeccenxxxxxC
\else
??????\fi
\fi
\fi
}
\newcommand{\hatcurSMEiiloggeccen}[1]{\ifnum#1=19 %
\hatcurSMEiiloggeccenxxxxxA
\else
\ifnum#1=20 %
\hatcurSMEiiloggeccenxxxxxB
\else
\ifnum#1=21 %
\hatcurSMEiiloggeccenxxxxxC
\else
??????\fi
\fi
\fi
}
\newcommand{\hatcurSMEiiteffeccen}[1]{\ifnum#1=19 %
\hatcurSMEiiteffeccenxxxxxA
\else
\ifnum#1=20 %
\hatcurSMEiiteffeccenxxxxxB
\else
\ifnum#1=21 %
\hatcurSMEiiteffeccenxxxxxC
\else
??????\fi
\fi
\fi
}
\newcommand{\hatcurSMEiivsineccen}[1]{\ifnum#1=19 %
\hatcurSMEiivsineccenxxxxxA
\else
\ifnum#1=20 %
\hatcurSMEiivsineccenxxxxxB
\else
\ifnum#1=21 %
\hatcurSMEiivsineccenxxxxxC
\else
??????\fi
\fi
\fi
}
\newcommand{\hatcurSMEiizfeheccen}[1]{\ifnum#1=19 %
\hatcurSMEiizfeheccenxxxxxA
\else
\ifnum#1=20 %
\hatcurSMEiizfeheccenxxxxxB
\else
\ifnum#1=21 %
\hatcurSMEiizfeheccenxxxxxC
\else
??????\fi
\fi
\fi
}
\newcommand{\hatcurSMEiizfehshorteccen}[1]{\ifnum#1=19 %
\hatcurSMEiizfehshorteccenxxxxxA
\else
\ifnum#1=20 %
\hatcurSMEiizfehshorteccenxxxxxB
\else
\ifnum#1=21 %
\hatcurSMEiizfehshorteccenxxxxxC
\else
??????\fi
\fi
\fi
}
\newcommand{\hatcurSMEiloggeccen}[1]{\ifnum#1=19 %
\hatcurSMEiloggeccenxxxxxA
\else
\ifnum#1=20 %
\hatcurSMEiloggeccenxxxxxB
\else
\ifnum#1=21 %
\hatcurSMEiloggeccenxxxxxC
\else
??????\fi
\fi
\fi
}
\newcommand{\hatcurSMEiteffeccen}[1]{\ifnum#1=19 %
\hatcurSMEiteffeccenxxxxxA
\else
\ifnum#1=20 %
\hatcurSMEiteffeccenxxxxxB
\else
\ifnum#1=21 %
\hatcurSMEiteffeccenxxxxxC
\else
??????\fi
\fi
\fi
}
\newcommand{\hatcurSMEivmaceccen}[1]{\ifnum#1=19 %
\hatcurSMEivmaceccenxxxxxA
\else
\ifnum#1=20 %
\hatcurSMEivmaceccenxxxxxB
\else
\ifnum#1=21 %
\hatcurSMEivmaceccenxxxxxC
\else
??????\fi
\fi
\fi
}
\newcommand{\hatcurSMEivmiceccen}[1]{\ifnum#1=19 %
\hatcurSMEivmiceccenxxxxxA
\else
\ifnum#1=20 %
\hatcurSMEivmiceccenxxxxxB
\else
\ifnum#1=21 %
\hatcurSMEivmiceccenxxxxxC
\else
??????\fi
\fi
\fi
}
\newcommand{\hatcurSMEivsineccen}[1]{\ifnum#1=19 %
\hatcurSMEivsineccenxxxxxA
\else
\ifnum#1=20 %
\hatcurSMEivsineccenxxxxxB
\else
\ifnum#1=21 %
\hatcurSMEivsineccenxxxxxC
\else
??????\fi
\fi
\fi
}
\newcommand{\hatcurSMEizfeheccen}[1]{\ifnum#1=19 %
\hatcurSMEizfeheccenxxxxxA
\else
\ifnum#1=20 %
\hatcurSMEizfeheccenxxxxxB
\else
\ifnum#1=21 %
\hatcurSMEizfeheccenxxxxxC
\else
??????\fi
\fi
\fi
}
\newcommand{\hatcurSMEizfehshorteccen}[1]{\ifnum#1=19 %
\hatcurSMEizfehshorteccenxxxxxA
\else
\ifnum#1=20 %
\hatcurSMEizfehshorteccenxxxxxB
\else
\ifnum#1=21 %
\hatcurSMEizfehshorteccenxxxxxC
\else
??????\fi
\fi
\fi
}
\newcommand{\hatcurXAveccen}[1]{\ifnum#1=19 %
\hatcurXAveccenxxxxxA
\else
\ifnum#1=20 %
\hatcurXAveccenxxxxxB
\else
\ifnum#1=21 %
\hatcurXAveccenxxxxxC
\else
??????\fi
\fi
\fi
}
\newcommand{\hatcurXdisteccen}[1]{\ifnum#1=19 %
\hatcurXdisteccenxxxxxA
\else
\ifnum#1=20 %
\hatcurXdisteccenxxxxxB
\else
\ifnum#1=21 %
\hatcurXdisteccenxxxxxC
\else
??????\fi
\fi
\fi
}
\newcommand{\hatcurXdistredeccen}[1]{\ifnum#1=19 %
\hatcurXdistredeccenxxxxxA
\else
\ifnum#1=20 %
\hatcurXdistredeccenxxxxxB
\else
\ifnum#1=21 %
\hatcurXdistredeccenxxxxxC
\else
??????\fi
\fi
\fi
}
\newcommand{\hatcurXEBVeccen}[1]{\ifnum#1=19 %
\hatcurXEBVeccenxxxxxA
\else
\ifnum#1=20 %
\hatcurXEBVeccenxxxxxB
\else
\ifnum#1=21 %
\hatcurXEBVeccenxxxxxC
\else
??????\fi
\fi
\fi
}
\newcommand{\hatcurXjhisoredeccen}[1]{\ifnum#1=19 %
\hatcurXjhisoredeccenxxxxxA
\else
\ifnum#1=20 %
\hatcurXjhisoredeccenxxxxxB
\else
\ifnum#1=21 %
\hatcurXjhisoredeccenxxxxxC
\else
??????\fi
\fi
\fi
}
\newcommand{\hatcurXjkisoredeccen}[1]{\ifnum#1=19 %
\hatcurXjkisoredeccenxxxxxA
\else
\ifnum#1=20 %
\hatcurXjkisoredeccenxxxxxB
\else
\ifnum#1=21 %
\hatcurXjkisoredeccenxxxxxC
\else
??????\fi
\fi
\fi
}
\newcommand{\hatcurXmhisoredeccen}[1]{\ifnum#1=19 %
\hatcurXmhisoredeccenxxxxxA
\else
\ifnum#1=20 %
\hatcurXmhisoredeccenxxxxxB
\else
\ifnum#1=21 %
\hatcurXmhisoredeccenxxxxxC
\else
??????\fi
\fi
\fi
}
\newcommand{\hatcurXmiisoredeccen}[1]{\ifnum#1=19 %
\hatcurXmiisoredeccenxxxxxA
\else
\ifnum#1=20 %
\hatcurXmiisoredeccenxxxxxB
\else
\ifnum#1=21 %
\hatcurXmiisoredeccenxxxxxC
\else
??????\fi
\fi
\fi
}
\newcommand{\hatcurXmjisoredeccen}[1]{\ifnum#1=19 %
\hatcurXmjisoredeccenxxxxxA
\else
\ifnum#1=20 %
\hatcurXmjisoredeccenxxxxxB
\else
\ifnum#1=21 %
\hatcurXmjisoredeccenxxxxxC
\else
??????\fi
\fi
\fi
}
\newcommand{\hatcurXmkisoredeccen}[1]{\ifnum#1=19 %
\hatcurXmkisoredeccenxxxxxA
\else
\ifnum#1=20 %
\hatcurXmkisoredeccenxxxxxB
\else
\ifnum#1=21 %
\hatcurXmkisoredeccenxxxxxC
\else
??????\fi
\fi
\fi
}
\newcommand{\hatcurXmvisoredeccen}[1]{\ifnum#1=19 %
\hatcurXmvisoredeccenxxxxxA
\else
\ifnum#1=20 %
\hatcurXmvisoredeccenxxxxxB
\else
\ifnum#1=21 %
\hatcurXmvisoredeccenxxxxxC
\else
??????\fi
\fi
\fi
}
\newcommand{\hatcurXsecdureccen}[1]{\ifnum#1=19 %
\hatcurXsecdureccenxxxxxA
\else
\ifnum#1=20 %
\hatcurXsecdureccenxxxxxB
\else
\ifnum#1=21 %
\hatcurXsecdureccenxxxxxC
\else
??????\fi
\fi
\fi
}
\newcommand{\hatcurXsecingdureccen}[1]{\ifnum#1=19 %
\hatcurXsecingdureccenxxxxxA
\else
\ifnum#1=20 %
\hatcurXsecingdureccenxxxxxB
\else
\ifnum#1=21 %
\hatcurXsecingdureccenxxxxxC
\else
??????\fi
\fi
\fi
}
\newcommand{\hatcurXsecondaryeccen}[1]{\ifnum#1=19 %
\hatcurXsecondaryeccenxxxxxA
\else
\ifnum#1=20 %
\hatcurXsecondaryeccenxxxxxB
\else
\ifnum#1=21 %
\hatcurXsecondaryeccenxxxxxC
\else
??????\fi
\fi
\fi
}
\newcommand{\hatcurXsecphaseeccen}[1]{\ifnum#1=19 %
\hatcurXsecphaseeccenxxxxxA
\else
\ifnum#1=20 %
\hatcurXsecphaseeccenxxxxxB
\else
\ifnum#1=21 %
\hatcurXsecphaseeccenxxxxxC
\else
??????\fi
\fi
\fi
}
\newcommand{\hatcurXviisoredeccen}[1]{\ifnum#1=19 %
\hatcurXviisoredeccenxxxxxA
\else
\ifnum#1=20 %
\hatcurXviisoredeccenxxxxxB
\else
\ifnum#1=21 %
\hatcurXviisoredeccenxxxxxC
\else
??????\fi
\fi
\fi
}
\newcommand{\hatcurXvkisoredeccen}[1]{\ifnum#1=19 %
\hatcurXvkisoredeccenxxxxxA
\else
\ifnum#1=20 %
\hatcurXvkisoredeccenxxxxxB
\else
\ifnum#1=21 %
\hatcurXvkisoredeccenxxxxxC
\else
??????\fi
\fi
\fi
}

\newcommand{\hatcurxxxxxA}{HATS-19}
\newcommand{\hatcurbxxxxxA}{HATS-19b}
\newcommand{\hatcurcxxxxxA}{HATS-19c}

\newcommand{\hatcurplanetnumxxxxxA}{19}

\newcommand{\hatcurRVgammaabsxxxxxA}{\hatcurRVgammaAeccen{\hatcurplanetnumxxxxxA}}                           

\newcommand{\hatcurRVgammarelxxxxxA}{***TBD***}                           

\newcommand{\hatcurCCtassvixxxxxA}{NULL}                  

\newcommand{\hatcurSMEversionxxxxxA}{ii}                                       

\newcommand{\hatcurisoshortxxxxxA}{YY}
\newcommand{\hatcurisofullxxxxxA}{Yonsei-Yale (YY)}
\newcommand{\hatcurisocitexxxxxA}{yi:2001}

\newcommand{\hatcurlumindxxxxxA}{\arstar}

\newcommand{\hatcurjhkfilsetxxxxxA}{ESO}

\newcommand{\hatcurSMEteffxxxxxA}{\ifthenelse{\equal{\hatcurSMEversionxxxxxA}{i}}{\hatcurSMEiteff{\hatcurplanetnumxxxxxA}}{\hatcurSMEiiteff{\hatcurplanetnumxxxxxA}}}
\newcommand{\hatcurSMEzfehxxxxxA}{\ifthenelse{\equal{\hatcurSMEversionxxxxxA}{i}}{\hatcurSMEizfeh{\hatcurplanetnumxxxxxA}}{\hatcurSMEiizfeh{\hatcurplanetnumxxxxxA}}}
\newcommand{\hatcurSMEzfehshortxxxxxA}{\ifthenelse{\equal{\hatcurSMEversionxxxxxA}{i}}{\hatcurSMEizfehshort{\hatcurplanetnumxxxxxA}}{\hatcurSMEiizfehshort{\hatcurplanetnumxxxxxA}}}
\newcommand{\hatcurSMEloggxxxxxA}{\ifthenelse{\equal{\hatcurSMEversionxxxxxA}{i}}{\hatcurSMEilogg{\hatcurplanetnumxxxxxA}}{\hatcurSMEiilogg{\hatcurplanetnumxxxxxA}}}
\newcommand{\hatcurSMEvsinxxxxxA}{\ifthenelse{\equal{\hatcurSMEversionxxxxxA}{i}}{\hatcurSMEivsin{\hatcurplanetnumxxxxxA}}{\hatcurSMEiivsin{\hatcurplanetnumxxxxxA}}}
\newcommand{\hatcurSMEvmacxxxxxA}{\ifthenelse{\equal{\hatcurSMEversionxxxxxA}{i}}{\hatcurSMEivmac{\hatcurplanetnumxxxxxA}}{\hatcurSMEiivmac{\hatcurplanetnumxxxxxA}}}
\newcommand{\hatcurSMEvmicxxxxxA}{\ifthenelse{\equal{\hatcurSMEversionxxxxxA}{i}}{\hatcurSMEivmic{\hatcurplanetnumxxxxxA}}{\hatcurSMEiivmic{\hatcurplanetnumxxxxxA}}}

\newcommand{\hatcurxxxxxB}{HATS-20}
\newcommand{\hatcurbxxxxxB}{HATS-20b}
\newcommand{\hatcurcxxxxxB}{HATS-20c}

\newcommand{\hatcurplanetnumxxxxxB}{20}

\newcommand{\hatcurRVgammaabsxxxxxB}{\hatcurRVgamma{\hatcurplanetnumxxxxxB}}                           

\newcommand{\hatcurRVgammarelxxxxxB}{\hatcurRVgamma{\hatcurplanetnumxxxxxB}}                           

\newcommand{\hatcurCCtassvixxxxxB}{***TBD***}                  

\newcommand{\hatcurSMEversionxxxxxB}{ii}                                       

\newcommand{\hatcurisoshortxxxxxB}{YY}
\newcommand{\hatcurisofullxxxxxB}{Yonsei-Yale (YY)}
\newcommand{\hatcurisocitexxxxxB}{yi:2001}

\newcommand{\hatcurlumindxxxxxB}{\arstar}

\newcommand{\hatcurjhkfilsetxxxxxB}{ESO}

\newcommand{\hatcurSMEteffxxxxxB}{\ifthenelse{\equal{\hatcurSMEversionxxxxxB}{i}}{\hatcurSMEiteff{\hatcurplanetnumxxxxxB}}{\hatcurSMEiiteff{\hatcurplanetnumxxxxxB}}}
\newcommand{\hatcurSMEzfehxxxxxB}{\ifthenelse{\equal{\hatcurSMEversionxxxxxB}{i}}{\hatcurSMEizfeh{\hatcurplanetnumxxxxxB}}{\hatcurSMEiizfeh{\hatcurplanetnumxxxxxB}}}
\newcommand{\hatcurSMEzfehshortxxxxxB}{\ifthenelse{\equal{\hatcurSMEversionxxxxxB}{i}}{\hatcurSMEizfehshort{\hatcurplanetnumxxxxxB}}{\hatcurSMEiizfehshort{\hatcurplanetnumxxxxxB}}}
\newcommand{\hatcurSMEloggxxxxxB}{\ifthenelse{\equal{\hatcurSMEversionxxxxxB}{i}}{\hatcurSMEilogg{\hatcurplanetnumxxxxxB}}{\hatcurSMEiilogg{\hatcurplanetnumxxxxxB}}}
\newcommand{\hatcurSMEvsinxxxxxB}{\ifthenelse{\equal{\hatcurSMEversionxxxxxB}{i}}{\hatcurSMEivsin{\hatcurplanetnumxxxxxB}}{\hatcurSMEiivsin{\hatcurplanetnumxxxxxB}}}
\newcommand{\hatcurSMEvmacxxxxxB}{\ifthenelse{\equal{\hatcurSMEversionxxxxxB}{i}}{\hatcurSMEivmac{\hatcurplanetnumxxxxxB}}{\hatcurSMEiivmac{\hatcurplanetnumxxxxxB}}}
\newcommand{\hatcurSMEvmicxxxxxB}{\ifthenelse{\equal{\hatcurSMEversionxxxxxB}{i}}{\hatcurSMEivmic{\hatcurplanetnumxxxxxB}}{\hatcurSMEiivmic{\hatcurplanetnumxxxxxB}}}

\newcommand{\hatcurxxxxxC}{HATS-21}
\newcommand{\hatcurbxxxxxC}{HATS-21b}
\newcommand{\hatcurcxxxxxC}{HATS-21c}

\newcommand{\hatcurplanetnumxxxxxC}{21}

\newcommand{\hatcurRVgammaabsxxxxxC}{\hatcurRVgammaC{\hatcurplanetnumxxxxxC}}                           

\newcommand{\hatcurRVgammarelxxxxxC}{\hatcurRVgammaC{\hatcurplanetnumxxxxxC}}                           

\newcommand{\hatcurCCtassvixxxxxC}{NULL}                  

\newcommand{\hatcurSMEversionxxxxxC}{ii}                                       

\newcommand{\hatcurisoshortxxxxxC}{YY}
\newcommand{\hatcurisofullxxxxxC}{Yonsei-Yale (YY)}
\newcommand{\hatcurisocitexxxxxC}{yi:2001}

\newcommand{\hatcurlumindxxxxxC}{\arstar}

\newcommand{\hatcurjhkfilsetxxxxxC}{ESO}

\newcommand{\hatcurSMEteffxxxxxC}{\ifthenelse{\equal{\hatcurSMEversionxxxxxC}{i}}{\hatcurSMEiteff{\hatcurplanetnumxxxxxC}}{\hatcurSMEiiteff{\hatcurplanetnumxxxxxC}}}
\newcommand{\hatcurSMEzfehxxxxxC}{\ifthenelse{\equal{\hatcurSMEversionxxxxxC}{i}}{\hatcurSMEizfeh{\hatcurplanetnumxxxxxC}}{\hatcurSMEiizfeh{\hatcurplanetnumxxxxxC}}}
\newcommand{\hatcurSMEzfehshortxxxxxC}{\ifthenelse{\equal{\hatcurSMEversionxxxxxC}{i}}{\hatcurSMEizfehshort{\hatcurplanetnumxxxxxC}}{\hatcurSMEiizfehshort{\hatcurplanetnumxxxxxC}}}
\newcommand{\hatcurSMEloggxxxxxC}{\ifthenelse{\equal{\hatcurSMEversionxxxxxC}{i}}{\hatcurSMEilogg{\hatcurplanetnumxxxxxC}}{\hatcurSMEiilogg{\hatcurplanetnumxxxxxC}}}
\newcommand{\hatcurSMEvsinxxxxxC}{\ifthenelse{\equal{\hatcurSMEversionxxxxxC}{i}}{\hatcurSMEivsin{\hatcurplanetnumxxxxxC}}{\hatcurSMEiivsin{\hatcurplanetnumxxxxxC}}}
\newcommand{\hatcurSMEvmacxxxxxC}{\ifthenelse{\equal{\hatcurSMEversionxxxxxC}{i}}{\hatcurSMEivmac{\hatcurplanetnumxxxxxC}}{\hatcurSMEiivmac{\hatcurplanetnumxxxxxC}}}
\newcommand{\hatcurSMEvmicxxxxxC}{\ifthenelse{\equal{\hatcurSMEversionxxxxxC}{i}}{\hatcurSMEivmic{\hatcurplanetnumxxxxxC}}{\hatcurSMEiivmic{\hatcurplanetnumxxxxxC}}}

\newcommand{\hatcur}[1]{\ifnum#1=19 %
\hatcurxxxxxA
\else
\ifnum#1=20 %
\hatcurxxxxxB
\else
\ifnum#1=21 %
\hatcurxxxxxC
\else
??????\fi
\fi
\fi
}
\newcommand{\hatcurb}[1]{\ifnum#1=19 %
\hatcurbxxxxxA
\else
\ifnum#1=20 %
\hatcurbxxxxxB
\else
\ifnum#1=21 %
\hatcurbxxxxxC
\else
??????\fi
\fi
\fi
}
\newcommand{\hatcurc}[1]{\ifnum#1=19 %
\hatcurcxxxxxA
\else
\ifnum#1=20 %
\hatcurcxxxxxB
\else
\ifnum#1=21 %
\hatcurcxxxxxC
\else
??????\fi
\fi
\fi
}
\newcommand{\hatcurCCtassvi}[1]{\ifnum#1=19 %
\hatcurCCtassvixxxxxA
\else
\ifnum#1=20 %
\hatcurCCtassvixxxxxB
\else
\ifnum#1=21 %
\hatcurCCtassvixxxxxC
\else
??????\fi
\fi
\fi
}
\newcommand{\hatcurisocite}[1]{\ifnum#1=19 %
\hatcurisocitexxxxxA
\else
\ifnum#1=20 %
\hatcurisocitexxxxxB
\else
\ifnum#1=21 %
\hatcurisocitexxxxxC
\else
??????\fi
\fi
\fi
}
\newcommand{\hatcurisofull}[1]{\ifnum#1=19 %
\hatcurisofullxxxxxA
\else
\ifnum#1=20 %
\hatcurisofullxxxxxB
\else
\ifnum#1=21 %
\hatcurisofullxxxxxC
\else
??????\fi
\fi
\fi
}
\newcommand{\hatcurisoshort}[1]{\ifnum#1=19 %
\hatcurisoshortxxxxxA
\else
\ifnum#1=20 %
\hatcurisoshortxxxxxB
\else
\ifnum#1=21 %
\hatcurisoshortxxxxxC
\else
??????\fi
\fi
\fi
}
\newcommand{\hatcurjhkfilset}[1]{\ifnum#1=19 %
\hatcurjhkfilsetxxxxxA
\else
\ifnum#1=20 %
\hatcurjhkfilsetxxxxxB
\else
\ifnum#1=21 %
\hatcurjhkfilsetxxxxxC
\else
??????\fi
\fi
\fi
}
\newcommand{\hatcurlumind}[1]{\ifnum#1=19 %
\hatcurlumindxxxxxA
\else
\ifnum#1=20 %
\hatcurlumindxxxxxB
\else
\ifnum#1=21 %
\hatcurlumindxxxxxC
\else
??????\fi
\fi
\fi
}
\newcommand{\hatcurplanetnum}[1]{\ifnum#1=19 %
\hatcurplanetnumxxxxxA
\else
\ifnum#1=20 %
\hatcurplanetnumxxxxxB
\else
\ifnum#1=21 %
\hatcurplanetnumxxxxxC
\else
??????\fi
\fi
\fi
}
\newcommand{\hatcurRVgammaabs}[1]{\ifnum#1=19 %
\hatcurRVgammaabsxxxxxA
\else
\ifnum#1=20 %
\hatcurRVgammaabsxxxxxB
\else
\ifnum#1=21 %
\hatcurRVgammaabsxxxxxC
\else
??????\fi
\fi
\fi
}
\newcommand{\hatcurRVgammarel}[1]{\ifnum#1=19 %
\hatcurRVgammarelxxxxxA
\else
\ifnum#1=20 %
\hatcurRVgammarelxxxxxB
\else
\ifnum#1=21 %
\hatcurRVgammarelxxxxxC
\else
??????\fi
\fi
\fi
}
\newcommand{\hatcurSMElogg}[1]{\ifnum#1=19 %
\hatcurSMEloggxxxxxA
\else
\ifnum#1=20 %
\hatcurSMEloggxxxxxB
\else
\ifnum#1=21 %
\hatcurSMEloggxxxxxC
\else
??????\fi
\fi
\fi
}
\newcommand{\hatcurSMEteff}[1]{\ifnum#1=19 %
\hatcurSMEteffxxxxxA
\else
\ifnum#1=20 %
\hatcurSMEteffxxxxxB
\else
\ifnum#1=21 %
\hatcurSMEteffxxxxxC
\else
??????\fi
\fi
\fi
}
\newcommand{\hatcurSMEversion}[1]{\ifnum#1=19 %
\hatcurSMEversionxxxxxA
\else
\ifnum#1=20 %
\hatcurSMEversionxxxxxB
\else
\ifnum#1=21 %
\hatcurSMEversionxxxxxC
\else
??????\fi
\fi
\fi
}
\newcommand{\hatcurSMEvmac}[1]{\ifnum#1=19 %
\hatcurSMEvmacxxxxxA
\else
\ifnum#1=20 %
\hatcurSMEvmacxxxxxB
\else
\ifnum#1=21 %
\hatcurSMEvmacxxxxxC
\else
??????\fi
\fi
\fi
}
\newcommand{\hatcurSMEvmic}[1]{\ifnum#1=19 %
\hatcurSMEvmicxxxxxA
\else
\ifnum#1=20 %
\hatcurSMEvmicxxxxxB
\else
\ifnum#1=21 %
\hatcurSMEvmicxxxxxC
\else
??????\fi
\fi
\fi
}
\newcommand{\hatcurSMEvsin}[1]{\ifnum#1=19 %
\hatcurSMEvsinxxxxxA
\else
\ifnum#1=20 %
\hatcurSMEvsinxxxxxB
\else
\ifnum#1=21 %
\hatcurSMEvsinxxxxxC
\else
??????\fi
\fi
\fi
}
\newcommand{\hatcurSMEzfeh}[1]{\ifnum#1=19 %
\hatcurSMEzfehxxxxxA
\else
\ifnum#1=20 %
\hatcurSMEzfehxxxxxB
\else
\ifnum#1=21 %
\hatcurSMEzfehxxxxxC
\else
??????\fi
\fi
\fi
}
\newcommand{\hatcurSMEzfehshort}[1]{\ifnum#1=19 %
\hatcurSMEzfehshortxxxxxA
\else
\ifnum#1=20 %
\hatcurSMEzfehshortxxxxxB
\else
\ifnum#1=21 %
\hatcurSMEzfehshortxxxxxC
\else
??????\fi
\fi
\fi
}

\newlength{\plotwidthtwo}

\shortauthors{W.~Bhatti, et al.}
\shorttitle{HATS-19\lowercase{b}--\hatcur{21}\lowercase{b}}

\begin{document}

\title{\hatcur{19}\lowercase{b}, \hatcur{20}\lowercase{b},
  \hatcur{21}\lowercase{b}: Three transiting hot-Saturns discovered by
  the HATSouth survey\altaffilmark{$\dagger$}}

\author{
  W.~Bhatti\altaffilmark{1},
  G.~\'A.~Bakos\altaffilmark{1},
  J.~D.~Hartman\altaffilmark{1},
  G.~Zhou\altaffilmark{2},
  K.~Penev\altaffilmark{1},
  D.~Bayliss\altaffilmark{3},
  A.~Jord\'an\altaffilmark{4,5},
  R.~Brahm\altaffilmark{4,5},
  N.~Espinoza\altaffilmark{4,5},
  M.~Rabus\altaffilmark{4,6},
  L.~Mancini\altaffilmark{6},
  M.~de Val-Borro\altaffilmark{1},
  J.~Bento\altaffilmark{7},
  S.~Ciceri\altaffilmark{6},
  Z.~Csubry\altaffilmark{1},
  T.~Henning\altaffilmark{6},
  B.~Schmidt\altaffilmark{7},
  P.~Arriagada\altaffilmark{8},
  R.~P.~Butler\altaffilmark{8},
  J.~Crane\altaffilmark{9},
  S.~Shectman\altaffilmark{9},
  I.~Thompson\altaffilmark{8},
  T.~G.~Tan\altaffilmark{10},
  V.~Suc\altaffilmark{4},
  J.~ L\'az\'ar\altaffilmark{11},
  I.~Papp\altaffilmark{11},
  P.~S\'ari\altaffilmark{11}
}
\altaffiltext{1}{Department of Astrophysical Sciences, 4 Ivy Ln.,
  Princeton, NJ 08544; \url{wbhatti@astro.princeton.edu}}
\altaffiltext{2}{Harvard-Smithsonian Center for Astrophysics,
 Cambridge, MA 02138, USA}
\altaffiltext{3}{Observatoire
 Astronomique de l'Universit\'e de Gen\`eve, 51 ch. des Maillettes,
 1290 Versoix, Switzerland}
\altaffiltext{4}{Instituto de Astrof\'isica, Facultad de F\'isica,
 Pontificia Universidad Cat\'olica de Chile, Av. Vicu\~na Mackenna
 4860, 7820436 Macul, Santiago, Chile}
\altaffiltext{5}{Millennium Institute of Astrophysics, Av. Vicu\~na
 Mackenna 4860, 7820436 Macul, Santiago, Chile}
\altaffiltext{6}{Max Planck Institute for Astronomy, Heidelberg, Germany}
\altaffiltext{7}{Research School of Astronomy and Astrophysics,
 Australian National University, Canberra, ACT 2611, Australia}
\altaffiltext{8}{Department of Terrestrial Magnetism, Carnegie
  Institution of Washington, 5241 Broad Branch Road, NW, Washington, DC
  20015, USA}
\altaffiltext{9}{The Observatories of the Carnegie Institution of
  Washington, 813 Santa Barbara Street, Pasadena, CA 91101, USA}
\altaffiltext{10}{Perth Exoplanet Survey Telescope, Perth, Australia}
\altaffiltext{11}{Hungarian Astronomical Association, Budapest, Hungary}
\altaffiltext{$\dagger$}{
  The HATSouth network is operated by a
  collaboration consisting of Princeton University (PU), the Max Planck
  Institute f\"ur Astronomie (MPIA), the Australian National University
  (ANU), and the Pontificia Universidad Cat\'olica de Chile (PUC).  The
  station at Las Campanas Observatory (LCO) of the Carnegie Institute is
  operated by PU in conjunction with PUC, the station at the High Energy
  Spectroscopic Survey (H.E.S.S.) site is operated in conjunction with
  MPIA, and the station at Siding Spring Observatory (SSO) is operated
  jointly with ANU.
  Based in part on observations made with the MPG~2.2\,m Telescope at
  the ESO Observatory in La Silla.}


\setcounter{footnote}{20}

\begin{abstract}

  We report the discovery by the HATSouth exoplanet survey of three
  hot-Saturn transiting exoplanets: \hatcurb{19}, \hatcurb{20}, and
  \hatcurb{21}. The planet host \hatcur{19} is a slightly evolved $V =
  \hatcurCCtassmvshort{19}$ G0 star with enhanced metallicity of $\feh =
  \hatcurSMEiizfeheccen{19}$, a mass of $\mstar =
  \hatcurISOmeccen{19}$\,\msun\ and a radius of $\rstar =
  \hatcurISOreccen{19}$\,\rsun. \hatcurb{19} is in an eccentric orbit
  ($e = \hatcurRVecceneccen{19}$) around this star with an orbital
  period of \hatcurLCPshorteccen{19} days, and has a mass of $\mpl =
  \hatcurPPmeccen{19}$\,\mjup\ and a highly inflated radius of $\rpl =
  \hatcurPPreccen{19}$\,\rjup. In contrast, the planet \hatcurb{20} has
  a Saturn-like mass and radius of $\mpl = \hatcurPPm{20}$\,\mjup\ and
  $\rpl = \hatcurPPr{20}$\,\rjup\ respectively. It orbits the less
  massive $V = \hatcurCCtassmvshort{20}$ G9V star \hatcur{20} ($\mstar =
  \hatcurISOm{20}$\,\msun; $\rstar = \hatcurISOr{20}$\,\rsun) with a
  period of \hatcurLCPshort{20}\ days. Finally, \hatcur{21} is a
  relatively bright G4V star ($V = \hatcurCCtassmvshort{21}$ mag) with
  super-Solar metallicity of $\feh = \hatcurSMEiizfeh{21}$, a mass of
  $\mstar = \hatcurISOm{21}$\,\msun, and a radius of $\rstar =
  \hatcurISOr{21}$\,\rsun. Its accompanying planet \hatcurb{21} has a
  \hatcurLCPshort{21}-day orbital period, a mass of $\mpl =
  \hatcurPPm{21}$\,\mjup, and a moderately inflated radius of $\rpl =
  \hatcurPPr{21}$\,\rjup. With the addition of these three very
  different planets to the growing sample of hot-Saturns, we re-examine
  the relations between the observed giant planet radii, stellar
  irradiation, and host metallicity. In agreement with earlier results,
  we find that there is a significant positive correlation between
  planet equilibrium temperature and radius, and a weak negative
  correlation between host metallicity and radius. To assess the
  relative influence of various physical parameters on the observed
  planet radii, we train and fit models using Random Forest
  regression. We find that for hot-Saturns ($0.1 < \mpl < 0.5\ M_{\rm
    J}$), the planetary mass and equilibrium temperature play dominant
  roles in determining planet radii. In contrast, for hot-Jupiters ($0.5
  < \mpl < 2.0\ M_{\rm J}$), the most important parameter appears to be
  equilibrium temperature alone. Finally, for irradiated higher-mass
  planets ($\mpl > 2.0\ M_{\rm J}$), we find that equilibrium
  temperature dominates in influence, with smaller contributions from
  the planet mass, and host metallicity.

\end{abstract}

\keywords{ planetary systems --- stars: individual (\hatcur{19}, GSC
  7172-01459, \hatcur{20}, GSC 8247-02184, \hatcur{21}, GSC 8770-00400) ---
  techniques: spectroscopic, photometric }

\section{Introduction}
\label{sec:intro}

The accelerating rate of discovery of transiting exoplanets in the past
decade has been driven by the continuing efforts of ground-based surveys
such as HATNet \citep{bakos:2004:hatnet}, HATSouth
\citep{bakos:2013:hatsouth}, WASP \citep{2006PASP..118.1407P}, KELT
\citep{2007PASP..119..923P}, and the important contributions of space
missions, including CoRoT \citep{2003AdSpR..31..345B} and {\it Kepler}
\citep{2010Sci...327..977B}. It is now becoming increasingly possible to
study populations of exoplanets, characterize trends in their
properties, and perform robust comparisons to theoretical models of
their formulation and evolution. Many of the giant exoplanets (with mass
$M_{\rm p} > 0.1\ M_{\rm J}$) discovered so far have measured radii that
are inflated with respect to Jupiter itself over a large range of
planetary masses. This is an expected outcome based on the small orbital
semi-major axes of the majority of these planets, which are all
significantly affected by stellar irradiation from their host stars. The
details of this mechanism, however, are not fully understood. Possible
candidates include tidal heating \citep{2008ApJ...681.1631J},
opacity-induced inefficiencies in energy transport in the planet
atmosphere \citep{2007ApJ...661..502B}, and several other methods of
depositing energy into the planet interior
\citep{2002A&A...385..156G,2011ApJ...738....1B}, thus inflating its
radius. The effects of these different mechanisms may differ over a
range of planet masses, host star metallicity and luminosity, and
orbital eccentricities, thus allowing us to distinguish between them.

The small number of transiting low mass giant planets, particularly
hot-Saturns ($0.1 < M_{\rm p} < 0.5\ M_{\rm J}$), however, makes the
determination of any definitive trends of planet radius with planet
mass, the level of stellar irradiation, or the host star metallicity,
rather difficult. With the focus of space-based transiting exoplanet
missions shifting to even lower mass Earth-like planets, ground-based
transit surveys have the unique opportunity to deeply explore this
parameter space.

In this paper, we report three transiting Saturn-mass exoplanets in
close orbit around G stars, \hatcurb{19}, \hatcurb{20}, and
\hatcurb{21}, discovered using HATSouth survey observations in 2011 and
2012, and confirmed via subsequent photometric and spectroscopic
follow-up observations.  The HATSouth
survey\footnote{\url{http://hatsouth.org}} achieved first light in 2009,
and has discovered many interesting transiting exoplanetary systems in
the Southern sky since then. Recent highlights include a transiting
hot-Saturn in orbit around an M-dwarf \citep{2015AJ....149..166H} and
the longest period transiting exoplanet discovered by a ground-based
survey so far \citep{2016AJ....151...89B}.

The planets discussed in this work are quite diverse in their
properties. \hatcurb{19} is one of the most highly inflated giant
planets discovered so far, despite its Saturn-like mass. \hatcurb{20} is
a planet much like Saturn itself in mass, density, and radius, despite
being in close orbit around and under significant irradiation from its
host star. Finally, \hatcurb{21} is a significantly inflated Saturn that
orbits a relatively metal-rich host star.

This paper is organized as follows. In \S~\ref{subsec:detection}, we
describe the initial observations leading to the detection of transits
of these three exoplanets. Follow up efforts are described in \S~
\ref{subsec:followup}, including reconnaissance spectroscopy,
high-precision follow-up light curves, lucky imaging to rule out close
companions, and finally, precise radial velocity measurements. We
present our analysis in \S~\ref{sec:analysis}, including determination
of the properties of the host stars (\S~\ref{subsec:hoststars}), ruling
out blends (\S~\ref{subsec:blending}), and final parameters for
\hatcurb{19}, \hatcurb{20}, and \hatcurb{21}
(\S~\ref{subsec:datamodeling}). Finally, in \S~\ref{sec:discussion}, we
discuss these newly discovered planets and investigate the relations
between planet radius and stellar irradiation and stellar metallicity
for a sample of well-characterized transiting giant exoplanets from the
literature, and resulting implications.

\section{Observations}
\label{sec:observations}

\subsection{Initial photometric detection}
\label{subsec:detection}

The six robotic HATSouth telescope units, distributed evenly in
longitude for near-continuous phase coverage, are located (two units per
site) at the Las Campanas Observatory in Chile (LCO), the High Energy
Stereoscopic System (HESS) site in Namibia, and the Siding Spring
Observatory in Australia (SSO). Each telescope unit consists of four
180-mm aperture f/2.8 Takahashi astrograph telescopes backed by 4K
$\times$ 4K Apogee U16M Alta CCDs on a common mount, with a $4^\circ
\times 4^\circ$ field of view per telescope (resulting in a per-unit
combined field-of-view of $8^\circ \times 8^\circ$) and pixel scale of
3$\farcs$7 pixel$^{-1}$. The units observe autonomously from dusk to
dawn, suspending operations as needed during bad weather. During more
than five years of HATSouth operations, we have collected $> 3$ million
frames for $\sim10$ million stars to a limiting $r < 16$ mag, covering
$\sim 13\%$ of the Southern sky.

\hatcur{19}, -20, and -21 are stars observed in the HATSouth primary
fields G606 centered at $\alpha = 09^{\rm h}36^{\rm m}$, $\delta =
-30^{\circ}00^{\prime}$, G700 centered at $\alpha = 13^{\rm h}12^{\rm
  m}$, $\delta = -45^{\circ}00^{\prime}$, and G777 centered at $\alpha =
18^{\rm h}24^{\rm m}$, $\delta = -60^{\circ}00^{\prime}$
respectively. Field G606 was observed by the HATSouth units HS-1 at LCO,
HS-3 at HESS, and HS-5 at SSO from 2011 January to 2012 June. Additional
observations of an overlapping field G607 were taken by the HATSouth
units HS-2 at LCO, HS-4 at HESS, and HS-6 at SSO during 2012
February--June. Field G700 was observed by HS-2, HS-4, and HS-6 from
2011 April to 2012 July. Field G777 was observed by HS-1, HS-3, and HS-5
from 2011 May to 2012 September, with additional observations of an
overlapping field G778 by HATSouth units HS-2, HS-4, and HS-6 during
2011 May to 2012 October.

All photometric observations were reduced to light curves following the
aperture photometry procedures detailed in \citet{bakos:2013:hatsouth}
and \citet{penev:2013:hats1}. Systematics in the light curves were
removed using the External Parameter Decorrelation (EPD;
\citealt{bakos:2010:hat11}) method and the Trend Filtering Algorithm
(TFA; \citealt{kovacs:2005:TFA}). These detrended light curves were then
searched for exoplanet transit signals using the Box-fitting Least
Squares algorithm (BLS; \citealt{kovacs:2002:BLS}).


\begin{figure*}[ht]
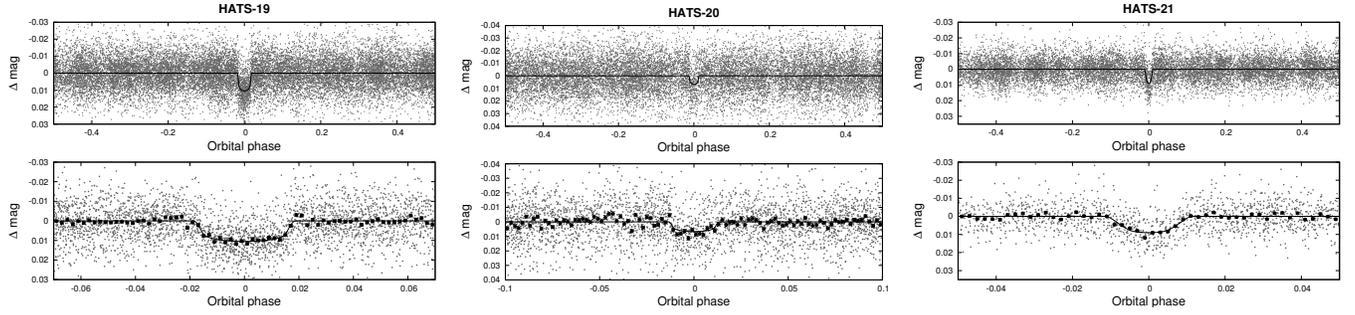

  {
    \centering
    \setlength{\plotwidthtwo}{0.315\linewidth}
    \includegraphics[width={\plotwidthtwo}]{\hatcurhtr{19}-hs-eccen-is}
    \hfil
    \includegraphics[width={\plotwidthtwo}]{\hatcurhtr{20}-hs-is}
    \hfil
    \includegraphics[width={\plotwidthtwo}]{\hatcurhtr{21}-hs-is}
  }
  \caption[]{ Phase-folded unbinned HATSouth light curves for \hatcur{19}
    (left), \hatcur{20} (middle) and \hatcur{21} (right). In each case we show
    two panels. The top panel shows the full light curve, while the
    bottom panel shows the light curve zoomed-in on the transit. The
    solid lines show the model fits to the light curves. The dark filled
    circles in the bottom panels show the light curves binned in phase
    with a bin size of 0.002. \label{fig:hslcs}}
\end{figure*}

We detected a transit signal in combined observations of HATSouth fields
G606 and G607 for the star \hatcur{19} ($\alpha = 09^{\rm h}49^{\rm
  m}37\fs63$, $\delta = -33^{\circ}13^{\prime}06\farcs6$, $V =
\hatcurCCtassmv{19}\ {\rm mag}$, also known as \hatcurCCtwomass{19})
with a depth of 10.3 milli-mag (mmag), a period of \hatcurLCPshort{19}
days, and duration of $4.005$ hours. In total, 19990 Sloan $r$-band
light curve points with 4-minute exposure time were obtained for this
object with a median cadence of $\sim 5$ minutes.

Similarly, a transit signal was detected for the star \hatcur{20}
($\alpha = 13^{\rm h}12^{\rm m}32\fs04$, $\delta =
-45^{\circ}35^{\prime}26\farcs0, V = \hatcurCCtassmv{20}\ {\rm mag}$,
also known as \hatcurCCtwomass{20}) during observations of field G700
with a depth of 8.3 mmag, a period of \hatcurLCPshort{20} days, and
transit duration of $2.453$ hours. The light curve for this object has
total of 16191 4-minute Sloan $r$-band exposures with a median cadence
of $\sim 5$ minutes.

Finally, we detected a transit signal in combined observations of field
G777 and G778 for the star \hatcur{21} ($\alpha = 18^{\rm h}40^{\rm
  m}44\fs40$, $\delta = -58^{\circ}27^{\prime}33\farcs3, V =
\hatcurCCtassmv{21}\ {\rm mag}$, also known as \hatcurCCtwomass{21})
with a depth of 11.1 mmag, duration of $1.978$ hours, and a period of
\hatcurLCPshort{21} days. The light curve for this object has 13106
4-minute Sloan $r$-band light curve points with a 5-minute median
cadence.

Table \ref{tab:photobs} presents a summary of the various HATSouth
photometric observations. Figure \ref{fig:hslcs} shows the discovery
light curves, phase-folded at their respective orbital periods, for all
three transiting systems.

All transit candidates from HATSouth observations of these fields were
then vetted by reconnaissance spectroscopy to determine stellar
parameters and rule out large radial velocity variations indicative of
eclipsing binaries (\S~\ref{subsubsec:reconspec}). \hatcur{19}, -20, and
-21 were identified as promising targets for further photometric
follow-up observations to obtain high-quality light curves and confirm
their transit signals (\S~\ref{subsubsec:followuplc}). Finally, high
precision radial velocity (RV) measurements were carried out for the
three transit candidates (\S~ \ref{subsubsec:preciserv}) to determine
their fundamental properties.


\begin{deluxetable*}{llrrrr}

  \tablewidth{0pc}
  \tabletypesize{\scriptsize}
  \tablecaption{
    Summary of photometric observations\label{tab:photobs}
  }
  \tablehead{
    \multicolumn{1}{c}{Instrument/Field\tablenotemark{a}} &
    \multicolumn{1}{c}{Date(s)} &
    \multicolumn{1}{c}{\# Images} &
    \multicolumn{1}{c}{Cadence\tablenotemark{b}} &
    \multicolumn{1}{c}{Filter} &
    \multicolumn{1}{c}{Precision\tablenotemark{c}} \\
    \multicolumn{1}{c}{} &
    \multicolumn{1}{c}{} &
    \multicolumn{1}{c}{} &
    \multicolumn{1}{c}{(sec)} &
    \multicolumn{1}{c}{} &
    \multicolumn{1}{c}{(mmag)}
  }
  \startdata
  \sidehead{\textbf{\hatcur{19}}}
  ~~~~HS-2.4/G606 & 2012 Feb--2012 Jun & 3702 & 291 & $r$ & 7.8 \\
  ~~~~HS-4.4/G606 & 2012 Mar--2012 Jun & 2154 & 300 & $r$ & 7.7 \\
  ~~~~HS-6.4/G606 & 2012 Feb--2012 Jun & 1164 & 299 & $r$ & 9.4 \\
  ~~~~HS-1.1/G607 & 2011 Jan--2012 Jun & 6735 & 289 & $r$ & 9.2 \\
  ~~~~HS-3.1/G607 & 2011 Jan--2012 Jun & 3180 & 289 & $r$ & 9.9 \\
  ~~~~HS-5.1/G607 & 2011 Jan--2012 Apr & 3055 & 288 & $r$ & 9.3 \\
  ~~~~Swope~1\,m/site3 & 2013 Nov 21 & 117 & 100 & $r$ & 1.6 \\
  ~~~~DK~1.54\,m/DFOSC & 2014 Mar 20 & 104 & 225 & $B$ & 1.2 \\
  ~~~~Swope~1\,m/e2v & 2014 Mar 20 & 168 & 161 & $i$ & 2.3 \\
  \sidehead{\textbf{\hatcur{20}}}
  ~~~~HS-2.1/G700 & 2011 Apr--2012 Jul & 2186 & 292 & $r$ & 19.7 \\
  ~~~~HS-4.1/G700 & 2011 Jul--2012 Jul & 3754 & 301 & $r$ & 14.4 \\
  ~~~~HS-6.1/G700 & 2011 May--2012 Jul & 854 & 300 & $r$ & 28.0 \\
  ~~~~HS-2.4/G700 & 2011 Apr--2012 Jul & 4428 & 292 & $r$ & 11.8 \\
  ~~~~HS-4.4/G700 & 2011 Jul--2012 Jul & 3553 & 301 & $r$ & 11.0 \\
  ~~~~HS-6.4/G700 & 2011 May--2012 Jul & 1416 & 300 & $r$ & 12.9 \\
  ~~~~PEST~0.3\,m & 2015 Apr 23 & 233 & 132 & $R_{C}$ & 4.6 \\
  ~~~~LCOGT~1\,m+SAAO/SBIG & 2015 May 12 & 86 & 145 & $i$ & 2.5 \\
  ~~~~LCOGT~1\,m+CTIO/sinistro & 2015 May 27 & 70 & 226 & $i$ & 2.0 \\
  ~~~~Swope~1\,m/e2v & 2015 May 27 & 113 & 159 & $i$ & 1.5 \\
  \sidehead{\textbf{\hatcur{21}}}
  ~~~~HS-1.3/G777 & 2011 May--2012 Sep & 1519 & 298 & $r$ & 8.2 \\
  ~~~~HS-3.3/G777 & 2011 Jul--2012 Sep & 1632 & 297 & $r$ & 7.1 \\
  ~~~~HS-5.3/G777 & 2011 May--2012 Sep & 1000 & 303 & $r$ & 7.2 \\
  ~~~~HS-2.2/G778 & 2011 May--2012 Nov & 3057 & 288 & $r$ & 6.0 \\
  ~~~~HS-4.2/G778 & 2011 Jul--2012 Nov & 3707 & 298 & $r$ & 6.0 \\
  ~~~~HS-6.2/G778 & 2011 Apr--2012 Oct & 2191 & 298 & $r$ & 7.8 \\
  ~~~~LCOGT~1\,m+SAAO/SBIG & 2015 Jul 15 & 90 & 143 & $i$ & 1.0 \\
  \enddata
  \tablenotetext{a}{
    For HATSouth data we list the HATSouth unit, CCD and field name
    from which the observations are taken. HS-1 and -2 are located at
    Las Campanas Observatory in Chile, HS-3 and -4 are located at the
    H.E.S.S. site in Namibia, and HS-5 and -6 are located at Siding
    Spring Observatory in Australia. Each unit has four CCDs. Each field
    corresponds to one of 838 fixed pointings used to cover the full
    4$\pi$ celestial sphere. All data from a given HATSouth field and
    CCD number are reduced together, while detrending through External
    Parameter Decorrelation (EPD) is done independently for each
    unique unit+CCD+field combination.
  }
  \tablenotetext{b}{
    The median time between consecutive images rounded to the nearest
    second. Due to factors such as weather, the day--night cycle,
    guiding and focus corrections the cadence is only approximately
    uniform over short timescales.
  }
  \tablenotetext{c}{
    The RMS of the residuals from the best-fit model.
  }

\end{deluxetable*}

\subsection{Follow-up observations}
\label{subsec:followup}

\subsubsection{Reconnaissance spectroscopy}
\label{subsubsec:reconspec}

Reconnaissance spectroscopy was performed for all three targets using
the Wide Field Spectrograph (WiFeS; \citealt{dopita:2007}) instrument on
the Australian National University (ANU) 2.3-m telescope at SSO. Details
of reductions for these data are presented in
\citet{bayliss:2013:hats3}; we briefly summarize the process:

The first stage of these observations involved a single spectrum taken
with modest resolution $R \equiv \lambda/\Delta\lambda = 3000$ to
determine if the transit candidate stars were dwarfs, as transit signals
for giant stars with the measured duration from HATSouth light curves
would not indicate possible planetary origin. To this end, we determined
the rough stellar parameters of the targets, including $T_{\rm eff}$,
log $g$, and [Fe/H] by a grid search to minimize $\chi^2$ differences
between the observed spectra and synthetic templates prepared using the
MARCS atmosphere models \citep{2008A&A...486..951G}. From these
observations, \hatcur{19} was found to be a G star with $T_{\rm eff} =
5695 \pm 300$ K, a surface gravity of log $g = 3.6 \pm 0.3$ that is
borderline between that of a dwarf and sub-giant star, and [Fe/H] $= 0.0
\pm 0.5$ dex. Similarly, \hatcur{20} was noted as a G-dwarf star with
$T_{\rm eff} = 5564 \pm 300$ K, log $g = 5.0 \pm 0.3$, and [Fe/H] $= 0.0
\pm 0.5$ dex. Finally, \hatcur{21} was found to be a G-dwarf star with
$T_{\rm eff} = 5322 \pm 300$ K, log $g = 4.3 \pm 0.3$, and [Fe/H] $= 0.0
\pm 0.5$ dex.

The second stage of reconnaissance spectroscopy involved obtaining
spectra at several points in orbital phase for all three targets with
the WiFeS instrument at a slightly higher resolution of $R = 7000$ to
rule out large radial velocity variations ($K > 2$ km
s$^{-1}$). Velocities of this order would indicate high-mass companions
to the target stars, thus invalidating the planetary origin for their
transit signals. No evidence for such variations was found for any of
the three transit candidates.

Details of the reconnaissance spectroscopy observations are presented in
Table \ref{tab:specobs}, while final stellar parameters derived after
global modeling including high-resolution spectra and precise
measurements of radial velocities (\S~\ref{subsubsec:preciserv}) are
listed in Table \ref{tab:starprop}.


\begin{deluxetable*}{llrrrrr}

  \tablewidth{0pc}
  \tabletypesize{\scriptsize}
  \tablecaption{
    Summary of spectroscopy observations\label{tab:specobs}
  }
  \tablehead{
    \multicolumn{1}{c}{Instrument} &
    \multicolumn{1}{c}{UT Date(s)} &
    \multicolumn{1}{c}{\# Spec.}   &
    \multicolumn{1}{c}{Res.}       &
    \multicolumn{1}{c}{S/N Range\tablenotemark{a}} &
    \multicolumn{1}{c}{$\gamma_{\rm RV}$\tablenotemark{b}} &
    \multicolumn{1}{c}{RV Precision\tablenotemark{c}} \\
    &
    &
    &
    \multicolumn{1}{c}{$\lambda$/$\Delta \lambda$/1000} &
    &
    \multicolumn{1}{c}{(\kms)} &
    \multicolumn{1}{c}{(\ms)}
  }
  \startdata
  %
  %
  \sidehead{\textbf{\hatcur{19}}}\\
  ANU~2.3\,m/WiFeS & 2013 Dec 26 & 1 & 3 & 57 & $\cdots$ & $\cdots$ \\
  ANU~2.3\,m/WiFeS & 2013 Dec--2014 Feb & 4 & 7 & 33--79 & 25.9 & 4000 \\
  Euler~1.2\,m/Coralie & 2014 Mar 11--16 & 6 & 60 & 19--23 & 27.456 & 24 \\
  MPG~2.2\,m/FEROS & 2014 Jun--2015 Feb & 12 & 48 & 47--74 & 27.544 & 20 \\
  Magellan~6.5\,m/PFS+I$_{\rm 2}$ & 2014 Dec--2015 Feb & 12 & 76 & 45--55 & $\cdots$ & 18 \\
  Magellan~6.5\,m/PFS & 2015 Jan & 3 & 76 & 59--61 & $\cdots$ & $\cdots$ \\
  \sidehead{\textbf{\hatcur{20}}}\\
  ANU~2.3\,m/WiFeS & 2014 Jun 3 & 1 & 3 & 24 & $\cdots$ & $\cdots$ \\
  ANU~2.3\,m/WiFeS & 2014 Jun 4--5 & 2 & 7 & 3--4 & 19.2 & 4000 \\
  MPG~2.2\,m/FEROS & 2014 Jun--2015 Jul & 10 & 48 & 21--49 & 22.116 & 14 \\
  ESO~3.6\,m/HARPS \tablenotemark{d} & 2015 Apr 6--8 & 3 & 115 & 9--15 & 22.115 & 27 \\
  \sidehead{\textbf{\hatcur{21}}}\\
  ANU~2.3\,m/WiFeS & 2015 Feb 3--8 & 3 & 7 & 24--63 & 31.3 & 4000 \\
  ANU~2.3\,m/WiFeS & 2015 Feb 10 & 1 & 3 & 45 & $\cdots$ & $\cdots$ \\
  ESO~3.6\,m/HARPS\tablenotemark{d} & 2015 Apr--Sep & 3 & 115 & 16--21 & 32.058 & 5.2 \\
  Euler~1.2\,m/Coralie\tablenotemark{d} & 2015 Jun--Sep & 3 & 60 & 13--17 & 32.016 & 40 \\
  Magellan~6.5\,m/PFS+I$_{\rm 2}$ & 2015 Jun--Jul & 7 & 76 & 45--55 & $\cdots$ & 6.9 \\
  Magellan~6.5\,m/PFS           & 2015 Jun & 3 & 76 & 59--61 & $\cdots$ & $\cdots$ \\
  MPG~2.2\,m/FEROS & 2015 Jul--Aug & 8 & 48 & 35--81 & 32.041 & 26 \\
  \enddata
  \tablenotetext{a}{
    S/N per resolution element near 5180\,\AA.
  }
  \tablenotetext{b}{
    For high-precision RV observations included in the orbit
    determination, excluding the PFS+I$_{2}$ observations, this is the
    zero-point RV from the best-fit orbit. For other instruments,
    excluding PFS, it is the mean value. We do not provide this quantity
    for the lower resolution WiFeS observations which were only used to
    measure stellar atmospheric parameters, or for the PFS observations
    for which only relative RVs are measured.
  }
  \tablenotetext{c}{
    For high-precision RV observations included in the orbit
    determination this is the scatter in the RV residuals from the
    best-fit orbit (which may include astrophysical jitter), for other
    instruments this is either an estimate of the precision (not
    including jitter), or the measured standard deviation. We do not
    provide this quantity for low-resolution observations from the
    ANU~2.3\,m/WiFeS, or for the I$_{2}$-free PFS template observations.
  }
  \tablenotetext{d}{
    We excluded the three ESO~3.6\,m/HARPS observations of
    \hatcur{20} from the analysis as these appeared to be unreliable due
    to low S/N and significant sky contamination.  One of the
    ESO~2.2\,m/HARPS observations of \hatcur{21} was excluded also from
    the analysis due to excessive sky contamination, while all three
    Euler~1.2\,m/Coralie observations of \hatcur{21} were excluded due
    to low S/N.
  }

\end{deluxetable*}

\subsubsection{Follow-up light curves}
\label{subsubsec:followuplc}


\begin{figure*}[!ht]
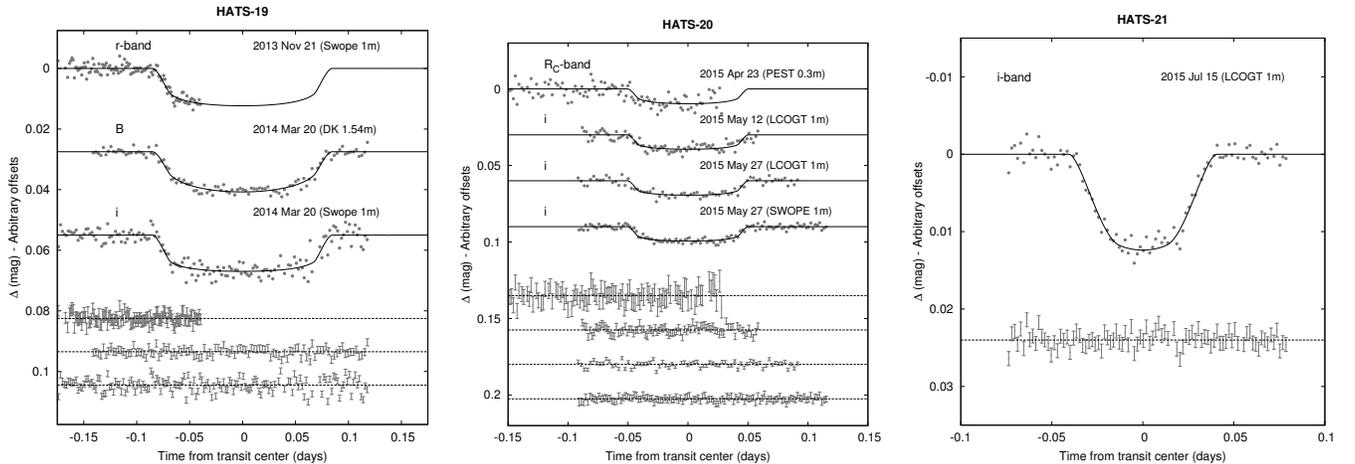

  {
    \centering
    \setlength{\plotwidthtwo}{0.31\linewidth}
    \includegraphics[width={\plotwidthtwo}]{\hatcurhtr{19}-lc-eccen-is}
    \hfil
    \includegraphics[width={\plotwidthtwo}]{\hatcurhtr{20}-lc-is}
    \hfil
    \includegraphics[width={\plotwidthtwo}]{\hatcurhtr{21}-lc-is}
  }
  \caption[]{Unbinned transit follow-up light curves for \hatcur{19} (left),
    \hatcur{20} (middle), and \hatcur{21} (right). The light curves have been
    corrected for quadratic trends in time fitted simultaneously with
    the transit model. The dates of the events, filters and instruments
    used are indicated. Light curves following the first are displaced
    vertically for clarity.  Our best fit from the global modeling
    described in \S~\ref{subsec:datamodeling} is shown by the solid
    lines. The residuals from the best-fit model are shown below in the
    same order as the original light curves.  The error bars represent
    the photon and background shot noise, plus the readout noise.
    \label{fig:followuplcs}}
\end{figure*}

High quality photometric observations with larger telescopes than the
HATSouth instruments were obtained for the three candidates to constrain
transit parameters and refine transit ephemerides. We briefly describe
these observations below. Table \ref{tab:photobs} lists the various
telescopes, instruments, filters, and observing parameters. Figure
\ref{fig:followuplcs} shows all follow-up light curves for \hatcur{19},
-20, and -21 along with transit model fits (discussed further in \S~
\ref{subsec:datamodeling}). Table \ref{tab:allphot} lists all
photometric observations of these three exoplanet candidates, including
the initial HATSouth light curves as well as the follow up observations.

\hatcur{19} was observed with the Swope 1-m telescope and the SITe3
camera on 2013 November 21 (Sloan $r$; ingress event) and 2014 March 20
(Sloan $i$; full transit). Aperture photometry was performed on the
frames following the procedure in \citet{deeg:2001} and
\citet{2016arXiv160302894R}, and differential magnitude light curves
were obtained. Another full transit event for \hatcur{19} was observed
using the Danish 1.54-m telescope, the DFOSC camera, and the $B$ filter,
on 2015 March 20. This observation was carried out with the telescope
defocused to obtain high photometric precision ($\sim1.2$ mmag per
point), and frames were reduced following the procedure described
above. All three follow-up observations showed no variation in transit
depth between the different filters used, greatly improving the
likelihood that this transit candidate is not affected by blending with
a foreground/background eclipsing binary system.

A transit ingress event for \hatcur{20} was observed by the 0.3-m Perth
Exoplanet Survey Telescope (PEST) on 2015 April 23 using a Cousins-$R$
filter. These observations were reduced to light curves following
\citet{zhou:2014:mebs}. After refining ephemerides based on this light
curve, a nearly full-transit event was observed by the 1-m telescope of
the Las Cumbres Observatory Global Telescope network (LCOGT;
\citealt{brown:2013:lcogt}) at the South African Astronomical
Observatory (SAAO) with the SBIG camera and the Sloan $i$ filter on 2015
May 12. Another LCOGT observation of the full transit event took place
on 2015 May 27 using the 1-m telescope at Cerro Tololo Inter-American
Observatory (CTIO) and the Sinistro camera with the Sloan $i$
filter. Calibrated science frames were delivered by the LCOGT pipeline,
which we then reduced to light curves following the procedures described
in \citet{bayliss:2013:hats3}. Finally, we covered yet another
full-transit event including out-of-transit observations using the Swope
1-m telescope and the SITe3 camera with the Sloan $i$ filter. As before,
all transit event depths were achromatic, increasing confidence in the
planetary nature of these observed transits.

Finally, a full-transit event for \hatcur{21} was observed using the
LCOGT 1-m telescope at SAAO and the SBIG camera with the Sloan $i$
filter on 2015 July 15. Due to the relatively deep transit ($\sim 10$
mmag), the ephemerides for this target were well-constrained by the
HATSouth light curve itself, thus a single follow-up light curve was
sufficient to characterize this candidate's transit parameters. These
observations were reduced in a similar manner to those obtained for
\hatcur{20}.


\begin{deluxetable*}{llrrrrl}

  \tablewidth{0pc}
  \tablecaption{
    Light curve data for \hatcur{19}, \hatcur{20}, and \hatcur{21}\label{tab:allphot}.
  }
  \tablehead{
    \colhead{Object\tablenotemark{a}} &
    \colhead{BJD\tablenotemark{b}} &
    \colhead{Mag\tablenotemark{c}} &
    \colhead{\ensuremath{\sigma_{\rm Mag}}} &
    \colhead{Mag(orig)\tablenotemark{d}} &
    \colhead{Filter} &
    \colhead{Instrument} \\
    \colhead{} &
    \colhead{\hbox{~~~~(2,400,000$+$)~~~~}} &
    \colhead{} &
    \colhead{} &
    \colhead{} &
    \colhead{} &
    \colhead{}
  }
  \startdata
  HATS-19 & $ 56045.41332 $ & $  -0.00045 $ & $   0.00431 $ & $ \cdots $ & $ r$ &         HS\\
HATS-19 & $ 55976.86851 $ & $  -0.00197 $ & $   0.00426 $ & $ \cdots $ & $ r$ &         HS\\
HATS-19 & $ 56022.56757 $ & $  -0.00323 $ & $   0.00465 $ & $ \cdots $ & $ r$ &         HS\\
HATS-19 & $ 56017.99833 $ & $   0.00750 $ & $   0.00392 $ & $ \cdots $ & $ r$ &         HS\\
HATS-19 & $ 56086.54423 $ & $   0.00190 $ & $   0.00499 $ & $ \cdots $ & $ r$ &         HS\\
HATS-19 & $ 56068.26668 $ & $  -0.00722 $ & $   0.00416 $ & $ \cdots $ & $ r$ &         HS\\
HATS-19 & $ 56018.00219 $ & $  -0.00090 $ & $   0.00403 $ & $ \cdots $ & $ r$ &         HS\\
HATS-19 & $ 56086.54798 $ & $  -0.00375 $ & $   0.00447 $ & $ \cdots $ & $ r$ &         HS\\
HATS-19 & $ 56022.57275 $ & $   0.01002 $ & $   0.00468 $ & $ \cdots $ & $ r$ &         HS\\
HATS-19 & $ 56068.27017 $ & $  -0.01594 $ & $   0.00415 $ & $ \cdots $ & $ r$ &         HS\\

  \enddata
  \tablenotetext{a}{
    Either \hatcur{19}, \hatcur{20}, or \hatcur{21}.
  }
  \tablenotetext{b}{
    Barycentric Julian Date is computed directly from the UTC time
    without correction for leap seconds.
  }
  \tablenotetext{c}{
    The out-of-transit level has been subtracted. For observations
    made with the HATSouth instruments (identified by ``HS'' in the
    ``Instrument'' column) these magnitudes have been corrected for
    trends using the EPD and TFA procedures applied {\em prior} to
    fitting the transit model. This procedure may lead to an artificial
    dilution in the transit depths. The blend factors for the HATSouth
    light curves are listed in Table~\ref{tab:planetparam}. For observations
    made with follow-up instruments (anything other than ``HS'' in the
    ``Instrument'' column), the magnitudes have been corrected for a
    quadratic trend in time, and for variations correlated with up to
    three PSF shape parameters, fit simultaneously with the transit.
  }
  \tablenotetext{d}{
    Raw magnitude values without correction for the quadratic trend in
    time, or for trends correlated with the seeing. These are only
    reported for the follow-up observations.
  }
  \tablecomments{
    This table is available in a machine-readable form in the online
    journal.  A portion is shown here for guidance regarding its form
    and content.
  }

\end{deluxetable*}

\subsubsection{Imaging to rule out close companions}
\label{subsubsec:lucky}

\begin{figure*}[!th]
  \centering
  \setlength{\plotwidthtwo}{0.8\linewidth}
  \includegraphics[width={\plotwidthtwo}]{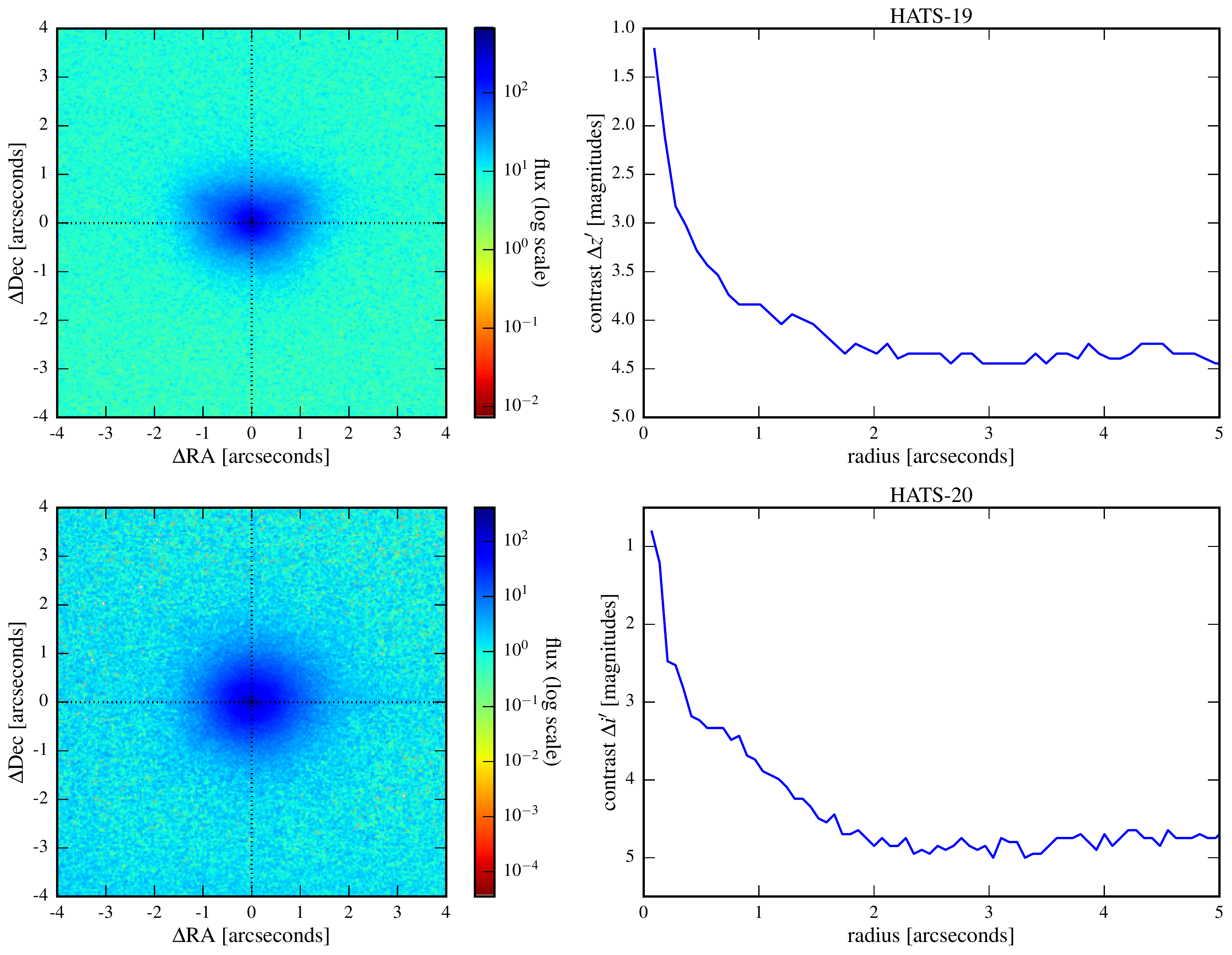}
  \caption{\emph{Upper-left}: Combined $z^{\prime}$-band frame from the
    AstraLux Sur lucky-imaging camera observation of HATS-19, showing no
    detected companions within 5 arcseconds of the target. The slightly
    elongated PSF is a known instrumental effect and was confirmed by
    observing other stars in the field. \emph{Upper-right}: 5$\sigma$
    detection contrast curve for HATS-19 obtained from the AstraLux
    combined frame. \emph{Lower-left}: Combined $i^{\prime}$-band frame
    from the AstraLux Sur lucky-imaging camera observation of HATS-20,
    showing no detected companions within 5 arcseconds of the
    target. \emph{Lower-right}: 5$\sigma$ detection contrast curve for
    HATS-20 obtained from the AstraLux combined
    frame.\label{fig:luckyimaging}}
\end{figure*}

Close companions to potential planet host stars can be a significant
source of extra light if they remain unresolved in photometric
observations of the target systems. These companion stars may in turn be
unresolved multiple star systems on their own. A close-by eclipsing
binary system may produce a diluted eclipse signal that mimics the
characteristic depth and shape of a planetary transit across the face of
the original target star. Photometric follow-up of transit candidates,
therefore, should include high-resolution imaging of the target stars to
rule out the possibility of a false-positive detection by the initial
survey (which generally has very wide-field low angular-resolution
images). For the HATSouth survey, in addition to careful inspection of
follow-up images, we started observing some of our transit candidates
with the AstraLux Sur lucky-imaging instrument
\citep{hippler2009astralux} on the ESO New Technology Telescope (NTT) at
La Silla Observatory in 2015.

We observed HATS-19 on 2015 December 22 with AstraLux Sur and obtained
$10^4$ frames with an exposure time of 70 milliseconds each in the SDSS
$z^\prime$ filter. Similarly, HATS-20 was observed on 2015 December 28
with the AstraLux Sur instrument; we obtained $10^4$ frames with an
exposure time of 30 milliseconds each in the SDSS $i^\prime$
filter. These observations were reduced following the procedure in
\cite{1742-6596-131-1-012051}; the frames with the best 10\% of the
measured Strehl ratio were selected and combined into a single frame
that is oversampled using a drizzle process resulting in a final pixel
scale of $\sim 23.7$ milliarcseconds (mas) pixel$^{-1}$. No companions
around HATS-19 or HATS-20 are detected by inspecting this final image
for both targets (see upper-left panel of Figure
\ref{fig:luckyimaging} for HATS-19, and lower-left panel for HATS-20).

We obtain 5$\sigma$ detection contrast curves for HATS-19 and -20
following the procedure in \citet{2016arXiv160600023E} and the
accompanying code\footnote{Available at
  \url{https://github.com/nespinoza/luckyimg-reduction}.}. We first fit
for the point-spread-function (PSF) of the target, then subtract this
from the image. On the residual image, we then place simulated sources
with PSFs of the derived full-width at half-maximum (FWHM) of the
original star and scaled fluxes corresponding to magnitude contrasts of
$\Delta z^{\prime} = 0$ to 10 mag at various positions on the image
around the location of the target star. Finally, we attempt to recover
these simulated sources, requiring that any detection be 5$\sigma$ above
the background. In this way, we generate a contrast curve placing upper
limits on the brightness of any close companions. We derived an
effective FWHM of the PSF of the HATS-19 observation of $4.84 \pm 0.35$
pixels, which corresponds to $111 \pm 8$ mas. The obtained 5$\sigma$
contrast curve following this procedure is shown in the upper-right
panel of Figure \ref{fig:luckyimaging}. Similarly, we derived an
effective FWHM of the PSF of the HATS-20 observation of $3.39 \pm 0.33$
pixels, which corresponds to $78 \pm 8$ mas. The obtained 5$\sigma$
contrast curve for HATS-20 following this procedure is shown in the
lower-right panel of Figure \ref{fig:luckyimaging}.


\subsubsection{Precise radial velocity measurements}
\label{subsubsec:preciserv}

We observed \hatcur{19}, -20, and -21 with high-resolution spectrographs
to obtain precise measurements of their RVs and stellar parameters, and
thus constrain the orbits and the fundamental properties of the
planetary companions to these stars. These observations are summarized
in Table \ref{tab:specobs}.  We briefly discuss each target's
observations below.

\hatcur{19} was observed extensively using three high-precision RV
instruments. During 2014 March 11--16, we observed this target using the
Coralie instrument on the 1.2-m Euler telescope at
the European Southern Observatory (ESO) at La Silla. Coralie is a high
resolution echelle spectrograph with $R = 60000$. We obtained six
spectra of \hatcur{19} during this observation run; these were reduced
following the procedures laid out in \citet{jordan:2014:hats4}. Twelve
more spectra were obtained for \hatcur{19} during observing runs taking
place in June 2014---Feb 2015 with the FEROS spectrograph
(\citealt{kaufer:1998}; $R = 48000$) on the ESO/MPG 2.2-m telescope at
La Silla. These were reduced using the Coralie pipeline described in
\citet{jordan:2014:hats4} adapted for use with FEROS data. Finally, we
obtained twelve more spectra using the Carnegie Planet Finder
Spectrograph (PFS; \citealt{crane:2010}; $R = 76000$) on the Magellan-II
6.5-m telescope at LCO. These spectra were obtained during observing
runs in December 2014---February 2015, and reduced following
\citet{butler:1996}.


\begin{figure*}[ht]
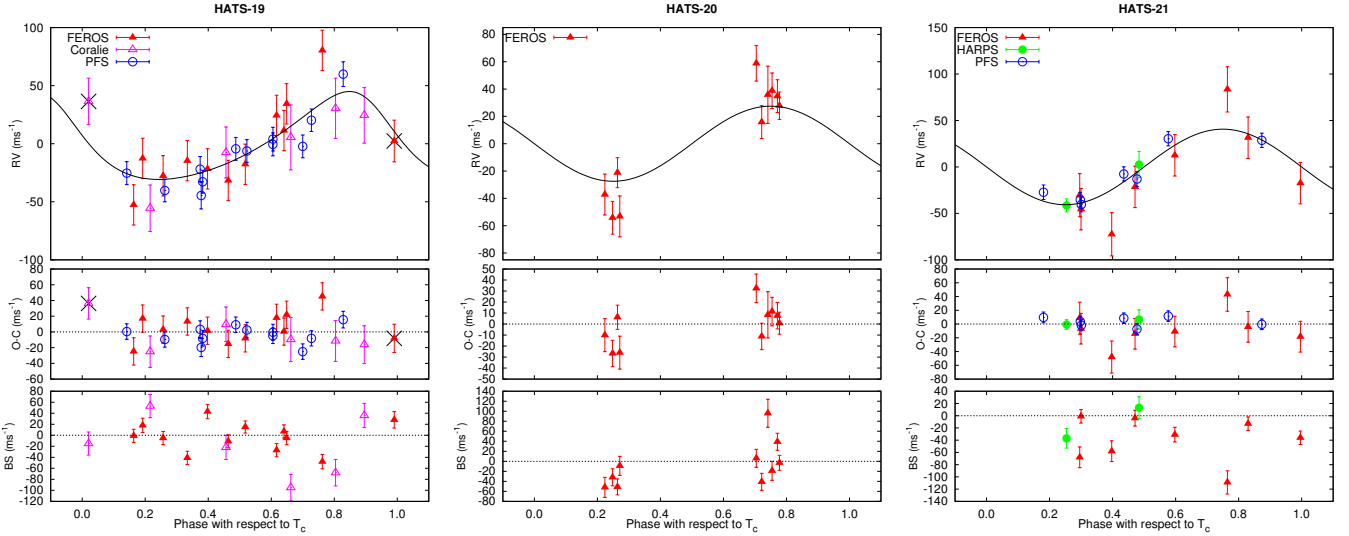

  {
    \centering
    \setlength{\plotwidthtwo}{0.315\linewidth}
    \includegraphics[width={\plotwidthtwo}]{\hatcurhtr{19}-rv-eccen-is}
    \hfil
    \includegraphics[width={\plotwidthtwo}]{\hatcurhtr{20}-rv-is}
    \hfil
    \includegraphics[width={\plotwidthtwo}]{\hatcurhtr{21}-rv-is}
  }
  \caption[]{Phased high-precision RV measurements for \hatcur{19}
    (left), \hatcur{20} (middle), and \hatcur{21} (right). The
    instruments used are labelled in the plots. For \hatcur{19}: two
    observations marked with an X were obtained (partially) in transit
    and have been excluded from the analysis. In each case we show three
    panels. The top panel shows the phased measurements together with
    our best-fit circular-orbit model (see Table \ref{tab:planetparam})
    for each system. Zero-phase corresponds to the time of
    mid-transit. The center-of-mass velocity has been subtracted. The
    second panel shows the velocity $O\!-\!C$ residuals from the best
    fit. The error bars include the jitter terms listed in Table
    \ref{tab:planetparam} added in quadrature to the formal errors for
    each instrument. The third panel shows the bisector spans (BS). Note
    the different vertical scales of the panels. \label{fig:rvcurves}}
\end{figure*}

We obtained ten spectra for \hatcur{20} using the FEROS instrument on
the ESO/MPG 2.2-m telescope at La Silla over the period of June---July
2015. These were reduced using the procedure outlined previously. In
addition, we observed this target using the High Accuracy Radial
Velocity Planet Searcher instrument (HARPS; \citealt{mayor:2003}; $R =
115000$) on the ESO 3.6-m telescope at La Silla during an observing run
on 2015 April 6---8, obtaining three more spectra. These observations
were reduced using the calibration pipeline provided by the instrument
facility. These spectra resulted in unreliable RV measurements due to
low S/N and significant sky contamination, thus were not suitable for
detection of low-amplitude radial velocity variations induced by
Saturn-mass companions. These were were not used for the subsequent
analysis of the system.

Finally, for \hatcur{21}, we obtained seven PFS/Magellan-II spectra
during June---July 2015 and eight FEROS/MPG 2.2-m spectra during
July---August 2015. These were reduced using the procedures outlined
above to obtain precise RV estimates. In addition to these measurements,
we also observed this target with HARPS/ESO 3.6-m during
April---September 2015 (three spectra obtained) and Coralie/Euler 1.2-m
during June---September 2015 (three more spectra obtained). One HARPS
observation and all three Coralie observations turned out to suffer from
low S/N and sky contamination; these were excluded from any further
analysis.

Radial velocity curves phased with the determined orbital periods for
\hatcurb{19}, -20b, and -21b are shown in Figure \ref{fig:rvcurves},
along with the associated bisector spans (BS) measured over the orbital
phases. Table \ref{tab:allrvs} lists all the measured radial velocities
and bisector spans for each of three targets.

\begin{figure*}[ht]
  {
    \centering
    \setlength{\plotwidthtwo}{0.32\linewidth}
    \includegraphics[width={\plotwidthtwo}]{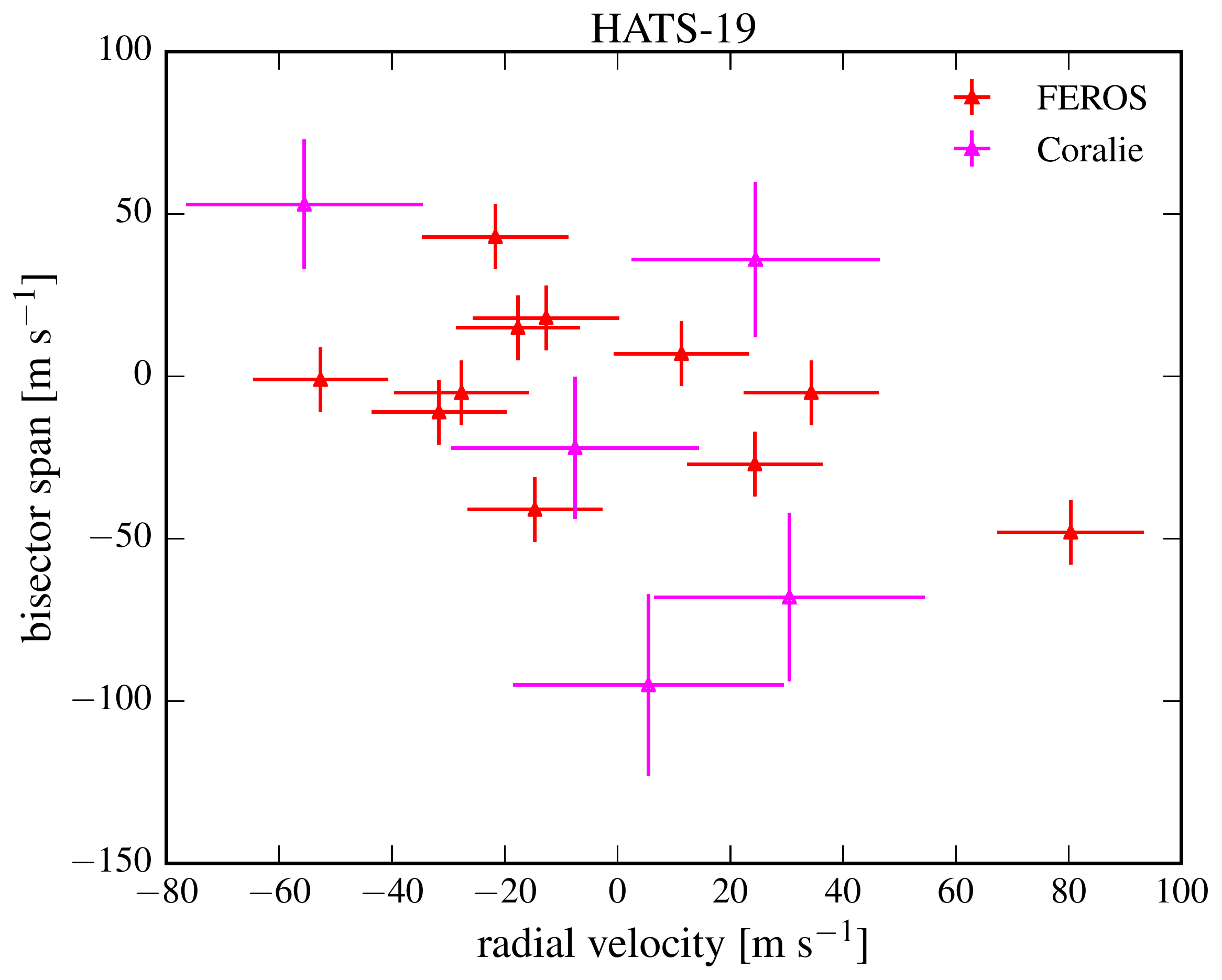}
    \hfil
    \includegraphics[width={\plotwidthtwo}]{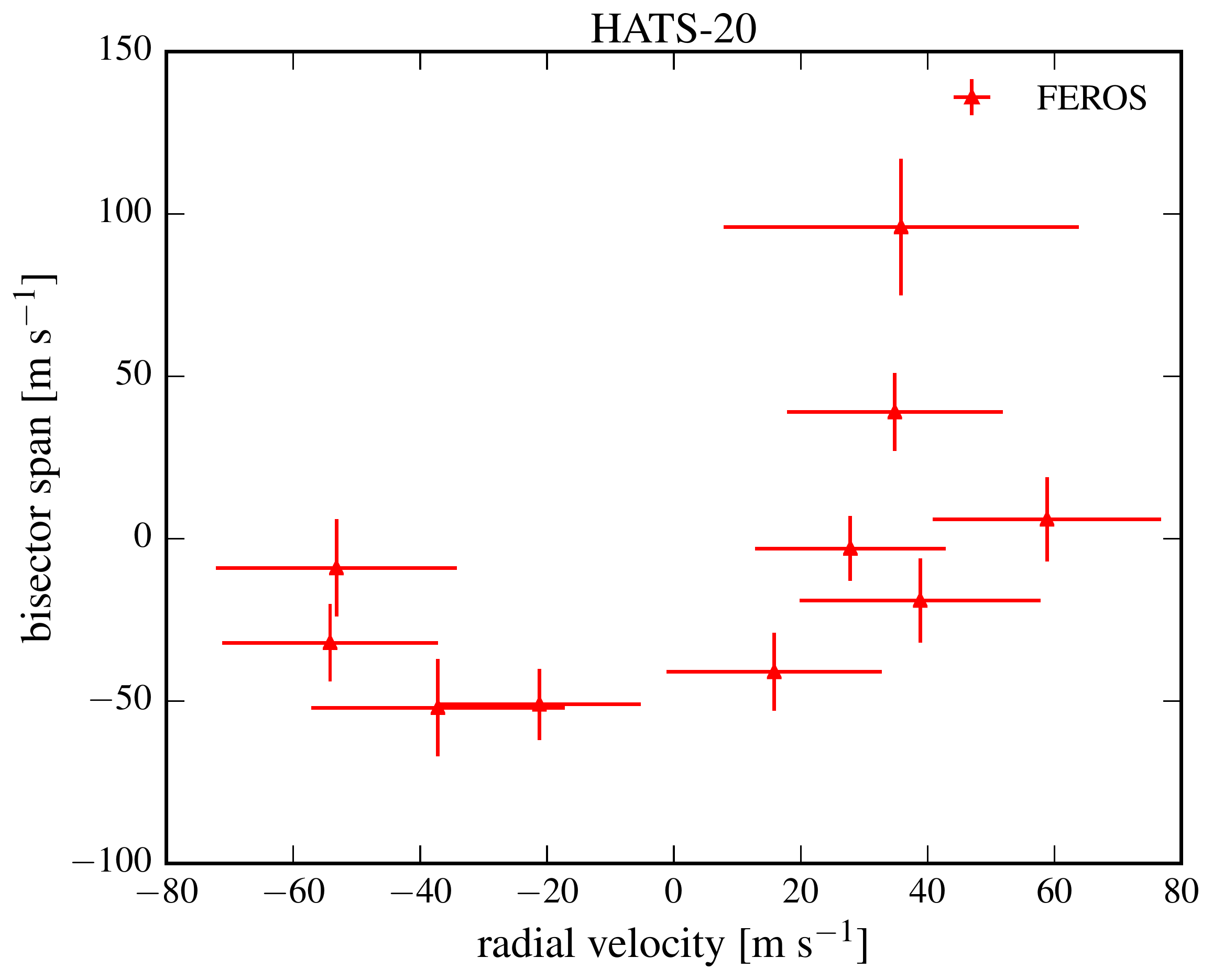}
    \hfil
    \includegraphics[width={\plotwidthtwo}]{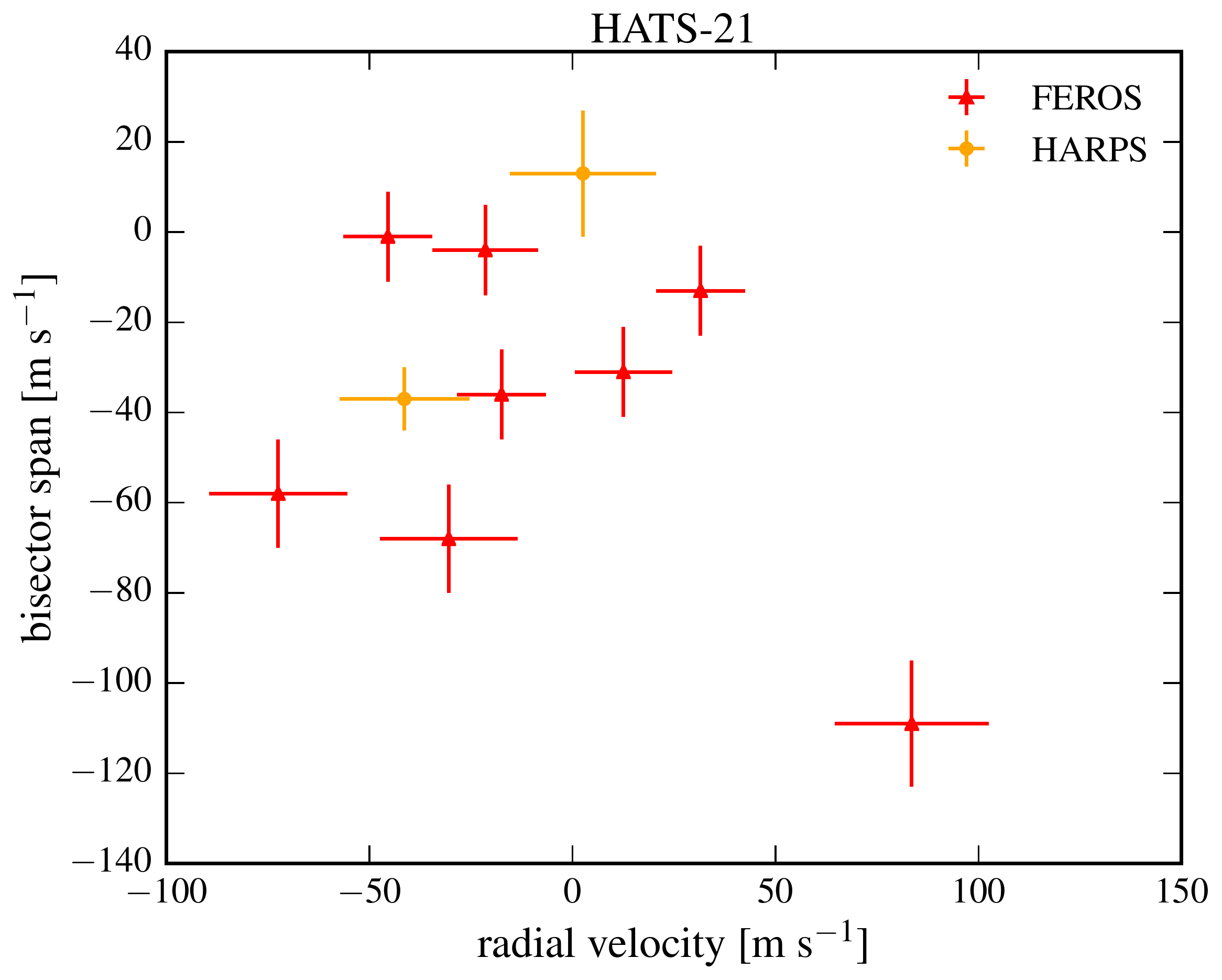}
  }
  \caption[]{Bisector-span and radial velocity correlation plots for
    \hatcur{19} (left), \hatcur{20} (middle), and \hatcur{21}
    (right). The instruments used are labeled in the plots. The center
    of mass radial velocities have been subtracted as in Figure
    \ref{fig:rvcurves}. \label{fig:bsrv}}
\end{figure*}

For \hatcur{19} and \hatcur{20}, there is a hint that the bisector spans
vary in phase with the orbital ephemeris. To check for possible
correlations between the radial velocities and bisector spans, we
calculated the Pearson correlation coefficient for these quantities. The
bootstrap sampling derived 95\%-confidence intervals for the correlation
coefficient are [-0.75, -0.05], [0.22, 0.81], and [-0.77, 0.69] for
\hatcur{19}, \hatcur{20}, and \hatcur{21} respectively. There is a weak
correlation seen between the RVs and BS values for \hatcur{19} and
\hatcur{20} (Figure \ref{fig:bsrv}), therefore detailed photometric
blend modeling is required to rule out the possibility that these
planetary RV signals are false-positives. We discuss this effort in
\S\ \ref{subsec:blending}, where we also discuss likely explanations for
the correlation.


\section{Analysis}
\label{sec:analysis}

\subsection{Stellar parameters for the planet hosts}
\label{subsec:hoststars}

Initial estimates of spectroscopic stellar parameters for \hatcur{19},
\hatcur{20}, and \hatcur{21} are obtained from the WiFeS $R = 3000$
reconnaissance spectra. More reliable estimates of these parameters are
obtained using the Zonal Atmospherical Stellar Parameter Estimator
(ZASPE; Brahm et al. 2016, in prep) method outlined in
\citet{2015AJ....150...33B} and \citet{2016AJ....151...89B}. Briefly:
the $\chi^{2}$ difference between the median-combined observed spectra
from the FEROS instrument and synthetic spectra, generated using the
SPECTRUM code \citep{1999ascl.soft10002G} and model atmospheres from
\citet{2004astro.ph..5087C}, is minimized over a grid to obtain
estimates of $T_{\rm eff}$, log $g$, [Fe/H], and $v \sin i$.
Measurements of these stellar parameters obtained from a run of ZASPE
are combined with constraints on the stellar mean density $\rho_\star$
obtained from the transit light curves (following
\citealt{sozzetti:2007}) and orbital parameters obtained from the RV
measurements during the global modeling of the data
(\S\ \ref{subsec:datamodeling}).

\begin{figure*}[ht]
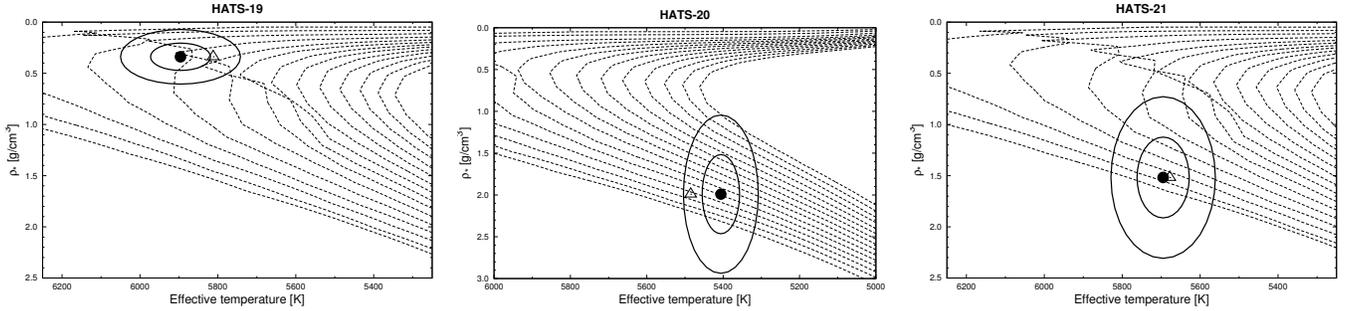

  {
    \centering
    \setlength{\plotwidthtwo}{0.315\linewidth}
    \includegraphics[width={\plotwidthtwo}]{\hatcurhtr{19}-iso-rho-eccen-is}
    \hfil
    \includegraphics[width={\plotwidthtwo}]{\hatcurhtr{20}-iso-rho-is}
    \hfil
    \includegraphics[width={\plotwidthtwo}]{\hatcurhtr{21}-iso-rho-is}
  }
  \caption[]{ Model isochrones from \citet{yi:2001} for the measured
    metallicities of \hatcur{19} (left), \hatcur{20} (middle), and \hatcur{21}
    (right). We show models for ages of 0.2\,Gyr and 1.0 to 14.0\,Gyr in
    1.0\,Gyr increments (ages increasing from left to right). The
    adopted values of T$_{{\rm eff} \star}$ and $\rho_\star$ are shown
    together with their 1$\sigma$ and 2$\sigma$ confidence ellipsoids.
    The initial values of T$_{{\rm eff} \star}$ and $\rho_\star$ from
    the first iteration of ZASPE and light curve/radial velocity
    analyses are represented with a triangle.\label{fig:isochrones}}
\end{figure*}

The physical parameters $R_\star$, $M_\star$ and stellar age, are
determined by comparing \teffstar, \feh, and \rhostar\ to Yonsei-Yale
(Y$^2$; \citealt{yi:2001}) stellar evolution models. This results in a
more precise determination of log $g_\star$. The log $g_\star$ is then
held fixed and another iteration of ZASPE and comparison to ${\rm Y}^2$
models performed to determine the final stellar parameters. See Figure
\ref{fig:isochrones} for comparisons between the resulting estimates of
T$_{{\rm eff} \star}$ and $\rho_\star$ for \hatcur{19}, \hatcur{20}, and
\hatcur{21} and the model isochrones. We derive distances to each of the
systems by comparing the measured magnitudes to those predicted by the
Y$^2$ models in several broadband filters, assuming a $R_{V} = 3.1$
extinction law from \citet{cardelli:1989}. All adopted parameters,
including derived stellar radii, masses, ages, and distances are
detailed in Table \ref{tab:starprop}.


\hatcur{19} is found to be an early G-dwarf (spectral type G0V; based on
tabulations by \citealt{2013ApJS..208....9P} and
\citealt{2007AJ....134.2398C}) with a mass of
$\hatcurISOmlongeccen{19}$\,\msun, radius of
$\hatcurISOrlongeccen{19}$\,\rsun, and log $g_{\star} =
\hatcurISOloggeccen{19}$. The stellar effective temperature $T_{\rm
  eff}$ measured is $\hatcurSMEiiteffeccen{19}$\,K, the measured
metallicity $\feh = \hatcurSMEiizfeheccen{19}$\,dex. The star is near
the end of its main-sequence lifetime; its derived age is
$\hatcurISOageeccen{19}$\,Gyr and its estimated distance is
$\hatcurXdistredeccen{19}$\,pc.

\hatcur{20} is a late G dwarf (G9V) with estimated mass of
$\hatcurISOmlong{20}$\,\msun, radius of $\hatcurISOrlong{20}$\,\rsun,
and log $g_{\star} = \hatcurISOlogg{20}$. The star has $T_{\rm eff} =
\hatcurSMEiiteff{20}$\,K and $\feh = \hatcurSMEiizfeh{20}$\,dex. The star
is $\hatcurISOage{20}$\,Gyr old, and is at an estimated distance of
$\hatcurXdistred{20}$\,pc.

Finally, \hatcur{21} is a mid G-dwarf star (G4V), with a measured mass
of $\hatcurISOmlong{21}$\,\msun\ and an estimated radius of
$\hatcurISOrlong{21}$\,\rsun. The effective temperature is measured to
be $T_{\rm eff} = \hatcurSMEiiteff{21}$\,K, $\feh =
\hatcurSMEiizfeh{21}$\,dex, and log $g_{\star} =
\hatcurISOlogg{21}$. The star is $\hatcurISOage{21}$\,Gyr old, and is at
an estimated distance of $\hatcurXdistred{21}$\,pc.

\subsection{Rotation of the host stars}
\label{subsec:hostrot}

We checked for signatures of stellar rotation in the HATSouth light
curves for \hatcur{19}, \hatcur{20}, and \hatcur{21} by looking for
sinusoidal modulation caused by star spots rotating through the line of
sight. We checked for photometric variability in the EPD de-trended
light curves as well as the TFA de-trended light curves. We checked both
because the non-reconstructive TFA procedure is known to suppress some
light curve modulation that may be astrophysical in nature. The
planetary transits were masked, and we searched for any remaining
periodic signals using the Generalized Lomb Scargle periodogram
\citep{2009A&A...496..577Z}. No significant peaks were found in
periodograms computed using the EPD and TFA light curves for any of the
objects. We note that the target stars appear to be quiet slow-rotating
G stars based on the lack of photometric variability, the relatively
small observed radial velocity jitter, and the long estimated stellar
rotation periods from the spectroscopic measured values of $v \sin i$:
18.5, 29.9, and 19.6 days for \hatcur{19}, -20, and -21 respectively.


\begin{deluxetable*}{lcccl}

  \tablewidth{0pc}
  \tabletypesize{\footnotesize}
  \tablecaption{
    Stellar parameters for \hatcur{19}, \hatcur{20}, and \hatcur{21}.
    \label{tab:starprop}
  }
  \tablehead{
    \multicolumn{1}{c}{} &
    \multicolumn{1}{c}{\bf \hatcur{19}} &
    \multicolumn{1}{c}{\bf \hatcur{20}} &
    \multicolumn{1}{c}{\bf \hatcur{21}} &
    \multicolumn{1}{c}{} \\
    \multicolumn{1}{c}{~~~~~~Parameter~~~~~~} &
    \multicolumn{1}{c}{Value} &
    \multicolumn{1}{c}{Value} &
    \multicolumn{1}{c}{Value} &
    \multicolumn{1}{c}{Source}
  }
  \startdata
  \noalign{\vskip -3pt}
\sidehead{Astrometric properties and cross-identifications}
~~~~2MASS-ID\dotfill               & \hatcurCCtwomasseccen{19}  & \hatcurCCtwomass{20} & \hatcurCCtwomass{21} & \\
~~~~GSC-ID\dotfill                 & \hatcurCCgsceccen{19}      & \hatcurCCgsc{20}     & \hatcurCCgsc{21}     & \\
~~~~R.A. (J2000)\dotfill            & \hatcurCCraeccen{19}       & \hatcurCCra{20}    & \hatcurCCra{21}    & 2MASS\\
~~~~Dec. (J2000)\dotfill            & \hatcurCCdececcen{19}      & \hatcurCCdec{20}   & \hatcurCCdec{21}   & 2MASS\\
~~~~$\mu_{\rm R.A.}$ (\masy)              & \hatcurCCpmraeccen{19}     & \hatcurCCpmra{20} & \hatcurCCpmra{21} & UCAC4\\
~~~~$\mu_{\rm Dec.}$ (\masy)              & \hatcurCCpmdececcen{19}    & \hatcurCCpmdec{20} & \hatcurCCpmdec{21} & UCAC4\\
  \sidehead{Spectroscopic properties}
~~~~$\teffstar$ (K)\dotfill         &  \hatcurSMEiiteff{19}   & \hatcurSMEiiteff{20} & \hatcurSMEiiteff{21} & ZASPE \tablenotemark{a}\\
~~~~$\feh$\dotfill                  &  \hatcurSMEiizfeh{19}   & \hatcurSMEiizfeh{20} & \hatcurSMEiizfeh{21} & ZASPE               \\
~~~~$\vsini$ (\kms)\dotfill         &  \hatcurSMEiivsin{19}   & \hatcurSMEiivsin{20} & \hatcurSMEiivsin{21} & ZASPE                \\
~~~~$\vmac$ (\kms)\dotfill          &  $4.17$   & $3.42$ & $3.86$ & Assumed              \\
~~~~$\vmic$ (\kms)\dotfill          &  $1.16$   & $0.89$ & $1.03$ & Assumed              \\
~~~~$\gamma_{\rm RV}$ (\ms)\dotfill&  \hatcurRVgammaabs{19}  & \hatcurRVgammaabs{20} & \hatcurRVgammaabs{21} & FEROS \tablenotemark{b}  \\
  \sidehead{Photometric properties}
~~~~$B$ (mag)\dotfill               &  \hatcurCCtassmBeccen{19}  & \hatcurCCtassmB{20} & \hatcurCCtassmB{21} & APASS \tablenotemark{c} \\
~~~~$V$ (mag)\dotfill               &  \hatcurCCtassmveccen{19}  & \hatcurCCtassmv{20} & \hatcurCCtassmv{21} & APASS \tablenotemark{c} \\
~~~~$g$ (mag)\dotfill               &  \hatcurCCtassmgeccen{19}  & \hatcurCCtassmg{20} & \hatcurCCtassmg{21} & APASS \tablenotemark{c} \\
~~~~$r$ (mag)\dotfill               &  \hatcurCCtassmreccen{19}  & \hatcurCCtassmr{20} & \hatcurCCtassmr{21} & APASS \tablenotemark{c} \\
~~~~$i$ (mag)\dotfill               &  \hatcurCCtassmieccen{19}  & \hatcurCCtassmi{20} & \hatcurCCtassmi{21} & APASS \tablenotemark{c} \\
~~~~$J$ (mag)\dotfill               &  \hatcurCCtwomassJmageccen{19} & \hatcurCCtwomassJmag{20} & \hatcurCCtwomassJmag{21} & 2MASS           \\
~~~~$H$ (mag)\dotfill               &  \hatcurCCtwomassHmageccen{19} & \hatcurCCtwomassHmag{20} & \hatcurCCtwomassHmag{21} & 2MASS           \\
~~~~$K_s$ (mag)\dotfill             &  \hatcurCCtwomassKmageccen{19} & \hatcurCCtwomassKmag{20} & \hatcurCCtwomassKmag{21} & 2MASS           \\
  \sidehead{Derived properties}
~~~~$\mstar$ ($\msun$)\dotfill      &  \hatcurISOmlongeccen{19}   & \hatcurISOmlong{20} & \hatcurISOmlong{21} & Y$^2$+$\rhostar$+ZASPE \tablenotemark{d}\\
~~~~$\rstar$ ($\rsun$)\dotfill      &  \hatcurISOrlongeccen{19}   & \hatcurISOrlong{20} & \hatcurISOrlong{21} & Y$^2$+$\rhostar$+ZASPE         \\
~~~~$\loggstar$ (cgs)\dotfill       &  \hatcurISOloggeccen{19}    & \hatcurISOlogg{20} & \hatcurISOlogg{21} & Y$^2$+$\rhostar$+ZASPE         \\
~~~~$\rhostar$ (\gcmc) \tablenotemark{e}\dotfill       &  \hatcurLCrhoeccen{19}    & \hatcurLCrho{20} & \hatcurLCrho{21} & Light curves         \\
~~~~$\rhostar$ (\gcmc) \tablenotemark{e}\dotfill       &  \hatcurISOrhoeccen{19}    & \hatcurISOrho{20} & \hatcurISOrho{21} & Y$^2$+Light curves+ZASPE         \\
~~~~$\lstar$ ($\lsun$)\dotfill      &  \hatcurISOlumeccen{19}     & \hatcurISOlum{20} & \hatcurISOlum{21} & Y$^2$+$\rhostar$+ZASPE         \\
~~~~$M_V$ (mag)\dotfill             &  \hatcurISOmveccen{19}      & \hatcurISOmv{20} & \hatcurISOmv{21} & Y$^2$+$\rhostar$+ZASPE         \\
~~~~$M_K$ (mag,\hatcurjhkfilset{19})\dotfill &  \hatcurISOMKeccen{19} & \hatcurISOMK{20} & \hatcurISOMK{21} & Y$^2$+$\rhostar$+ZASPE         \\
~~~~Age (Gyr)\dotfill               &  \hatcurISOageeccen{19}     & \hatcurISOage{20} & \hatcurISOage{21} & Y$^2$+$\rhostar$+ZASPE         \\
~~~~$A_{V}$ (mag)\dotfill               &  \hatcurXAveccen{19}     & \hatcurXAv{20} & \hatcurXAv{21} & Y$^2$+$\rhostar$+ZASPE         \\
~~~~Distance (pc)\dotfill           &  \hatcurXdistredeccen{19}\phn  & \hatcurXdistred{20} & \hatcurXdistred{21} & Y$^2$+$\rhostar$+ZASPE\\
  \enddata

  \tablecomments{
    For each system we adopt the class of model which has the highest
    Bayesian evidence from among those tested. For \hatcur{20} and
    \hatcur{21}, the adopted parameters come from a fit in which the
    orbit is assumed to be circular. For \hatcur{19}, the eccentricity is
    allowed to vary.
  }
  \tablenotetext{a}{
    ZASPE = Zonal Atmospherical Stellar Parameter Estimator routine for
    the analysis of high-resolution spectra (Brahm et al.~2016, in
    preparation), applied to the FEROS spectra. These parameters rely
    primarily on ZASPE, but have a small dependence also on the
    iterative analysis incorporating the isochrone search and global
    modeling of the data.
  }
  \tablenotetext{b}{
    The error on $\gamma_{\rm RV}$ is determined from the orbital fit to
    the FEROS RV measurements, and does not include the systematic
    uncertainty in transforming the velocities from FEROS to the IAU
    standard system. The velocities have not been corrected for
    gravitational redshifts.
  } \tablenotetext{c}{
    From APASS DR6 \citep{henden:2009} as listed in the UCAC 4 catalog
    \citep{zacharias:2012:ucac4}.
  }
  \tablenotetext{d}{
    Y$^{2}$+\rhostar+ZASPE = Based on the Yonsei-Yale isochrones
    \citep{yi:2001}, \rhostar\ as a luminosity indicator, and
    the ZASPE results.
  }
  \tablenotetext{e}{
    In the case of $\rhostar$ we list two values. The first value is
    determined from the global fit to the light curves and RV data,
    without imposing a constraint that the parameters match the
    stellar evolution models. The second value results from
    restricting the posterior distribution to combinations of
    $\rhostar$+$\teffstar$+$\feh$ that match to a Y$^{2}$
    stellar model.
  }

\end{deluxetable*}

\subsection{Ruling out blend scenarios}
\label{subsec:blending}

We carried out an analysis following \citet{hartman:2012:hat39hat41} to
rule out the possibility that our transit detections are instead
unresolved foreground/background stellar eclipsing binary systems
blended with the target systems, thus masquerading as planetary transit
signals in either their light curves or the radial velocity
measurements. We attempt to model the available photometric data
(including light curves and catalog broad-band photometric measurements)
for each object as a blend between an eclipsing binary star system and a
third star along the line of sight. The physical properties of the stars
are constrained using the Padova isochrones \citep{girardi:2000}, while
we also require that the brightest of the three stars in the blend have
atmospheric parameters consistent with those measured with ZASPE for the
target stars \hatcur{19}, -20, and -21. We simulate composite
cross-correlation functions (CCFs) and use them to predict radial
velocity (RV) signals and bisector spans (BS) for each blend scenario
considered.

For all three objects, we find that blend models that cannot be rejected
with at least 4$\sigma$ confidence based on the photometry alone would
have produced RV and/or BS variations in excess of 1\,\kms, or would
have been easily identified as having composite CCFs. The source of the
observed BS--RV correlations remains unclear. The small radial velocity
jitter measured for all three systems, coupled with no photometric
detection of stellar rotation, points to low levels of stellar activity,
likely ruling it out as the cause. Alternatively, any remaining
unsubtracted signal from the sky background inside the spectroscopic
aperture might lead to small variations in BS and RVs that
coincidentally end up being correlated. Radial velocity follow-up
observing runs tend to be clustered in time; the effective radial
velocity signal of the scattered moonlight is similar for many of the
measurements, so it is not uncommon for this effect to result in BS--RV
correlations \citep{2009ApJ...706..785H}. We are especially sensitive to
these small signals due to the low masses and thus smaller radial
velocity amplitudes of these Saturn-mass transit candidates.

Based on our blend analysis, we conclude that all three objects are
indeed transiting planet systems. We cannot, however, exclude the
possibility that one or more of these objects is an unresolved binary
stellar system with one component hosting a short period transiting
planet. The presence of a still unresolved binary star companion to
either \hatcur{19} or \hatcur{20}, despite the null results from lucky
imaging observations presented in \S\ \ref{subsubsec:lucky}, could also
explain the slight BS--RV correlations observed for these systems. For
the remainder of the paper we assume that these are all single stars
with transiting planets, but we note that the radii, and potentially the
masses, of the planets would be larger than what we infer here if
subsequent observations reveal binary star companions.

\subsection{Modeling of the data and resulting planet parameters}
\label{subsec:datamodeling}

We modeled the HATSouth photometry, the follow-up photometry, and the
high-precision RV measurements following
\citet{pal:2008:hat7,bakos:2010:hat11,hartman:2012:hat39hat41}. We fit
\citet{mandel:2002} transit models to the light curves, allowing for a
dilution of the HATSouth transit depth as a result of blending from
neighboring stars and over-correction by the trend-filtering method. To
correct for systematic errors in the follow-up light curves, we include
in our model for each event a quadratic trend in time, and linear trends
with up to three parameters describing the shape of the instrument
point-spread-function (PSF). We fit Keplerian orbits to the RV curves
allowing the zero-point for each instrument to vary independently in the
fit, and allowing for RV jitter which we we also vary as a free
parameter for each instrument. We used a Differential Evolution Markov
Chain Monte Carlo procedure \citep{terbraak:2006, eastman:2013} to
explore the fitness landscape and to determine the posterior
distribution of the parameters. We fit both fixed circular orbits and
free-eccentricity models to the data for all three systems, and then
used the method of \citet{weinberg:2013} to estimate the Bayesian
evidence for each scenario. The final resulting parameters for each
system are listed in Table \ref{tab:planetparam}, and we discuss each
exoplanet briefly below.

We find that for \hatcurb{19}, the free-eccentricity model has the
higher Bayesian evidence (it is 500 times greater), and that this system
has a significant non-zero eccentricity of $e =
\hatcurRVecceneccen{19}$. \hatcurb{19} is more massive than Saturn, with
estimated planetary mass $\mpl = \hatcurPPmlong{19}$\,\mjup, a
rather-inflated planetary radius $\rpl = \hatcurPPrlong{19}$\,\rjup, and
a density $\rho_{p} = \hatcurPPrho{19}$\,\gcmc. The planet's equilibrium
surface temperature $T_{\rm eq}$ (averaged over the orbit, assuming zero
albedo and full redistribution of heat in the planet's atmosphere) is
$\hatcurPPteffeccen{19}$\,K.

For \hatcurb{20}, the fixed circular orbit model has Bayesian evidence
$3$ times greater than a free-eccentricity model, with a 95\%-confidence
upper limit on the eccentricity of
$e\hatcurRVeccentwosiglimeccen{20}$. This planet is less massive than
Saturn, with $\mpl = \hatcurPPmlong{20}$\,\mjup, and a measured
planetary radius $R_{p}$ of $\hatcurPPrlong{20}$\,\rjup. \hatcurb{20}
has a density comparable to that of Saturn itself, with $\rho_{p} =
\hatcurPPrho{20}$\,\gcmc, despite being in close orbit around the host
star (the planet's equilibrium surface temperature $T_{\rm eq}$ is $1147
\pm 36$ K).

Finally, for \hatcurb{21}, the fixed circular orbit model has Bayesian
evidence 120000 times that of an eccentric orbit model and a
95\%-confidence upper limit on the eccentricity of
$e\hatcurRVeccentwosiglimeccen{21}$. This planet is slightly more
massive than Saturn, with $\mpl = \hatcurPPmlong{21}$\,\mjup, an
inflated planetary radius $\rpl = \hatcurPPrlong{21}$\,\rjup, and a
density $\rho_{p} = \hatcurPPrho{21}$\,\gcmc. Its equilibrium surface
temperature $T_{\rm eq}$ is $\hatcurPPteff{21}$\,K.


\begin{deluxetable*}{lccc}

  \tabletypesize{\scriptsize}

  \tablecaption{Orbital and planetary parameters for \hatcurb{19}, \hatcurb{20}
    and \hatcurb{21}\label{tab:planetparam}}

  \tablehead{
    \multicolumn{1}{c}{} &
    \multicolumn{1}{c}{\bf \hatcurb{19}} &
    \multicolumn{1}{c}{\bf \hatcurb{20}} &
    \multicolumn{1}{c}{\bf \hatcurb{21}} \\
    \multicolumn{1}{c}{~~~~~~~~~~~~~Parameter~~~~~~~~~~~~~} &
    \multicolumn{1}{c}{Value} &
    \multicolumn{1}{c}{Value} &
    \multicolumn{1}{c}{Value}
  }
  \startdata
  \noalign{\vskip -3pt}
  \sidehead{\Lc{} parameters}
~~~$P$ (days)             \dotfill    & $\hatcurLCPeccen{19}$ & $\hatcurLCP{20}$ & $\hatcurLCP{21}$ \\
~~~$T_c$ (${\rm BJD}$)
      \tablenotemark{a}   \dotfill    & $\hatcurLCTeccen{19}$ & $\hatcurLCT{20}$ & $\hatcurLCT{21}$ \\
~~~$T_{12}$ (days)
      \tablenotemark{a}   \dotfill    & $\hatcurLCdureccen{19}$ & $\hatcurLCdur{20}$ & $\hatcurLCdur{21}$ \\
~~~$T_{12} = T_{34}$ (days)
      \tablenotemark{a}   \dotfill    & $\hatcurLCingdureccen{19}$ & $\hatcurLCingdur{20}$ & $\hatcurLCingdur{21}$ \\
~~~$\arstar$              \dotfill    & $\hatcurPPareccen{19}$ & $\hatcurPPar{20}$ & $\hatcurPPar{21}$ \\
~~~$\zrstar$ \tablenotemark{b}             \dotfill    & $\hatcurLCzetaeccen{19}$\phn & $\hatcurLCzeta{20}$\phn & $\hatcurLCzeta{21}$\phn \\
~~~$\rpl/\rstar$          \dotfill    & $\hatcurLCrprstareccen{19}$ & $\hatcurLCrprstar{20}$ & $\hatcurLCrprstar{21}$ \\
~~~$b^2$                  \dotfill    & $\hatcurLCbsqeccen{19}$ & $\hatcurLCbsq{20}$ & $\hatcurLCbsq{21}$ \\
~~~$b \equiv a \cos i/\rstar$
                          \dotfill    & $\hatcurLCimpeccen{19}$ & $\hatcurLCimp{20}$ & $\hatcurLCimp{21}$ \\
~~~$i$ (deg)              \dotfill    & $\hatcurPPieccen{19}$\phn & $\hatcurPPi{20}$\phn & $\hatcurPPi{21}$\phn \\

  \sidehead{HATSouth blend factors \tablenotemark{c}}
~~~Blend factor 1 \dotfill & $\hatcurLCiblendA{19}$ & $\hatcurLCiblendA{20}$ & $\hatcurLCiblendA{21}$ \\
~~~Blend factor 2 \dotfill & $\hatcurLCiblendB{19}$ & $\hatcurLCiblendB{20}$ & $\hatcurLCiblendB{21}$ \\

  \sidehead{Limb-darkening coefficients \tablenotemark{d}}
~~~$c_1,B$ (linear term)    \dotfill    & $0.6289$ & $\cdots$ & $\cdots$ \\
~~~$c_2,B$ (quadratic term) \dotfill    & $0.1892$ & $\cdots$ & $\cdots$ \\
~~~$c_1,R$                  \dotfill    & $\cdots$ & $\hatcurLBiR{20}$ & $\cdots$ \\
~~~$c_2,R$                  \dotfill    & $\cdots$ & $\hatcurLBiiR{20}$ & $\cdots$ \\
~~~$c_1,r$                  \dotfill    & $\hatcurLBireccen{19}$ & $\hatcurLBir{20}$ & $\hatcurLBir{21}$ \\
~~~$c_2,r$                  \dotfill    & $\hatcurLBiireccen{19}$ & $\hatcurLBiir{20}$ & $\hatcurLBiir{21}$ \\
~~~$c_1,i$                  \dotfill    & $\hatcurLBiieccen{19}$ & $\hatcurLBii{20}$ & $\hatcurLBii{21}$ \\
~~~$c_2,i$                  \dotfill    & $\hatcurLBiiieccen{19}$ & $\hatcurLBiii{20}$ & $\hatcurLBiii{21}$ \\

  \sidehead{RV parameters}
~~~$K$ (\ms)              \dotfill    & $\hatcurRVKeccen{19}$\phn\phn & $\hatcurRVK{20}$\phn\phn & $\hatcurRVK{21}$\phn\phn \\
~~~$e$ \tablenotemark{e}               \dotfill    & $\hatcurRVecceneccen{19}$ & $\hatcurRVeccentwosiglimeccen{20}$ & $\hatcurRVeccentwosiglimeccen{21}$ \\
~~~$\omega$ (deg) \dotfill    & $\hatcurRVomegaeccen{19}$ & $\cdots$ & $\cdots$ \\
~~~$\sqrt{e}\cos\omega$               \dotfill    & $\hatcurRVrkeccen{19}$ & $\cdots$ & $\cdots$ \\
~~~$\sqrt{e}\sin\omega$               \dotfill    & $\hatcurRVrheccen{19}$ & $\cdots$ & $\cdots$ \\
~~~$e\cos\omega$               \dotfill    & $\hatcurRVkeccen{19}$ & $\cdots$ & $\cdots$ \\
~~~$e\sin\omega$               \dotfill    & $\hatcurRVheccen{19}$ & $\cdots$ & $\cdots$ \\
~~~RV jitter FEROS (\ms) \tablenotemark{f}       \dotfill    & \hatcurRVjitterAeccen{19} & \hatcurRVjittertwosiglim{20} & \hatcurRVjitterA{21} \\
~~~RV jitter HARPS (\ms)        \dotfill    & $\cdots$ & $\cdots$ & \hatcurRVjittertwosiglimB{21} \\
~~~RV jitter Coralie (\ms)        \dotfill    & \hatcurRVjittertwosiglimBeccen{19} & $\cdots$ & $\cdots$ \\
~~~RV jitter PFS (\ms)        \dotfill    & \hatcurRVjitterCeccen{19} & $\cdots$ & \hatcurRVjitterC{21} \\

  \sidehead{Planetary parameters}
~~~$\mpl$ ($\mjup$)       \dotfill    & $\hatcurPPmlongeccen{19}$ & $\hatcurPPmlong{20}$ & $\hatcurPPmlong{21}$ \\
~~~$\rpl$ ($\rjup$)       \dotfill    & $\hatcurPPrlongeccen{19}$ & $\hatcurPPrlong{20}$ & $\hatcurPPrlong{21}$ \\
~~~$C(\mpl,\rpl)$
    \tablenotemark{g}     \dotfill    & $\hatcurPPmrcorreccen{19}$ & $\hatcurPPmrcorr{20}$ & $\hatcurPPmrcorr{21}$ \\
~~~$\rhopl$ (\gcmc)       \dotfill    & $\hatcurPPrhoeccen{19}$ & $\hatcurPPrho{20}$ & $\hatcurPPrho{21}$ \\
~~~$\log g_p$ (cgs)       \dotfill    & $\hatcurPPloggeccen{19}$ & $\hatcurPPlogg{20}$ & $\hatcurPPlogg{21}$ \\
~~~$a$ (AU)               \dotfill    & $\hatcurPPareleccen{19}$ & $\hatcurPParel{20}$ & $\hatcurPParel{21}$ \\
~~~$T_{\rm eq}$ (K)        \dotfill   & $\hatcurPPteffeccen{19}$ & $\hatcurPPteff{20}$ & $\hatcurPPteff{21}$ \\
~~~$\Theta$ \tablenotemark{h} \dotfill & $\hatcurPPthetaeccen{19}$ & $\hatcurPPtheta{20}$ & $\hatcurPPtheta{21}$ \\
  %
~~~$\log_{10}\langle F \rangle$ (cgs) \tablenotemark{i}
                          \dotfill    & $\hatcurPPfluxavglogeccen{19}$ & $\hatcurPPfluxavglog{20}$ & $\hatcurPPfluxavglog{21}$ \\ [-1.5ex]
  \enddata
  \tablenotetext{a}{
    Times are in Barycentric Julian Date calculated directly from UTC
    {\em without} correction for leap seconds.
    \ensuremath{T_c}: Reference epoch of
    mid transit that minimizes the correlation with the orbital
    period.
    \ensuremath{T_{14}}: total transit duration, time
    between first to last contact;
    \ensuremath{T_{12}=T_{34}}: ingress/egress time, time between first
    and second, or third and fourth contact.
  }
  \tablecomments{
    For each system we adopt the class of model which has the highest
    Bayesian evidence from among those tested. For \hatcurb{20} and
    \hatcurb{21} the adopted parameters come from a fit in which the orbit
    is assumed to be circular. For \hatcurb{19} the eccentricity is allowed
    to vary.
  }
  \tablenotetext{b}{
    Reciprocal of the half duration of the transit used as a jump
    parameter in our MCMC analysis in place of $\arstar$. It is related to
    $\arstar$ by the expression $\zrstar =
    \arstar(2\pi(1+e\sin\omega))/(P\sqrt{1-b^2}\sqrt{1-e^2})$
    \citep{bakos:2010:hat11}.
  }
  \tablenotetext{c}{
    Scaling factor applied to the model transit that is fit to the
    HATSouth light curves to account for dilution of the
    transit due to blending from neighboring stars and over-filtering of
    the light curve. These factors are varied in the fit, and we allow
    independent factors for observations obtained with different HATSouth
    camera and field combinations. For \hatcur{19}, blend factors 1 and 2
    are used for the G606.4 and G607.1 observations, respectively. For
    \hatcur{20}, they are used for the G700.1 and G700.4 observations,
    respectively. For \hatcur{21}, they are used for the G777.3 and
    G778.2 observations, respectively.
  }
  \tablenotetext{d}{
    Values for a quadratic law, adopted from \citet{claret:2004} according
    to the spectroscopic (ZASPE) parameters listed in Table \ref{tab:starprop}.
  }
  \tablenotetext{e}{
    For fixed circular orbit models we list
    the 95\% confidence upper limit on the eccentricity determined
    when $\sqrt{e}\cos\omega$ and $\sqrt{e}\sin\omega$ are allowed to
    vary in the fit.
  }
  \tablenotetext{f}{
    Term added in quadrature to the formal RV uncertainties for each
    instrument and treated as a free parameter in the fitting
    routine. In cases where the jitter is consistent with zero, we
    list its 95\% confidence upper limit.
}
  \tablenotetext{g}{
    Correlation coefficient between the planetary mass \mpl\ and radius
    \rpl\ estimated from the posterior parameter distribution.
  }
  \tablenotetext{h}{
    The Safronov number is given by $\Theta = \frac{1}{2}(V_{\rm
      esc}/V_{\rm orb})^2 = (a/\rpl)(\mpl / \mstar )$
    \citep[see][]{hansen:2007}.
  }
  \tablenotetext{i}{
    Incoming flux per unit surface area, averaged over the orbit.
  }

\end{deluxetable*}

\section{Discussion}
\label{sec:discussion}

\begin{figure}[t]
  {
    \centering
    \setlength{\plotwidthtwo}{1.0\linewidth}
    \includegraphics[width=\plotwidthtwo]{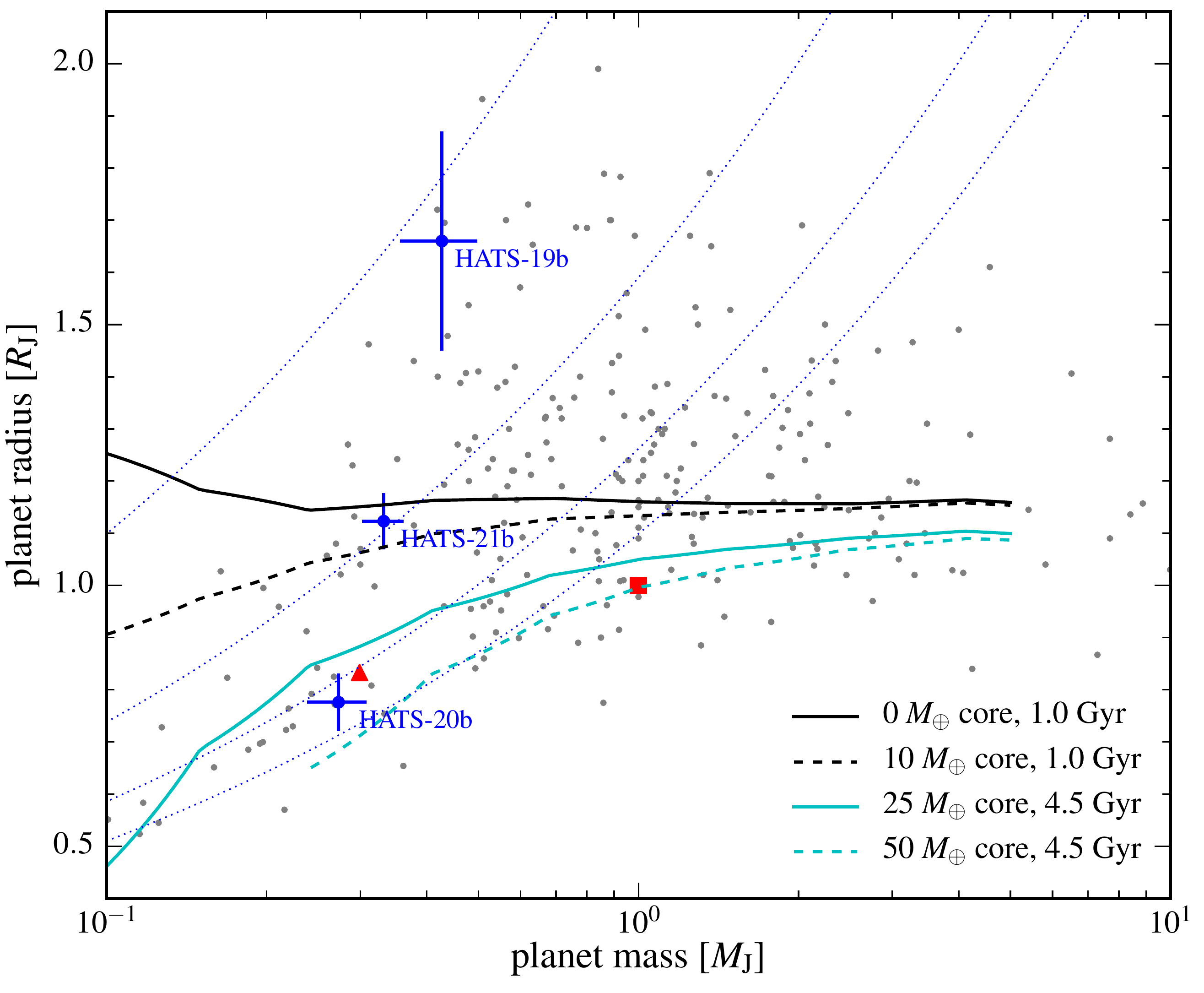}
  }
  \caption[]{The planet mass in $M_{\rm J}$ and planet radius in $R_{\rm
      J}$ plotted for 252 well-characterized transiting giant exoplanets
    selected from exoplanets.org (see text for method; errorbars
    suppressed for clarity) with mass $M_{\rm p} > 0.1 M_{\rm J}$. In
    addition to this sample, \hatcurb{19}, -20b, and -21b are shown as
    filled blue circles. The lines depict the mass-radius relations from
    \citet{2007ApJ...659.1661F}, for irradiated planets at 0.045 AU from
    the host star at 1.0 Gyr (black lines) and 4.5 Gyr (cyan lines). For
    the 1.0 Gyr relations, solid lines depict core masses of 0
    $M_\oplus$ and dashed lines depict core masses of 10 $M_\oplus$. For
    the 4.5 Gyr relations, the solid lines depict core masses of 25
    $M_\oplus$ and dashed lines depict core masses of 50 $M_\oplus$. The
    position of Saturn in this mass-radius plane is shown as the filled
    red triangle, while that of Jupiter is shown as the filled red
    square. The blue dotted iso-density lines are for $\rho = 0.10$,
    $0.33$, $0.66$, and ${1.0\ \rm g\ cm^{-3}}$ from left to
    right. \label{fig:massradius}}
\end{figure}

We plot the derived masses and radii for \hatcurb{19}, -20b, and -21b in
Figure \ref{fig:massradius}, along with 252 other transiting giant
exoplanets from the literature (taken from exoplanets.org on 2016 April
12) with measured masses $M_{\rm p} > 0.1\ M_{\rm J}$. \hatcurb{19} is
immediately seen as an outlier due to its highly inflated radius
compared to the theoretical mass-radius relations from
\citet{2007ApJ...659.1661F}. Other highly-inflated Saturn-mass planets
with radii comparable to \hatcurb{19} include WASP-31b
(\citealt{2011A&A...531A..60A}; $R = 1.55\ R_{\rm J}$), WASP-94 A b
(\citealt{2014A&A...572A..49N}; $R = 1.72\ R_{\rm J}$), and Kepler-12b
(\citealt{2011ApJS..197....9F}; $R = 1.70\ R_{\rm J}$). These planets
orbit host stars with ${\rm [Fe/H]} = -0.20$, $+0.26$, and $+0.07$,
respectively, while \hatcurb{19} has a host star with ${\rm [Fe/H]} =
+0.24$. The highly inflated radius observed for \hatcurb{19} despite its
host star's enhanced metallicity and increased heavy element fraction
may be explained by the combination of kinetic heating in the planet
interior \citep{2002A&A...385..156G}, opacity-induced energy transport
inefficiencies \citep{2007ApJ...661..502B}, and additional energy input
by tidal heating due to the planet's eccentric orbit
\citep{2008ApJ...681.1631J}.

\hatcurb{20} is a dense planet orbiting an older star with nearly-Solar
metallicity. \hatcurb{20}'s position in the mass-radius plane despite
being in close orbit around its host star is likely to be a result of a
large inferred core mass ($> 25\ M_{\oplus}$) combined with its
relatively low level of insolation receiving much less energy deposited
into its atmosphere, and perhaps thermal contraction over the long
main-sequence life-time of the host star.

Finally, \hatcurb{21} is another Saturn-mass planet orbiting a star that
has enhanced metallicity ${\rm [Fe/H]} = +0.30$, yet has a significantly
inflated radius. Models from \citet{2007ApJ...659.1661F} calculated at
the level of stellar irradiation expected (near 0.045 AU from the star)
and an age of 1.0 Gyr appear to encompass the observed radius of the
planet, but these require small core masses ($< 10\ M_{\oplus}$). The
small core mass, combined with the high level of insolation may serve to
inflate the planetary radius to the observed value.

\begin{figure*}[!ht]
  {
    \centering
    \setlength{\plotwidthtwo}{1.0\linewidth}
    \includegraphics[width=\plotwidthtwo]{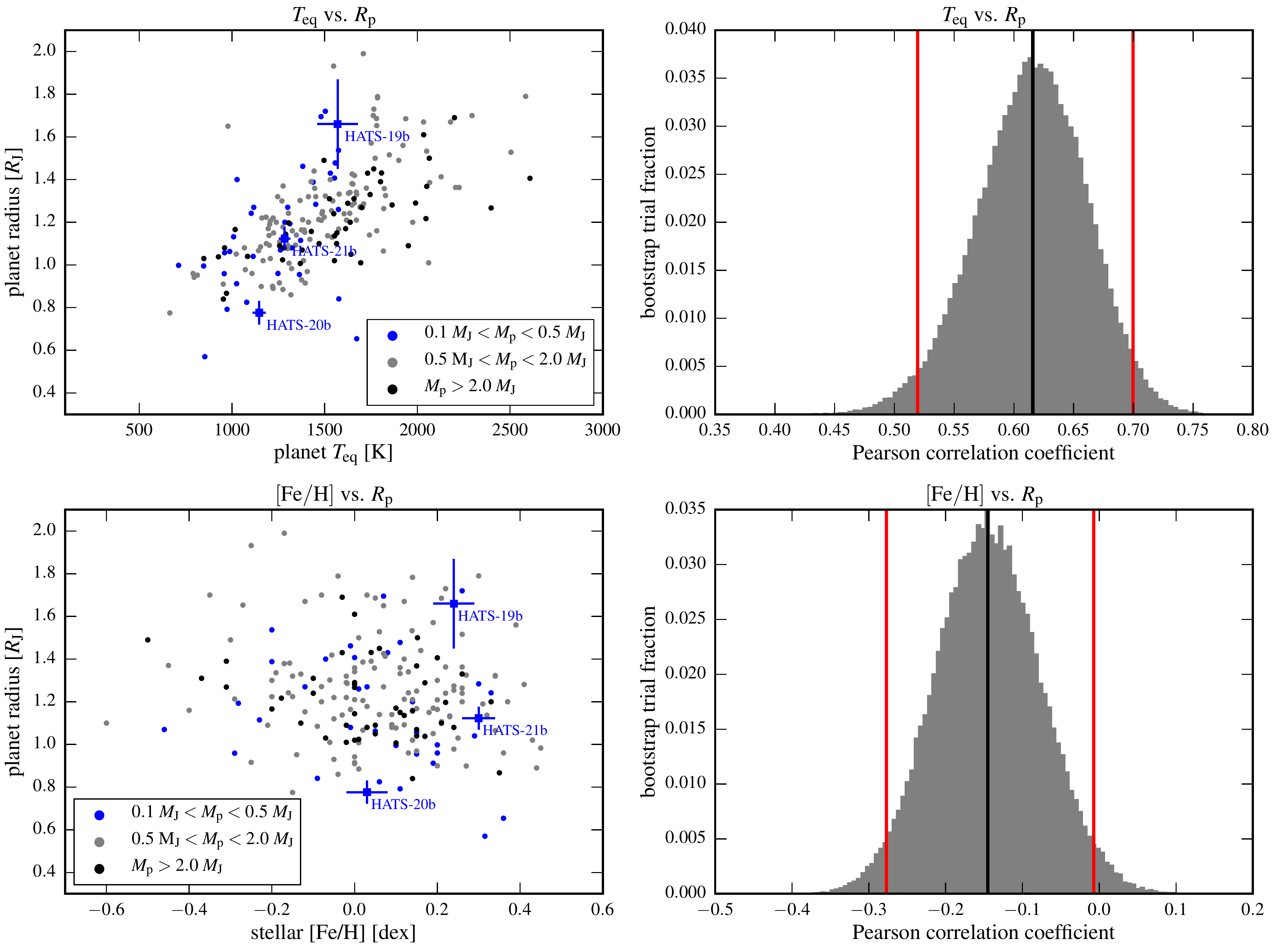}
  }
  \caption[]{\emph{Upper-left}: The planet equilibrium temperature
    $T_{\rm eq}$ and planet radius $R_{\rm p}$ plotted for 205
    transiting exoplanets selected in the same way as for Figure
    \ref{fig:massradius}, but restricted to orbital periods $P < 10$
    days to select irradiated planets. Blue filled circles represent
    hot-Saturns ($0.1 < M_{\rm p} < 0.5$ $M_J$), grey filled circles
    represent hot-Jupiters planets ($0.5 < M_{\rm p} < 2.0$ $M_J$), and
    black filled circles represent high-mass planets ($M_{\rm p} > 2.0$
    $M_J$). \hatcurb{19}, -20b, and -21b are shown as filled blue
    squares. \emph{Upper-right}: The median Pearson correlation
    coefficient (black line) for the relation between $T_{\rm eq}$ and
    $R_{\rm p}$ for all planets in this sample determined after $10^5$
    bootstrap trials along with its 95\% confidence limits (red
    lines). \emph{Lower-left}: The stellar metallicity ${\rm [Fe/H]}$
    and planet radius $R_{\rm p}$ for the same sample. \hatcurb{19},
    -20b, and -21b are shown as filled blue squares. \emph{Lower-right}:
    The median Pearson correlation coefficient (black line) for the
    relation between ${\rm [Fe/H]}$ and $R_{\rm p}$ of all planets in
    this sample determined after $10^5$ bootstrap trials along with its
    95\% confidence limits (red lines).\label{fig:planetradius}}
\end{figure*}

Relations between various physical parameters and the observed giant
planet radii have been investigated by \citet{2011ApJ...729L...7L},
\citet{2011ApJ...734..109B}, \citet[hereafter E12]{2012A&A...540A..99E},
and \citet{zhou:2014:hats5}, among others. In particular, E12 fit
empirical relations to the radii of Saturn- and Jupiter-mass giant
planets as functions of stellar irradiation parameterized by $T_{\rm
  eq}$, planet host metallicity $\feh$, planet mass $M_{p}$, orbital
semi-major axis $a$, and the tidal heating rate $H$. Significant
correlations were found between planet radius and planet equilibrium
temperature $T_{\rm eq}$, as well as between planet radius and host star
metallicity $\feh$. Figure \ref{fig:planetradius} summarizes the
relations between the planet radius $R_{\rm p}$, planetary equilibrium
temperature $T_{\rm eq}$, and stellar metallicity $\feh$ for 204
irradiated transiting giant exoplanets with {\em measured} masses
$M_{\rm p} > 0.1 M_{\rm J}$ and periods $P < 10$ days taken from
exoplanets.org (accessed on 2016 April 12) plus HATS-19b, -20b, and
-21b. In addition to these criteria, the host stars for these selected
planets were required to have measured values of metallicity and
effective temperature. There is a significant strong positive
correlation seen between the planetary radius and equilibrium
temperature, with a bootstrap 95\% confidence interval of the
correlation coefficient of [+0.52, +0.70]. On the other hand, there is
only a weak negative correlation seen between the planetary radius and
stellar metallicity with a bootstrap 95\%-confidence interval of the
correlation coefficient of [-0.28, -0.02], but it still appears
statistically significant. Both results are in line with the conclusions
in E12.

Characterizing the empirical relations of planetary and host star
parameters to the observed planet radii may help distinguish between
various proposed models of the input, internal transport, and loss of
energy from planetary atmospheres, and may explain the observed radius
distributions. We investigate the relative importance of these
parameters by fitting a model relating them to the observed planet
radius using regression carried out with Random Forests
\citep{brieman:randomforests}. Random forest regression does not require
explicit functional forms for the dependence between model parameters
and the explained variable. It also includes a way to determine the
relative importance of the regression model parameters. We used the
implementation of this method in the Python \emph{scikit-learn} library
\citep{scikit-learn}; see Appendix \ref{app:randomforest} for details.

Using the random forest regression method, we fit the following model
for the observed planetary radius $\rpl$, which is described as a
function of the predictor variables $T_{\rm eq}$ (the equilibrium
temperature), ${\rm [Fe/H]}$ (planet host metallicity), $M_{p}$ (planet
mass), $a$ (orbital semi-major axis), $e$ (orbital eccentricity), and
$M_\star$ (planet host mass):

\begin{equation}
  \rpl \sim f(T_{\rm eq}, {\rm [Fe/H]}, M_{p}, a, e, M_\star)
\end{equation}

We break up the sample of 207 irradiated giant transiting planets ($P <
10$ days) that have measured values of radius, mass, semi-major axis,
host metallicity, host mass, and orbital eccentricity into three sets:
hot-Saturns ($0.1 < \mpl < 0.5\ M_{\rm J}$; 37 members), hot-Jupiters
($0.5 < \mpl < 2.0\ M_{\rm J}$; 125 members), and higher mass planets
($\mpl > 2.0\ M_{\rm J}$; 45 members). This allows us to investigate
which model parameters are more important for each class of planet, and
enables a comparison with E12, which used an identical scheme to break
up their sample of planets and characterize relations between their
physical parameters and observed planet radii.

\begin{figure*}[ht]
  {
    \centering
    \setlength{\plotwidthtwo}{1.0\linewidth}
    \includegraphics[width=\plotwidthtwo]{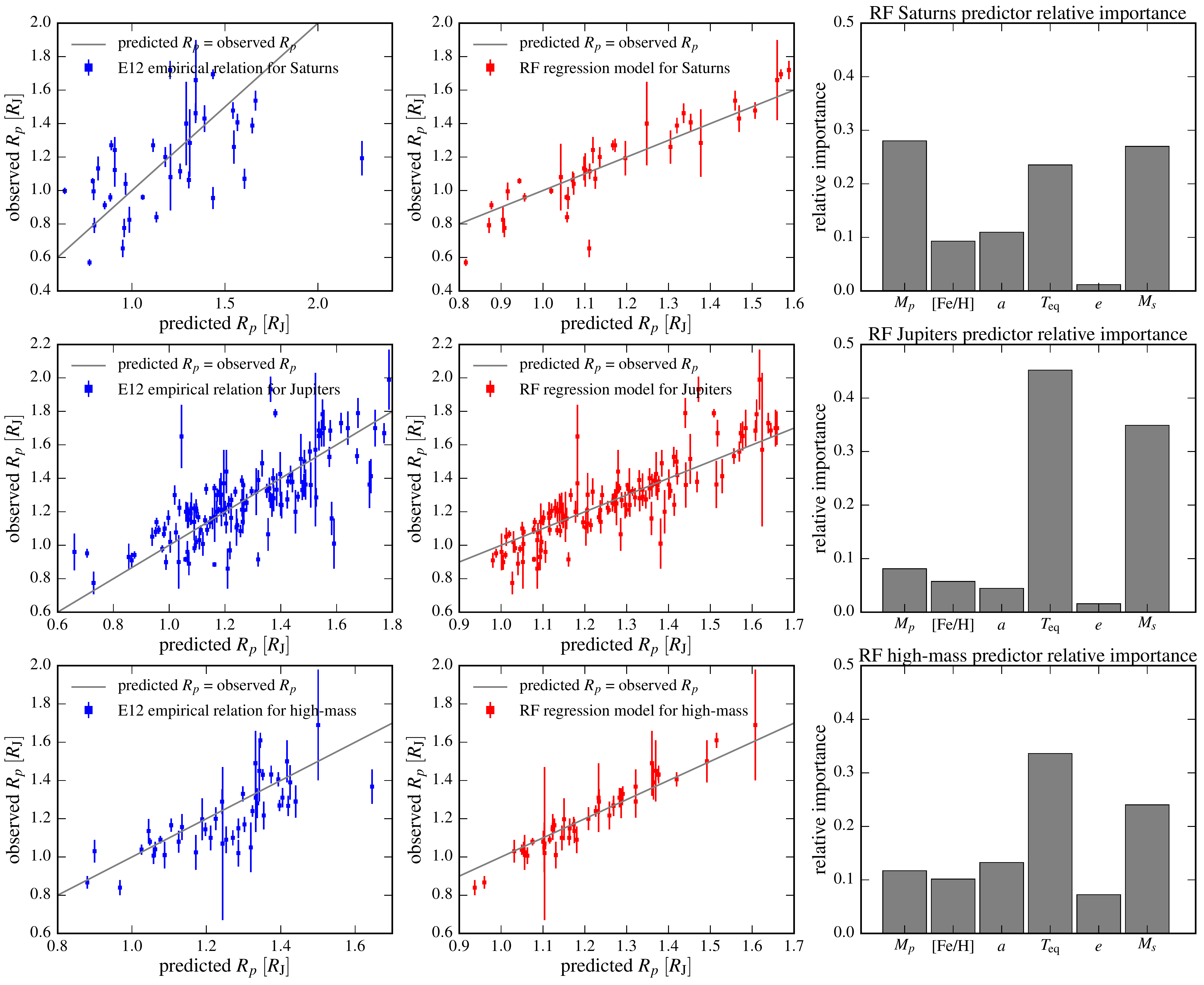}
  }
  \caption[]{\emph{Upper-left}: Observed and predicted radii from the
    E12 relation for Saturn-mass planets ($0.1 < \mpl < 0.5\ M_{\rm
      J}$). The gray line indicates $R_{\rm predicted} = R_{\rm
      observed}$. \emph{Upper-middle}: Observed and predicted radii from
    the Random Forest regression (RF) model for Saturn-mass
    planets. \emph{Upper-right}: Relative importance for six predictor
    variables used in the RF model for Saturn-mass planets. These have
    been normalized so that they sum to 1.0. \emph{Center-left}:
    Observed and predicted radii from the E12 relation for Jupiter-mass
    planets ($0.5 < \mpl < 2.0\ M_{\rm J}$). \emph{Center-middle}:
    Observed and predicted radii from the RF model for Jupiter-mass
    planets. \emph{Center-right}: Relative importance for six predictor
    variables used in the RF model for Jupiter-mass
    planets. \emph{Lower-left}: Observed and predicted radii from the
    E12 relation for high-mass planets ($\mpl > 2.0\ M_{\rm
      J}$). \emph{Lower-middle}: Observed and predicted radii from the
    RF model for high-mass planets.  \emph{Lower-right}: Relative
    importance for six predictor variables used in the RF model for
    high-mass planets. \label{fig:rfradius}}
\end{figure*}

Results of the model training and fitting process\footnote{See
  \url{https://github.com/waqasbhatti/hats19to21} for trained model
  estimators.} are shown in Figure \ref{fig:rfradius} for each set of
planets. The median absolute difference between the predicted and
observed radii is 0.08, 0.06, and 0.04 $R_{\rm J}$ for Saturn-,
Jupiter-, and high-mass planets respectively. The findings are broadly
similar to those in E12, and are summarized below. We note that these
conclusions are based on the observed sample of transiting giant
exoplanets, without correcting for observational completeness. In
particular, the bias towards brighter stars in the observed target
population prefers main sequence stars with higher luminosity, mass, and
radius, which results in a preference for transiting giant planets with
larger radii that are more easily detectable. As such, the radii of all
giant planet classes show a significant dependency on the planet host
star mass. We assume that observational biases do not affect the
correlations measured between the planetary radius and the other
physical parameters. Taking these caveats into account:

\begin{itemize}

\item The radii of hot-Saturns appear to be largely dependent on planet
  mass and then on equilibrium temperature, with a small dependence on
  the planet host metallicity. This suggests that these planets are
  core-dominated \citep{2011ApJ...736L..29M}, and thus require inflation
  mechanisms more sensitive to heavy element content in the core, such
  as kinetic heating \citep{2002A&A...385..156G}, as opposed to
  mechanisms that rely on atmospheric opacity induced energy transport
  inefficiency \citep{2007ApJ...661..502B}.

\item The radii of hot-Jupiters are far more dependent on the
  equilibrium temperature than on either planet mass or planet host
  metallicity, unlike the lower mass hot-Saturns. This suggests an
  inflation mechanism strongly tied to the radiation incident on the
  planets, such as Ohmic heating
  \citep{2011ApJ...729L...7L,2011ApJ...738....1B}.

\item The radii of irradiated higher mass planets appear to be largely
  dependent on the equilibrium temperature, with smaller dependencies on
  the planet mass and host metallicity of comparable magnitude. In this
  mass regime, the inflation mechanism may be a combination of Ohmic
  heating and energy transport inefficiency caused by the increasing
  presence of heavy elements in the envelope. Interestingly, the planet
  radius shows a small but significant dependence on the orbital
  eccentricity as well, indicating that tidal heating may play an
  increasing role in this mass regime.

\item Eccentricity of the orbit appears to play nearly no role in
  determining the planet radii, except for the high mass planets. Note
  that the observed eccentricity in most cases is an upper limit, and is
  usually set to zero if not fully determined when fitting photometric
  and RV data to obtain planet parameters. The true eccentricity for
  these assumed-circular orbits may be up to 0.03
  \citep{2008ApJ...681.1631J}; its effect on the planetary radius via
  tidal heating may thus be under-estimated in this sample.

\end{itemize}

Much of the uncertainty in characterizing the relations between giant
planet radii and other physical parameters arises due to the small
number of lower mass planets; there are only 37 Saturn-mass vs. 170
Jupiter-mass and higher-mass planets in the sample discussed in this
work. A larger sample with directly measured masses, radii, and other
physical parameters is required to effectively determine correlations
that may distinguish between inflation mechanisms for these irradiated
planets. To this end, one of the most important contributions that
exoplanet transit surveys will make in the next few years will be to
fill out this parameter space by discovering and characterizing with
high precision populations of giant planets with masses $M_{\rm p} <
1.0\ M_{\rm J}$ over a wide range in orbital semi-major axis and planet
host metallicity and luminosity.

\begin{deluxetable*}{lrrrrrl}

  \tablewidth{0pc}
  \tablecaption{
    Relative radial velocities and bisector spans for \hatcur{19}, \hatcur{20},
    and \hatcur{21}.
    \label{tab:allrvs}
  }
  \tablehead{
    \colhead{BJD} &
    \colhead{RV\tablenotemark{a}} &
    \colhead{\ensuremath{\sigma_{\rm RV}}\tablenotemark{b}} &
    \colhead{BS} &
    \colhead{\ensuremath{\sigma_{\rm BS}}} &
    \colhead{Phase} &
    \colhead{Instrument}\\
    \colhead{\hbox{(2,450,000$+$)}} &
    \colhead{(\ms)} &
    \colhead{(\ms)} &
    \colhead{(\ms)} &
    \colhead{(\ms)} &
    \colhead{} &
    \colhead{}
  }
  \startdata
  \multicolumn{7}{c}{\bf \hatcur{19}} \\
  \hline\\
  \input{\hatcurhtr{19}_rvtable.tex}
  \cutinhead{\bf \hatcur{20}}
  \input{\hatcurhtr{20}_rvtable.tex}
  \cutinhead{\bf \hatcur{21}}
  \input{\hatcurhtr{21}_rvtable.tex}
  \enddata
  \tablenotetext{a}{
    The zero-point of these velocities is arbitrary. An overall offset
    $\gamma_{\rm rel}$ fitted independently to the velocities from
    each instrument has been subtracted.
  }
  \tablenotetext{b}{
    Internal errors excluding the component of astrophysical jitter
    considered in \S\ \ref{subsec:datamodeling}.
  }
  \tablenotetext{c}{
    These observations were excluded from the analysis because the
    observations were (partially) obtained with the planet in transit,
    and thus may be affected by the Rossiter-McLaughlin effect.
  }
  \tablecomments{
    The PFS observations of \hatcur{19} and \hatcur{21} without a BS
    measurement have too low S/N in the I$_{2}$-free blue spectral region
    to pass our quality threshold for calculating accurate BS values.
  }

\end{deluxetable*}

\acknowledgements Development of the HATSouth project was funded by NSF
MRI grant NSF/AST-0723074, operations have been supported by NASA grants
NNX09AB29G and NNX12AH91H, and follow-up observations receive partial
support from grant NSF/AST-1108686.
A.J.\ acknowledges support from FONDECYT project 1130857, BASAL CATA
PFB-06, and project IC120009 ``Millennium Institute of Astrophysics
(MAS)'' of the Millennium Science Initiative, Chilean Ministry of
Economy. R.B.\ and N.E.\ are supported by CONICYT-PCHA/Doctorado
Nacional. R.B.\ and N.E.\ acknowledge additional support from project
IC120009 ``Millennium Institute of Astrophysics (MAS)'' of the
Millennium Science Initiative, Chilean Ministry of Economy.
V.S.\ acknowledges support form BASAL CATA PFB-06. M.R.\ acknowledges
support from FONDECYT postdoctoral fellowship 3120097.
This work is based on observations made with ESO Telescopes at the La
Silla Observatory, the Swope telescope at the Las Campanas Observatory,
and the Danish 1.54-m telescope at La Silla Observatory.
This paper also uses observations obtained with facilities of the Las
Cumbres Observatory Global Telescope.
Work at the Australian National University is supported by ARC Laureate
Fellowship Grant FL0992131.
We acknowledge the use of the AAVSO Photometric All-Sky Survey (APASS),
funded by the Robert Martin Ayers Sciences Fund, and the SIMBAD
database, operated at CDS, Strasbourg, France.
Operations at the MPG~2.2\,m Telescope are jointly performed by the Max
Planck Gesellschaft and the European Southern Observatory. We thank the
MPG 2.2\,m telescope support crew for their technical assistance during
observations. Observing times were obtained through proposals
CN2013A-171, CN2013B-55, CN2014A-104, CN2014B-57, CN2015A-51 and ESO
096.C-0544.
We are grateful to P.~Sackett for her help in the early phase of the
HATSouth project.

\bibliographystyle{apj}
\bibliography{hatsbib}

\begin{thebibliography}{66}
\expandafter\ifx\csname natexlab\endcsname\relax\def\natexlab#1{#1}\fi

\bibitem[{{Anderson} {et~al.}(2011){Anderson}, {Collier Cameron}, {Hellier},
  {Lendl}, {Lister}, {Maxted}, {Queloz}, {Smalley}, {Smith}, {Triaud}, {West},
  {Brown}, {Gillon}, {Pepe}, {Pollacco}, {S{\'e}gransan}, {Street}, \&
  {Udry}}]{2011A&A...531A..60A}
{Anderson}, D.~R., {Collier Cameron}, A., {Hellier}, C., {et~al.} 2011, \aap,
  531, A60

\bibitem[{{Baglin}(2003)}]{2003AdSpR..31..345B}
{Baglin}, A. 2003, Advances in Space Research, 31, 345

\bibitem[{{Bakos} {et~al.}(2004){Bakos}, {Noyes}, {Kov{\'a}cs}, {Stanek},
  {Sasselov}, \& {Domsa}}]{bakos:2004:hatnet}
{Bakos}, G., {Noyes}, R.~W., {Kov{\'a}cs}, G., {et~al.} 2004, \pasp, 116, 266

\bibitem[{{Bakos} {et~al.}(2010){Bakos}, {Torres}, {P{\'a}l}, {Hartman},
  {Kov{\'a}cs}, {Noyes}, {Latham}, {Sasselov}, {Sip{\H o}cz}, {Esquerdo},
  {Fischer}, {Johnson}, {Marcy}, {Butler}, {Isaacson}, {Howard}, {Vogt},
  {Kov{\'a}cs}, {Fernandez}, {Mo{\'o}r}, {Stefanik}, {L{\'a}z{\'a}r}, {Papp},
  \& {S{\'a}ri}}]{bakos:2010:hat11}
{Bakos}, G.~{\'A}., {Torres}, G., {P{\'a}l}, A., {et~al.} 2010, \apj, 710, 1724

\bibitem[{{Bakos} {et~al.}(2013){Bakos}, {Csubry}, {Penev}, {Bayliss},
  {Jord{\'a}n}, {Afonso}, {Hartman}, {Henning}, {Kov{\'a}cs}, {Noyes},
  {B{\'e}ky}, {Suc}, {Cs{\'a}k}, {Rabus}, {L{\'a}z{\'a}r}, {Papp}, {S{\'a}ri},
  {Conroy}, {Zhou}, {Sackett}, {Schmidt}, {Mancini}, {Sasselov}, \&
  {Ueltzhoeffer}}]{bakos:2013:hatsouth}
{Bakos}, G.~{\'A}., {Csubry}, Z., {Penev}, K., {et~al.} 2013, \pasp, 125, 154

\bibitem[{{Batygin} {et~al.}(2011){Batygin}, {Stevenson}, \&
  {Bodenheimer}}]{2011ApJ...738....1B}
{Batygin}, K., {Stevenson}, D.~J., \& {Bodenheimer}, P.~H. 2011, \apj, 738, 1

\bibitem[{{Bayliss} {et~al.}(2013){Bayliss}, {Zhou}, {Penev}, {Bakos},
  {Hartman}, {Jord{\'a}n}, {Mancini}, {Mohler-Fischer}, {Suc}, {Rabus},
  {B{\'e}ky}, {Csubry}, {Buchhave}, {Henning}, {Nikolov}, {Cs{\'a}k}, {Brahm},
  {Espinoza}, {Noyes}, {Schmidt}, {Conroy}, {Wright}, {Tinney}, {Addison},
  {Sackett}, {Sasselov}, {L{\'a}z{\'a}r}, {Papp}, \&
  {S{\'a}ri}}]{bayliss:2013:hats3}
{Bayliss}, D., {Zhou}, G., {Penev}, K., {et~al.} 2013, \aj, 146, 113

\bibitem[{{B{\'e}ky} {et~al.}(2011){B{\'e}ky}, {Bakos}, {Hartman}, {Torres},
  {Latham}, {Jord{\'a}n}, {Arriagada}, {Bayliss}, {Kiss}, {Kov{\'a}cs},
  {Quinn}, {Marcy}, {Howard}, {Fischer}, {Johnson}, {Esquerdo}, {Noyes},
  {Buchhave}, {Sasselov}, {Stefanik}, {Perumpilly}, {L{\'a}z{\'a}r}, {Papp}, \&
  {S{\'a}ri}}]{2011ApJ...734..109B}
{B{\'e}ky}, B., {Bakos}, G.~{\'A}., {Hartman}, J., {et~al.} 2011, \apj, 734,
  109

\bibitem[{Bergstra \& Bengio(2012)}]{bergstra2012random}
Bergstra, J., \& Bengio, Y. 2012, The Journal of Machine Learning Research, 13,
  281

\bibitem[{{Borucki} {et~al.}(2010){Borucki}, {Koch}, {Basri}, {Batalha},
  {Brown}, {Caldwell}, {Caldwell}, {Christensen-Dalsgaard}, {Cochran},
  {DeVore}, {Dunham}, {Dupree}, {Gautier}, {Geary}, {Gilliland}, {Gould},
  {Howell}, {Jenkins}, {Kondo}, {Latham}, {Marcy}, {Meibom}, {Kjeldsen},
  {Lissauer}, {Monet}, {Morrison}, {Sasselov}, {Tarter}, {Boss}, {Brownlee},
  {Owen}, {Buzasi}, {Charbonneau}, {Doyle}, {Fortney}, {Ford}, {Holman},
  {Seager}, {Steffen}, {Welsh}, {Rowe}, {Anderson}, {Buchhave}, {Ciardi},
  {Walkowicz}, {Sherry}, {Horch}, {Isaacson}, {Everett}, {Fischer}, {Torres},
  {Johnson}, {Endl}, {MacQueen}, {Bryson}, {Dotson}, {Haas}, {Kolodziejczak},
  {Van Cleve}, {Chandrasekaran}, {Twicken}, {Quintana}, {Clarke}, {Allen},
  {Li}, {Wu}, {Tenenbaum}, {Verner}, {Bruhweiler}, {Barnes}, \&
  {Prsa}}]{2010Sci...327..977B}
{Borucki}, W.~J., {Koch}, D., {Basri}, G., {et~al.} 2010, Science, 327, 977

\bibitem[{{Brahm} {et~al.}(2015){Brahm}, {Jord{\'a}n}, {Hartman}, {Bakos},
  {Bayliss}, {Penev}, {Zhou}, {Ciceri}, {Rabus}, {Espinoza}, {Mancini}, {de
  Val-Borro}, {Bhatti}, {Sato}, {Tan}, {Csubry}, {Buchhave}, {Henning},
  {Schmidt}, {Suc}, {Noyes}, {Papp}, {L{\'a}z{\'a}r}, \&
  {S{\'a}ri}}]{2015AJ....150...33B}
{Brahm}, R., {Jord{\'a}n}, A., {Hartman}, J.~D., {et~al.} 2015, \aj, 150, 33

\bibitem[{{Brahm} {et~al.}(2016){Brahm}, {Jord{\'a}n}, {Bakos}, {Penev},
  {Espinoza}, {Rabus}, {Hartman}, {Bayliss}, {Ciceri}, {Zhou}, {Mancini},
  {Tan}, {de Val-Borro}, {Bhatti}, {Csubry}, {Bento}, {Henning}, {Schmidt},
  {Rojas}, {Suc}, {L{\'a}z{\'a}r}, {Papp}, \& {S{\'a}ri}}]{2016AJ....151...89B}
{Brahm}, R., {Jord{\'a}n}, A., {Bakos}, G.~{\'A}., {et~al.} 2016, \aj, 151, 89

\bibitem[{Breiman(2001)}]{brieman:randomforests}
Breiman, L. 2001, Machine Learning, 45, 5

\bibitem[{Breiman {et~al.}(1984)Breiman, Friedman, Stone, \&
  Olshen}]{breiman1984classification}
Breiman, L., Friedman, J., Stone, C.~J., \& Olshen, R.~A. 1984, Classification
  and regression trees (CRC press)

\bibitem[{{Brown} {et~al.}(2013){Brown}, {Baliber}, {Bianco}, {Bowman},
  {Burleson}, {Conway}, {Crellin}, {Depagne}, {De Vera}, {Dilday}, {Dragomir},
  {Dubberley}, {Eastman}, {Elphick}, {Falarski}, {Foale}, {Ford}, {Fulton},
  {Garza}, {Gomez}, {Graham}, {Greene}, {Haldeman}, {Hawkins}, {Haworth},
  {Haynes}, {Hidas}, {Hjelstrom}, {Howell}, {Hygelund}, {Lister}, {Lobdill},
  {Martinez}, {Mullins}, {Norbury}, {Parrent}, {Paulson}, {Petry}, {Pickles},
  {Posner}, {Rosing}, {Ross}, {Sand}, {Saunders}, {Shobbrook}, {Shporer},
  {Street}, {Thomas}, {Tsapras}, {Tufts}, {Valenti}, {Vander Horst}, {Walker},
  {White}, \& {Willis}}]{brown:2013:lcogt}
{Brown}, T.~M., {Baliber}, N., {Bianco}, F.~B., {et~al.} 2013, \pasp, 125, 1031

\bibitem[{{Burrows} {et~al.}(2007){Burrows}, {Hubeny}, {Budaj}, \&
  {Hubbard}}]{2007ApJ...661..502B}
{Burrows}, A., {Hubeny}, I., {Budaj}, J., \& {Hubbard}, W.~B. 2007, \apj, 661,
  502

\bibitem[{{Butler} {et~al.}(1996){Butler}, {Marcy}, {Williams}, {McCarthy},
  {Dosanjh}, \& {Vogt}}]{butler:1996}
{Butler}, R.~P., {Marcy}, G.~W., {Williams}, E., {et~al.} 1996, \pasp, 108, 500

\bibitem[{{Cardelli} {et~al.}(1989){Cardelli}, {Clayton}, \&
  {Mathis}}]{cardelli:1989}
{Cardelli}, J.~A., {Clayton}, G.~C., \& {Mathis}, J.~S. 1989, \apj, 345, 245

\bibitem[{{Castelli} \& {Kurucz}(2004)}]{2004astro.ph..5087C}
{Castelli}, F., \& {Kurucz}, R.~L. 2004, ArXiv Astrophysics e-prints

\bibitem[{{Claret}(2004)}]{claret:2004}
{Claret}, A. 2004, \aap, 428, 1001

\bibitem[{{Covey} {et~al.}(2007){Covey}, {Ivezi{\'c}}, {Schlegel},
  {Finkbeiner}, {Padmanabhan}, {Lupton}, {Ag{\"u}eros}, {Bochanski}, {Hawley},
  {West}, {Seth}, {Kimball}, {Gogarten}, {Claire}, {Haggard}, {Kaib},
  {Schneider}, \& {Sesar}}]{2007AJ....134.2398C}
{Covey}, K.~R., {Ivezi{\'c}}, {\v Z}., {Schlegel}, D., {et~al.} 2007, \aj, 134,
  2398

\bibitem[{{Crane} {et~al.}(2010){Crane}, {Shectman}, {Butler}, {Thompson},
  {Birk}, {Jones}, \& {Burley}}]{crane:2010}
{Crane}, J.~D., {Shectman}, S.~A., {Butler}, R.~P., {et~al.} 2010, in Society
  of Photo-Optical Instrumentation Engineers (SPIE) Conference Series, Vol.
  7735, Society of Photo-Optical Instrumentation Engineers (SPIE) Conference
  Series

\bibitem[{{Deeg} \& {Doyle}(2001)}]{deeg:2001}
{Deeg}, H.~J., \& {Doyle}, L.~R. 2001, in Third Workshop on Photometry, ed.
  W.~J. {Borucki} \& L.~E. {Lasher}, 85

\bibitem[{{Dopita} {et~al.}(2007){Dopita}, {Hart}, {McGregor}, {Oates},
  {Bloxham}, \& {Jones}}]{dopita:2007}
{Dopita}, M., {Hart}, J., {McGregor}, P., {et~al.} 2007, \apss, 310, 255

\bibitem[{{Eastman} {et~al.}(2013){Eastman}, {Gaudi}, \& {Agol}}]{eastman:2013}
{Eastman}, J., {Gaudi}, B.~S., \& {Agol}, E. 2013, \pasp, 125, 83

\bibitem[{{Enoch} {et~al.}(2012){Enoch}, {Collier Cameron}, \&
  {Horne}}]{2012A&A...540A..99E}
{Enoch}, B., {Collier Cameron}, A., \& {Horne}, K. 2012, \aap, 540, A99

\bibitem[{{Espinoza} {et~al.}(2016){Espinoza}, {Bayliss}, {Hartman}, {Bakos},
  {Jord{\'a}n}, {Zhou}, {Mancini}, {Brahm}, {Ciceri}, {Bhatti}, {Csubry},
  {Rabus}, {Penev}, {Bento}, {de Val-Borro}, {Henning}, {Schmidt}, {Suc},
  {Wright}, {Tinney}, {Tan}, \& {Noyes}}]{2016arXiv160600023E}
{Espinoza}, N., {Bayliss}, D., {Hartman}, J.~D., {et~al.} 2016, ArXiv e-prints,
  1606.00023

\bibitem[{{Fortney} {et~al.}(2007){Fortney}, {Marley}, \&
  {Barnes}}]{2007ApJ...659.1661F}
{Fortney}, J.~J., {Marley}, M.~S., \& {Barnes}, J.~W. 2007, \apj, 659, 1661

\bibitem[{{Fortney} {et~al.}(2011){Fortney}, {Demory}, {D{\'e}sert}, {Rowe},
  {Marcy}, {Isaacson}, {Buchhave}, {Ciardi}, {Gautier}, {Batalha}, {Caldwell},
  {Bryson}, {Nutzman}, {Jenkins}, {Howard}, {Charbonneau}, {Knutson}, {Howell},
  {Everett}, {Fressin}, {Deming}, {Borucki}, {Brown}, {Ford}, {Gilliland},
  {Latham}, {Miller}, {Seager}, {Fischer}, {Koch}, {Lissauer}, {Haas}, {Still},
  {Lucas}, {Gillon}, {Christiansen}, \& {Geary}}]{2011ApJS..197....9F}
{Fortney}, J.~J., {Demory}, B.-O., {D{\'e}sert}, J.-M., {et~al.} 2011, \apjs,
  197, 9

\bibitem[{{Girardi} {et~al.}(2000){Girardi}, {Bressan}, {Bertelli}, \&
  {Chiosi}}]{girardi:2000}
{Girardi}, L., {Bressan}, A., {Bertelli}, G., \& {Chiosi}, C. 2000, \aaps, 141,
  371

\bibitem[{{Gray}(1999)}]{1999ascl.soft10002G}
{Gray}, R.~O. 1999, {SPECTRUM: A stellar spectral synthesis program},
  Astrophysics Source Code Library

\bibitem[{Gr{\"o}mping(2012)}]{gromping2012variable}
Gr{\"o}mping, U. 2012, The American Statistician

\bibitem[{{Guillot} \& {Showman}(2002)}]{2002A&A...385..156G}
{Guillot}, T., \& {Showman}, A.~P. 2002, \aap, 385, 156

\bibitem[{{Gustafsson} {et~al.}(2008){Gustafsson}, {Edvardsson}, {Eriksson},
  {J{\o}rgensen}, {Nordlund}, \& {Plez}}]{2008A&A...486..951G}
{Gustafsson}, B., {Edvardsson}, B., {Eriksson}, K., {et~al.} 2008, \aap, 486,
  951

\bibitem[{{Hansen} \& {Barman}(2007)}]{hansen:2007}
{Hansen}, B.~M.~S., \& {Barman}, T. 2007, \apj, 671, 861

\bibitem[{{Hartman} {et~al.}(2009){Hartman}, {Bakos}, {Torres}, {Kov{\'a}cs},
  {Noyes}, {P{\'a}l}, {Latham}, {Sip{\H o}cz}, {Fischer}, {Johnson}, {Marcy},
  {Butler}, {Howard}, {Esquerdo}, {Sasselov}, {Kov{\'a}cs}, {Stefanik},
  {Fernandez}, {L{\'a}z{\'a}r}, {Papp}, \& {S{\'a}ri}}]{2009ApJ...706..785H}
{Hartman}, J.~D., {Bakos}, G.~{\'A}., {Torres}, G., {et~al.} 2009, \apj, 706,
  785

\bibitem[{{Hartman} {et~al.}(2012){Hartman}, {Bakos}, {B{\'e}ky}, {Torres},
  {Latham}, {Csubry}, {Penev}, {Shporer}, {Fulton}, {Buchhave}, {Johnson},
  {Howard}, {Marcy}, {Fischer}, {Kov{\'a}cs}, {Noyes}, {Esquerdo}, {Everett},
  {Szklen{\'a}r}, {Quinn}, {Bieryla}, {Knox}, {Hinz}, {Sasselov}, {F{\H
  u}r{\'e}sz}, {Stefanik}, {L{\'a}z{\'a}r}, {Papp}, \&
  {S{\'a}ri}}]{hartman:2012:hat39hat41}
{Hartman}, J.~D., {Bakos}, G.~{\'A}., {B{\'e}ky}, B., {et~al.} 2012, \aj, 144,
  139

\bibitem[{{Hartman} {et~al.}(2015){Hartman}, {Bayliss}, {Brahm}, {Bakos},
  {Mancini}, {Jord{\'a}n}, {Penev}, {Rabus}, {Zhou}, {Butler}, {Espinoza}, {de
  Val-Borro}, {Bhatti}, {Csubry}, {Ciceri}, {Henning}, {Schmidt}, {Arriagada},
  {Shectman}, {Crane}, {Thompson}, {Suc}, {Cs{\'a}k}, {Tan}, {Noyes},
  {L{\'a}z{\'a}r}, {Papp}, \& {S{\'a}ri}}]{2015AJ....149..166H}
{Hartman}, J.~D., {Bayliss}, D., {Brahm}, R., {et~al.} 2015, \aj, 149, 166

\bibitem[{{Henden} {et~al.}(2009){Henden}, {Welch}, {Terrell}, \&
  {Levine}}]{henden:2009}
{Henden}, A.~A., {Welch}, D.~L., {Terrell}, D., \& {Levine}, S.~E. 2009, in
  American Astronomical Society Meeting Abstracts, Vol. 214, American
  Astronomical Society Meeting Abstracts \#214, \#407.02

\bibitem[{Hippler {et~al.}(2009)Hippler, Bergfors, Brandner, Daemgen, Henning,
  Hormuth, Huber, Janson, Rochau, Rohloff, {et~al.}}]{hippler2009astralux}
Hippler, S., Bergfors, C., Brandner, W., {et~al.} 2009, The Messenger, 137, 14

\bibitem[{Hormuth {et~al.}(2008)Hormuth, Brandner, Hippler, \&
  Henning}]{1742-6596-131-1-012051}
Hormuth, F., Brandner, W., Hippler, S., \& Henning, T. 2008, Journal of
  Physics: Conference Series, 131, 012051

\bibitem[{{Jackson} {et~al.}(2008){Jackson}, {Greenberg}, \&
  {Barnes}}]{2008ApJ...681.1631J}
{Jackson}, B., {Greenberg}, R., \& {Barnes}, R. 2008, \apj, 681, 1631

\bibitem[{{Jord{\'a}n} {et~al.}(2014){Jord{\'a}n}, {Brahm}, {Bakos}, {Bayliss},
  {Penev}, {Hartman}, {Zhou}, {Mancini}, {Mohler-Fischer}, {Ciceri}, {Sato},
  {Csubry}, {Rabus}, {Suc}, {Espinoza}, {Bhatti}, {Borro}, {Buchhave},
  {Cs{\'a}k}, {Henning}, {Schmidt}, {Tan}, {Noyes}, {B{\'e}ky}, {Butler},
  {Shectman}, {Crane}, {Thompson}, {Williams}, {Martin}, {Contreras},
  {L{\'a}z{\'a}r}, {Papp}, \& {S{\'a}ri}}]{jordan:2014:hats4}
{Jord{\'a}n}, A., {Brahm}, R., {Bakos}, G.~{\'A}., {et~al.} 2014, \aj, 148, 29

\bibitem[{{Kaufer} \& {Pasquini}(1998)}]{kaufer:1998}
{Kaufer}, A., \& {Pasquini}, L. 1998, in Society of Photo-Optical
  Instrumentation Engineers (SPIE) Conference Series, Vol. 3355, Optical
  Astronomical Instrumentation, ed. S.~{D'Odorico}, 844--854

\bibitem[{{Kov{\'a}cs} {et~al.}(2005){Kov{\'a}cs}, {Bakos}, \&
  {Noyes}}]{kovacs:2005:TFA}
{Kov{\'a}cs}, G., {Bakos}, G., \& {Noyes}, R.~W. 2005, \mnras, 356, 557

\bibitem[{{Kov{\'a}cs} {et~al.}(2002){Kov{\'a}cs}, {Zucker}, \&
  {Mazeh}}]{kovacs:2002:BLS}
{Kov{\'a}cs}, G., {Zucker}, S., \& {Mazeh}, T. 2002, \aap, 391, 369

\bibitem[{{Laughlin} {et~al.}(2011){Laughlin}, {Crismani}, \&
  {Adams}}]{2011ApJ...729L...7L}
{Laughlin}, G., {Crismani}, M., \& {Adams}, F.~C. 2011, \apjl, 729, L7

\bibitem[{{Mandel} \& {Agol}(2002)}]{mandel:2002}
{Mandel}, K., \& {Agol}, E. 2002, \apjl, 580, L171

\bibitem[{{Mayor} {et~al.}(2003){Mayor}, {Pepe}, {Queloz}, {Bouchy},
  {Rupprecht}, {Lo Curto}, {Avila}, {Benz}, {Bertaux}, {Bonfils}, {Dall},
  {Dekker}, {Delabre}, {Eckert}, {Fleury}, {Gilliotte}, {Gojak}, {Guzman},
  {Kohler}, {Lizon}, {Longinotti}, {Lovis}, {Megevand}, {Pasquini}, {Reyes},
  {Sivan}, {Sosnowska}, {Soto}, {Udry}, {van Kesteren}, {Weber}, \&
  {Weilenmann}}]{mayor:2003}
{Mayor}, M., {Pepe}, F., {Queloz}, D., {et~al.} 2003, The Messenger, 114, 20

\bibitem[{{Miller} \& {Fortney}(2011)}]{2011ApJ...736L..29M}
{Miller}, N., \& {Fortney}, J.~J. 2011, \apjl, 736, L29

\bibitem[{{Neveu-VanMalle} {et~al.}(2014){Neveu-VanMalle}, {Queloz},
  {Anderson}, {Charbonnel}, {Collier Cameron}, {Delrez}, {Gillon}, {Hellier},
  {Jehin}, {Lendl}, {Maxted}, {Pepe}, {Pollacco}, {S{\'e}gransan}, {Smalley},
  {Smith}, {Southworth}, {Triaud}, {Udry}, \& {West}}]{2014A&A...572A..49N}
{Neveu-VanMalle}, M., {Queloz}, D., {Anderson}, D.~R., {et~al.} 2014, \aap,
  572, A49

\bibitem[{{P{\'a}l} {et~al.}(2008){P{\'a}l}, {Bakos}, {Torres}, {Noyes},
  {Latham}, {Kov{\'a}cs}, {Marcy}, {Fischer}, {Butler}, {Sasselov}, {Sip{\H
  o}cz}, {Esquerdo}, {Kov{\'a}cs}, {Stefanik}, {L{\'a}z{\'a}r}, {Papp}, \&
  {S{\'a}ri}}]{pal:2008:hat7}
{P{\'a}l}, A., {Bakos}, G.~{\'A}., {Torres}, G., {et~al.} 2008, \apj, 680, 1450

\bibitem[{{Pecaut} \& {Mamajek}(2013)}]{2013ApJS..208....9P}
{Pecaut}, M.~J., \& {Mamajek}, E.~E. 2013, \apjs, 208, 9

\bibitem[{Pedregosa {et~al.}(2011)Pedregosa, Varoquaux, Gramfort, Michel,
  Thirion, Grisel, Blondel, Prettenhofer, Weiss, Dubourg, Vanderplas, Passos,
  Cournapeau, Brucher, Perrot, \& Duchesnay}]{scikit-learn}
Pedregosa, F., Varoquaux, G., Gramfort, A., {et~al.} 2011, Journal of Machine
  Learning Research, 12, 2825

\bibitem[{{Penev} {et~al.}(2013){Penev}, {Bakos}, {Bayliss}, {Jord{\'a}n},
  {Mohler}, {Zhou}, {Suc}, {Rabus}, {Hartman}, {Mancini}, {B{\'e}ky}, {Csubry},
  {Buchhave}, {Henning}, {Nikolov}, {Cs{\'a}k}, {Brahm}, {Espinoza}, {Conroy},
  {Noyes}, {Sasselov}, {Schmidt}, {Wright}, {Tinney}, {Addison},
  {L{\'a}z{\'a}r}, {Papp}, \& {S{\'a}ri}}]{penev:2013:hats1}
{Penev}, K., {Bakos}, G.~{\'A}., {Bayliss}, D., {et~al.} 2013, \aj, 145, 5

\bibitem[{{Pepper} {et~al.}(2007){Pepper}, {Pogge}, {DePoy}, {Marshall},
  {Stanek}, {Stutz}, {Poindexter}, {Siverd}, {O'Brien}, {Trueblood}, \&
  {Trueblood}}]{2007PASP..119..923P}
{Pepper}, J., {Pogge}, R.~W., {DePoy}, D.~L., {et~al.} 2007, \pasp, 119, 923

\bibitem[{{Pollacco} {et~al.}(2006){Pollacco}, {Skillen}, {Collier Cameron},
  {Christian}, {Hellier}, {Irwin}, {Lister}, {Street}, {West}, {Anderson},
  {Clarkson}, {Deeg}, {Enoch}, {Evans}, {Fitzsimmons}, {Haswell}, {Hodgkin},
  {Horne}, {Kane}, {Keenan}, {Maxted}, {Norton}, {Osborne}, {Parley}, {Ryans},
  {Smalley}, {Wheatley}, \& {Wilson}}]{2006PASP..118.1407P}
{Pollacco}, D.~L., {Skillen}, I., {Collier Cameron}, A., {et~al.} 2006, \pasp,
  118, 1407

\bibitem[{{Rabus} {et~al.}(2016){Rabus}, {Jord{\'a}n}, {Hartman}, {Bakos},
  {Espinoza}, {Brahm}, {Penev}, {Ciceri}, {Zhou}, {Bayliss}, {Mancini},
  {Bhatti}, {de Val-Borro}, {Csbury}, {Sato}, {Tan}, {Henning}, {Schmidt},
  {Bento}, {Suc}, {Noyes}, {L{\'a}z{\'a}r}, {Papp}, \&
  {S{\'a}ri}}]{2016arXiv160302894R}
{Rabus}, M., {Jord{\'a}n}, A., {Hartman}, J.~D., {et~al.} 2016, ArXiv e-prints,
  1603.02894

\bibitem[{{Sozzetti} {et~al.}(2007){Sozzetti}, {Torres}, {Charbonneau},
  {Latham}, {Holman}, {Winn}, {Laird}, \& {O'Donovan}}]{sozzetti:2007}
{Sozzetti}, A., {Torres}, G., {Charbonneau}, D., {et~al.} 2007, \apj, 664, 1190

\bibitem[{{ter Braak}(2006)}]{terbraak:2006}
{ter Braak}, C.~J.~F. 2006, Statistics and Computing, 16, 239

\bibitem[{{Weinberg} {et~al.}(2013){Weinberg}, {Yoon}, \&
  {Katz}}]{weinberg:2013}
{Weinberg}, M.~D., {Yoon}, I., \& {Katz}, N. 2013, ArXiv e-prints, 1301.3156

\bibitem[{{Yi} {et~al.}(2001){Yi}, {Demarque}, {Kim}, {Lee}, {Ree}, {Lejeune},
  \& {Barnes}}]{yi:2001}
{Yi}, S., {Demarque}, P., {Kim}, Y.-C., {et~al.} 2001, \apjs, 136, 417

\bibitem[{{Zacharias} {et~al.}(2012){Zacharias}, {Finch}, {Girard}, {Henden},
  {Bartlett}, {Monet}, \& {Zacharias}}]{zacharias:2012:ucac4}
{Zacharias}, N., {Finch}, C.~T., {Girard}, T.~M., {et~al.} 2012, VizieR Online
  Data Catalog, 1322, 0

\bibitem[{{Zechmeister} \& {K{\"u}rster}(2009)}]{2009A&A...496..577Z}
{Zechmeister}, M., \& {K{\"u}rster}, M. 2009, \aap, 496, 577

\bibitem[{{Zhou} {et~al.}(2014{\natexlab{a}}){Zhou}, {Bayliss}, {Penev},
  {Bakos}, {Hartman}, {Jord{\'a}n}, {Mancini}, {Mohler}, {Csubry}, {Ciceri},
  {Brahm}, {Rabus}, {Buchhave}, {Henning}, {Suc}, {Espinoza}, {B{\'e}ky},
  {Noyes}, {Schmidt}, {Butler}, {Shectman}, {Thompson}, {Crane}, {Sato},
  {Cs{\'a}k}, {L{\'a}z{\'a}r}, {Papp}, {S{\'a}ri}, \&
  {Nikolov}}]{zhou:2014:hats5}
{Zhou}, G., {Bayliss}, D., {Penev}, K., {et~al.} 2014{\natexlab{a}}, \aj, 147,
  144

\bibitem[{{Zhou} {et~al.}(2014{\natexlab{b}}){Zhou}, {Bayliss}, {Hartman},
  {Bakos}, {Penev}, {Csubry}, {Tan}, {Jord{\'a}n}, {Mancini}, {Rabus}, {Brahm},
  {Espinoza}, {Mohler-Fischer}, {Ciceri}, {Suc}, {Cs{\'a}k}, {Henning}, \&
  {Schmidt}}]{zhou:2014:mebs}
{Zhou}, G., {Bayliss}, D., {Hartman}, J.~D., {et~al.} 2014{\natexlab{b}},
  \mnras, 437, 2831

\end{thebibliography}

\appendix

\section{Random forest regression}
\label{app:randomforest}

Given a set of $m$ predictors for a sample of $n$ observations of
$\mathrm{Y}$, $\mathbf{X} = \{ (x_{1,1} \ldots x_{m,1}), (x_{1,2} \ldots
x_{m,2}), \ldots, (x_{1,n} \ldots x_{m,n}) \}$, a \emph{decision tree}
splits $\mathbf{X}$ and the accompanying $\mathbf{Y}$ recursively until
the \emph{leaf} nodes are reached (with a single sample per node), or
the tree construction is halted at a certain depth. For each node in the
tree, a set of predictors is chosen so that a split based on a threshold
applied to these predictors minimizes the ``impurity'' of the samples in
the subsequent left and right child nodes (this is the ``best'' split;
see \citealt{breiman1984classification}). The ``impurity'' $I$ in
decision trees used for regression is defined as:

\begin{equation}
  I = \frac{n_{\mathrm{left}}}{N_{\mathrm{node}}} H_{\mathrm{left}} +
  \frac{n_{\mathrm{right}}}{N_{\mathrm{node}}} H_{\mathrm{right}},
\end{equation}

where $n_{\mathrm{left}}$ is the number of items after a proposed split
in the left child node, $n_{\mathrm{right}}$ is the number of items
after a proposed split in the right child node, and $N_{\mathrm{node}}$
is the total number of items in the current node. $H_{\mathrm{left}}$
and $H_{\mathrm{right}}$ are the variances of predicted values in the
left and right child nodes, respectively:

\begin{equation}
  H_{\mathrm{z}} = \frac{1}{N_{\mathrm{node}}}
  \sum_{i,\mathrm{z} \in N_{\mathrm{node}}} (y_{i,\mathrm{z}} - c_{\mathrm{node}})^2,
\end{equation}

where z refers to either the left or right child node, and
$c_{\mathrm{node}}$ is defined as:

\begin{equation}
  c_{\mathrm{node}} = \frac{1}{N_{\mathrm{node}}}\sum_{{i} \in
    N_{\mathrm{node}}} y_{i}.
\end{equation}

\begin{figure}[ht]
  {
    \centering
    \setlength{\plotwidthtwo}{1.0\linewidth}
    \includegraphics[width=\plotwidthtwo]{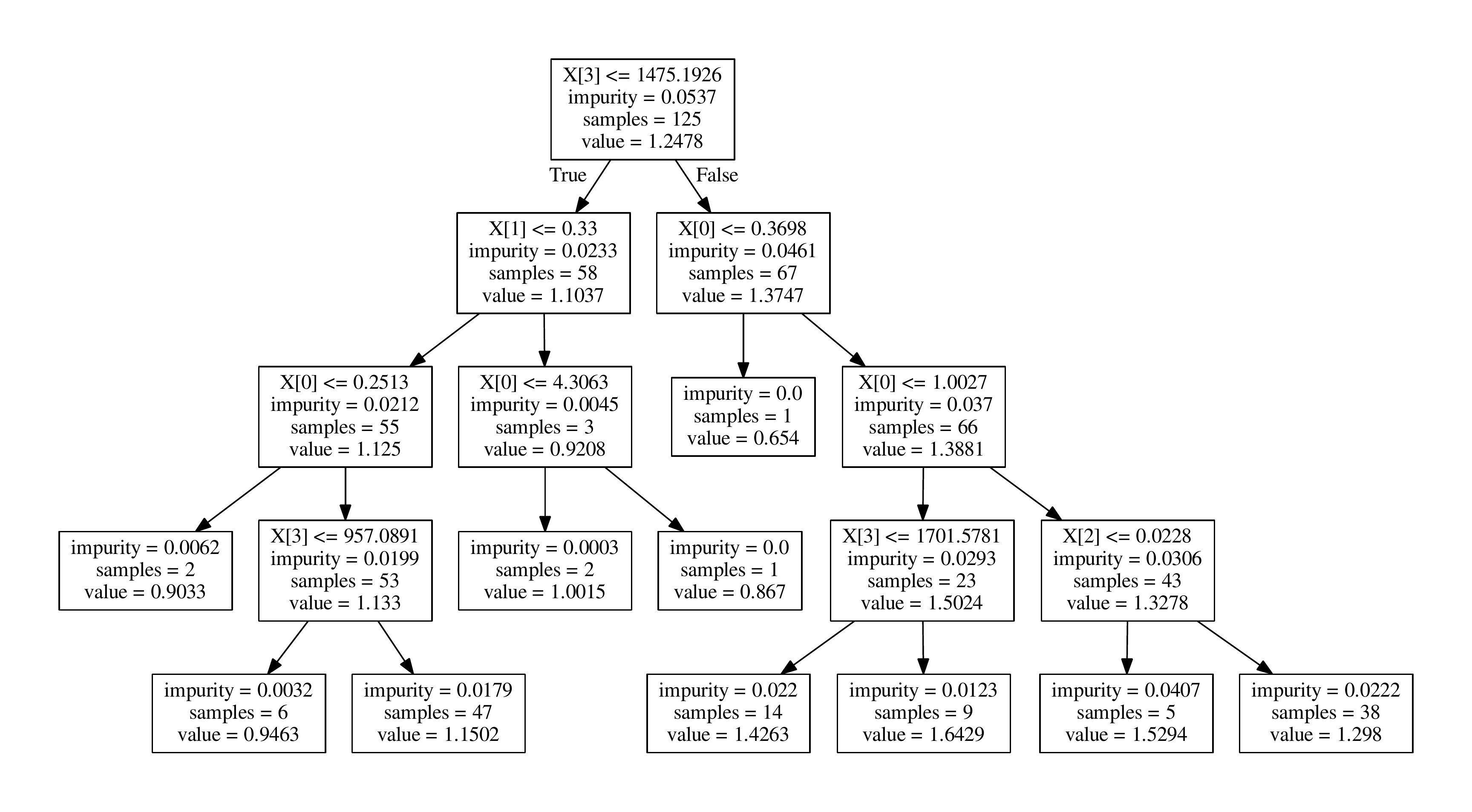}
  }
  \caption[]{The structure for a single decision tree from the random
    forest used to fit a regression model to the observed planet radii
    (``value'' above) using various predictors (the vector $\mathbf{X}$
    above). The ``impurity'' above refers to the criterion used to split
    the sample in the tree node at each level (see text for
    definition).\label{fig:graph}}
\end{figure}

The prediction of an observed value is obtained by following the
decision tree along these splits in the predictor values to the deepest
level of the tree. The relative importance of a predictor is calculated
by determining how much it contributes to reducing the ``impurity'' of
the node samples at each level of the tree. Predictors that are used to
split the nodes near the top of the tree are thus deemed more important
than those that contribute to splits near the bottom of the tree. Note,
however, that degeneracies between the model parameters themselves may
result in calculated importances that do not reflect their true values
\citep{gromping2012variable}.

Figure \ref{fig:graph} shows how this works for a decision tree used in
regression between planet radius and several predictor variables
(discussed in \S~\ref{sec:discussion}). At the top level, the sample is
split based on $X[3]$ (the equilibrium temperature in this case). The
predicted value of the radius for this node is simply the mean of all
radius values in the node sample. The two resulting nodes at the next
level of the tree are themselves split based on $X[1]$ (metallicity) for
the left node, and on $X[0]$ (planet mass) respectively. This splitting
process continues until either there is only one element left in the
node sample, or the construction of the tree is halted.

Decision tree regression in this way amounts to fitting sums of
successive step functions based on the ``best'' splits for each
predictor at each level of the tree. The process is non-parametric and
does not rely on a prior functional form of the model. On the other
hand, decision tree regression is very prone to overfitting if the tree
is allowed to grow all the way to leaf nodes. In addition, small changes
in the sample used to construct the tree can lead to very different
trees being constructed. Ensembles of decision trees overcome these
limitations, especially if different sets of predictors are chosen to
split each tree. In particular, \emph{random forests}
\citep{brieman:randomforests} carry out bootstrap sampling of the full
set of predictors, use the chosen samples of predictors to construct
decision trees, and then average the predictions of these trees to
estimate the final predicted values of the observations. The final
relative importance of predictors may also be calculated as the average
of the determined values of predictor importance for each tree.

Practical implementations of random forest regression involve tuning of
so-called \emph{hyper-parameters}. These are variables associated with
the construction process of the ensemble of decision trees and include:
$d_{\rm max}$ (the maximum depth of each tree),
$n_{\rm trees}$ (the total number of decision trees),
$m_{\rm split}$ (the maximum number of predictors to consider when
calculating the ``best'' split for a node),
$s_{\rm split}$ (the minimum number of samples in a node required to
consider a split),
and $s_{\rm leaf}$ (the minimum number of samples required for a node to
consider it a leaf node).

Optimizing these hyper-parameters results in better performance of the
trained random forest regression model. This usually involves a
grid-search among the parameters listed above, seeking to minimize a
performance metric (in our case, the median absolute deviation of the
predicted values from the observed values). In lieu of an exhaustive
grid search, a random search in the hyper-parameter space may be
performed, and appears to return hyper-parameter values that perform
just as well \citep{bergstra2012random}. The search for optimal
hyper-parameters is carried out repeatedly using subsets of the training
sample, training the model on one subset, validating it on the next
subset, and testing it on the rest of the subsets. The whole procedure
is referred to as \emph{cross-validation}. In this way, the regression
model with the best performance on the majority of the test subsets can
be determined, and then used for all subsequent predictions.

We utilize the Python library \emph{scikit-learn} \citep{scikit-learn}
for the entire procedure described above. The
\texttt{RandomForestRegressor}
class\footnote{\url{http://scikit-learn.org/stable/modules/ensemble.html\#random-forests}}
is used. We tune the hyper-parameters by using a random search
cross-validation\footnote{\url{http://scikit-learn.org/stable/modules/grid_search.html\#randomized-parameter-optimization}}
  among the following distributions:

\begin{itemize}

\item $d_{\rm max} = [3, 4, 5, 10, 20, {\rm full}]$,
\item $n_{\rm trees} = {\rm Uniform}(100, 2000)$,
\item $m_{\rm split} = {\rm Uniform}(1, 6)$,
\item $s_{\rm split} = {\rm Uniform}(1, 11)$,
\item and $s_{\rm leaf} = {\rm Uniform}(1, 11)$.

\end{itemize}

We conduct a 3-fold cross-validation to optimize the model further as
part of the hyper-parameter search described above. This involves
breaking each sample of predictors and observations into three subsets;
we train and validate the model on the first two subsets, and test it on
the final subset.

As the regression models are non-parametric, no simple functional
relations can be written down. Instead, we provide the trained models as
Python pickles that may readily be imported, and an accompanying Jupyter
notebook explaining their use at:
\url{https://github.com/waqasbhatti/hats19to21}.

\end{document}